\documentclass{aa} 
\usepackage{graphicx}
\usepackage{longtable,lscape}
\usepackage{txfonts}
\usepackage{natbib}
\bibpunct{(}{)}{;}{a}{}{,}
\newcommand{\micron}{\,$\mu$m }
\begin{document}
   \title{Silicate features in Galactic and extragalactic post-AGB discs.
\thanks{Based on observations obtained at the European Southern Observatory (ESO),
La Silla, observing program 072.D-0263 and 077.D-0555, 
and on observations made with the Spitzer Space Telescope (program id 3274 and 50092), which is operated 
by the Jet Propulsion Laboratory, California Institute of Technology under a contract with NASA.
}}

   \author{
          C. Gielen\inst{1,2}\fnmsep \thanks{Postdoctoral Fellow of
the Fund for Scientific Research, Flanders}
		\and
		J. Bouwman\inst{2}
		\and
	    H. Van Winckel\inst{1}
		\and
		T. Lloyd Evans\inst{3}
		\and
		P.~M. Woods\inst{4,5}
		\and
		F. Kemper\inst{6,4}
		\and
		M. Marengo\inst{7}
		\and
		M. Meixner\inst{8}
		\and
		G.~C. Sloan\inst{9}
		\and
		A.~G.~G.~M. Tielens\inst{10}
		   }


   \institute{Instituut voor Sterrenkunde,
              Katholieke Universiteit Leuven, Celestijnenlaan 200D, 3001 Leuven, Belgium\\ 
	      \email{clio.gielen@ster.kuleuven.be}
	\and
	Max Planck Institut f\"{u}r Astronomie, K\"{o}nigstuhl 17, 69117 Heidelberg, Germany 
		 \and
	 SUPA, School of Physics and Astronomy, University of St Andrews, North Haugh, St Andrews, Fife KY16 9SS, UK 
	\and
    Jodrell Bank Centre for Astrophysics, Alan Turing Building, School of Physics and Astronomy, The University of Manchester, 
    Oxford Road, Manchester M13 9PL, UK 
    \and
    Department of Physics and Astronomy, University College London, Gower Street, London, WC1E 6BT, UK 
    \and
    Institute of Astronomy and Astrophysics, Academia Sinica, P.O. Box 23-141, Taipei 10617, Taiwan, R. O. C 
	\and
	Department of Physics and Astronomy, Iowa State University, A313E Zaffarano, Ames, IA 50010, USA 
	\and
	Space Telescope Science Institute, 3700 San Martin Drive, Baltimore, MD 21218, USA 
	 \and
	 Department of Astronomy, Cornell University, Ithaca, NY 14853, USA 
	 \and
	 Leiden Observatory, Leiden University, P.O. Box 9513, NL-2300 RA Leiden, The Netherlands 
	}

   \date{Received ; accepted }

  \abstract
   {}
   {In this paper we study the Spitzer and TIMMI2 infrared spectra of post-AGB disc sources, both in the Galaxy and the LMC.
   Using the observed infrared spectra we determine the mineralogy and dust parameters of the discs, and look for possible differences between the
   Galactic and extragalactic sources.}
   {Modelling the full spectral range observed allows us to determine the dust species present in the disc 
   and different physical parameters such as 
   grain sizes, dust abundance ratios, and the dust and continuum temperatures. }
   {We find that all the discs are dominated by emission features of crystalline and amorphous silicate dust. 
   Only a few sample sources show features due to CO$_{2}$ gas or carbonaceous molecules such as PAHs and C$_{60}$ fullerenes.     
   Our analysis shows that dust grain processing in these discs is strong, resulting in large average grain sizes and a very high crystallinity fraction. 
   However, we do not find any correlations between the derived dust parameters and properties of the central source. There also does not seem to be
   a noticeable difference between the mineralogy of the Galactic and LMC sources.
   Even though the observed spectra are very similar to those of protoplanetary discs around young stars, showing similar mineralogy
   and strong grain processing, we do find evidence for differences in the physical and chemical processes of the dust processing.}
   {}
   \titlerunning{Silicate features in Galactic and extragalactic post-AGB disks.}
   \maketitle
%

\section{Introduction}

Studies of the chemistry and geometry of circumstellar discs have, so far, mainly focussed
on the protoplanetary discs around young stars \citep[e.g.][]{meeus01,bouwman08,juhasz10}. However, in recent years it became clear
that circumstellar discs are present in nearly all stages of stellar evolution,
going from first-ascent giants \citep{jura03,verhoelst07,melis10}, B[e] supergiants \citep{kastner10},
asymptotic giant branch (AGB) stars \citep[e.g.][]{yamamura00,chiu06,deroo07a}, (proto-)planetary nebulae \citep[e.g.][]{chesneau06,chesneau07,lykou11}
to white dwarves \citep[e.g.][]{becklin05,dong10}. 
Even though circumstellar discs appear common throughout the Hertzsprung-Russell diagram, it is still unclear
what links the different disc-bearing objects throughout all the late evolutionary stages.
It is likely that there are different formation channels depending on the evolutionary status of the central object.

Whereas for young stars the disc is a by-product of the star formation, there is evidence
that for the majority of the evolved stars the disc is newly formed. The exact formation mechanisms
are unknown, and will most likely differ for different evolutionary stages. For example, disc formation has been linked
to binary mergers, wind capture or Roche-lobe overflow (see references above). However,
in most cases, binarity appears to be the key ingredient to the formation of discs in later stages
of stellar evolution.

In this work we study a particular class of evolved binary post-AGB stars surrounded by stable dusty discs.
These sources were initially selected on the basis of their very strong near-infrared excess. 
Follow-up studies confirmed the binarity,
and showed that the companion star is most likely a main-sequence star, with a typical separation of about 1\,AU \citep{vanwinckel09}. 
The presence of a disc was already proposed to explain the presence of hot dust in the system \citep{deruyter06}
and later resolved by interferometric observations \citep{bujarrabal01,bujarrabal07,deroo06,deroo07c}.
The discs also explain the observed depletion process in the photospheric abundances of the central post-AGB star \citep{waters92,maas05,gielen09b}, 
Since the dust sublimation radii for these sources are well beyond the orbit, all the discs are circumbinary.

Our previous studies have shown that the discs are ideal environments for strong
dust processing, in the form of grain growth and crystallisation \citep{gielen08,gielen09,gielen09b}. 
This dust composition is very similar to what is observed for protoplanetary discs around young stars,
even though the disc formation mechanisms, and probably also the initial dust species, are very different.

In the Galaxy, around 80 such systems are now known \citep{deruyter06}.
Recently, large programmes, such as the Spitzer SAGE (Surveying the Agents of Galaxy Evolution) photometric \citep{meixner06}, 
and follow-up SAGE-Spec spectroscopic \citep{kemper10}, programme indicate
that also in the Large Magellanic Cloud (LMC) post-AGB disc sources are common: The study of \citet{vanaarle11} 
lists about 650 probable post-AGB disc candidates in the LMC, and about the same number for 
post-AGB stars surrounded by a cool expanding dust shell, using SAGE photometric data.

In this paper we look in more detail to the mineralogy of the circumbinary discs, both
for sources in the Galaxy and in the LMC. For this we use high- and low-resolution
Spitzer and TIMMI2 infrared spectra. These spectra allow us to study dust and gas emission
features in the $5-35$\micron region.

The outline of the paper is as follows: In Sections~\ref{programmestars} and \ref{observations}
we describe the selected samples Galactic and LMC stars and the data reduction process.
In Sect.~\ref{complexes} we take a first look at the different emission features in individual sources,
and compare the Galactic and LMC sample. The results on the dust parameters using a 
more detailed model to fit the full Spitzer wavelength range are described in Sect.~\ref{fullspecfit}.
Finally, we end with a discussion and conclusions in Sects.~\ref{discussion} and ~\ref{conclusions}.

\section{Programme stars}
\label{programmestars}

In this paper we study a total of 57 post-AGB stars with evidence for the presence of a stable 
circumbinary disc, located in the Galaxy and the LMC.
The Galactic sample consists of 33 stars from the larger sample discussed in \citet{deruyter06}.
Of these stars, 21 sources are already discussed in \citet{gielen08} and \citet{gielen09}.
To complement these 21 sources we obtained Spitzer high- and low-resolution spectra of 13 additional 
suspected post-AGB disc sources. 
The LMC sample consists of 24 sources, of which 3 are already discussed briefly in \citet{gielen09b}.
These sources were observed in low-resolution mode,
either as part of the larger SAGE-Spec programme of \citet{kemper10},
a follow-up to the photometric SAGE legacy programme \citep{meixner06}, or as part of Spitzer programmes 
3274 (PI: Hans Van Winckel) and 50092 (PI: Clio Gielen) (http://irsa.ipac.caltech.edu/data/SPITZER/docs/).

From \citet{woods11}, we selected the stars which are classified as oxygen-rich post-AGB or RV\,Tauri sources in the SAGE-Spec catalogue. We removed the sources for which only a small part of the Spitzer wavelength range was observed (LH$\alpha$ \,120-N\,145 and MACHO\,81.9728.14). After this, 16 stars remained.
To increase this LMC sample, we searched the SAGE photometric catalogue for the presence of other possible disc bearing post-AGB sources. All objects with 24\,$\mu$m fluxes between 2\,mJy and 1\,Jy were selected, in order to exclude young stellar objects and supergiants. Other selection criteria were chosen to distinguish between post-AGB stars with an expanding shell (F$_{24} > \rm{F}_8$) and binary post-AGB sources with a circumbinary disc (F$_{24} > 0.5$\,F$_8$ and J $-$ K $< 1$). For a detailed description of the selection criteria we refer to \citet{vanaarle11}.
After cross-correlation with optical photometric catalogues and the SIMBAD Astronomical Database, 650 sources remained.
Of this larger sample, the 8 brightest stars were selected and observed with the Spitzer infrared spectrograph.
For 18 of the 24 LMC sources additional ground-based optical spectra were obtained at Siding Spring Observatory, the South African Astronomical Observatory (SAAO) or with the UVES spectrograph in Paranal. This allows us to determine a spectral type and assign an effective temperature \citep{gielen09b,vanaarle11}.

For 15 Galactic sources the binarity has been confirmed by radial velocity monitoring, resulting in orbital periods between 
200 and 1800 days \citep{vanwinckel09}. For the other Galactic sources, binarity can already be suspected from the monitoring programme but
not enough data are available to derive the exact orbital parameters.
Unfortunately, such long-term radial velocity monitoring programme for the LMC sources is difficult, since it requires several years
of observations with a high-resolution optical spectrograph on a large telescope, such as UVES on the VLT. 
But, given the strong resemblance of the LMC disc candidates to the Galactic disc sources, in chemistry of the central star, spectral energy distribution
and mineralogy of the circumstellar environment \citep{reyniers07a,gielen09b}, we postulate that these sources will also be part of a binary system.

\subsection{Spectral energy distribution}

\begin{figure*}[ht]
\hspace{-1cm}
\includegraphics[width=19cm]{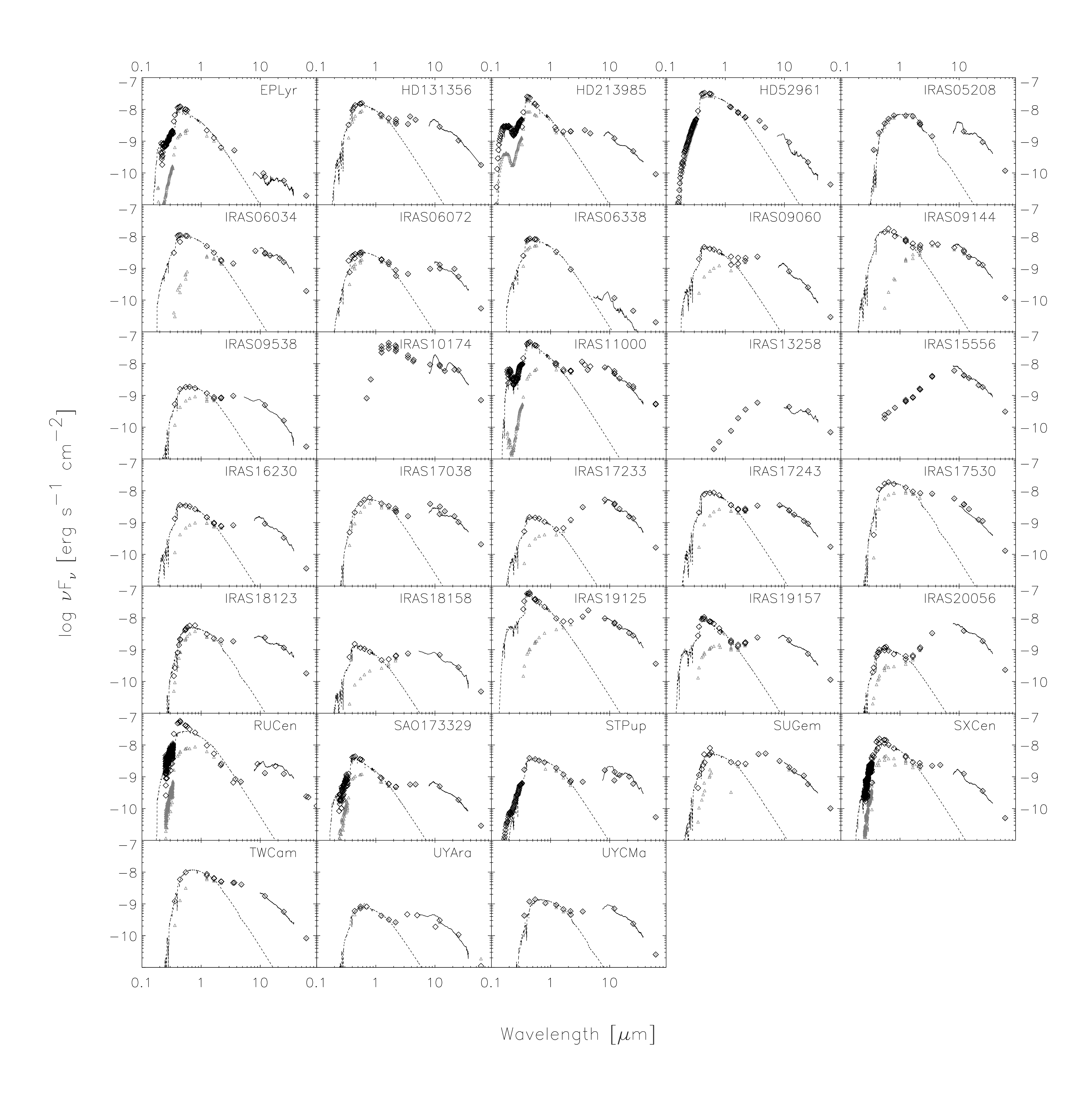}
\vspace{-1cm}
\caption{The spectral energy distributions of our sample stars. The dereddened fluxes (diamonds), reddened fluxes (gray triangles) and Spitzer spectra (solid line) are given together with the scaled photospheric Kurucz model (dashed line). For the sources where we lack the stellar parameters to determine the underlying
Kurucz model, we only plot the reddened data.}
\label{allseds1}
\end{figure*}

\begin{figure*}[ht]
\includegraphics[width=16cm]{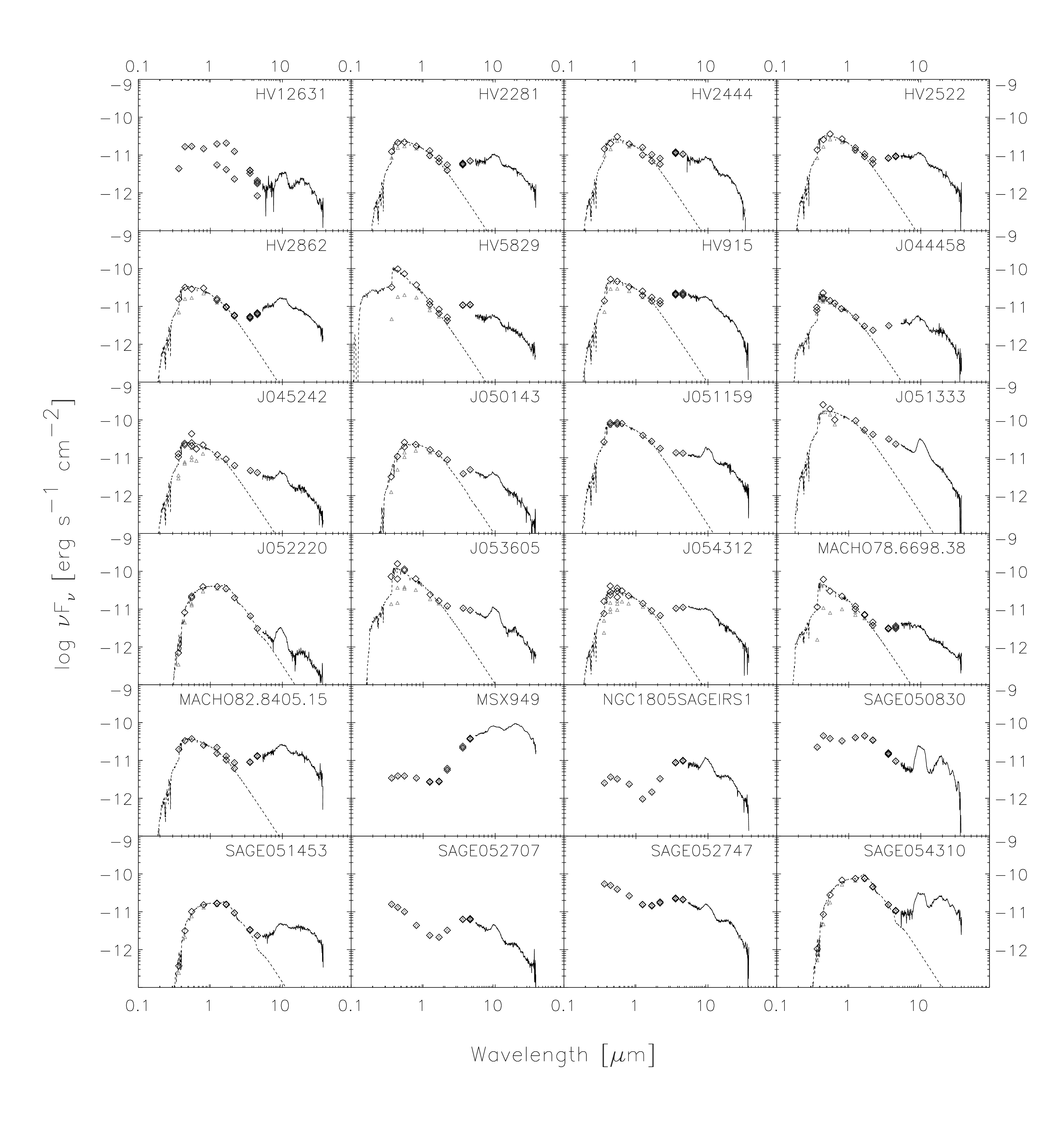}
\vspace{-1cm}
\caption{Same as previous figure.}
\label{allseds2}
\end{figure*}

For all Galactic sample stars, spectral energy distributions (SEDs) were calculated from the photometric data and stellar parameters 
as given in \citet{deruyter06}, the SAGE photometric catalogue and/or the Vizier database. The resulting SEDs can be seen in Figures~\ref{allseds1} and ~\ref{allseds2}. From the SED we also calculated the luminosity ratio $L_{IR}/L_*$.
The total extinction $E(B-V)_{tot}$ was determined by dereddening the observed photometry and infrared spectra, using the average
extinction law of \citet{savage79} extended with the theoretical extinction law of \citet{steenman89,steenman91}. 
Minimising the difference between the dereddened observed optical fluxes and the
appropriate Kurucz model \citep{kurucz79} gives the total colour excess $E(B-V)_{tot}$ (Tables~\ref{galsterren}-\ref{lmcsterren}).
This is done under
the assumption that the extinction is fully due to interstellar extinction, or that the circumstellar component follows the same extinction law. Since the total extinction probably
consists of both an interstellar and a circumstellar component, the applied dereddening is thus a maximal correction.
The errors on the value for $E(B-V)_{tot}$ are calculated using a Monte-Carlo simulation
on the photometric data. We use an error of 0.05 for the photometric measurements in a Gaussian distribution.
Since we do not know the distances to the Galactic sources, we adopt a likely luminosity for evolved low-gravity objects of $L_*=5000\pm2000$\,L$_{\odot}$. For the LMC sources we calculate the luminosity assuming a typical LMC distance of 50\,000\,pc \citep{kemper10}.
For the LMC sources we use the effective temperatures as given in \citet{vanaarle11}, if available. Since the metallicity and $\log g$ values
for these sources are not determined, we used values of [Fe/H]$=-1.0$ and $\log g=1.0$ for all stars.
These values are consistent with those found for the Galactic objects, and have only minimal impact on the derived
total reddening and infrared energy ratio.
Of the LMC sample, 6 sources lack optical spectra, and thus effective temperatures, and we could not determine
the total reddening.

For some sources there is evidence that the discs are seen close to edge on \citep{menzies88,lloydevans97}. The visible light of the central source
is then seen in reflection which makes an accurate determination of the total extinction, the luminosity ratio, and the distance very difficult.
These sources are marked with an asterisk in Table~\ref{galsterren}.

\begin{table*}
\caption{The name, equatorial coordinates $\alpha$ and $\delta$ (J2000), effective
temperature $T_{\rm eff}$, surface gravity $\log g$ and metallicity [Fe/H] of our Galactic sample stars.
For the model parameters we refer to \citet{deruyter06}. Also given is the orbital period \citep[see references in][]{deruyter06,gielen07,vanwinckel09}.
The total reddening $E(B-V)_{tot}$, the energy ratio $L_{IR}/L_*$ and the calculated distance,
assuming a luminosity of $L_*=5000\pm2000$\,L$_{\odot}$. Stars marked with * are seen in reflection only, resulting in unreliable $E(B-V)_{tot}$ values and luminosity ratios, and upper limits for the distances. The last column lists whether the spectra are part of the SAGE-Spec catalogue,
Spitzer programme 3274 or 50092.}
\label{galsterren}
\centering
\begin{tabular}{llrrcccrrrcr}
\hline \hline
N$^\circ$ & Name & $\alpha$ (J2000) & $\delta$ (J2000)  & $T_{\rm eff}$ & $\log g$ & [Fe/H] & $P_{\rm orbit}$ & $E(B-V)_{tot}$ & $L_{IR}/L_*$ & d & Prog. ID\\ 
 & & (h m s) & ($^\circ$ ' '')  & (K) & (cgs) & & (days) & & (\%) & (kpc)  &\\
\hline
1 & \object{EP\,Lyr}  & 19 18 17.5 & $+$27 50 38  & 7000 & 2.0 & -1.5   &               & 0.52$\pm$0.01 & 3$\pm$0    & 4.1$\pm$0.8   & 3274   \\ 
2 & \object{HD\,131356}   & 14 57 00.7 & $-$68 50 23  & 6000 & 1.0 & -0.5   & 1490          & 0.20$\pm$0.01 & 50$\pm$2   & 3.0$\pm$0.6   & 3274   \\ 
3 & \object{HD\,213985}   & 22 35 27.5 & $-$17 15 27  & 8250 & 1.5 & -1.0   & 259           & 0.27$\pm$0.01 & 24$\pm$1   & 3.1$\pm$0.6   & 3274   \\ 
4 & \object{HD\,52961}    & 07 03 39.6 & $+$10 46 13  & 6000 & 0.5 & -4.8   & 1310          & 0.06$\pm$0.01 & 12$\pm$1   & 2.1$\pm$0.4   & 3274   \\ 
5 & \object{IRAS\,05208$-$2035}  & 05 22 59.4 & $-$20 32 53  & 4000 & 0.5 & 0.0    & 236    & 0.00$\pm$0.00 & 38$\pm$2   & 3.9$\pm$0.8  & 3274    \\ 
6 & \object{IRAS\,06034$+$1354} & 06 06 12.3 & $+$13 53 09 & 6000 & 1.5 & -2.0 &  & 0.97$\pm$0.02 & 48$\pm$3 & 3.4$\pm$0.7 & 50092\\
7 & \object{IRAS\,06072$+$0953} & 06 09 57.4 & $+$09 52 35 & 5500 & 1.0 & -2.0 & & 0.20$\pm$0.01 & 54$\pm$3 & 5.9$\pm$1.2 & 50092\\
8 & \object{IRAS\,06338$+$5333} & 06 37 52.4 & $+$53 31 02 & 6250 & 1.0 & -1.5 & & 0.16$\pm$0.02 & 3$\pm$0 & 3.9$\pm$0.8 & 50092\\
9 & \object{IRAS\,09060$-$2807}  & 09 08 10.1 & $-$28 19 10  & 6500 & 1.5 & -0.5   & 371    & 0.57$\pm$0.02 & 63$\pm$3   & 5.4$\pm$1.1 & 3274   \\ 
10 & \object{IRAS\,09144$-$4933}  & 09 16 09.1 & $-$49 46 06  & 5750 & 0.5 & -0.5   & 1770   & 1.99$\pm$0.05 & 53$\pm$5   & 2.7$\pm$0.6  & 3274\\ 
11 & \object{IRAS\,09538$-$7622} & 09 53 58.5 & $-$76 36 53 & 5500 & 1.0 & -0.5 &  & 0.35$\pm$0.02 & 64$\pm$5 & 7.8$\pm$1.6 & 50092\\
12 & \object{IRAS\,10174$-$5704}  & 10 19 18.1 & $-$57 19 36  &  G8IaO   &     &        & 323    &  &  &     & 3274                                    \\ 
13 & \object{IRAS\,11000$-$6153} & 11 02 04.3 & $-$62 09 43 & 7600 & 2.0 & 0.1 & & 0.63$\pm$0.01 & 42$\pm$2 & 1.9$\pm$0.4 & 50092 \\
14 & \object{IRAS\,13258$-$8103}* & 13 31 07.1 & $-$81 18 30 & F4Ib-G0Ib & & & & & & & 50092\\ 
15 & \object{IRAS\,15556$-$5444} & 15 59 32.1 & $-$54 53 18 & F8 & & & & & &  & 50092\\
16 & \object{IRAS\,16230$-$3410}  & 16 26 20.3 & $-$34 17 12  & 6250 & 1.0 & -0.5   &        & 0.56$\pm$0.02 & 60$\pm$3   & 6.1$\pm$1.2 &3274   \\ 
17 & \object{IRAS\,17038$-$4815} & 17 07 36.3 & $-$48 19 08  & 4750 & 0.5 & -1.5   & 1381   & 0.22$\pm$0.02 & 69$\pm$5   & 4.5$\pm$1.0  & 3274 \\ 
18 & \object{IRAS\,17233$-$4330}* & 17 26 57.7 & $-$43 33 13 & 6250 & 1.5 & -1.0 & & 0.53$\pm$0.02 & 548$\pm$32 & 9.2$\pm$2.0 & 50092\\
19 & \object{IRAS\,17243$-$4348} & 17 27 56.1 & $-$43 50 48  & 6250 & 0.5 & 0.0    & 484    & 0.59$\pm$0.02 & 68$\pm$4   & 3.8$\pm$0.8 & 3274 \\ 
20 & \object{IRAS\,17530$-$3348} & 17 56 18.5 & $-$33 48 47 & 5000 & 0.0 & 0.0 & & 0.38$\pm$0.02 & 57$\pm$4 & 2.6$\pm$0.5 & 50092\\
21 & \object{IRAS\,18123$+$0511} & 18 14 49.4 & $+$05 12 55 & 5000 & 0.5 & 0.0 & & 0.24$\pm$0.02 & 89$\pm$6 & 4.9$\pm$1.0 & 50092\\
22 & \object{IRAS\,18158$-$3445} & 18 19 13.6 & $-$34 44 32 & 6500 & 1.5 & 0.0 & & 0.78$\pm$0.03 & 13$\pm$9 & 10$\pm$2.3 & 50092\\
23 & \object{IRAS\,19125$+$0343} & 19 15 00.8 & $+$03 48 41  & 7750 & 1.0 & -0.5   & 517    & 1.08$\pm$0.02 & 52$\pm$3   & 1.8$\pm$0.4  & 3274 \\ 
24 & \object{IRAS\,19157$-$0247} & 19 18 22.5 & $-$02 42 09  & 7750 & 1.0 & 0.0    & 120.5  & 0.68$\pm$0.01 & 63$\pm$2   & 4.2$\pm$0.9 &  3274\\ 
25 & \object{IRAS\,20056$+$1834}* & 20 07 54.8 & $+$18 42 57  & 5850 & 0.7 & -0.4   &        & 0.51$\pm$0.02 & 905$\pm$42 & 10.9$\pm$2.3 & 3274 \\ 
26 & \object{RU\,Cen}     & 12 09 23.7 & $-$45 25 35  & 6000 & 1.5 & -2.0   & 1489          & 0.55$\pm$0.01 & 13$\pm$1   & 2.3$\pm$0.5  & 3274\\ 
27 & \object{SAO\,173329} & 07 16 08.3 & $-$23 27 02  & 7000 & 1.5 & -0.8   & 115.9         & 0.39$\pm$0.01 & 36$\pm$1   & 6.5$\pm$1.3    & 3274\\ 
28 & \object{ST\,Pup}     & 06 48 56.4 & $-$37 16 33  & 5750 & 0.5 & -1.5   & 410           & 0.00$\pm$0.00 & 55$\pm$1   & 5.7$\pm$1.2   & 3274 \\ 
29 & \object{SU\,Gem}*     & 06 14 00.8 & $+$27 42 12  & 5750 & 1.125 & -0.7 &               & 0.58$\pm$0.02 & 111$\pm$7  & 4.8$\pm$1.0   & 3274\\ 
30 & \object{SX\,Cen}     & 12 21 12.6 & $-$49 12 41  & 6000 & 1.0 & -1.0   & 600           & 0.32$\pm$0.02 & 34$\pm$2   & 3.8$\pm$0.7  & 3274    \\ 
31 & \object{TW\,Cam}     & 04 20 48.1 & $+$57 26 26  & 4800 & 0.0 & -0.5   &               & 0.40$\pm$0.02 & 42$\pm$3   & 3.2$\pm$0.6   & 3274 \\ 
32 & \object{UY\,Ara}* & 17 29 28.9 & $-$59 54 02 & 5500 & 0.5 & -1.0 & & 0.00$\pm$0.00 & 72$\pm$3 & 12$\pm$2.5 & 50092\\
33 & \object{UY\,CMa}*     & 06 18 16.4 & $-$17 02 35  & 5500 & 1.0 & 0.0    &               & 0.00$\pm$0.00 & 89$\pm$3   & 9.6$\pm$2.0 & 3274  \\
\hline                                                                             
\end{tabular}
\end{table*}

\begin{table*}
\caption{The name, equatorial coordinates $\alpha$ and $\delta$ (J2000), and effective
temperature $T_{\rm eff}$ our LMC sample stars, taken from \citep{vanaarle11}.
The total reddening $E(B-V)_{tot}$, the energy ratio $L_{IR}/L_*$, and the luminosity as calculated from our SED modelling.
The last column lists whether the spectra are part of the SAGE-Spec catalogue,
Spitzer programme 3274 or 50092.}
\label{lmcsterren}
\centering
\begin{tabular}{llrrcrrrr}
\hline \hline
N$^\circ$ & Name & $\alpha$ (J2000) & $\delta$ (J2000)  & $T_{\rm eff}$ & $L_*$                  & $E(B-V)_{tot}$ & $L_{IR}/L_*$ & Prog. ID\\ 
                 &          & (h m s)                  & ($^\circ$ ' '')        & (K)           & L$_{\odot}$  &                       & (\%)   &\\
\hline
34 & HV\,12631 & 05 39 33.1  & $-$71 21 55 & & & &  & SAGE-Spec\\
35 & HV\,2281 & 05 03 05.0 & $-$68 40 25 & 5750 & 2000 & 0.02$\pm$0.02 & 63$\pm$2 & SAGE-Spec\\
36 & HV\,2444 & 05 18 46.0 & $-$69 03 22& 6750 & 4000 & 0.26$\pm$0.02 & 32$\pm$2 & SAGE-Spec\\
37 & HV\,2522 & 05 26 27.2 & $-$66 42 59& 6250 & 3700 & 0.17$\pm$0.02 & 45$\pm$3 & SAGE-Spec \\
38 & HV\,2862 & 05 51 21.1 & $-$69 53 47& 5750 & 2700 & 0.09$\pm$0.02 & 54$\pm$2 & SAGE-Spec\\
39 & HV\,5829 &05 25 19.3 & $-$70 54 07 & 5500 & 1800 & 0.00$\pm$0.02 & 60$\pm$2 & SAGE-Spec\\
40 & HV\,915 & 05 14 18.0 & $-$69 12 35 & 6250 & 4600 & 0.21$\pm$0.01 & 64$\pm$3 & SAGE-Spec\\
41 & J044458.18-703522.8 & 04 44 58.4 & $-$70 35 23 & 7000 & 1400 & 0.04$\pm$0.02 & 40$\pm$2 & 50092\\ 
42 & J045242.93-704737.4 & 04 52 43.2 & $-$70 47 37 & 5500 & 2500 & 0.30$\pm$0.03 & 22$\pm$3 & 50092\\
43 & J050143.18-694048.7 & 05 01 43.5 & $-$69 40 48 & 5000 & 2700 & 0.22$\pm$0.02  & 19$\pm$2 & 50092\\
44 & J051159.11-692532.8 & 05 11 59.4 & $-$69 25 33 & 6250 & 8500 & 0.02$\pm$0.02 & 23$\pm$2 & 50092\\
45 & J051333.74-663419.1 & 05 13 33.7 & $-$66 43 19 & 6500 & 17000 & 0.12$\pm$0.03 & 18$\pm$2 & 50092\\  
46 & J052220.87-655551.6 & 05 22 21.1 & $-$65 55 52 & 4250 & 5000 & 0.16$\pm$0.03 & 2$\pm$2 & 50092 \\
47 & J053605.56-695802.9 & 05 36 05.9 & $-$69 58 03 & 6750 & 8500 & 0.38$\pm$0.03 & 15$\pm$3 & 50092\\
48 & J054312.52-683356.9 & 05 43 12.9 & $-$68 33 57 & 6250 & 3000 & 0.27$\pm$0.02 & 49$\pm$5 & 50092\\
49 & MACHO\,78.6698.38 & 05 21 49.1 & $-$70 04 34 & 7000 & 3300 & 0.46$\pm$0.03 & 27$\pm$2 & SAGE-Spec\\
50 & MACHO\,82.8405.15 & 05 31 50.9 & $-$69 11 46 & 6000 & 3600 & 0.05$\pm$0.01 & 84$\pm$3 & SAGE-Spec\\
51 & MSX\,949 & 05 40 14.8 & $-$69 28 49 & & & &  & SAGE-Spec\\
52 & NGC\,1805\,SAGE\,IRS1 & 05 02 24.2 & $-$66 06 37 & & & &  & SAGE-Spec\\     
53 & SAGE050830 & 05 08 30.6 & $-$69 22 37 &  &  &  & & SAGE-Spec\\
54 & SAGE051453 & 05 14 18.2 & $-$69 17 24 & 4250 & 2100 & 0.10$\pm$0.02 & 28$\pm$2 & SAGE-Spec\\
55 & SAGE052707 & 05 27 07.2 & $-$70 20 02 & & & &  & SAGE-Spec\\
56 & SAGE052747 & 05 27 47.6 & $-$71 48 53 & & & &  & SAGE-Spec\\
57 & SAGE054310 & 05 43 10.9 & $-$67 27 28 & 4000 & 10500 &  0.32$\pm$0.02 & 27$\pm$2 & SAGE-Spec\\
\hline
\end{tabular}
\end{table*}

\section{Observations and data reduction}
\label{observations}

\subsection{Spitzer}

The spectra were obtained using the SL ($\lambda = 5.3-14.5\,\mu$m), LL ($\lambda = 14-38\,\mu$m), SH ($\lambda = 9.9-19.5\,\mu$m) and LH ($\lambda = 19.3-37\,\mu$m) staring modes on the Spitzer-IRS instrument \citep{werner04,houck04}. 
For the Galactic objects, exposure times were chosen to achieve a S/N ratio of 400. For the extragalactic objects in our own observing proposal,
exposure times were chosen to give a S/N ratio of 100 for the SL mode and 20 for the LL mode.
The SAGE-Spec LMC objects have a S/N ratio $\sim 60$ in SL mode and $\sim 30$ in LL mode.

The newly obtained spectra from our own Spitzer observations were extracted from the SSC raw data pipeline S18.0 version products, using the c2d and feps data reduction packages.  For a detailed description of these reduction packages, we refer to \citet{lahuis06} and \citet{hines05}. The reduction includes background and bad-pixel correction, extraction, defringing and order matching. Individual orders are corrected for offsets, if necessary, by applying small scaling corrections to match the bluer order.
For a detailed description of the target selection, observing strategy and reduction of the SAGE-Spec objects, we refer to \citet{kemper10}.

\subsection{TIMMI2}

For some stars we lack the Spitzer IRS-SH observations and we obtained additional ground-based N-band infrared
spectra with the Thermal Infrared Multi Mode Instrument 2 (TIMMI2, \citealp{reimann00,kaufl03}), mounted on the 3.6\,m telescope at the ESO La Silla Observatory. The low-resolution ($ R \sim 160$) N band grism
was used in combination with a 1.2\,arcsec slit; the pixel scale in the
spectroscopic mode of TIMMI2 is 0.45\,arcsec.
For the reduction of the spectra we used the method described in \citet{vanboekel05}.
We scaled the TIMMI2 spectra to the Spitzer spectra and found a very good agreement in spectral shape between the two data sets.

\begin{figure*}
\hspace{1cm}
\resizebox{8cm}{!}{\includegraphics{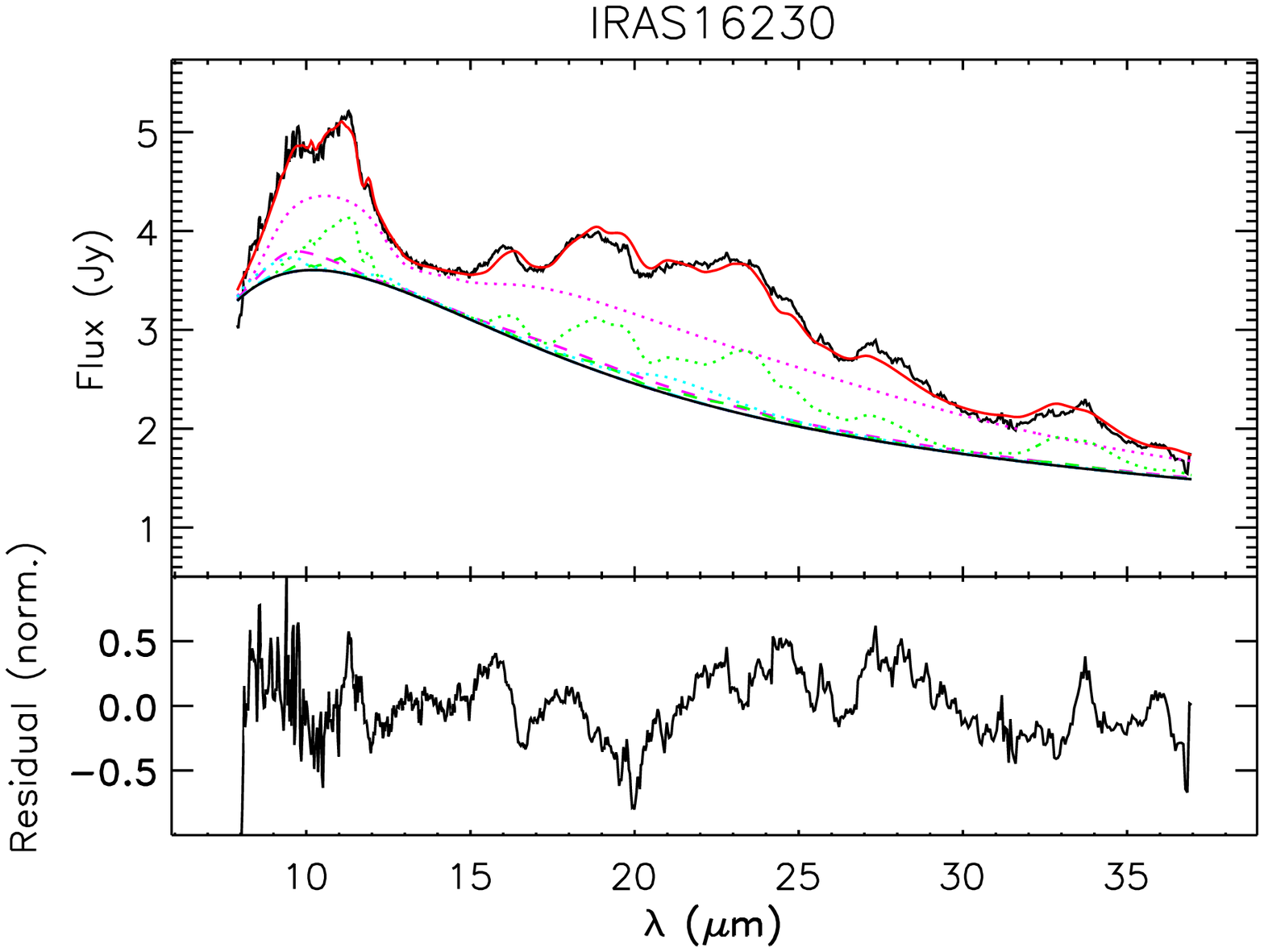}}
\hspace{1cm}
\resizebox{8cm}{!}{\includegraphics{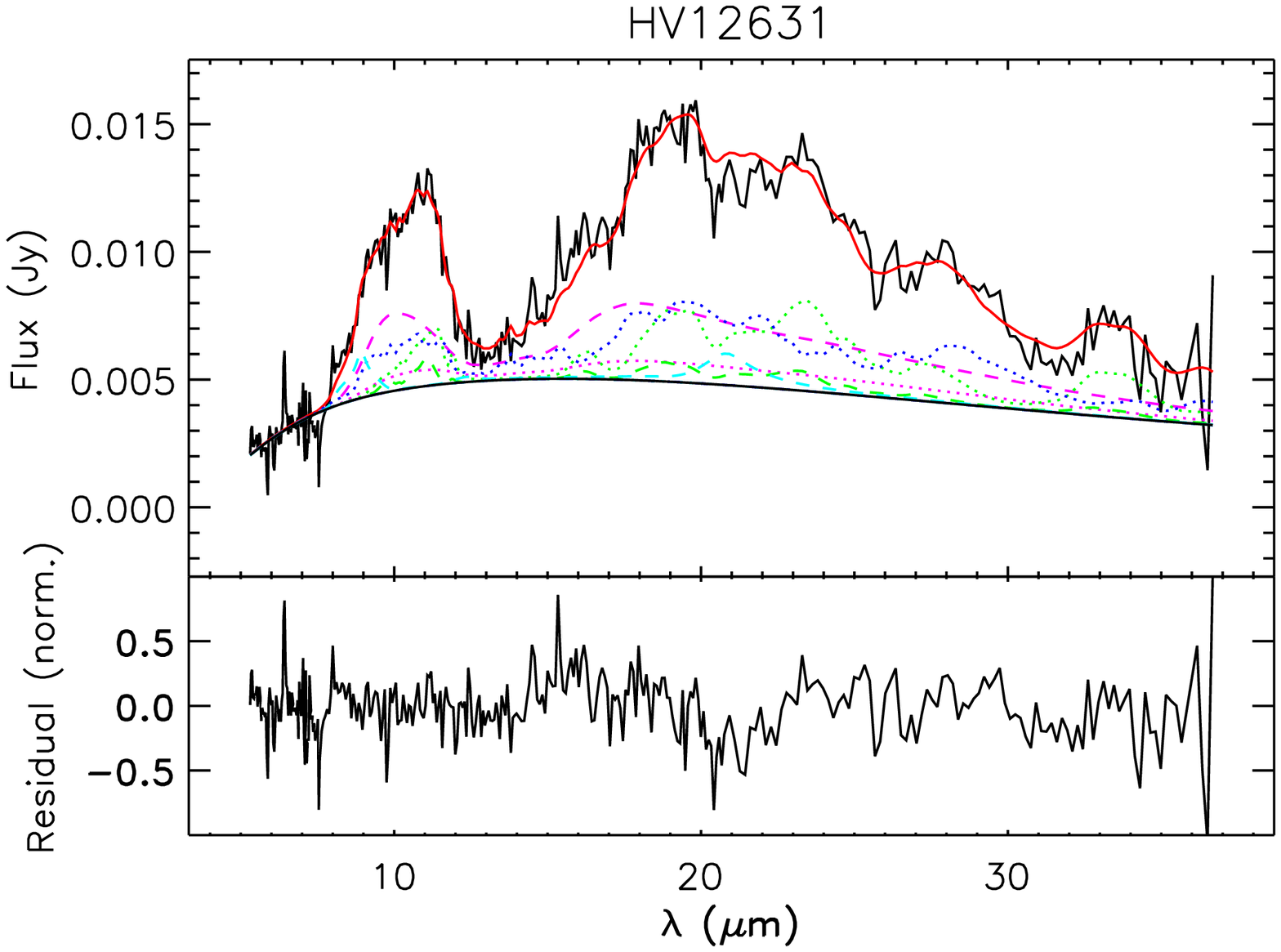}}
\vspace{0.5cm}
\caption{Best model fits for two of our sample stars, showing the contribution of the different dust species.
Top: The observed spectrum (black curve) is plotted together with the best model fit (red curve) and the continuum (black solid line).
Forsterite is plotted in green, enstatite in blue, silica in cyan and amorphous olivine and pyroxene in magenta.
Small grains (0.1\,$\mu$m) are plotted as dashed lines and larger grains (2 and 4\,$\mu$m) as dotted lines.
Bottom: The normalised residuals after subtraction of our best model of the observed spectra. The models for the other sample stars can be found in Figs.~\ref{fits1}-\ref{fits5}.}
\label{fits}
\end{figure*}

The resulting spectra can be found in Figure~\ref{fits} and Figs.~\ref{fits1}-\ref{fits5}.

\section{First inspection of emission features}
\label{complexes}

Looking at the spectra of the Galactic and LMC sources (Figs.~\ref{fits1}-\ref{fits5}), we find that all sources show clear
silicate emission. For nearly all sources the prominent broad amorphous silicate features at 
10 and 20\micron stand out. Furthermore, most spectra show additional narrower features at
$11.3,6,19,23,27$ and 33\,$\mu$m, which are due to crystalline silicate emission.

Even though all the spectra are dominated by oxygen-rich dust species, some stars do show evidence for the presence of carbonaceous molecules. Clear PAH emission can be seen in EP\,Lyr, IRAS\,06338 and IRAS\,13258, with peaks at 8 and 11.2\micron. The PAH features of EP\,Lyr were already discussed in \citet{gielen09}. The peculiar spectrum of IRAS\,06338 not only shows the typical PAH bands, but several smaller features between 6 and 8\,$\mu$m, most likely resulting from very small PAH grains. In this star, the strong narrow peaks between 13 and 18\,$\mu$m are due to CO$_2$ gas emission, which can also be seen in EP\,Lyr and HD\,52961.
HD\,52961 and IRAS\,06338 both show a strong feature at 18.7\,$\mu$m, which can be identified as C$_{60}$ fullerene emission \citep{cami10}.
The detection of these carbonaceous molecules in our sample stars will be further discussed in an upcoming paper.

\begin{figure*}[ht]
\centering
\resizebox{15cm}{!}{\includegraphics{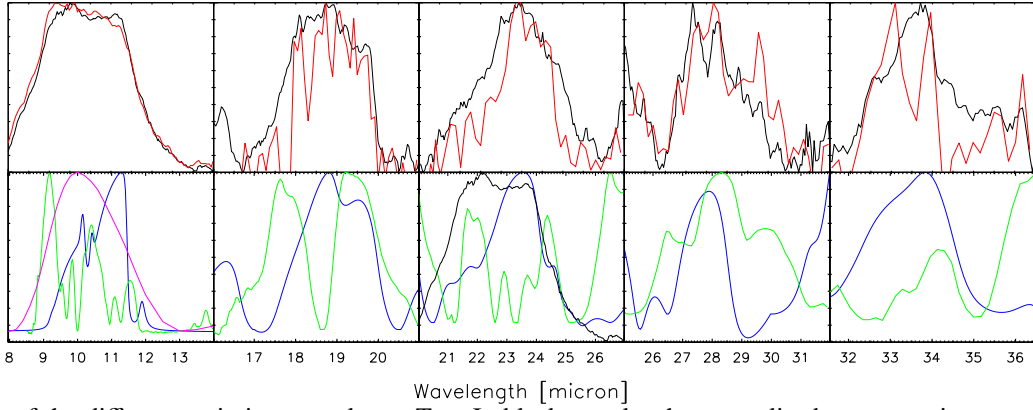}}
\vspace{0.3cm}
\caption{Overview of the different emission complexes. Top: In black we plot the normalised mean continuum-subtracted Galactic spectrum, in red the normalised mean LMC spectrum. We did not include the 14 and 16\micron complexes, since the noise level for the LMC sources made it impossible to
determine a mean spectrum.
Bottom: The different normalised continuum-subtracted spectra of forsterite, enstatite and amorphous olivine
are given in respectively blue, green and magenta.}
\label{allmeans}
\end{figure*}

To study the silicate signatures in the infrared spectra, we divided the full spectrum into 7 different complexes where
strong silicate emission is seen, more specifically at $10-14-16-19-23-27$ and 33\,$\mu$m.
To compare the Galactic stars to the LMC sources, we calculated for each group a mean continuum-subtracted spectrum in these 7 complexes. The continuum was determined by linearly interpolating between the beginning and end of the studied regions. These mean spectra are then normalised to the maximum flux in the wavelength interval.
An overview of the different mean spectra can be seen in Figure~\ref{allmeans}, together with synthetic spectra
of crystalline and amorphous silicates.
Below we discuss the different complexes in more detail, and results can be seen in Figure~\ref{mean10} and Figs.~\ref{mean1416} to ~\ref{mean33}.

\subsection{The 10\,$\mu$m complex ( $8-13\,\mu$m )}

It is clear from Fig.~\ref{mean10} that the mean spectrum in this region is very similar for the
Galactic and LMC sources. A flat-topped feature is seen, where the two peaks come from the emission of 
amorphous and crystalline olivine. The observed 10\micron complex seems to be broader at the left shoulder,
compared to the emission feature of amorphous olivine. Additional emission near 9\micron could point
to the presence of amorphous pyroxene or silica, which peak at shorter wavelengths.

Some individual sources do not follow the calculated mean complex.
IRAS\,13258  and EP\,Lyr show no silicate features, but exhibit emission due to PAHs. PAH emission 
probably also contributes to features seen in IRAS\,06338 and HD\,52961. Also note that in IRAS\,06338, the strong feature at 9\,$\mu$m seems to be shifted bluewards in comparison to the mean. This could point to the dominance of silica in this source. 
IRAS\,10174 shows almost no emission of crystalline species,
not only at 10\,$\mu$m but along its entire wavelength range, and is very similar to the extragalactic source J\,051333. 
These two sources also do not show the broadening at the left shoulder of the complex, and are thus expected to be 
devoid of silica.

\begin{figure}[ht]
\vspace{0cm}
\hspace{-0.5cm}
\resizebox{10cm}{!}{\includegraphics{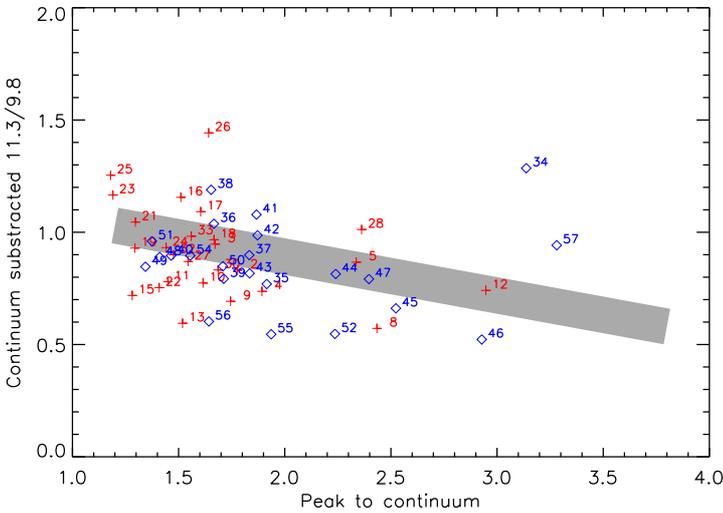}}
\caption{Ratio of the continuum-subtracted flux at 9.8 and 11.3\,$\mu$m versus the peak-to-continuum ratio of the
10\,$\mu$m silicate feature. Galactic sources are plotted in red plus signs, LMC sources in blue diamonds. The gray area shows typical values
found for protoplanetary discs around young stars. The numbers correspond to numbers given in Tables~\ref{galsterren} and \ref{lmcsterren}.}
\label{peakcont}
\end{figure}

\begin{figure*}
\hspace{0cm}
\includegraphics[width=9.5cm,height=10.5cm]{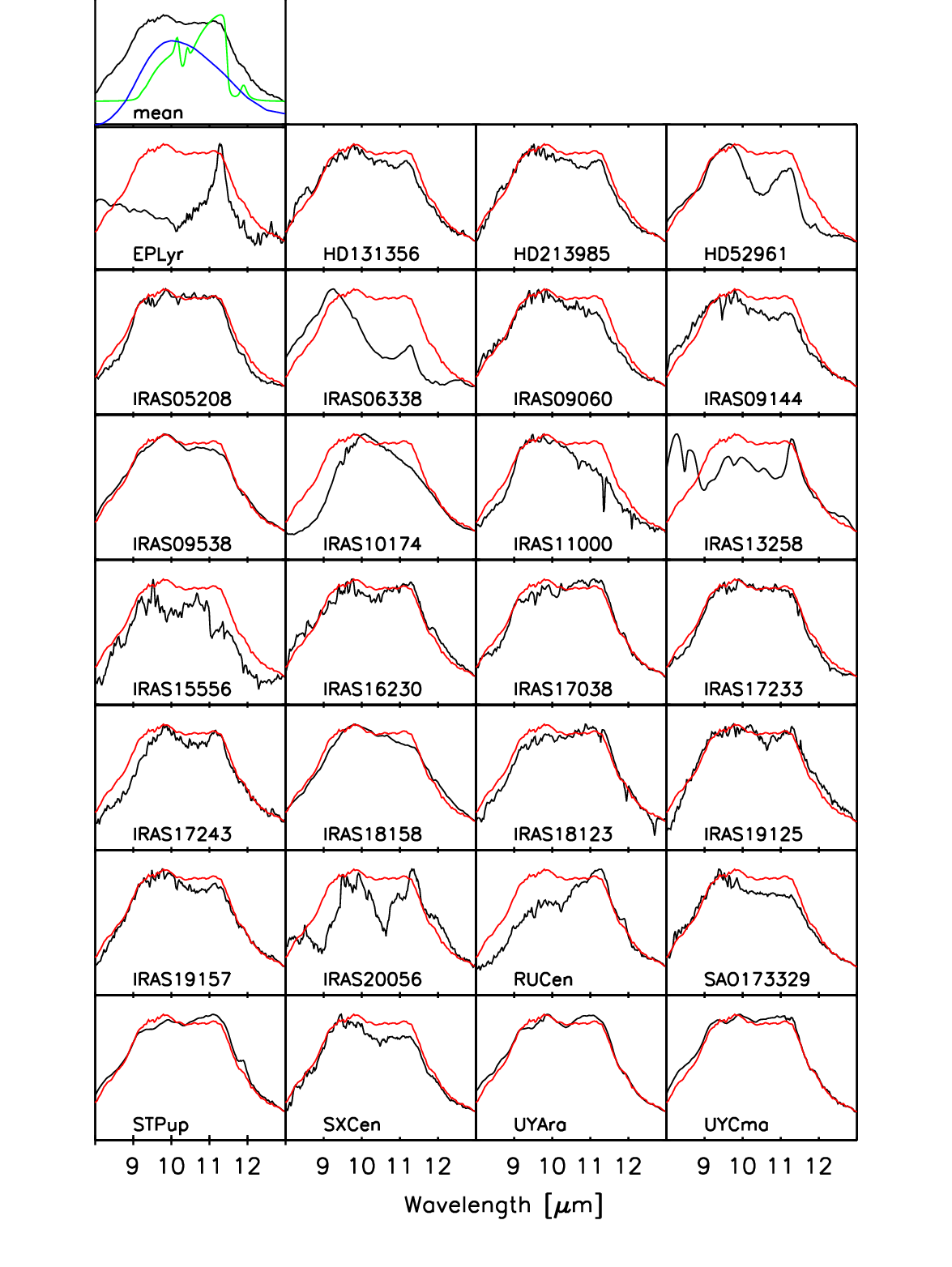}
\includegraphics[width=9.5cm,height=10.5cm]{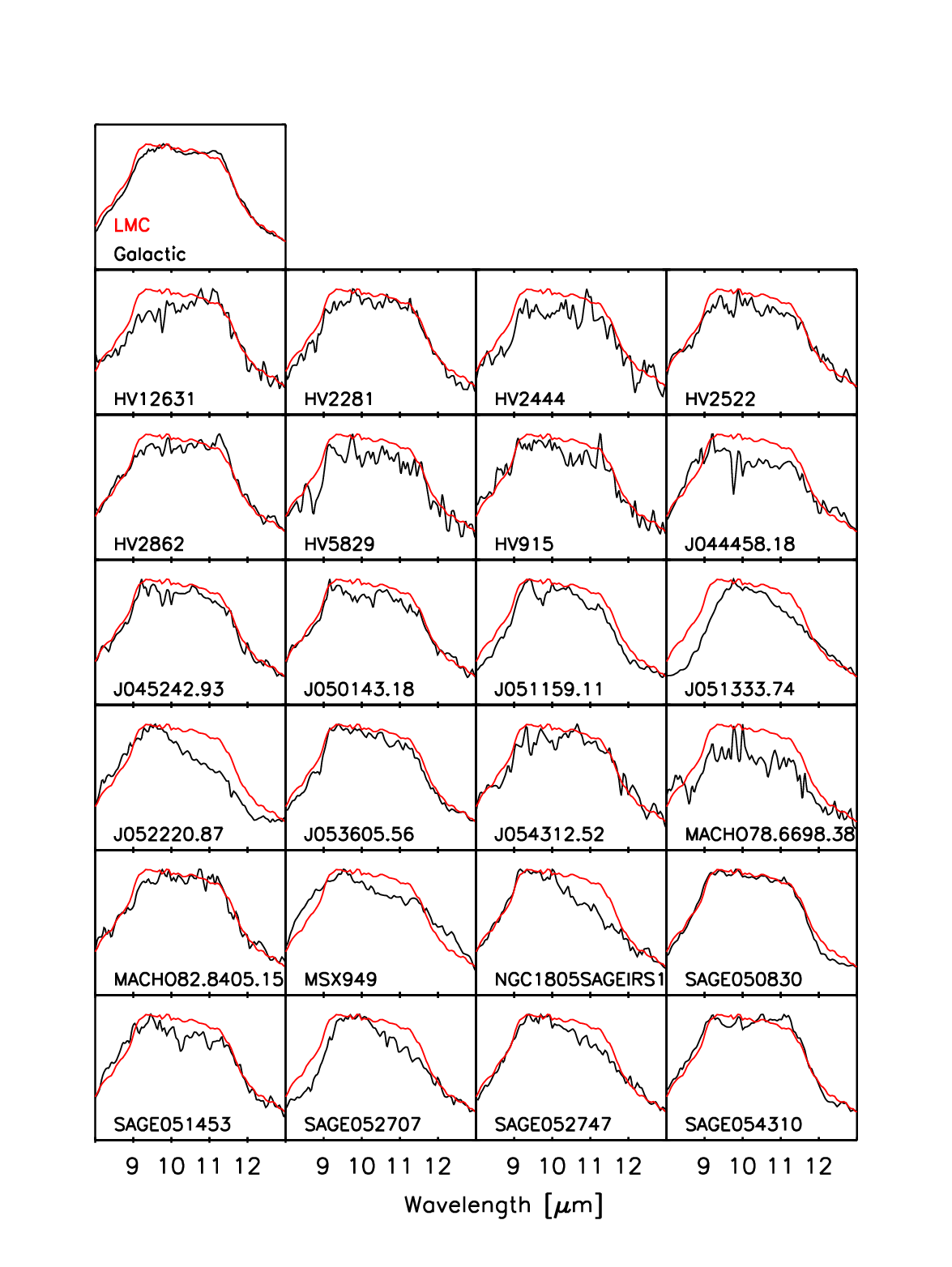}
\caption{Left: The 10\,$\mu$m complex for the Galactic sources, continuum subtracted and normalised. Overplotted in red the mean spectrum. The mass absorption coefficients of amorphous olivine and forsterite are plotted in blue and green. Right: Same as on the left, but for the LMC sources.
The top panel shows the comparison between the calculated means for the LMC and Galactic sources.}
\label{mean10}
\end{figure*}

The 10\,$\mu$m complex is a good tracer of grain processing, in the form of grain size and crystallisation \citep{vanboekel03,vanboekel05,juhasz10}.
Since amorphous and crystalline silicates peak at two distinct wavelengths, respectively 9.8 and 11.3\,$\mu$m,
the continuum-subtracted 11.3/9.8\,$\mu$m flux can be used as a measure for the amount for the crystallisation
the dust has undergone. Furthermore, the peak-to-continuum ratio of the 10\,$\mu$m complex can be used as a tracer for grain growth, since larger grains will result in a less pronounced feature. In Figure~\ref{peakcont} we plot these two ratios. We do not plot EP\,Lyr and IRAS\,13258, since they show strong PAH emission at 11.3\,$\mu$m, contaminating the crystalline emission at this wavelength. SAGE\,050830 has a very high peak-to-continuum ratio of 5.05 
( with a 11.3/9.8\,$\mu$m ratio of 0.81), and falls outside our plot range.
Most of our sources show rather high 11.3/9.8\,$\mu$m ratios, with low peak-to-continuum values, showing that the crystallinity fraction is high, and
the average grain sizes relatively large. 

A very weak correlation (Kendall rank correlation $\tau = -0.23$), can be seen. This is in 
contrast to the strong correlation seen in discs around young stars between grain growth and crystallisation processes \citep{vanboekel03,vanboekel05,juhasz10}. 
The gray area in Figure~\ref{peakcont} shows typical values found for protoplanetary discs, and it is clear that our sources show a much larger
spread in values for the continuum-subtracted 11.3/9.8\,$\mu$m flux ratio.
This could mean that in the case of the post-AGB discs, the dust might not consist of very small (0.1\micron) amorphous grains, but may already have a higher crystallinity or larger grain size.
It could also mean that different grain processes are at play, resulting in a slightly different dust grain evolution.
We find no evidence for different behaviour between our Galactic and LMC samples.

\subsection{The 14\,$\mu$m and 16\,$\mu$m complexes ( $13.5-15\,\mu$m and $15-17\,\mu$m)}
\label{15micronregion}

This region is dominated by two different emission complexes, respectively around 14 and 16\micron, as can be seen in Figure~\ref{mean1416}. 
Because of the high noise level in the LMC sources for this region, we could only calculate a mean spectrum for the Galactic stars. 

\begin{figure}[ht]
\vspace{-0.5cm}
\hspace{0cm}
\resizebox{8cm}{!}{\includegraphics{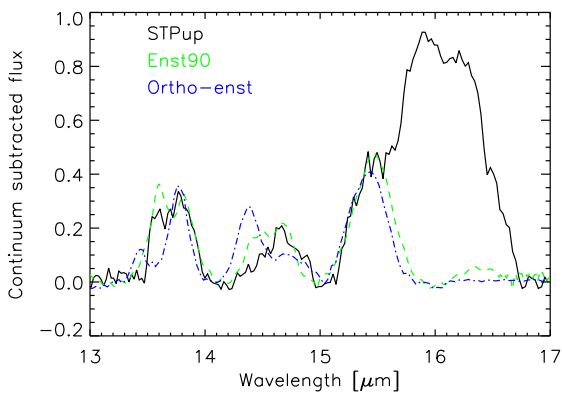}}
\caption{Comparison between the normalised and continuum-subtracted spectrum of one 
of our sample stars ST\,Pup and the laboratory spectra of ortho-enstatite (dot-dashed line) and clino-enstatite (dashed line)
with a 10\% iron content, as presented by \citet{chihara02}. The strong observed feature in ST\,Pup around
16\micron is due to forsterite.}
\label{stpup_enst}
\end{figure}

The 14\micron complex is sensitive to the emission of enstatite, which shows a clear feature around 13.8\,$\mu$m 
in our observed sources. The predicted feature at 14.4\,$\mu$m, however, is not seen.
Instead, we do see a clear signature around 14.7\,$\mu$m. The synthetic spectra of enstatite are known to be 
sensitive to the refractory indices used and the adopted grain size (see for example Fig.~10 in \citet{molster02a} and Fig.~20 in \citet{juhasz10}). 
\citet{chihara02} present an overview of the shift of peak position of crystalline pyroxenes with different iron contributions and we
find that the peak positions found in our spectra are better modelled with enstatite with a small iron contribution of about 10\%.
In Figure~\ref{stpup_enst} we show the continuum-subtracted spectrum of ST\,Pup, which has the strongest enstatite features and best S/N ratio in 
this region of our sample stars, together with the laboratory spectra of ortho-enstatite and clino-enstatite
with a 10\% iron content, as presented by \citet{chihara02}. Unfortunately, the 14\micron complex is the only
wavelength region where the enstatite features are not blended with forsterite emission.
This makes it impossible to study the enstatite iron content using other complexes.

In the Galactic sample the 16\micron complex is clearly visible in most sources, the outliers being EP\,Lyr and IRAS\,10174.
The feature seen in IRAS\,15556 is strongly deviating from the mean complex, and shows a stronger contribution of enstatite emission. IRAS\,06338 shows strong emission of CO$_2$ gas in the 16\micron region. 

As was already discussed in \citet{gielen08}, the strength of the 16\,$\mu$m feature seems to correspond to the emission of forsterite,
shifted bluewards in central wavelength. Our new spectra follow this trend. This shift of the 16\micron feature is also seen in
the infrared spectra of protoplanetary discs \citep{juhasz10}, and is probably an effect of the adopted synthetic spectrum of forsterite.

\subsection{The 19\,$\mu$m and 23\,$\mu$m complexes ( $17-21\,\mu$m and $21-26\,\mu$m)}

This region shows two strong emission complexes, around 19 and 23\micron, and a good agreement between the two samples is found (see Figs~\ref{mean19} and \ref{mean23}). 
Although the LMC sample has strong noise, the mean spectrum is very similar to the Galactic mean.

The 19\micron feature seems to be more pronounced in the Galactic sources, which could mean that the LMC sample is less crystalline,
since the feature is mainly formed by forsterite emission. However, the strong noise level of the LMC sources could also 
hamper the detection of the feature.
The 23\micron feature is dominated by the emission of forsterite, and is clearly seen in most Galactic and several LMC stars.
The mean Galactic and LMC complexes are again very similar. In the Galactic sources, the same outliers appear again:
EP\,Lyr and HD\,52961, which have a very particular mineralogy \citep{gielen09b}; IRAS\,10174, which is almost completely
amorphous, and IRAS\,15556 which shows no emission at 23\,$\mu$m.

\begin{table*}
\caption{Overview of the adopted dust species. For each component we list its chemical composition, whether it has an amorphous (A) or crystalline (C) structure,
density, adopted grain shape and grain sizes, and reference to the refractory indices used.}
\label{dustoverview}
\centering
\begin{tabular}{llccccl}
\hline \hline
Dust species & Composition & Structure & Density & Shape & Grain size & Reference\\
\hline
Olivine & Mg$_2$SiO$_4$ & A & 3.71\,g/cm$^3$ & GRF & $0.1-2-4$\micron & \citet{dorschner95} \\
Pyroxene & MgSiO$_3$ & A & 3.20\,g/cm$^3$ & GRF & $0.1-2-4$\micron & \citet{dorschner95} \\
Olivine & MgFeSiO$_4$ & A & 3.71\,g/cm$^3$ & GRF & $0.1-2-4$\micron & \citet{dorschner95} \\
Pyroxene & MgFeSi$_2$O$_6$ & A & 3.20\,g/cm$^3$ & GRF & $0.1-2-4$\micron & \citet{dorschner95} \\
Forsterite & Mg$_2$SiO$_4$ & C & 3.33\,g/cm$^3$ & GRF & $0.1-2-4$\micron &  \citet{servoin73} \\
Ortho-Enstatite & MgSiO$_3$ & C & 2.80\,g/cm$^3$ & GRF & $0.1-2-4$\micron & \citet{jaeger98} \\
Silica & SiO$_2$ & A &  2.20\,g/cm$^3$ & GRF & $0.1-2-4$\micron & \citet{henning97} \\
\hline                                                                             
\end{tabular}
\end{table*}

\subsection{The 27\,$\mu$m and 33\,$\mu$m complexes ( $25.5-30\,\mu$m and $32-36\,\mu$m)}

The two samples show a similar observed mean features, peaking around 27 and 33\micron (see Figs.~\ref{mean27} and \ref{mean33}).

The 27\micron complex peaks at the forsterite 27.5\micron feature, but with an additional shoulder around
29.3\,$\mu$m, which is due to enstatite emission.
In the Galactic sample, strong deviation can again be seen for IRAS\,06338, which is clearly a source with atypical dust emission features.
Some sources show a somewhat broader 27\,$\mu$m feature, such as IRAS\,15556, which could point to a larger enstatite contribution.
The LMC sample is again compromised by the strong noise, but the stars with strong observed emission feature do show a similar feature as observed in the Galactic sample. Only MSX 949 seems to deviate from the observed mean complex, with a very broad feature which peaks at 29.5\,$\mu$m.

Both samples show a clear 33\micron feature, due to the emission of forsterite crystals at lower temperature.
For the LMC sample the spectrum around the 33\micron complex is strongly hampered by high noise,
but the feature is still visible in the mean spectrum. A few sources have very strong emission at 33\,$\mu$m,
such as HV\,12631, J044458, MACHO\,78.6698.38, MSX\,949, SAGE\,054310, and SAGE\,050830.

\section{Full spectral model}
\label{fullspecfit}

To study the characteristics of the silicate emission observed in these sources, we constructed a
basic model to fit the full Spitzer wavelength range. The observed emission features will depend on the
chemical composition of the dust, the grain sizes and the grain shapes.
In \citet{gielen08} we constructed a model that takes all the above properties into account.
Note that a bug was present in the modelling routine used in \citet{gielen08,gielen09}, which we describe in \citet{gielen_err}.
For this paper, we also extended the routine to include an additional dust species, namely amorphous silica (SiO$_2$).

Assuming that the dust features are formed in an optically thin upper part of the disc, the spectrum can be 
approximated as a linear combination of dust absorption profiles.
The model emission is then given by
$$F_\lambda \sim (\sum_i \alpha_i\kappa_i)\times(\sum_j \beta_j B_\lambda(T_j))+F_{cont}$$
where $\kappa_i$ is the mass absorption coefficient of dust component $i$ and $\alpha_i$
gives the fraction of that dust component, $B_\lambda(T_j)$ denotes the
Planck function at temperature $T_j$ and $\beta_j$ a scaling factor for the Planck functions. 
A sum of two Planck functions is also used to represent the continuum flux $F_{cont}$.
Following \citet{gielen08}, we use two different dust and continuum temperatures,
ranging from 100\,K to 1000\,K. 

The dust species we included are amorphous olivine/pyroxene (Mg$_{2x}$Fe$_{2(1-x)}$SiO$_4$/Mg$_x$Fe$_{1-x}$SiO$_3$),
crystalline olivine/pyroxene (forsterite/enstatite) and amorphous silica. Silica has different polymorphs, such as quartz, cristobalite and tridimite, with similar emission profiles \citep[e.g.][]{sargent09}, and we cannot rule out that some of these other polymorphs contribute to the silica fraction. To keep the number of free parameters to a minimum, we opted to use only amorphous silica in our modelling.
In section~\ref{15micronregion} we showed that (part of) the enstatite content in our discs might be in the form of
clino-enstatite with a 10\% iron content. Unfortunately, the laboratory data of this enstatite species does not allow to 
calculate synthetic spectra for different grain sizes, so we opted to use the more commonly used iron-free ortho-enstatite.
As discussed above (Sect.~\ref{complexes}), our study of the different complexes show that these dust species are present, and 
that there is no strong evidence for the presence of other dust species.

Mass absorption coefficients for the different dust species are calculated from refractory indices in gaussian random fields 
(GRF) dust approximation \citep{shkuratov05}. The details of the different refractory indices that we used can be found in Table~\ref{dustoverview}. From our previous spectral studies we know that the observed emission features
are reproduced using a non-spherical grain shape. Even though the continuous distribution of ellipsoids approximation \citep[CDE,][]{bohren83}
is widely used, it is unfortunately only valid in the Rayleigh limit, and does not allow us to study grain growth effects. For this reason we prefer
the GRF approximation. We also tested the distribution of hollow spheres approximation \citep[DHS,][]{min05a}, but this did not result in a better fit to the observed emission features. 

To study the grain size distribution inferred from the modelling, we use three discrete dust grain sizes in the model: 0.1, 2.0 and 4.0\,$\mu$m.
The emission features of grains with larger sizes become too weak to distinguish from the continuum emission.
In \citet{gielen08} we already found that the presence of Mg-rich amorphous grains cannot be ruled out, and thus here also we use both
purely Mg-rich amorphous silicates ($x=1$) and amorphous silicates with an equal amount of Mg and Fe ($x=0.5$).
The ratio of magnesium and iron in the amorphous silicates mainly changes the peak position of the 10 and 18\,$\mu$m emission features
\citep{dorschner95}.

The best model was calculated using standard $\chi^2$ minimalisation. Errors on the model parameters were calculated using a 100 step
Monte Carlo simulation with gaussian noise distribution.
Even though this model is only a first approximation,
the model clearly succeeds in giving an overall good fit to the observed spectra (see Figs.~\ref{fits1}-\ref{fits5}).

For 6 sample (Galactic) sources the Spitzer spectrum only starts at 9.9\,$\mu$m, which means we lack information on the dust
composition in the 10\,$\mu$m wavelength range. Since this could influence the derived dust parameters, we depict these sources 
in a different color in our correlation plots (Figs~\ref{cryst_meansize}-\ref{cryst_meansizecryst}).

\subsection{Results}

\begin{figure}[h]
\vspace{0cm}
\hspace{-0.5cm}
\resizebox{9.5cm}{!}{\includegraphics{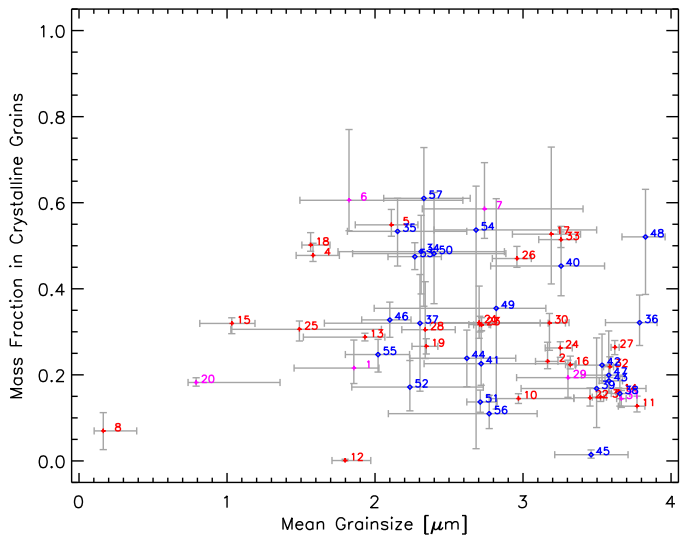}}
\caption{The mass fraction in crystalline grains versus the mean grain size of our spectral modelling. 
Galactic sources are
given in red plus signs and LMC sources in blue diamonds. The magenta symbols depict Galactic sources for which the infrared spectra
only start from 9.9\,$\mu$m. The numbers correspond to numbers given in Tables~\ref{galsterren} and \ref{lmcsterren}.}
\label{cryst_meansize}
\end{figure}

We find that, for most sources, the dust is dominated by large grains. We define the mass-weighted mean grain size of the dust as 
$$a_{mean} = \Sigma m_ia_i,$$ with $a_i$ the grain size, and $m_i$ the mass fraction of dust in that grain size. 
For 47/57 of our sample stars, the mean grain size is larger than 2\,$\mu$m (see Fig.~\ref{cryst_meansize}). 

\begin{figure}[h]
\vspace{0cm}
\hspace{0cm}
\resizebox{8cm}{!}{\includegraphics{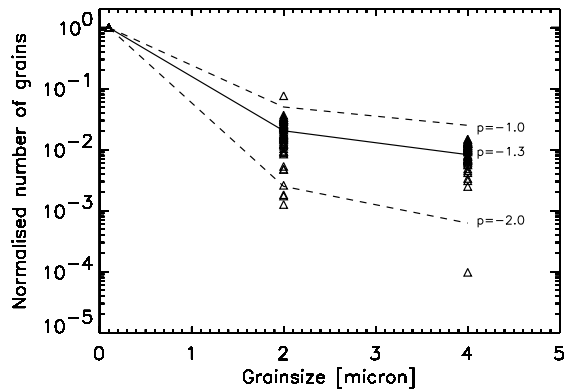}}
\caption{The normalised number of grains versus the adopted grain sizes. The triangles represent all the sample stars.
The solid line gives the best power-law distribution to the mean of all stars. The dashed lines represent different power-law indices, given for comparison. }
\label{grainslope}
\end{figure}

Since our model routine uses three different grain sizes, we can use them to determine a grain size distribution. The Spitzer spectra probably only trace the upper layers of the disc, and so the calculated distribution could not be valid for the entire disc. The grain size distribution is usually approximated by $n(a) \propto a^{p}$, with $n(a)$ the number of grains with grain size $a$, and $p$ a power-law index. For the interstellar medium a value of $p=-3.5$ is found \citep{mathis77} for typical ISM grains up to 0.3\,$\mu$m, rolling over exponentially for larger grains
\citep{zubko04}. To calculate the number of grains in a given grain size, we compute the mass fraction of these grains from our modelling and divide it by the corresponding volume of the grains. We then normalise all the grain numbers, such that $n(0.1\,\mu$m$) = 1$. The results of this calculation can be seen in Figure~\ref{grainslope}. We find a good fit to our results is achieved with a power-law index $p=-1.3^{0.1}_{0.2}$, for grain sizes between 0.1 and 4\micron.
It is clear that our grain size distribution is not ISM like, larger grains are much more abundant.  

From Figure~\ref{meansizecryst_meansizeamorf} it is clear that for about half the stars the crystalline grains are 
larger than the amorphous grains. For nearly all stars the mean grain size of the crystalline grains lies above 2\,$\mu$m, whereas the 
amorphous grains show a larger spread in grain sizes. This is in contrast to what is found for the dust in discs around Herbig Ae stars, where the crystalline grains are significantly smaller than the amorphous grains \citep{juhasz10}. We do not find any correlation between the size of crystalline and amorphous material. It is unclear what causes this difference in grain size between the amorphous and crystalline dust.
An effect that could come into play here is the apparent spectral signature of large dust aggregates. \citet{min08} showed
that aggregates with a very low abundance appear spectroscopically as very small grains, 
while more abundant materials appear spectroscopically to reside in larger grains. Since for our sources the amorphous dust is in most cases more abundant than the crystalline dust, this could mean that the amorphous grains reside in large fluffy aggregates, which have 
spectral signatures that are very similar to those of small grains \citep{min06,min08}. 

Similar to what is found in \citet{juhasz10} we find that the size of the enstatite grains is on average slightly larger than that of the forsterite grains. There seems to be a weak trend between the crystallinity and the mean size of the crystalline grains: sources with a higher crystallinity have on average larger crystalline grain sizes (see Fig.~\ref{cryst_meansizecryst}).

For discs around young stars a strong correlation is found between the mean grain size of the amorphous grains and the disc flaring.
This disc flaring is determined by the ratio of the 24\,$\mu$m and 8\,$\mu$m flux. \citet{juhasz10} find that sources with flatter discs
have larger amorphous grains in their disc atmosphere. This trend is not seen in our sample. Our sample sources seem to be centred around 
$F_{24}/F_8 = 0.83$, which shows that these discs are not strongly flared and that there is no large spread
in disc flaring. Of the 57 sources, 8 show higher values of the disc flaring, going from 
$F_{24}/F_8 = 2$ up to 4.3. We find no correlation between the disc flaring and any other dust parameter.


Figure~\ref{cryst_meansize} also shows the high crystallinity fraction derived from the Spitzer modelling. 
The crystallinity can reach values of 60\%, which is among the highest seen in astronomical environments.
High values of crystalline dust are also found for protoplanetary discs around young stars \citep[e.g.][]{bouwman08,juhasz10},
where crystalline fractions up to about 30\% are found.
For the crystalline dust, forsterite is almost always the dominant species: the forsterite fraction of the crystalline material has values between
20 and 100\% (Fig.~\ref{forst_cryst}). 
There does not seem to be a strong correlation between the crystallinity and the forsterite/enstatite fraction of the crystalline material.
The same holds for the forsterite/enstatite fraction of the crystalline material and the mean grain size of the crystalline grains (see Fig.~\ref{enstfrac_meansizecryst}).

The derived silica fractions are of the order of 5\%, but can go up to 20\%. We find no correlation between the silica mass fraction and any other dust parameters. Thermal annealing of amorphous dust produces both forsterite and silica, and so a relation is to be expected if this process is responsible for the crystallisation. 

In our modelling we used amorphous silicates with an equal Mg-Fe content and pure Mg-rich grains. 
We find that 25/57 sources show a dominance of iron-free amorphous dust, whereas the other sources are clearly dominated by
Mg-Fe amorphous silicates. On average we find that the Galactic sources have a slightly higher fraction (53\%) of the Mg-Fe rich amorphous silicates,
whereas the LMC sources have a higher fraction (56\%) of purely Mg-rich amorphous silicates. The derived differences are very minimal, so we cannot make strong statements on the iron content of the amorphous silicates. For the sources where we find a high fraction of iron-free dust, the iron grains could be stored as metallic inclusions in the grains, which would be very hard to detect.

We do not find any correlation between the derived dust parameters and central binary parameters such as the effective temperature or the orbit.

\subsection{Atypical sources: intruders?}

Some sources clearly deviate from the mean observed spectrum, by showing no crystalline grains (IRAS\,10174 and  J\,05133) or
carbonaceous molecules and/or gas emission (EP\,Lyr, HD\,52961, IRAS\,06338 and IRAS\,13258).
Other sources, such as MSX\,949, show less obvious differences, but are still not reproduced as well by the model as the other sources.
Since these sources were mainly selected on the basis of their infrared colours, we cannot exclude that 
non-post-AGB disc sources are present in the sample. Possible intruders could be young stars with protoplanetary discs,
red super giants or AGB stars. However, for most sources we have additional observations of the central star, such as optical spectroscopy,
which corroborate their post-AGB evolutionary phase.
We discuss several doubtful (or anomalous) cases below.

\subsubsection{SAGE\,050830}

For SAGE\,050830, the optical spectra show some evidence for a carbon-rich chemistry \citep{vanaarle11}.
However, the photometry and infrared spectral information for this source is suspected to be contaminated by a
foreground star of spectral type A0-1IV. Unfortunately, the angular resolution does not allow us to 
discriminate between the A star and the carbon star as the identification of the Spitzer source.
Still, if the carbon-rich spectrum truly belongs to the Spitzer source,
the strong oxygen-rich spectrum seems surprising. This source is one of the more crystalline objects of our
sample, and even has the most extreme 10\micron feature-to-continuum ratio of all sources!

The carbon-rich classification of the central star, together with the presence of crystalline silicates in its
circumstellar environment would make this star an ideal candidate to be a silicate J-type carbon star.
These are carbon-rich AGB stars, but with a very low $^{12}$C/$^{13}$C ratio and detection of crystalline silicates in 
their infrared spectrum \citep{lloydevans90,abia00}.
The sources are believed to be binary stars, with an unseen companion, surrounded by a circumbinary disc \citep{morris90,jura99,yamamura00,deroo07a}.
This scenario could explain the dual chemistry, since the disc could then be formed while the central star
was still oxygen rich, and has now evolved to be carbon rich. However, it does not explain the 
low $^{12}$C/$^{13}$C ratio, usually seen in J-type silicate carbon stars.

Unfortunately, due to the confusion with the foreground star, we cannot determine the stellar parameters,
which would shed light on the evolutionary status of this object. Also, the low-resolution optical spectrum
does not allow to determine the $^{12}$C/$^{13}$C ratio of the carbon star, hence corroborating the J-type nature.

\subsubsection{J\,05133 and IRAS\,10174}

\begin{figure}[ht]
\vspace{0cm}
\hspace{0cm}
\resizebox{10cm}{!}{\includegraphics{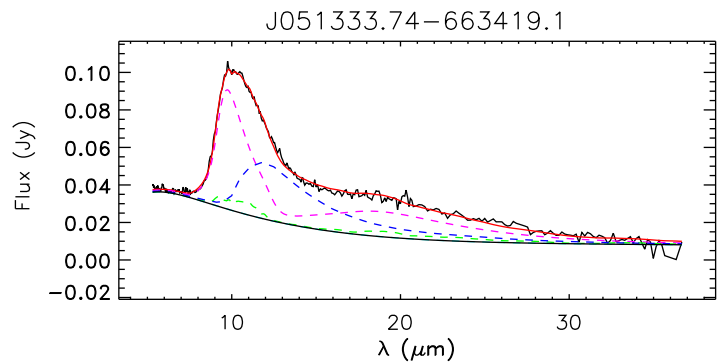}}
\caption{The results of our modelling of J\,05133.74, with the addition of alumina grains. 
The observed spectrum (black curve) is plotted together with the best model fit (red curve) and the continuum (black solid line).
Crystalline silicates are plotted in green, amorphous silicates in magenta and alumina in blue.}
\label{alumina}
\end{figure}

Two of our sample sources, IRAS\,10174 and J\,05133, clearly deviate from the rest of the sample by showing no strong evidence for crystalline features in
their spectra. However, our modelling shows that purely amorphous silicate dust is not sufficient in reproducing the observed features, especially around 13\,$\mu$m. The spectra of these stars are actually very similar to the observed spectra of AGB outflows, characterised by small amorphous grains. In these sources, emission from additional dust species, such as alumina (Al$_2$O$_3$), can influence the 13\,$\mu$m region.
To see if alumina could also be present in these sources, we remodelled the spectra, now including alumina grains.
For both stars we find an improvement when including Al$_2$O$_3$, especially for J\,05133. For IRAS\,10174 the improvement is only minor,
with an amount of alumina in the new model of 2\%. However, for J\,05133 the fit is improved drastically when including 30\% alumina (Fig.~\ref{alumina}). The bulk of the other dust ($\sim$50\%) is stored in small 0.1\,$\mu$m Mg-rich olivine in this new model. Less then 10\% of the mass fraction of dust is in crystalline form, and then mainly forsterite. This type of dust composition is more indicative of an outflow and not a disc.

The optical spectrum of J\,05133 \citep{vanaarle11} also indicates the peculiar nature of this source.
The spectrum points to a F8-G0Ip spectral classification, but shows very strong H\,I (6563\,\AA) and He\,I (5876\,\AA) emission,
and broad Ca\,II absorption lines, which are not expected in a star of this type, but point to the 
presence of a hotter source. One possibility would be that the system is actually a binary with an unseen
hot companion. The SED modelling gives, for this source, a luminosity of around 17\,000\,L$_\odot$.
This, combined with the spectral type as derived from the optical spectrum, shows that the source
cannot be an AGB star, but also shows that it is probably not a post-AGB disc source as normally understood.

The unusual chemistry of IRAS\,10174 compared to the rest of the sample made us re-investigate the optical spectra for this source.
The source was originally classified as a post-AGB source \citep{lloydevans99,deruyter06} from a blue spectrum which suffered badly from dust extinction in the violet. A new optical spectrum ($6400 - 9000$\AA), taken with the Cassegrain grating spectrograph of the 1.9\,m Radcliffe Telescope at the South African Astronomical Observatory, shows that the star is a luminous supergiant, of type G8Ia-O. This is in better agreement with the observed chemistry of the infrared spectrum. Note that even though the star is not a post-AGB disc source, the binarity of the star is confirmed \citep{maas03}.

Even though some non-post-AGB disc sources might be present in the sample, these sources do not change
the overall conclusions of this study. The removal of these sources does not introduce correlations
between the different stellar and dust parameters, that are currently not observed.

\section{Discussion}
\label{discussion}

For all sources, the infrared spectra are dominated by emission features due to oxygen-rich
dust species. For some of the LMC sources (such as J051159.11, J053605.56, and SAGE054310) this is surprising, since they have luminosities
which would put them in the peak of the carbon star luminosity function \citep{stancliffe05,groenewegen07,srinivasan11}.
Optical spectroscopy for these sources does not point to a carbon-rich chemistry of the central star \citep{vanaarle11}.
This shows that the AGB evolution for these binary post-AGB stars was shortcut, possibly under the influence of strong
binary interaction, preventing them to evolve into carbon stars. 

Our study shows that even a relatively simple model succeeds in reproducing the observed infrared spectra.
The model assumes only two dust temperatures, and uses the same dust abundances
for the cool and warm dust. This shows that the dust in the disc is relatively well mixed, in that cooler and hotter regions
in the disc have a similar dust composition. This is very different from the results for protoplanetary discs around young stars,
where a difference in dust composition is needed for the inner and outer disc regions \citep{juhasz10}.

The strong observed crystalline bands at longer wavelengths show that at least a significant fraction of the crystalline
grains are located at cooler temperatures. For some sources our model even underestimates the forsterite flux at 33\,$\mu$m,
showing that, for some sources at least, the forsterite fraction between the two temperatures might not be evenly distributed, but
dominated by the cooler temperature. The problem of reproducing the features at longer wavelengths might also be due
to the adopted synthetic spectra of forsterite. As can be seen in Fig.~3.14 of \citet{gielen08}, the 33\micron is best
reproduced by DHS grain shapes. The GRF grain shape gives a feature which is much broader and flat topped.
However, since on average our features were slightly better reproduced with GRF shapes, we used this approximation 
in our modelling.
Another effect that can influence the observed features in the optical depth. At different wavelengths we would look
at different depths in the disc, with a different temperature distribution. At longer wavelengths, we would then 
look deeper in the disc, where the cooler temperature might enhance the features at these longer wavelengths.
To study this effect in detail, a full radiative transfer model is needed, which goes beyond the scope of this paper.

Since the exact formation mechanism of these circumbinary discs is still uncertain, it is
difficult the relate the different observed dust characteristics to disc evolution.
One possibility is that the discs are formed after a common-envelope phase, with some dust formation already forming in the outflow phase.
Crystalline grains can then be formed directly out of the gas phase, at high temperatures \citep{gail04,petaev05}.
But this will probably not give rise to the very high amounts of crystalline material we see,
which shows that another crystallisation process is still active afterwards, such as thermal annealing \citep{wooden05}. 
Another formation mechanism is Roche-Lobe overflow through an outer Langrangian point, 
where the material is already confined to the midplane. Here one could expect the dust at the 
hot and denser inner regions to be more efficient in producing crystalline species.
Both gas-phase condensation and thermal annealing might be important to explain the
high crystallinity in these discs.

The condensation models predict forsterite to condense first, followed by the formation of enstatite through
reactions between forsterite and SiO$_2$ gas. 
In contrast to what is found for protoplanetary discs, our results show that 
forsterite is almost always the dominant crystalline dust species. This could point to a deviation from 
equilibrium conditions during condensation. 
If the forsterite grains reach large grain sizes quickly, the formation of enstatite
might be complicated, since it will become increasingly harder to infuse SiO$_2$ in the forserite lattice.
The resulting dust may the be in the form of a large forsterite grain, surrounded by a small layer of enstatite.
The formation of enstatite can be further weakened
if the material is allowed to cool very quickly after the condensation of forsterite.
The formation of forsterite through annealing
is especially efficient if the starting material has an olivine stoichiometry. A high forsterite fraction would
then go together with a higher pyroxene fraction of the amorphous material, which is not supported by our results.

Since crystallisation requires high temperatures above 1000\,K \citep{fabian00}, one would expect the
crystalline dust to be confined to the hot, inner regions of the disc. This is in clear contrast to our findings of 
cool crystalline material and a homogeneous dust composition throughout the disc.
This shows that mixing must be efficient is transporting the crystalline material to cooler regions which were initially dominated by amorphous material,
or a crystallisation process at lower temperatures is occurring in the discs.
\citet{molster02a,molster02c} already showed that the crystallinity fraction in disc sources is much higher
than that observed in typical outflow sources. This shows that the crystalline component in the disc sources is
most likely determined by subsequent dust grain processing in the discs, and not by cooling processes in the outflow
of the material forming the discs.

Dust formation models also show that iron will preferably condense out
as metallic iron, rather than be included in silicate formation. This could explain the presence of 
Mg-rich amorphous silicates in our results. The grain sizes of the crystalline dust, formed through condensation,
will not be correlated with the grain sizes of the amorphous material, which is in line with our results.
If the crystallisation occurs through annealing, we would expect
a relation between the initial amorphous material and final crystalline grains. This does not explain the
observed difference in crystalline and amorphous grain sizes, unless a subsequent process can be invoked
that would grow the crystalline material, but not the amorphous dust. Our results on the difference in 
crystalline and amorphous grain sizes is again in clear contrast to what is found for the dust in 
protoplanetary discs, where the crystalline grains are found to be significantly smaller than the 
amorphous grains \citep{juhasz10}. Clearly, different dust processes are responsible for the grain growth
and crystallisation in the discs around young and evolved stars. 

The derived large grain sizes show that there seems to be an efficient removal of the smallest grains. 
The question remains whether this lack of small grains is an effect of grain growth \citep{dullemond04} or whether the initial grain population already consisted of large grains. In that case the small-grain fraction could be a result of grain collision and subsequent break-up.
An effect which might also be important to the observed grain sizes is the strong radiation of the central source.
The central post-AGB stars are highly luminous, and radiation pressure could be responsible for the removal
of the smallest grains in the upper layers of the disc. Since our results show that the amorphous grains
tend to be smaller than the crystalline grains, radiation pressure might be (partly) responsible for the
large fraction of crystalline grains observed in the upper layers of the disc.

Surprisingly, we find no correlations between the derived dust parameters, such as crystallinity, grain size and abundances.
Also, no correlation between the dust parameters and parameters of the central binary system is found. 
The lack of correlation raises the question whether the optically thin upper layers traced by the Spitzer spectra are
a good representative of the global dust composition.

Except for the amorphous silicate dust, we find no evidence for the presence of dust species
usually associated with AGB outflows or single-star post-AGB shells, such as simple oxides or Al/Ca-bearing dust species. 
The theoretical oxygen-rich dust condensation sequence for dusty outflows starts with the formation 
of alumina (Al$_2$O$_3$) around 1760\,K, followed by formation of
gehlenite (Ca$_2$Al$_2$SiO$_7$) at slightly lower temperatures \citep{tielens90,tielens98}. Further interactions with magnesium will produce
species like spinel (MgAl$_2$O$_4$), akermanite (Ca$_2$MgSi$_2$O$_7$), diopside (CaMgSi$_2$O$_6$) and finally anorthite (Ca$_2$Al$_2$Si$_2$O$_8$)
around 1360\,K. 
A second condensation sequence, involving mostly magnesium and silicon, starts with the formation of forsterite around 1500\,K, 
followed by enstatite around 1300\,K. Only at temperatures below the glass temperature can iron interact to form amorphous iron-containing silicates. Not only the temperature plays a role, also the densities involved will determine which dust species can be formed.

From observations it is found that AGB stars start by forming Al- and Mg-rich oxides in their outflows, followed by an increase 
of amorphous silicates bands with increasing mass-loss rate, which start to grow on the
Al-rich oxides \citep{lebzelter06}. A similar trend is seen in outflows of red supergiants \citep{verhoelst09}.
A similar scenario might explain the lack of Al/Na/Ca-rich dust species in the discs around the post-AGB stars.
Since the photospheres of the central stars are strongly depleted in these elements \citep{maas05,hrivnak08,gielen09b}, we know these refractory elements must be present in the disc. 
Of course, since densities associated with these discs are much higher than for typical outflows, and dust might be subject to
a different temperature gradient, which could result in a different condensation sequence to that observed in AGB stars.

\section{Conclusions}
\label{conclusions}

We analysed the Spitzer infrared spectra of 33 Galactic and 24 LMC (candidate) post-AGB binaries
surrounded by a dusty circumbinary disc. For nearly all Galactic sources, previous studies have
already confirmed the binarity and post-AGB status. The LMC sources were taken from a list
of probable post-AGB disc candidates.
Our main focus was to determine the dust composition of the discs, but also
to look for possible differences between the Galactic and LMC sample.
Our study shows that: 

\begin{itemize}

\item
The Spitzer spectra are all dominated by emission features of oxygen-rich dust species, namely amorphous
and crystalline silicates of olivine and pyroxene stoichiometry.
\item
The observed silicate dust has a high crystallinity factor: most sources have crystalline mass fractions 
between $20-60$\%.
\item
Most of the dust is stored in larger grains ($> 2$\micron). This results in an average grain
size distribution of $n(a) \propto a^{-1.3^{0.1}_{0.2}}$ for grain sizes between 0.1 and 4\micron. 
\item
We find no correlations between the dust, stellar, and/or orbital parameters,
which makes it difficult to constrain the dust grain processes that are causing the
observed dust properties such as the grain sizes and crystallinity.
\item
We find no differences between the dust parameters of the Galactic and LMC sources.
\item
Although the observed spectra are very similar to those of protoplanetary dics, we find
evidence for a fundamental difference in the dust processing occurring in the two disc types,
more specifically in the homogeneous dust compostion throughout the disc, the observed degree of crystallinity, the crystalline grain sizes and, the strong dominance of forsterite in the crystalline grain fraction.

\end{itemize}

\bibliographystyle{aa}
\bibliography{referenties.bib}

\listofobjects

\Online

\begin{appendix}
\onecolumn
\section{Figures and Tables}

\begin{figure}[ht]
\vspace{-1.5cm}
\hspace{0cm}
\includegraphics[width=9cm,height=10cm]{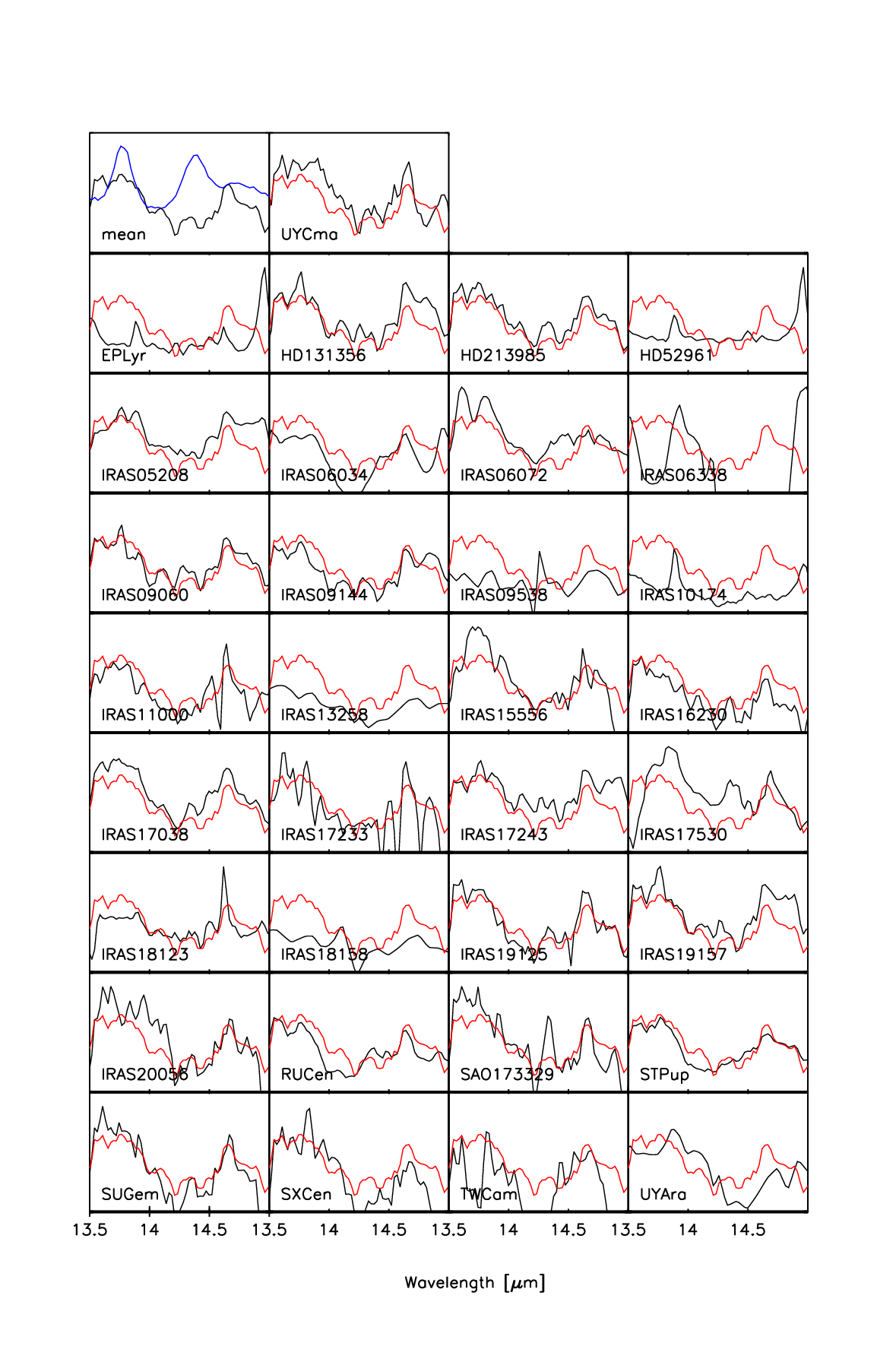}
\includegraphics[width=9cm,height=10cm]{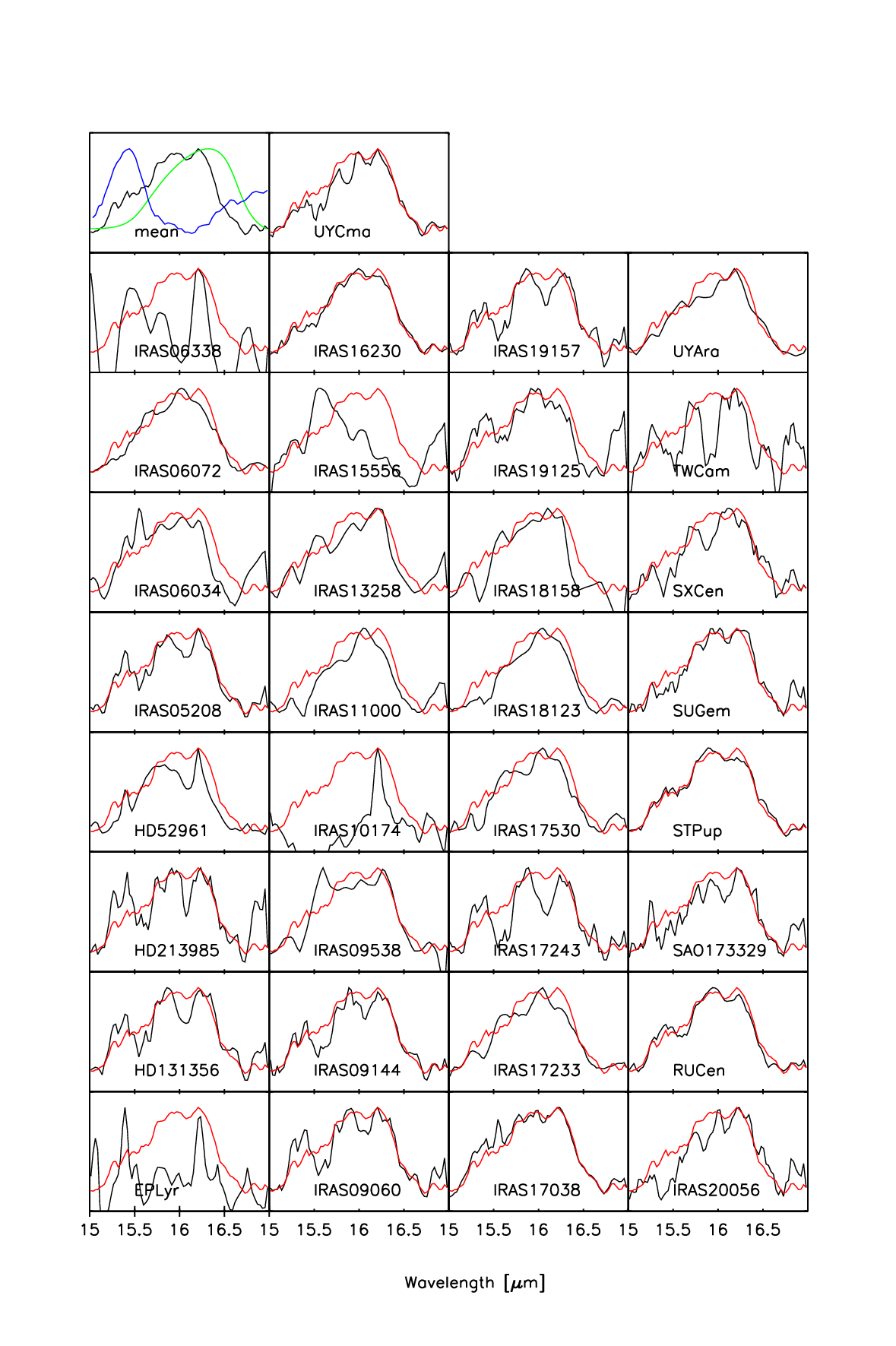}
\vspace{-0.2cm}
\caption{Left: The 14\,$\mu$m complex for the Galactic sources, continuum subtracted and normalised. Overplotted in red the mean spectrum. The mass absorption coefficients of forsterite and enstatite are plotted in green and blue. The high noise level on the LMC sources did not allow for a mean spectrum determination. Right: Same as on the left, but for the 16\micron complex.}
\label{mean1416}
\end{figure}

\begin{figure*}[ht]
\vspace{-1.0cm}
\hspace{0cm}
\includegraphics[width=9cm,height=10cm]{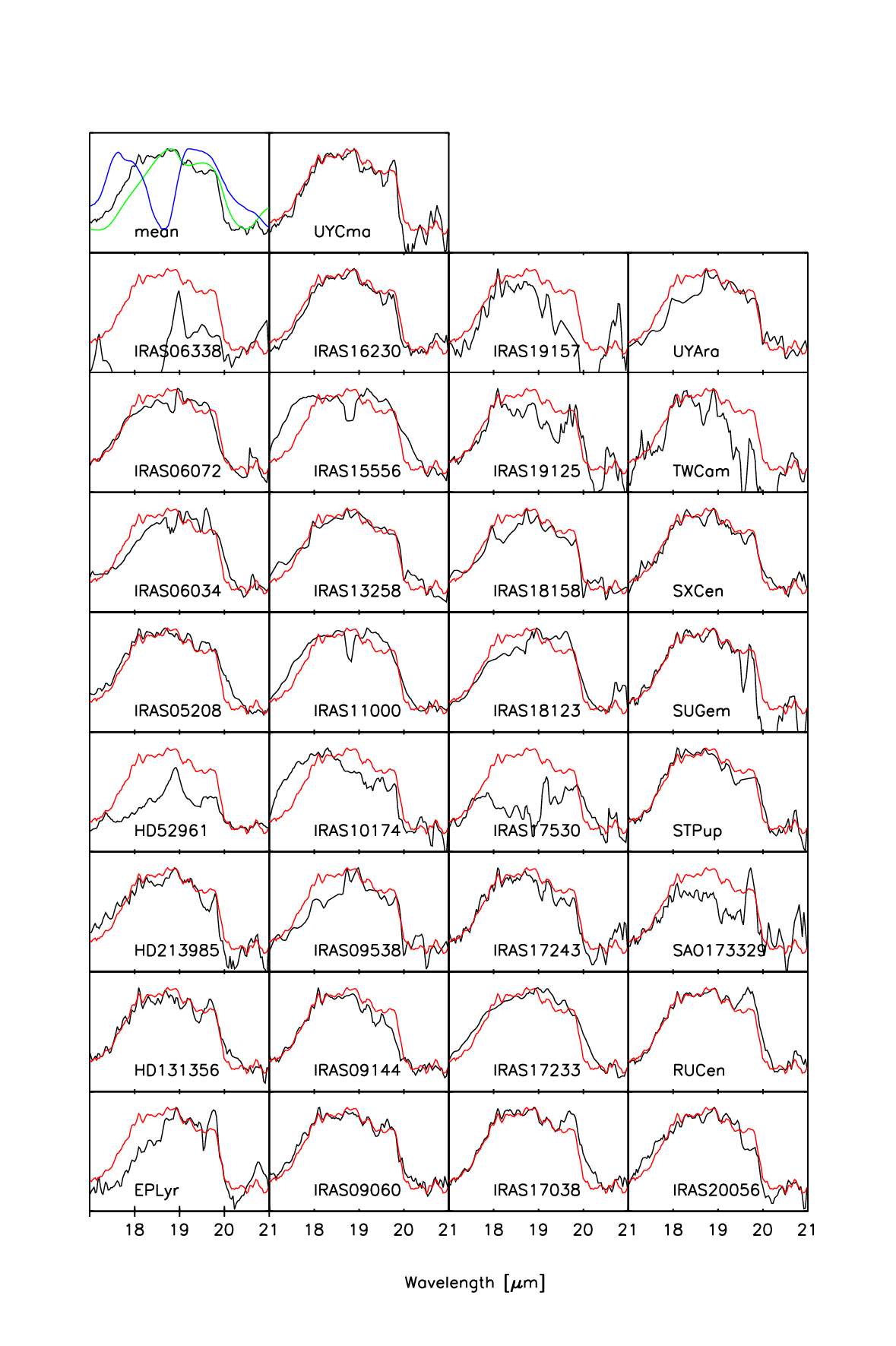}
\includegraphics[width=9cm,height=10cm]{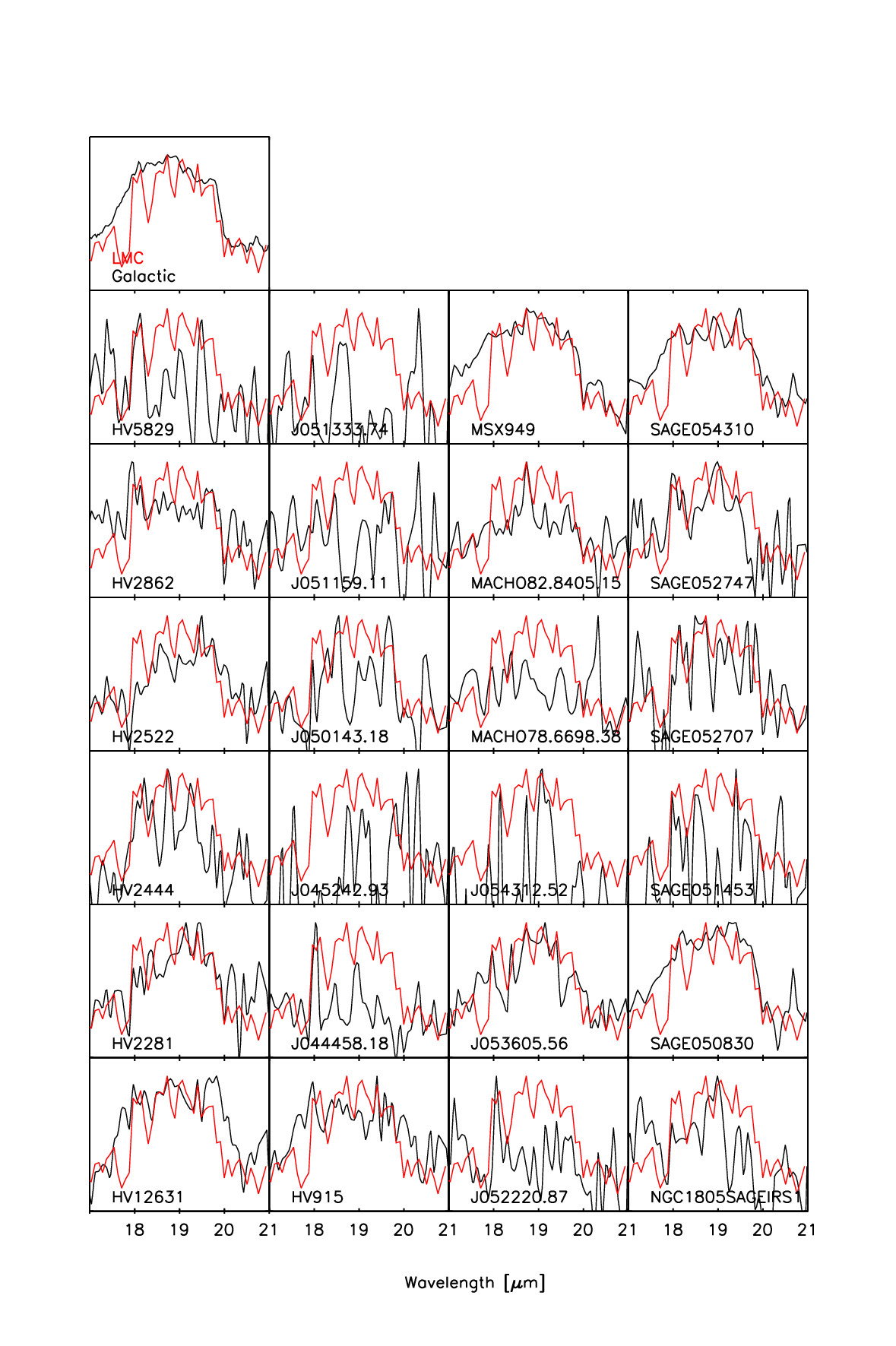}
\vspace{-0.2cm}
\caption{Left: The 19\,$\mu$m complex for the Galactic sources, continuum subtracted and normalised. Overplotted in red the mean spectrum. The mass absorption coefficients of forsterite and enstatite are plotted in green and blue. Right: Same as on the left, but for the LMC sources.
The top panel shows the comparison between the calculated mean for the LMC and Galactic sources.}
\label{mean19}
\end{figure*}

\begin{figure*}[ht]
\vspace{0cm}
\hspace{0cm}
\includegraphics[width=9cm,height=10cm]{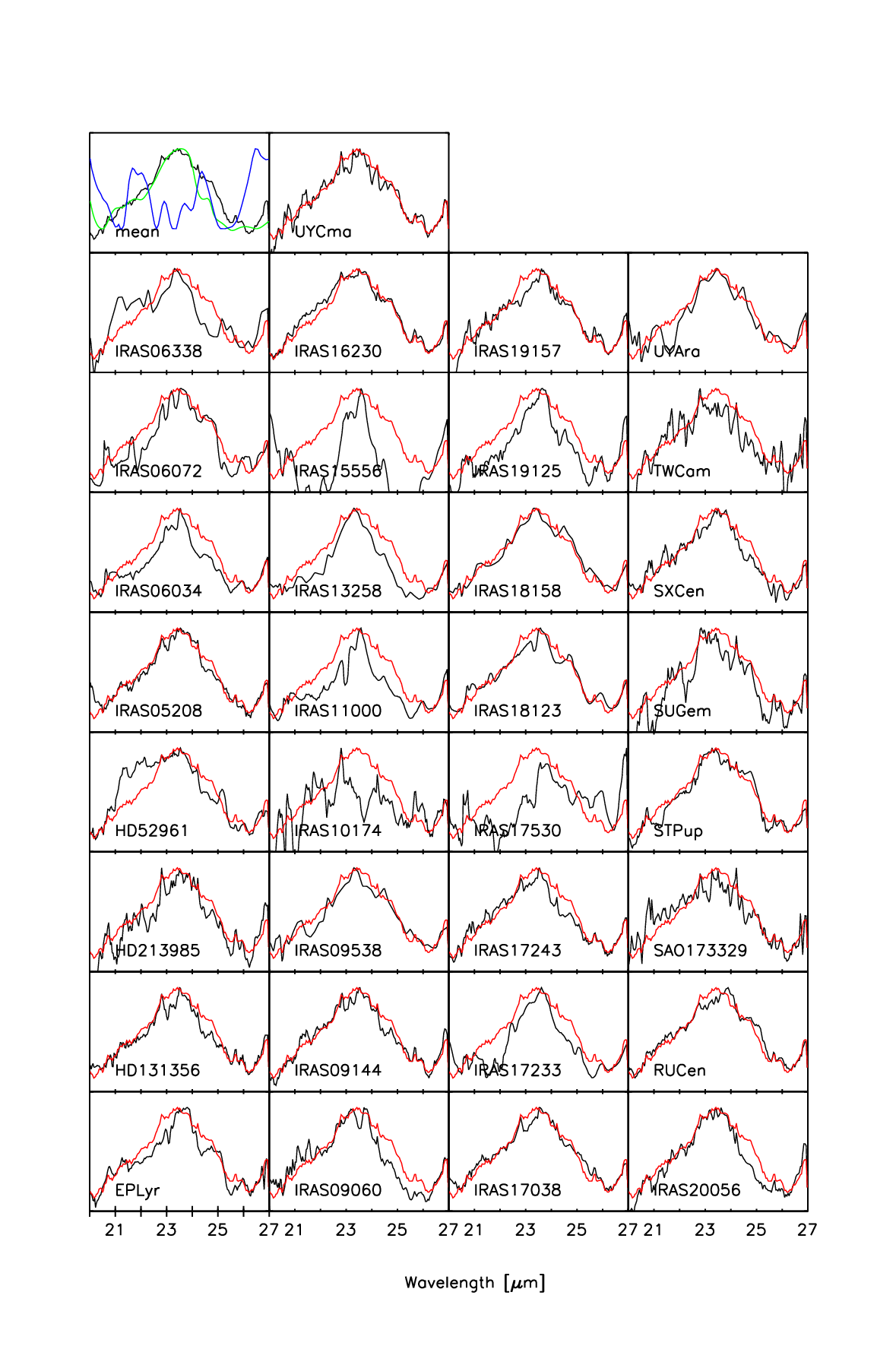}
\includegraphics[width=9cm,height=10cm]{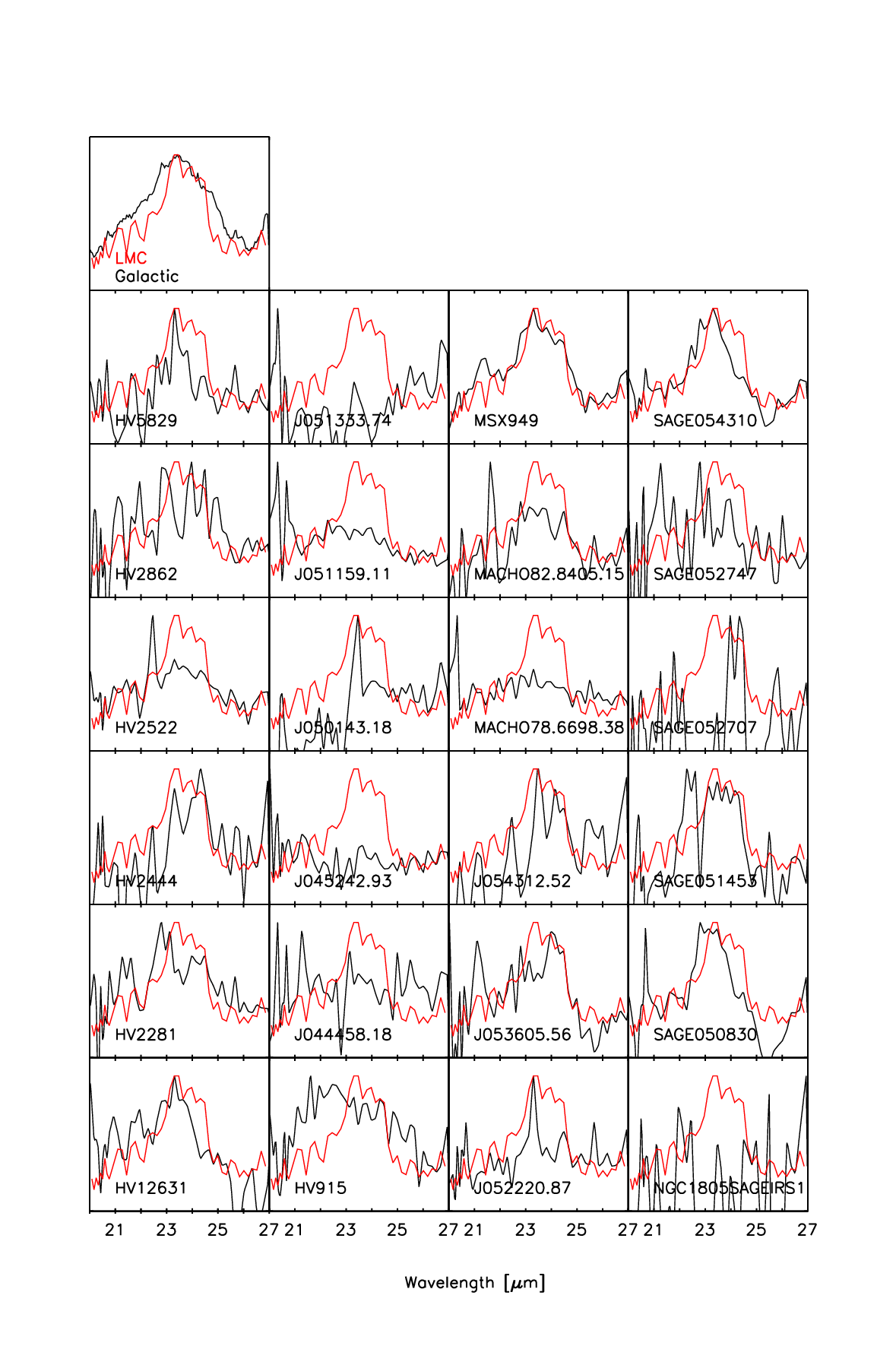}
\vspace{-0.2cm}
\caption{Same as Figure~\ref{mean19}, but for the 23\,$\mu$m complex.}
\label{mean23}
\end{figure*}

\begin{figure*}[ht]
\vspace{-1.cm}
\hspace{0cm}
\includegraphics[width=9cm,height=10cm]{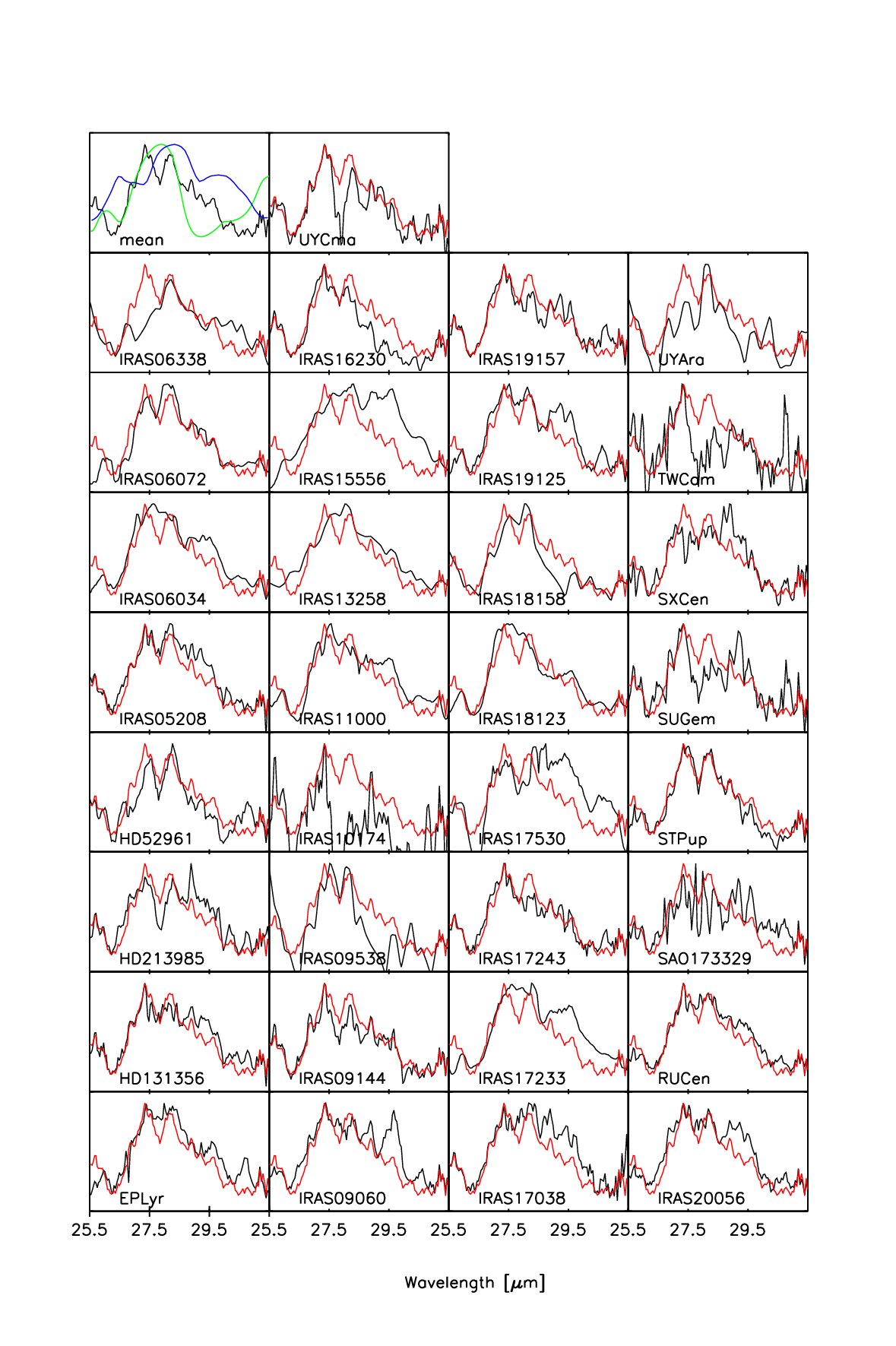}
\includegraphics[width=9cm,height=10cm]{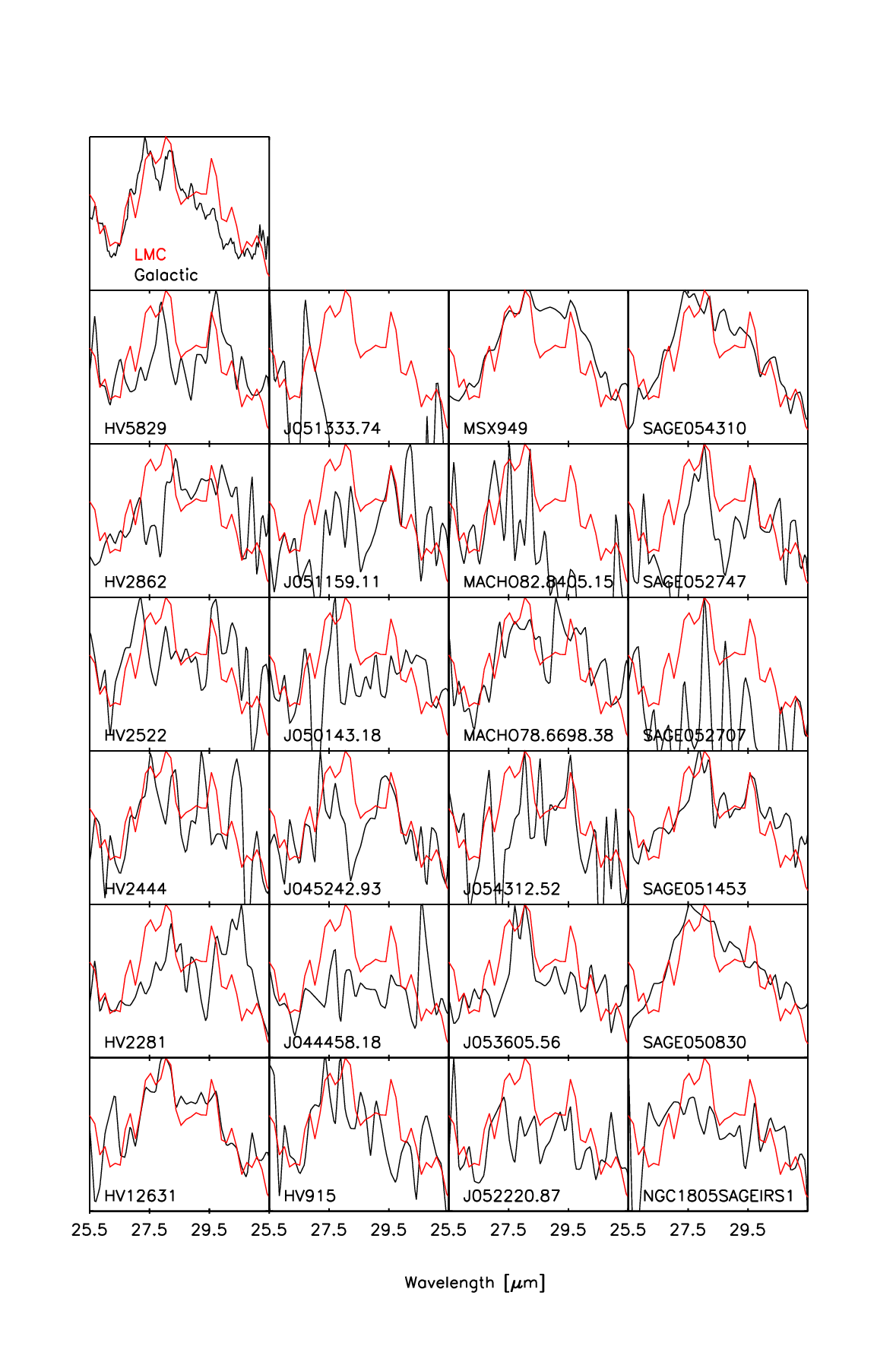}
\vspace{-0.15cm}
\caption{Same as Figure~\ref{mean19}, but for the 27\,$\mu$m complex.}
\label{mean27}
\end{figure*}

\begin{figure*}[ht]
\vspace{0cm}
\hspace{0cm}
\includegraphics[width=9cm,height=10cm]{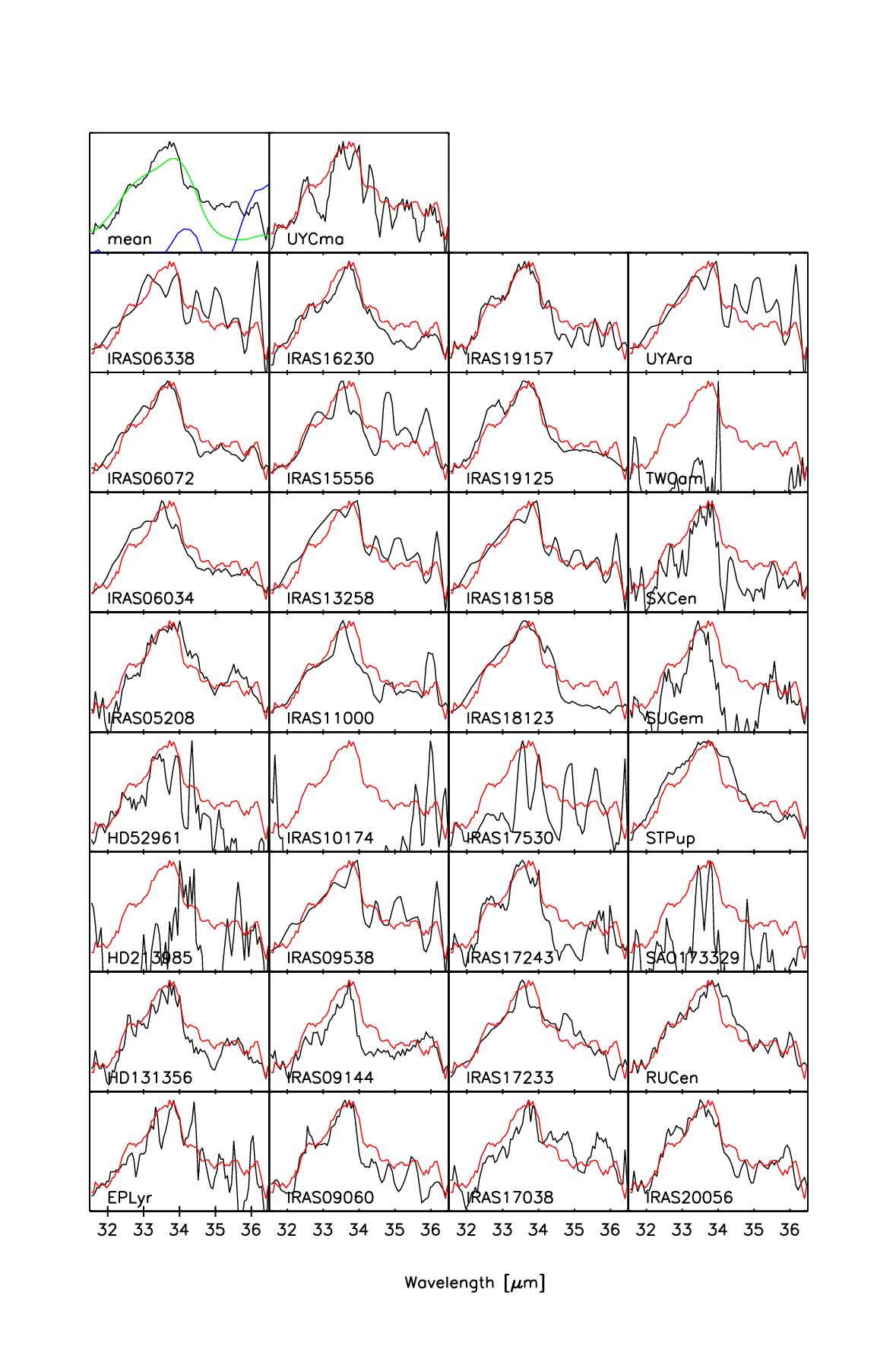}
\includegraphics[width=9cm,height=10cm]{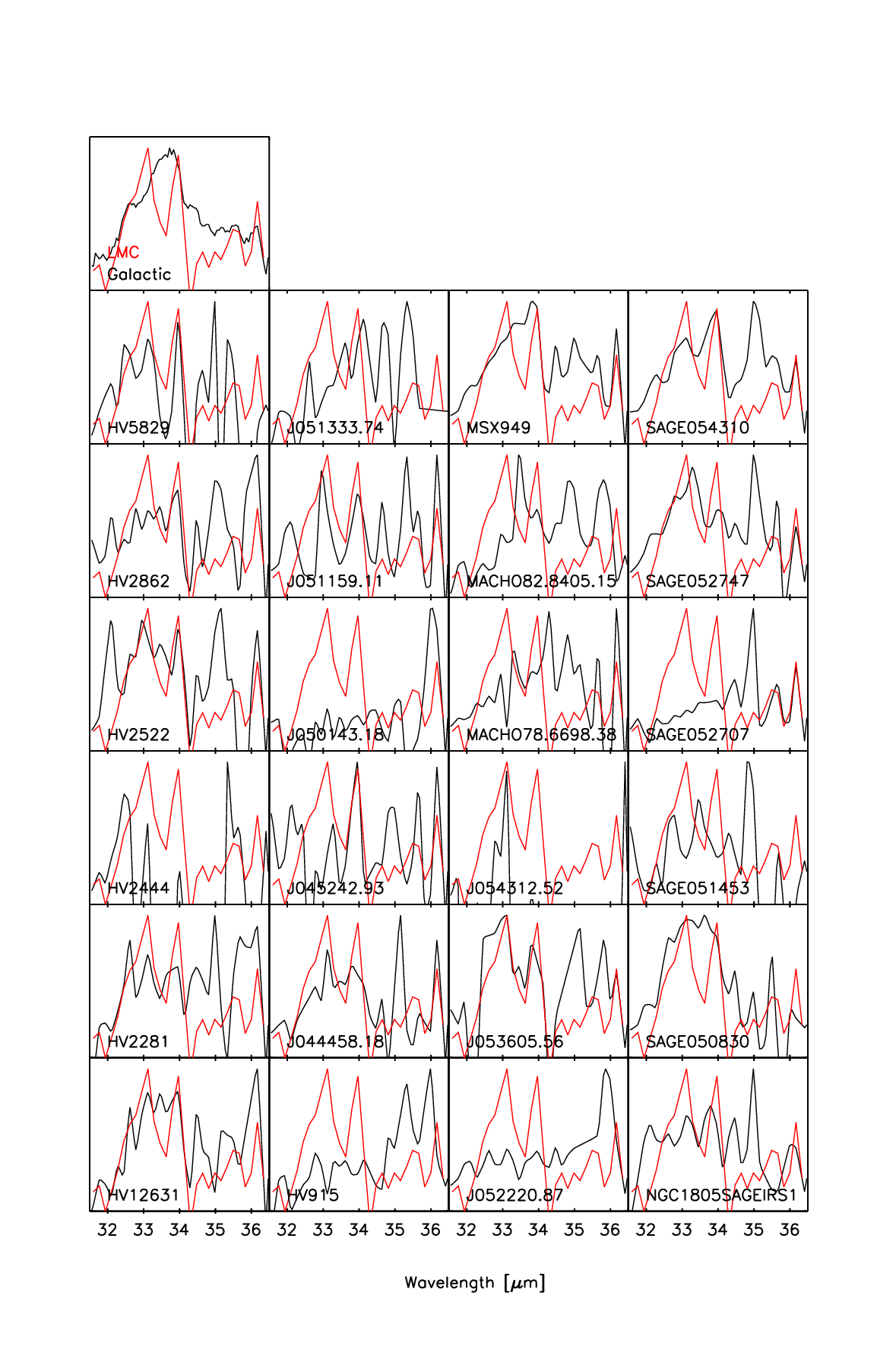}
\vspace{-0.2cm}
\caption{Same as Figure~\ref{mean19}, but for the 33\,$\mu$m complex.}
\label{mean33}
\end{figure*}

\begin{figure}[ht]
\resizebox{6cm}{!}{\includegraphics{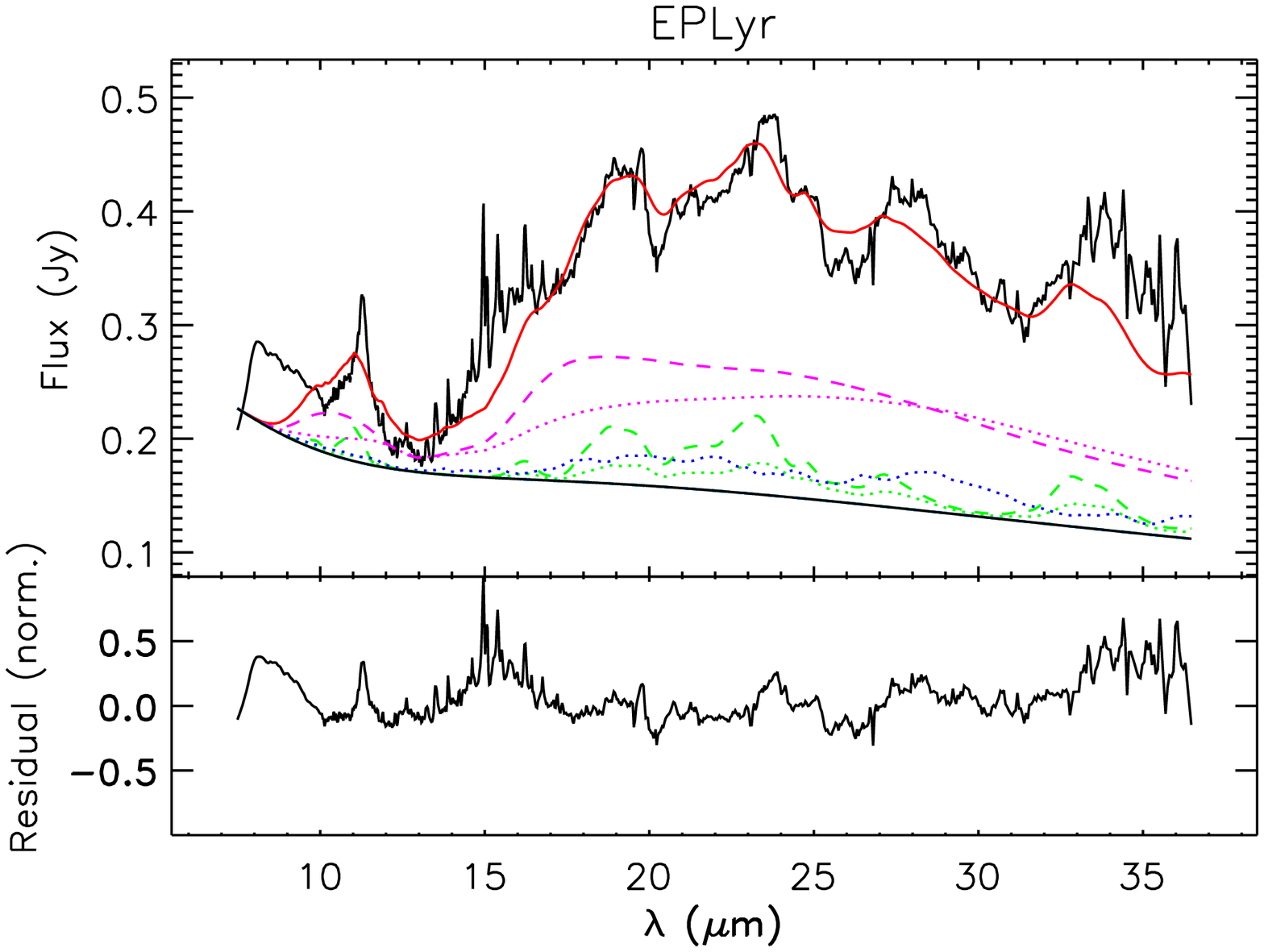}}
\vspace{0.3cm}
\hspace{0.3cm}
\resizebox{6cm}{!}{\includegraphics{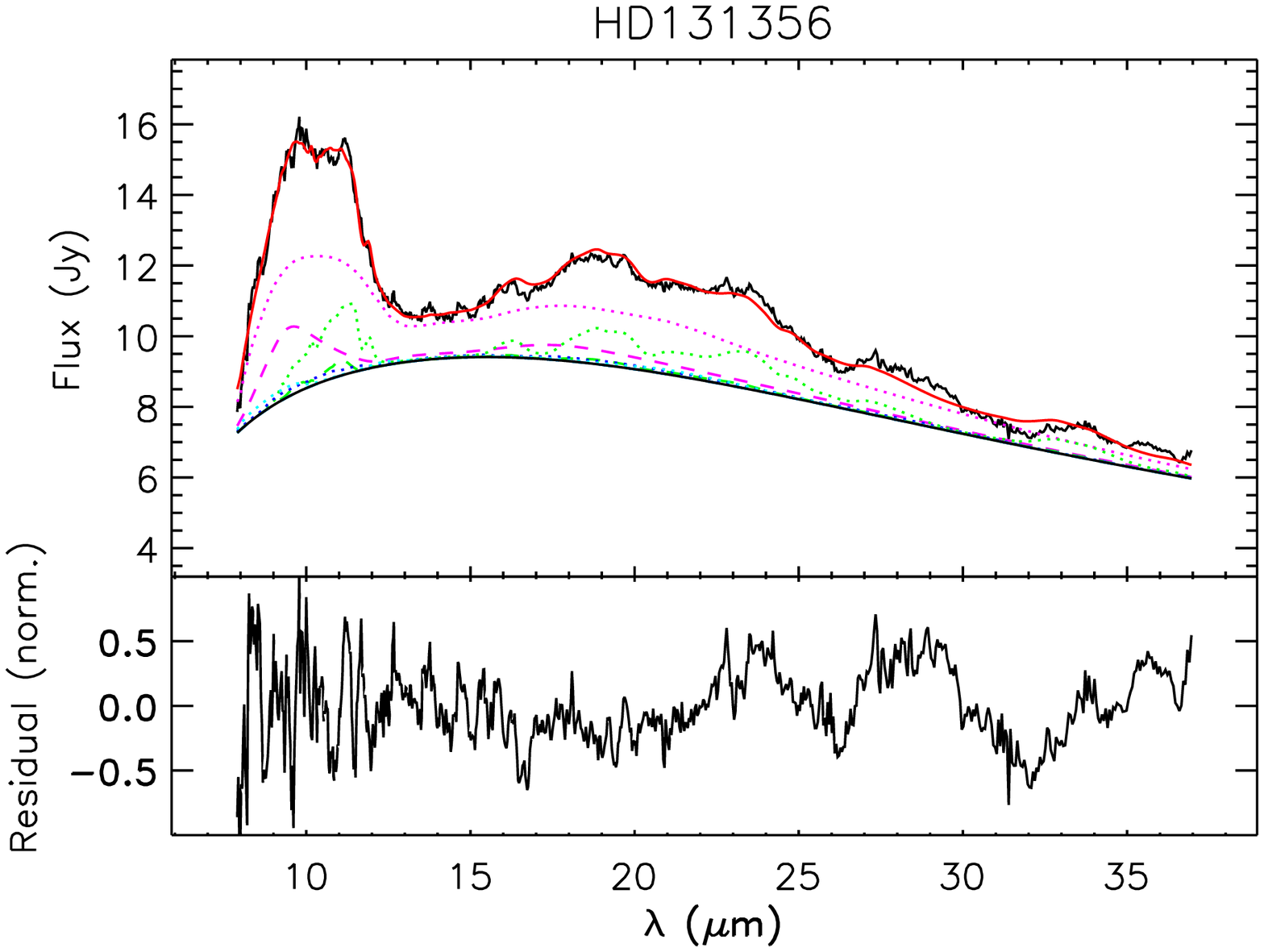}}
\vspace{0.3cm}
\hspace{0.3cm}
\resizebox{6cm}{!}{\includegraphics{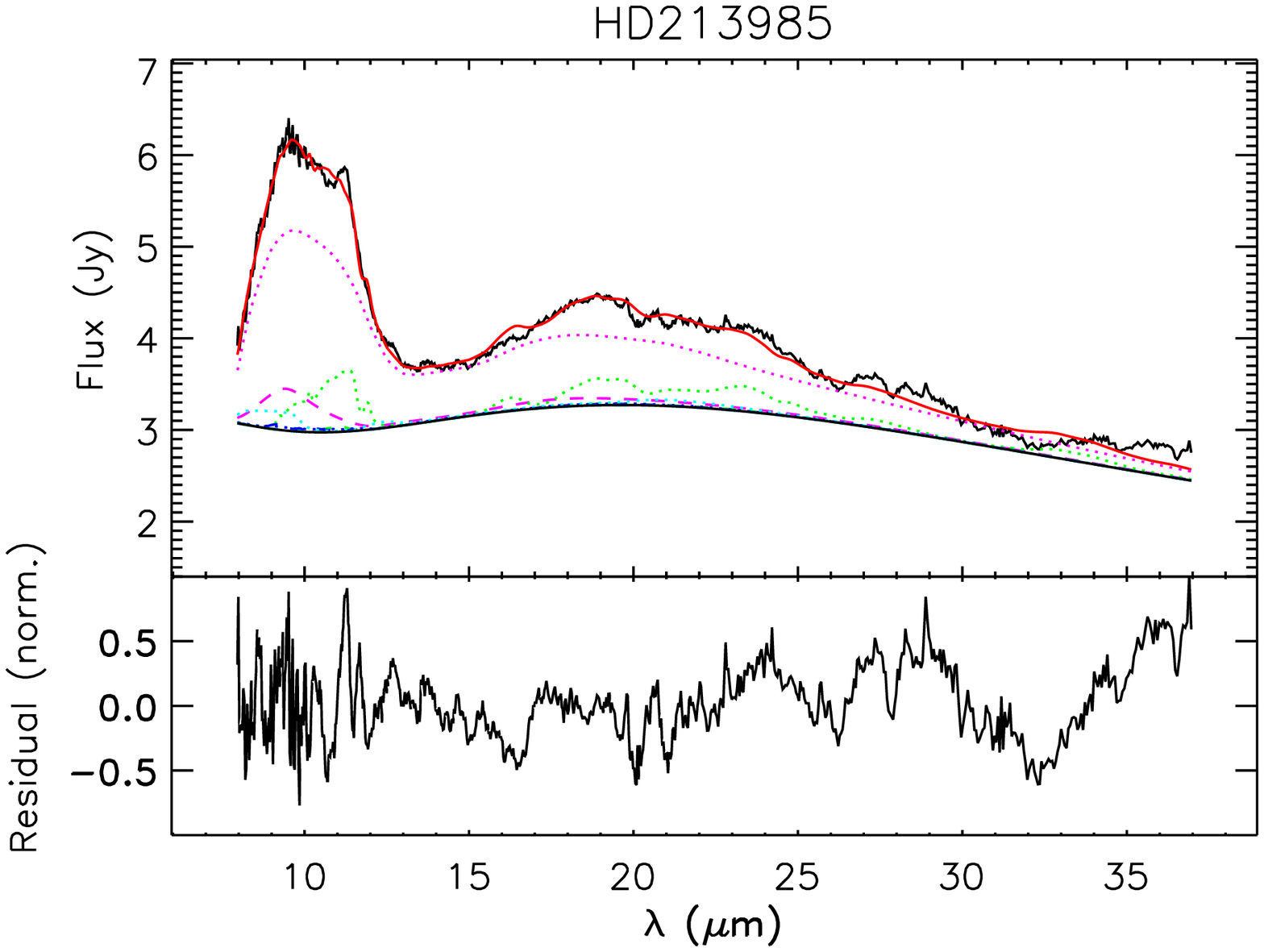}}
\resizebox{6cm}{!}{\includegraphics{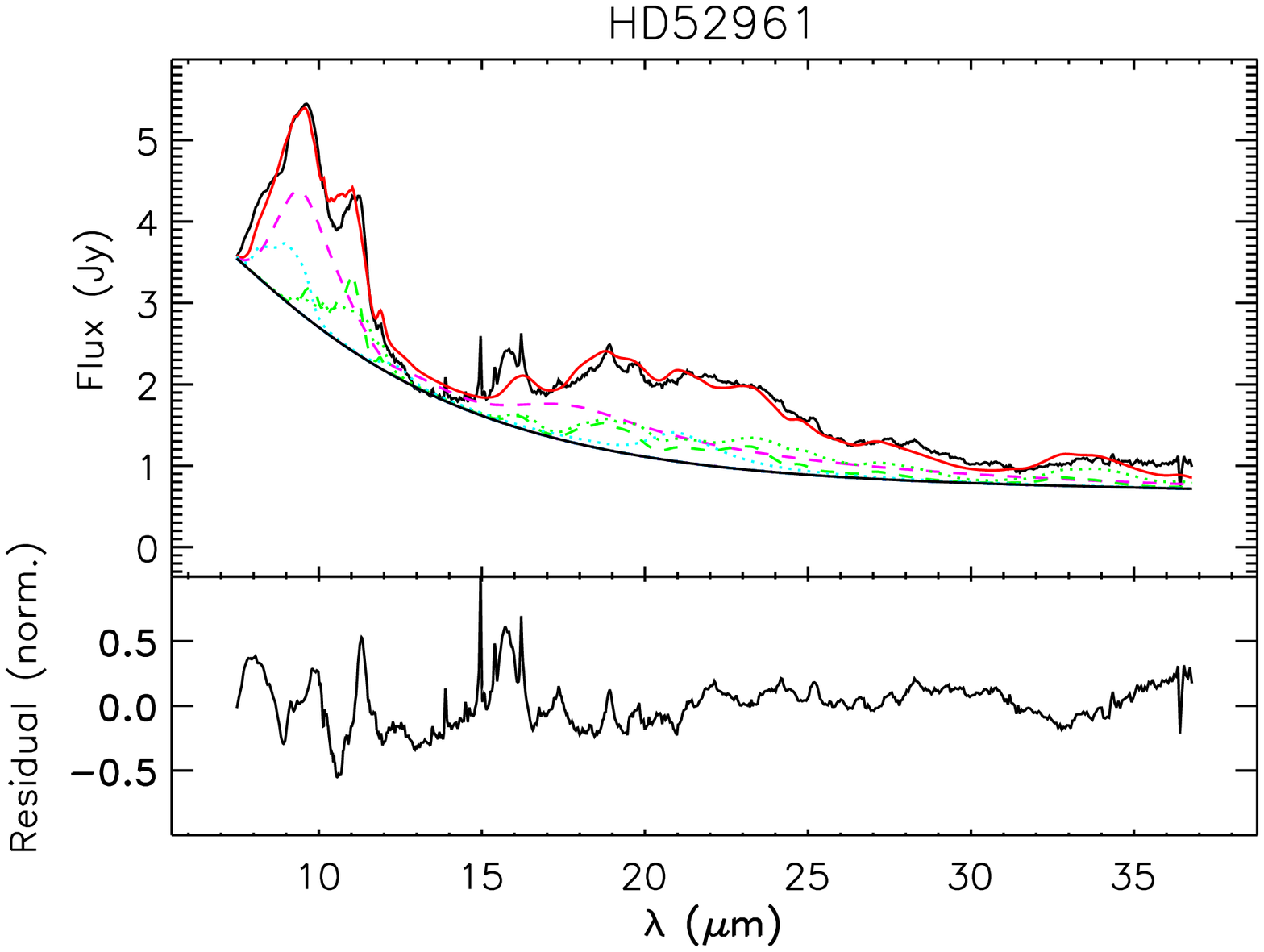}}
\vspace{0.3cm}
\hspace{0.3cm}
\resizebox{6cm}{!}{\includegraphics{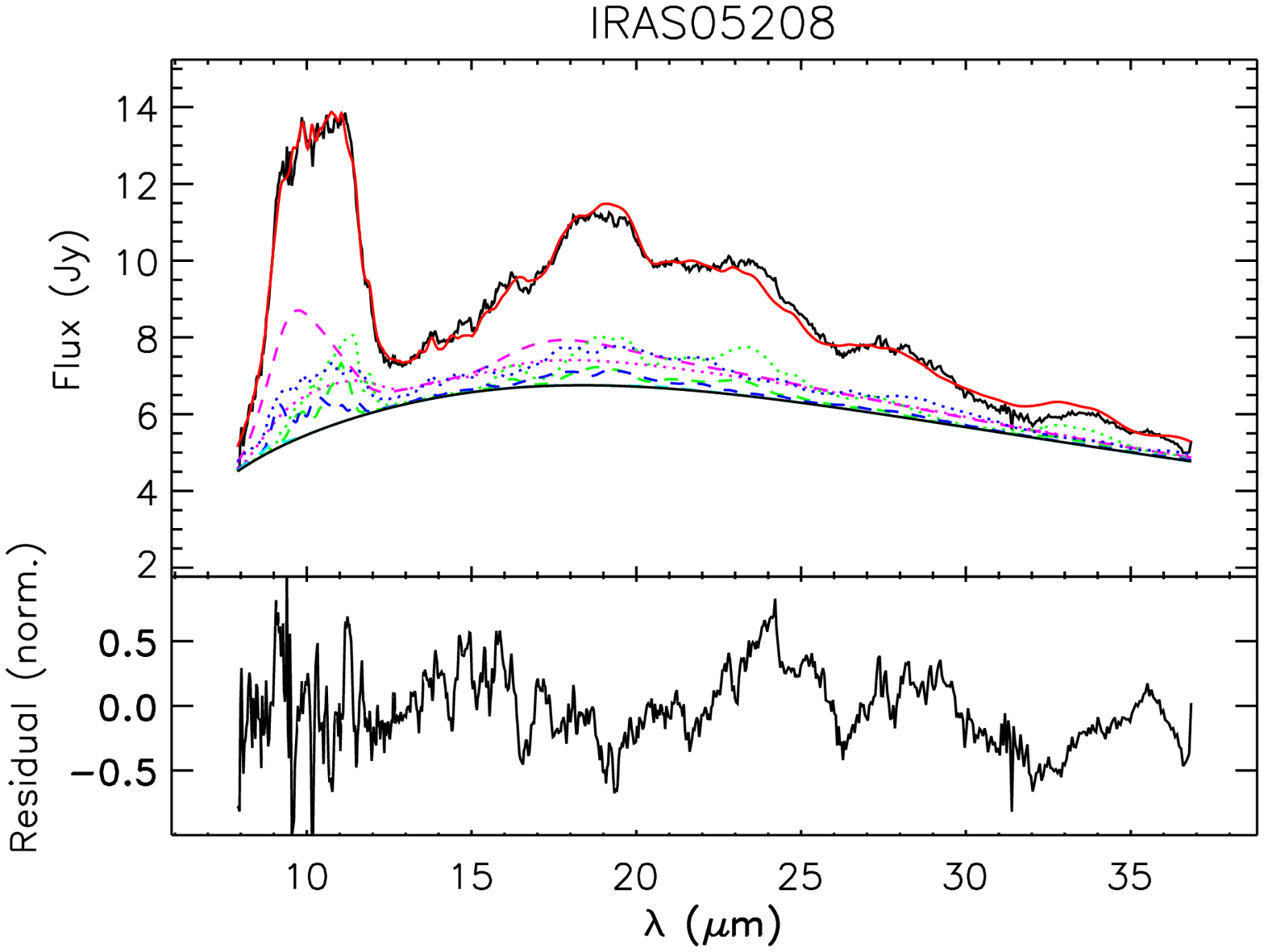}}
\vspace{0.3cm}
\hspace{0.3cm}
\resizebox{6cm}{!}{\includegraphics{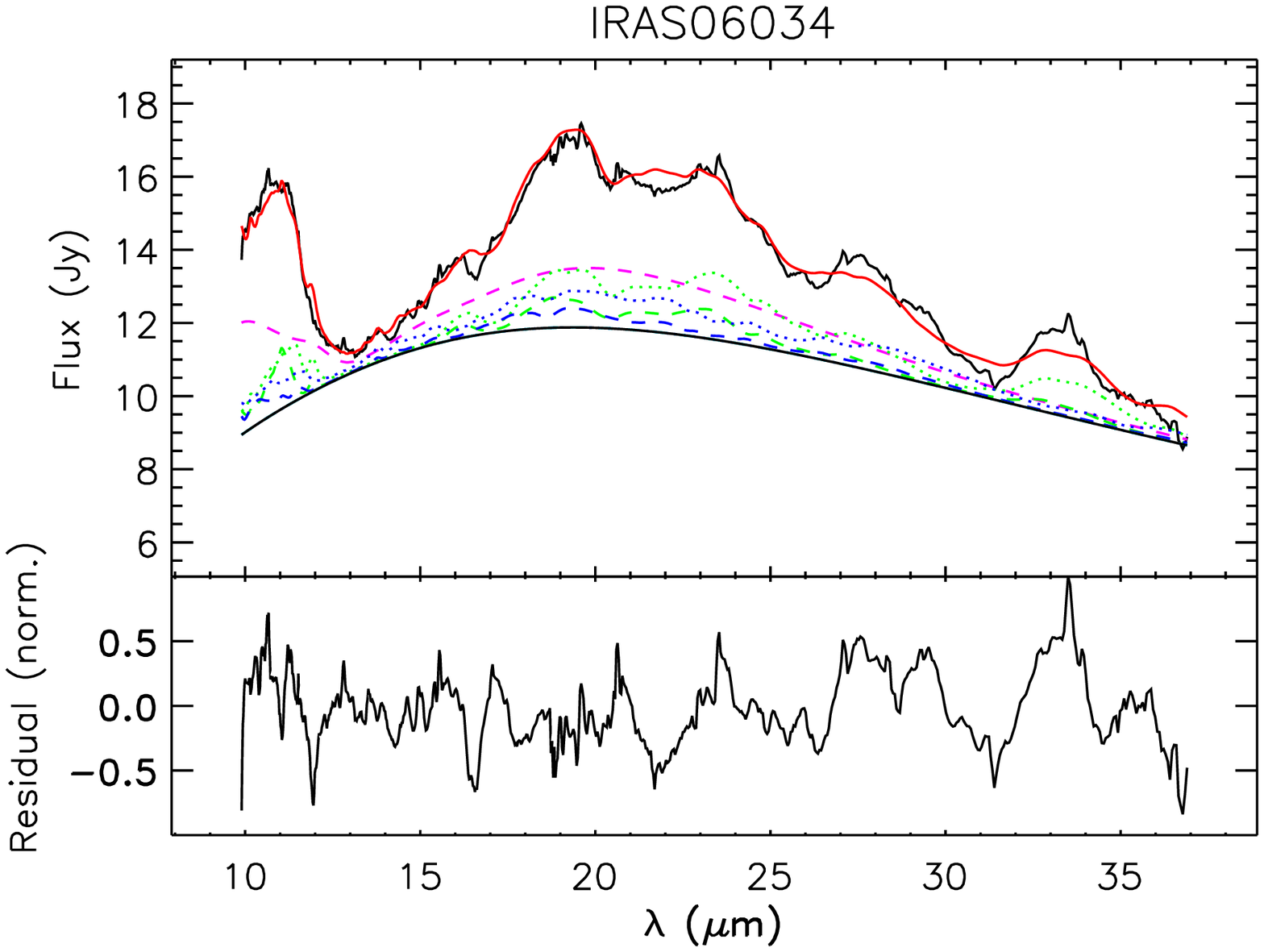}}
\caption{Best model fits for our Galactic sample stars, showing the contribution of the different dust species.
Top: The observed spectrum (black curve) is plotted together with the best model fit (red curve) and the continuum (black solid line).
Forsterite is plotted in green, enstatite in blue, silica in cyan and amorphous olivine and pyroxene in magenta.
Small grains (0.1\,$\mu$m) are plotted as dashed lines and larger grains (2 and 4\,$\mu$m) as dotted lines.
Bottom: The normalised residuals after subtraction of our best model of the observed spectra.}
\label{fits1}
\end{figure}
\begin{figure}[ht]

\resizebox{6cm}{!}{\includegraphics{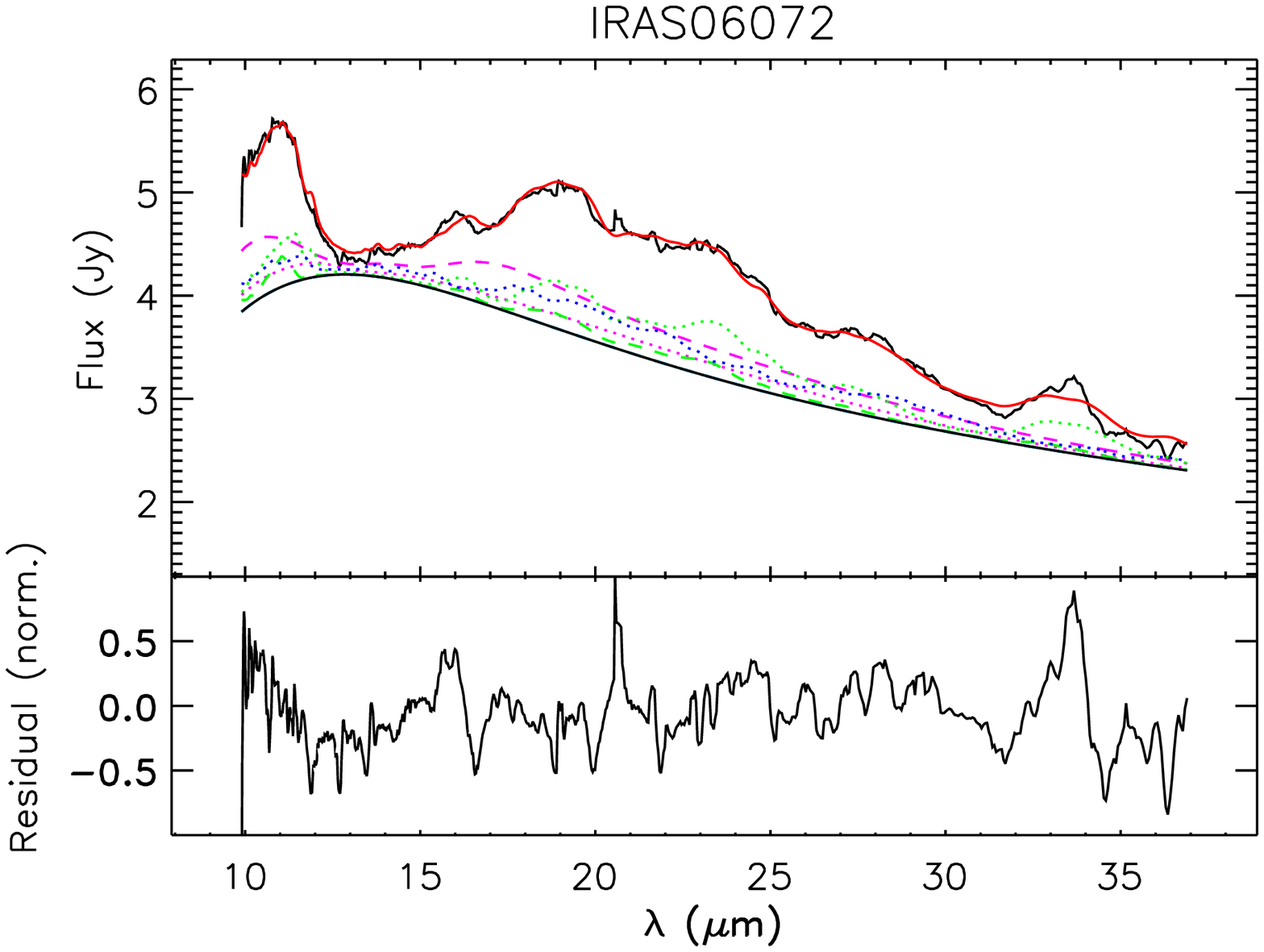}}
\vspace{0.3cm}
\hspace{0.3cm}
\resizebox{6cm}{!}{\includegraphics{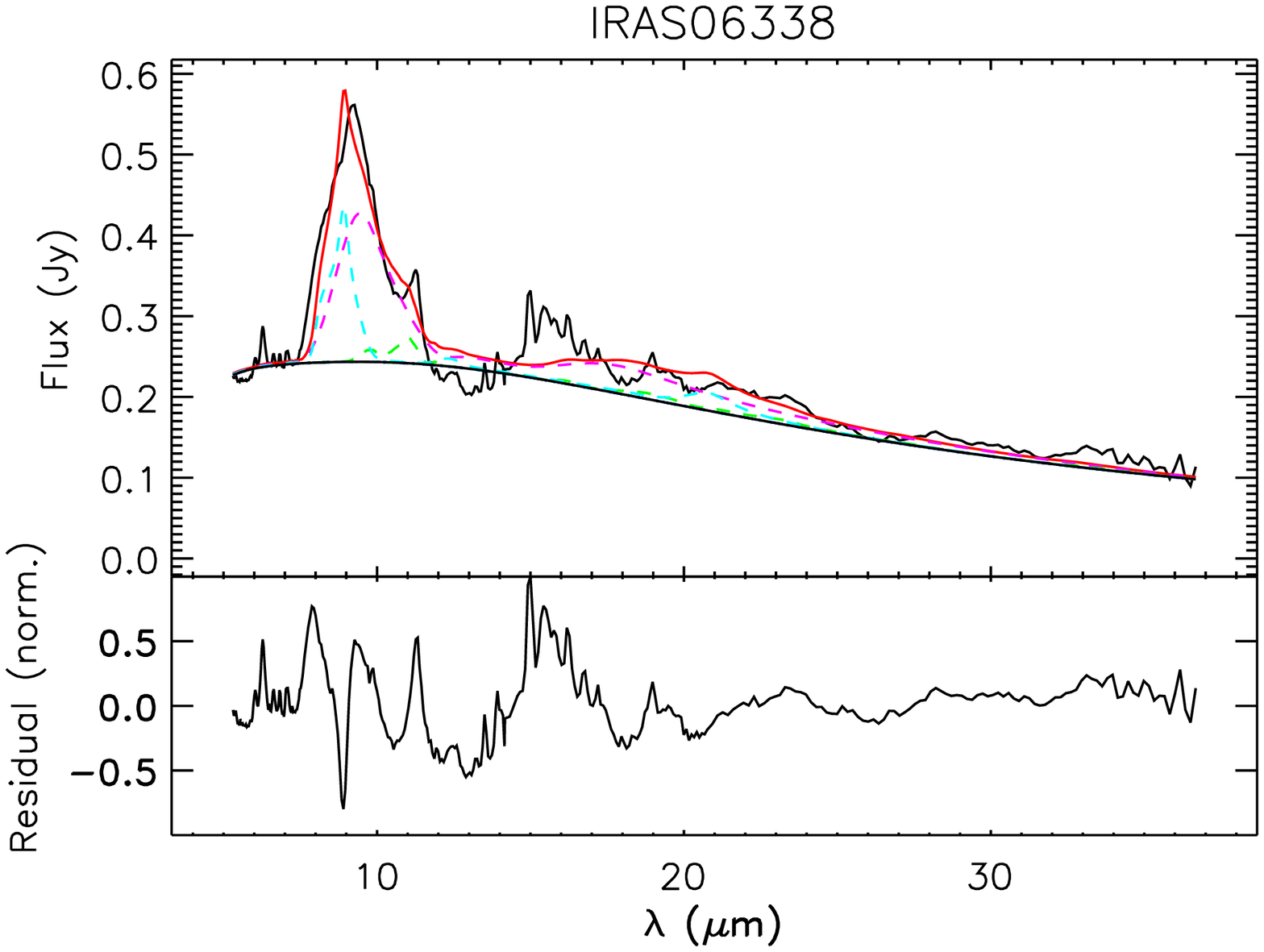}}
\vspace{0.3cm}
\hspace{0.3cm}
\resizebox{6cm}{!}{\includegraphics{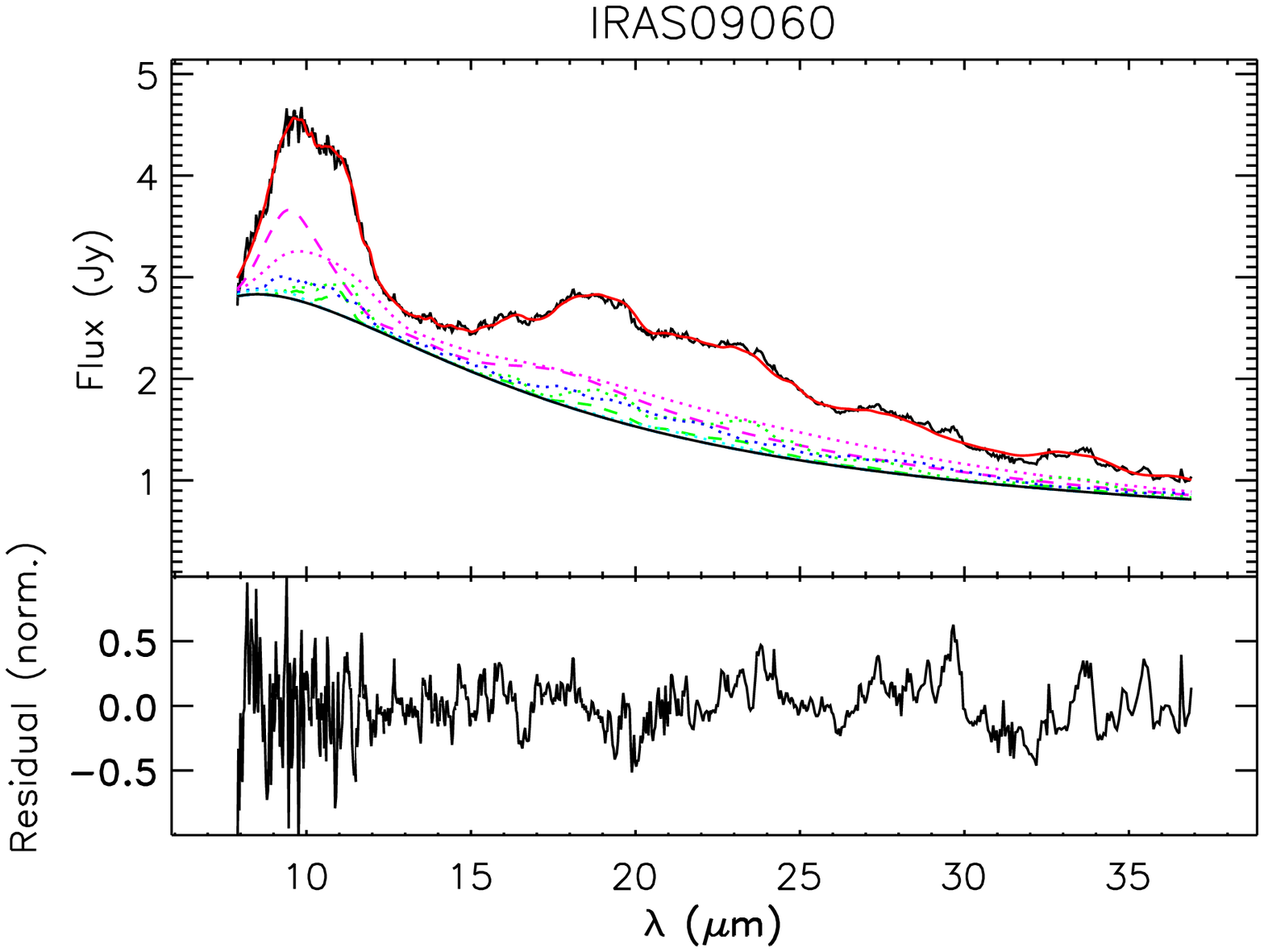}}

\resizebox{6cm}{!}{\includegraphics{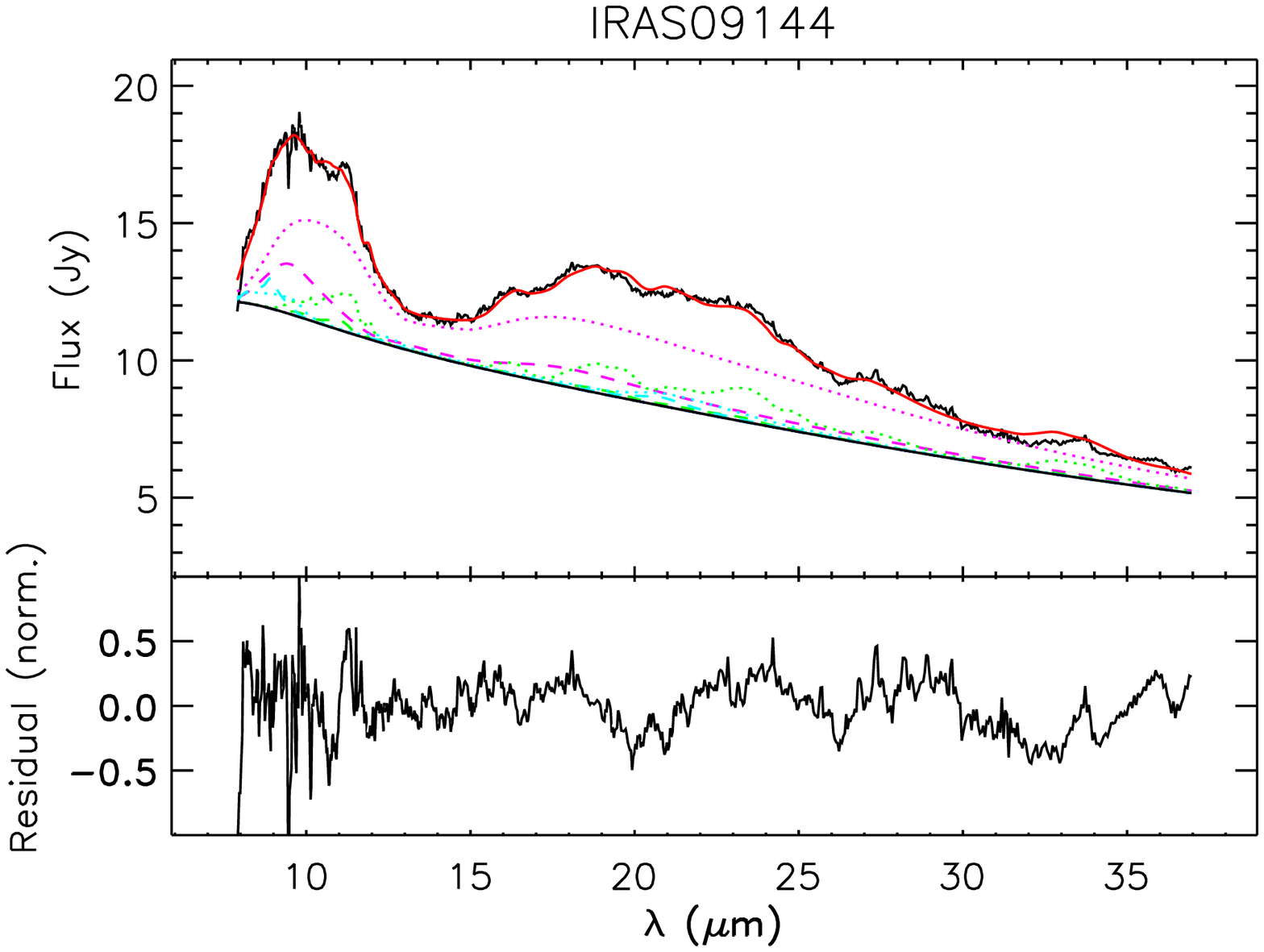}}
\vspace{0.3cm}
\hspace{0.3cm}
\resizebox{6cm}{!}{\includegraphics{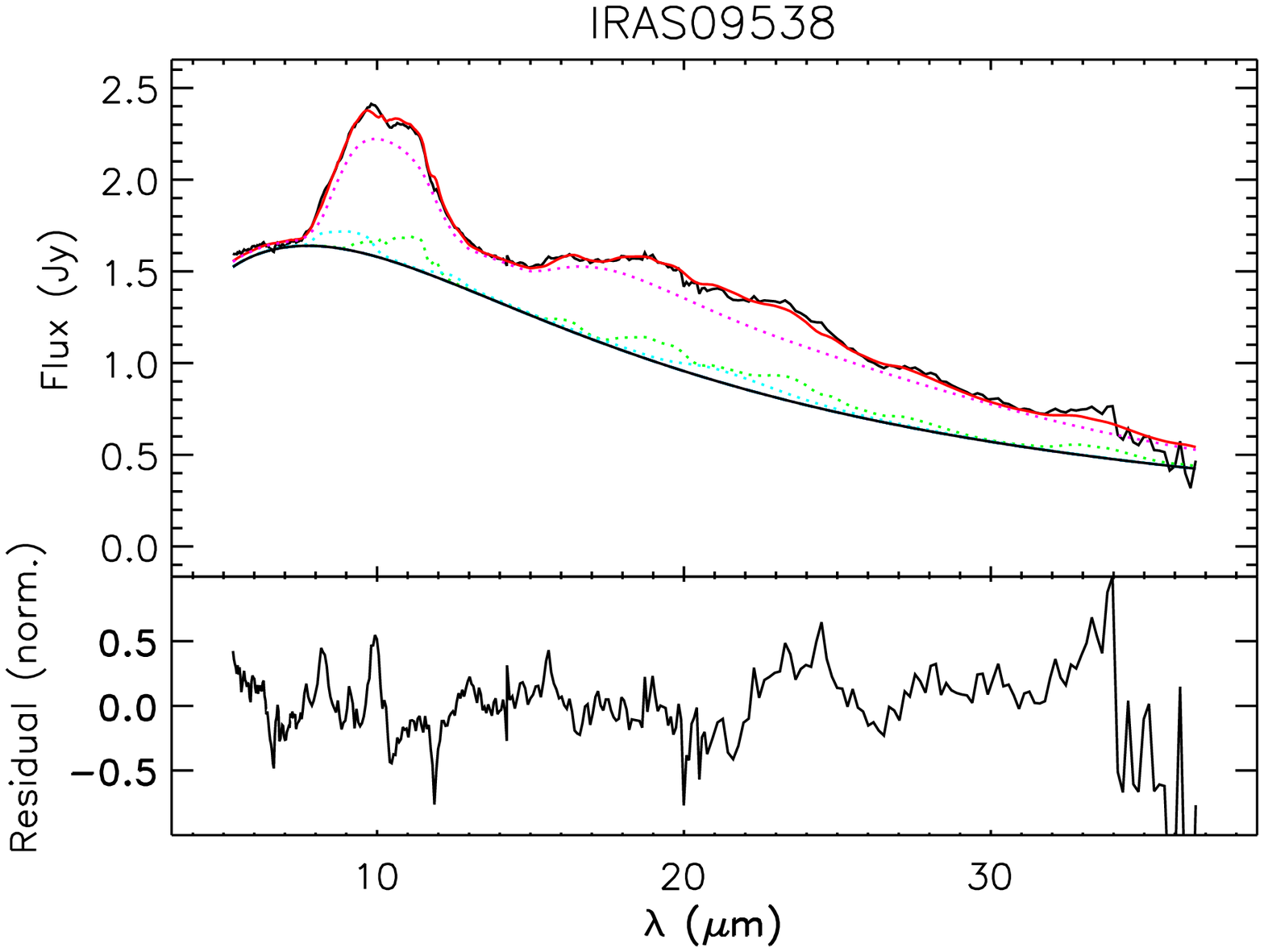}}
\vspace{0.3cm}
\hspace{0.3cm}
\resizebox{6cm}{!}{\includegraphics{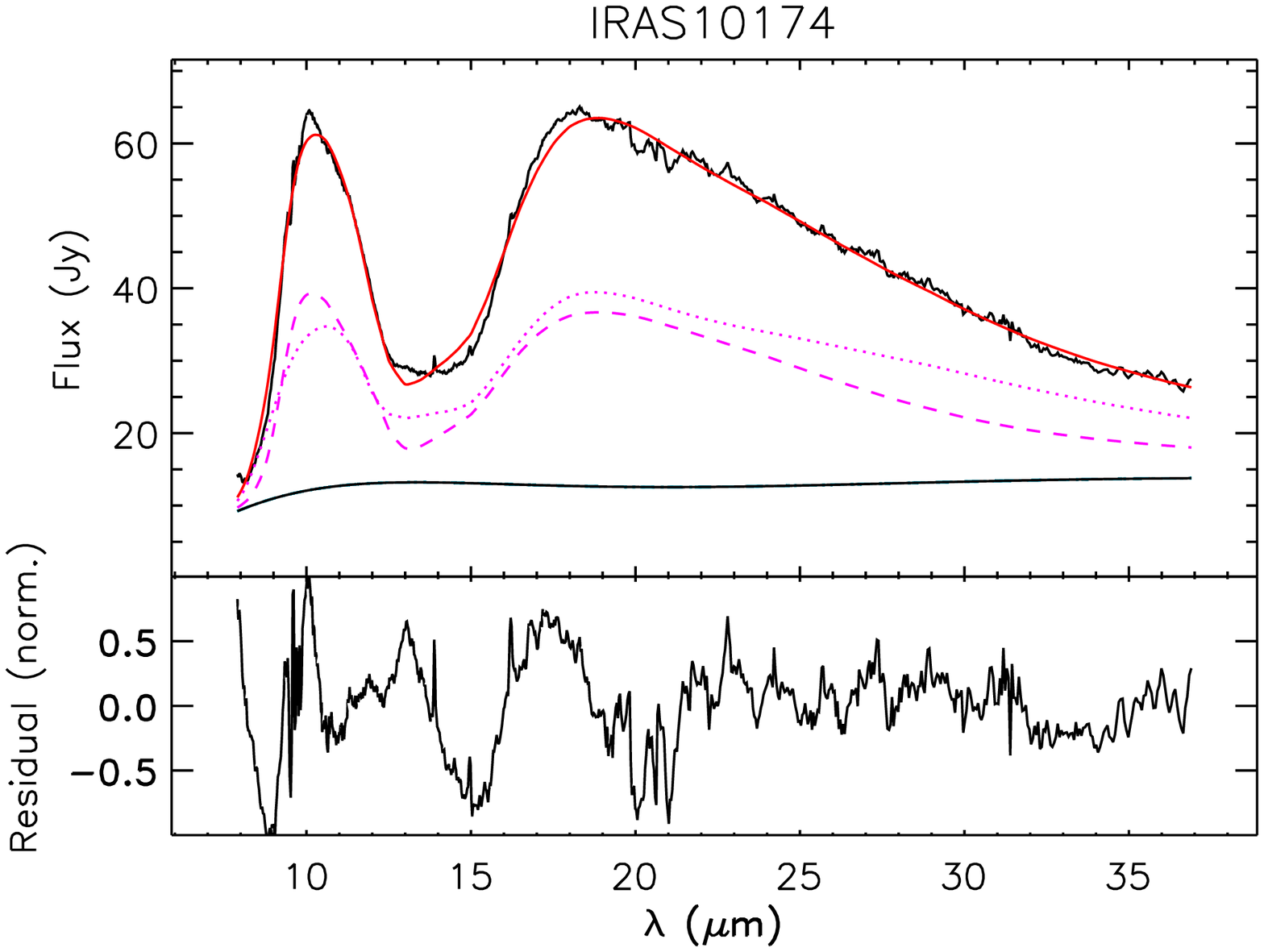}}

\resizebox{6cm}{!}{\includegraphics{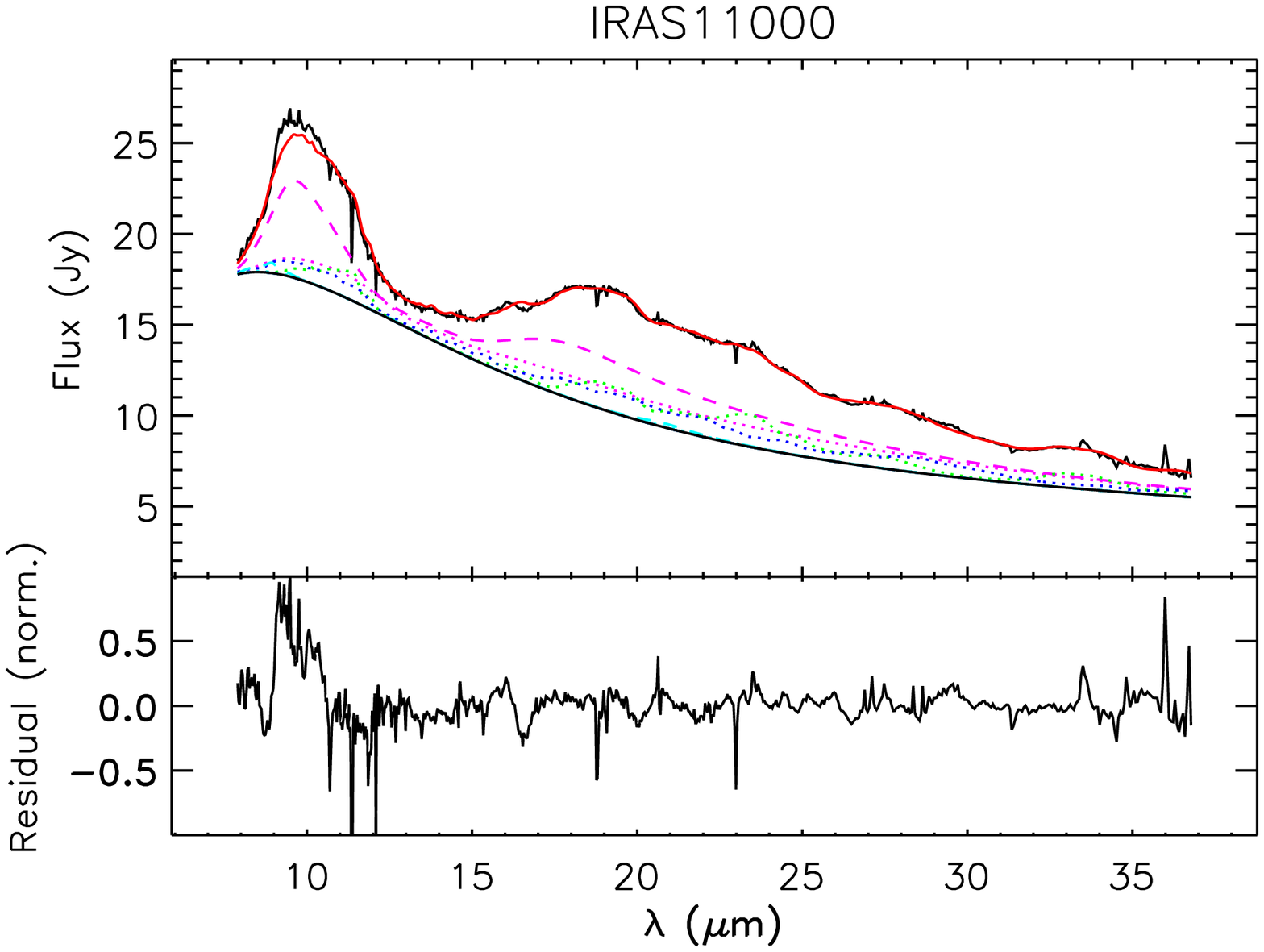}}
\vspace{0.3cm}
\hspace{0.3cm}
\resizebox{6cm}{!}{\includegraphics{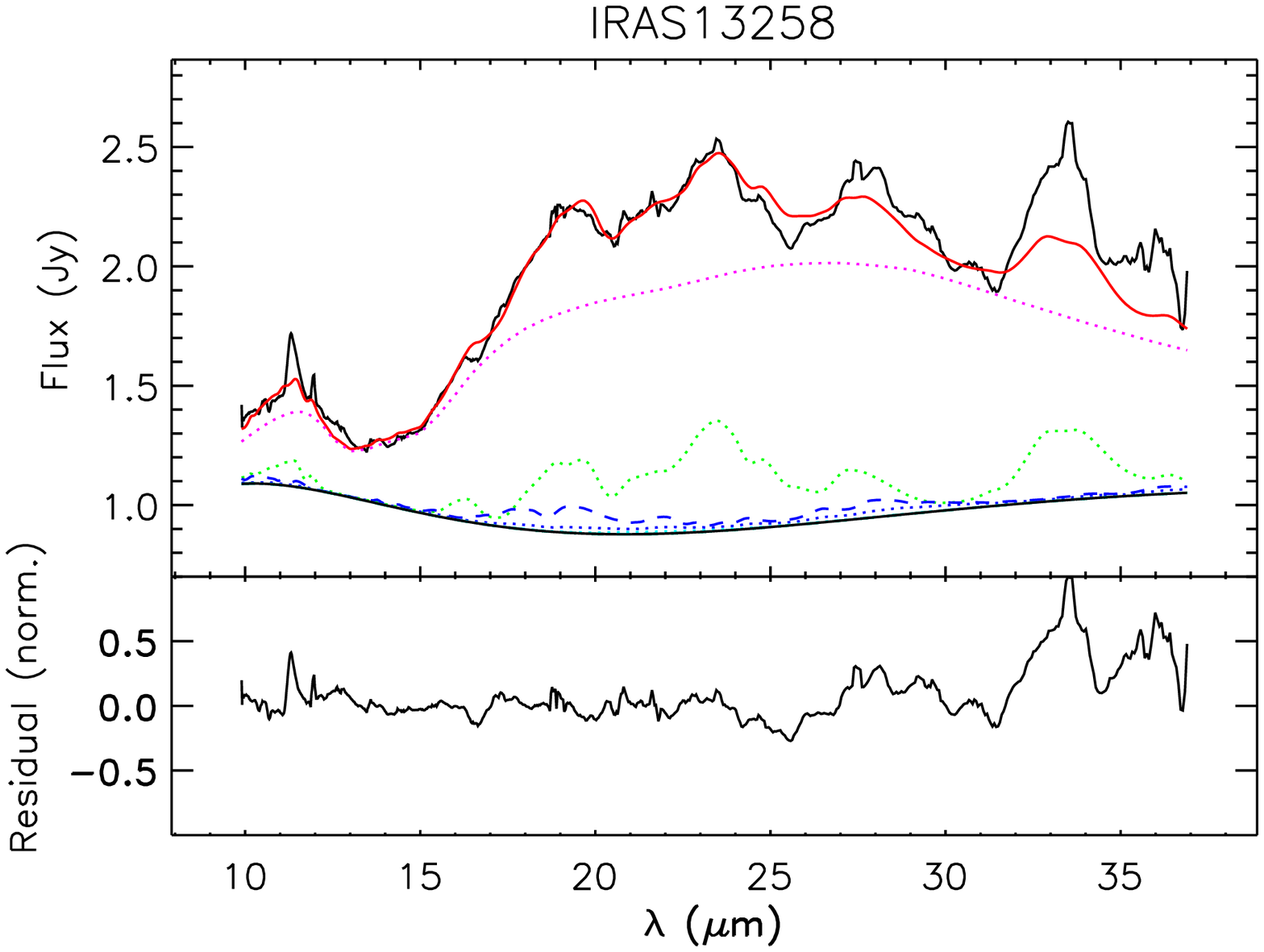}}
\vspace{0.3cm}
\hspace{0.3cm}
\resizebox{6cm}{!}{\includegraphics{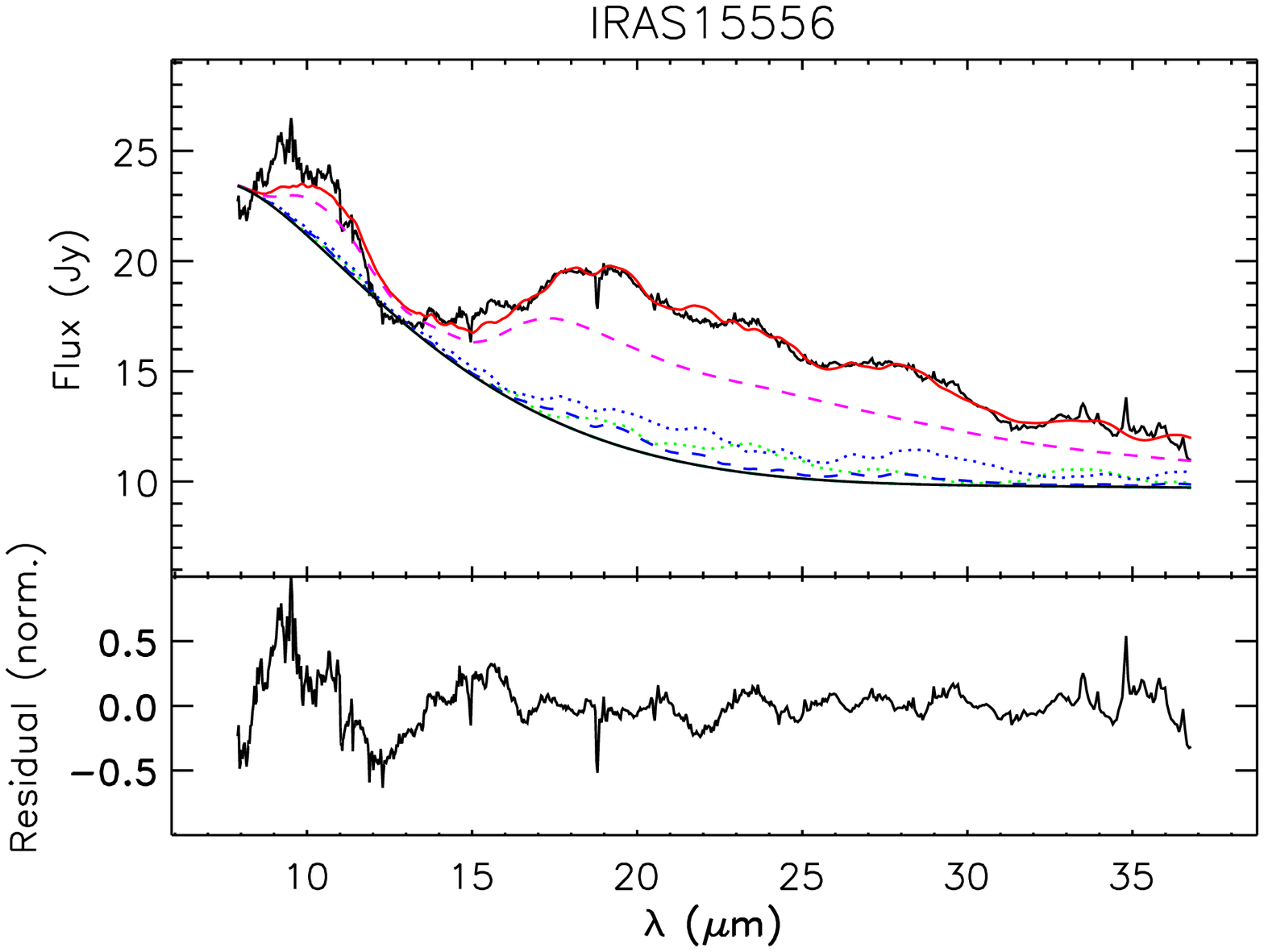}}

\resizebox{6cm}{!}{\includegraphics{plots/bestfit/IRAS16230.ps}}
\vspace{0.3cm}
\hspace{0.3cm}
\resizebox{6cm}{!}{\includegraphics{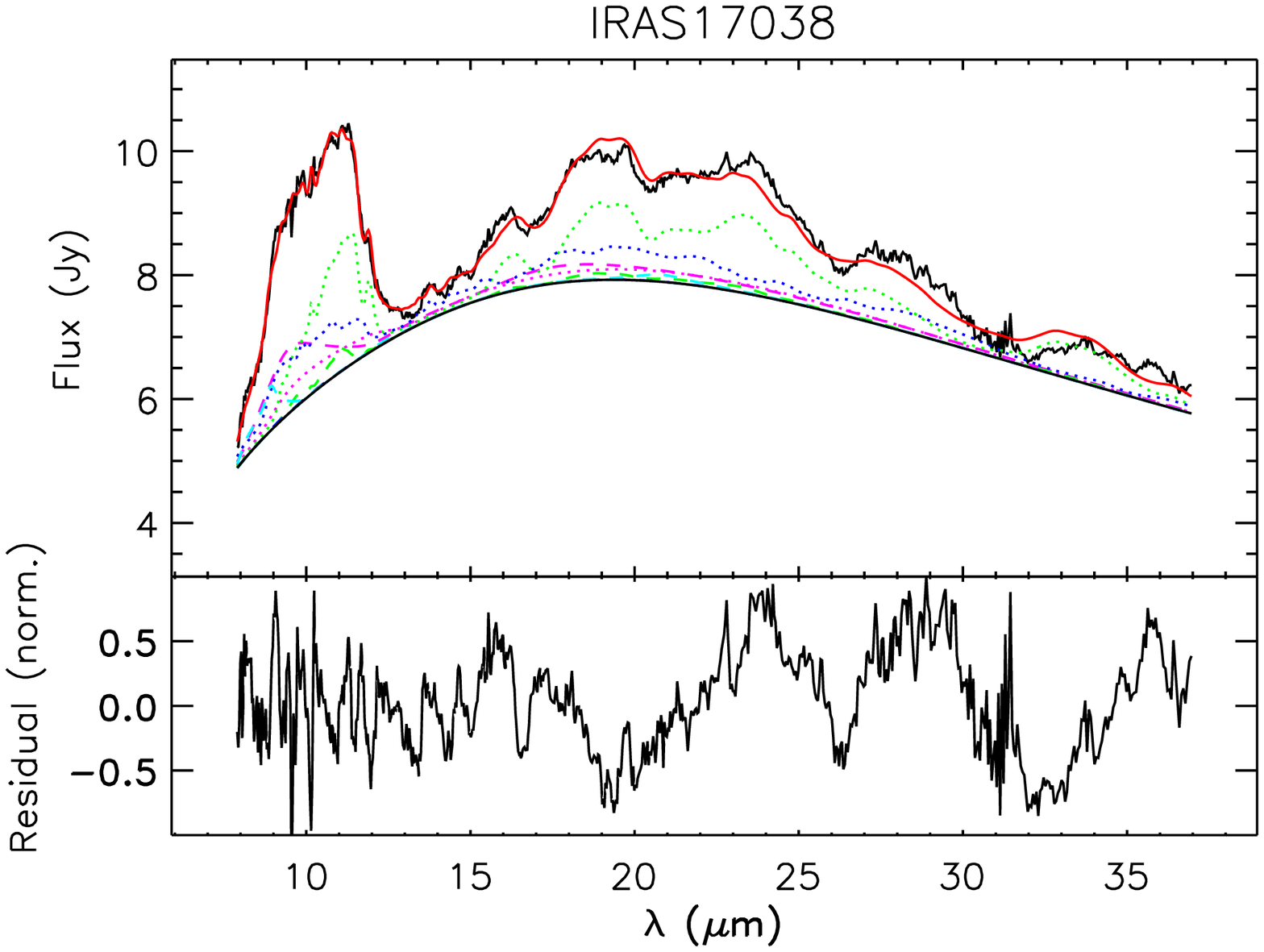}}
\vspace{0.3cm}
\hspace{0.3cm}
\resizebox{6cm}{!}{\includegraphics{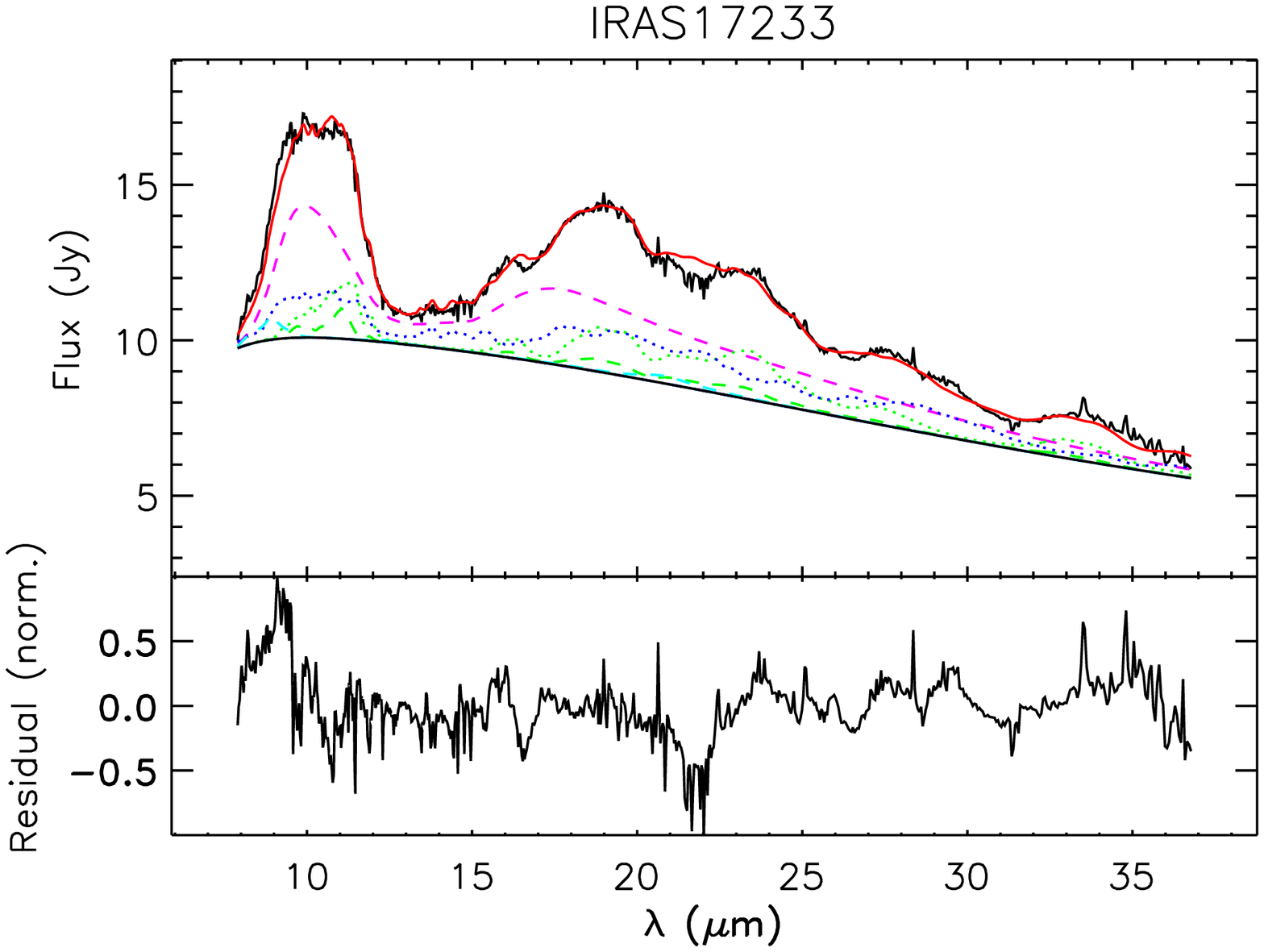}}

\resizebox{6cm}{!}{\includegraphics{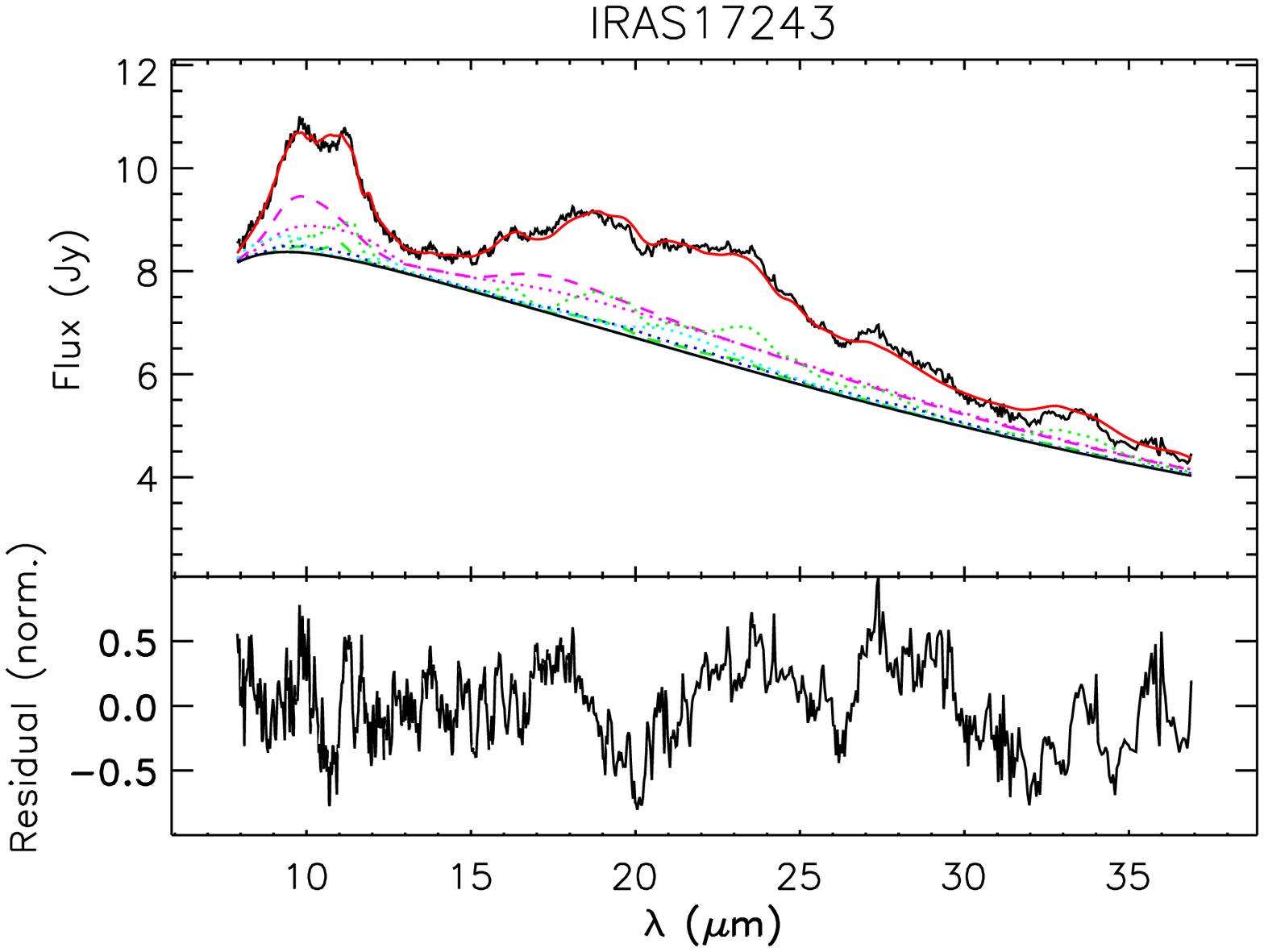}}
\vspace{0.3cm}
\hspace{0.3cm}
\resizebox{6cm}{!}{\includegraphics{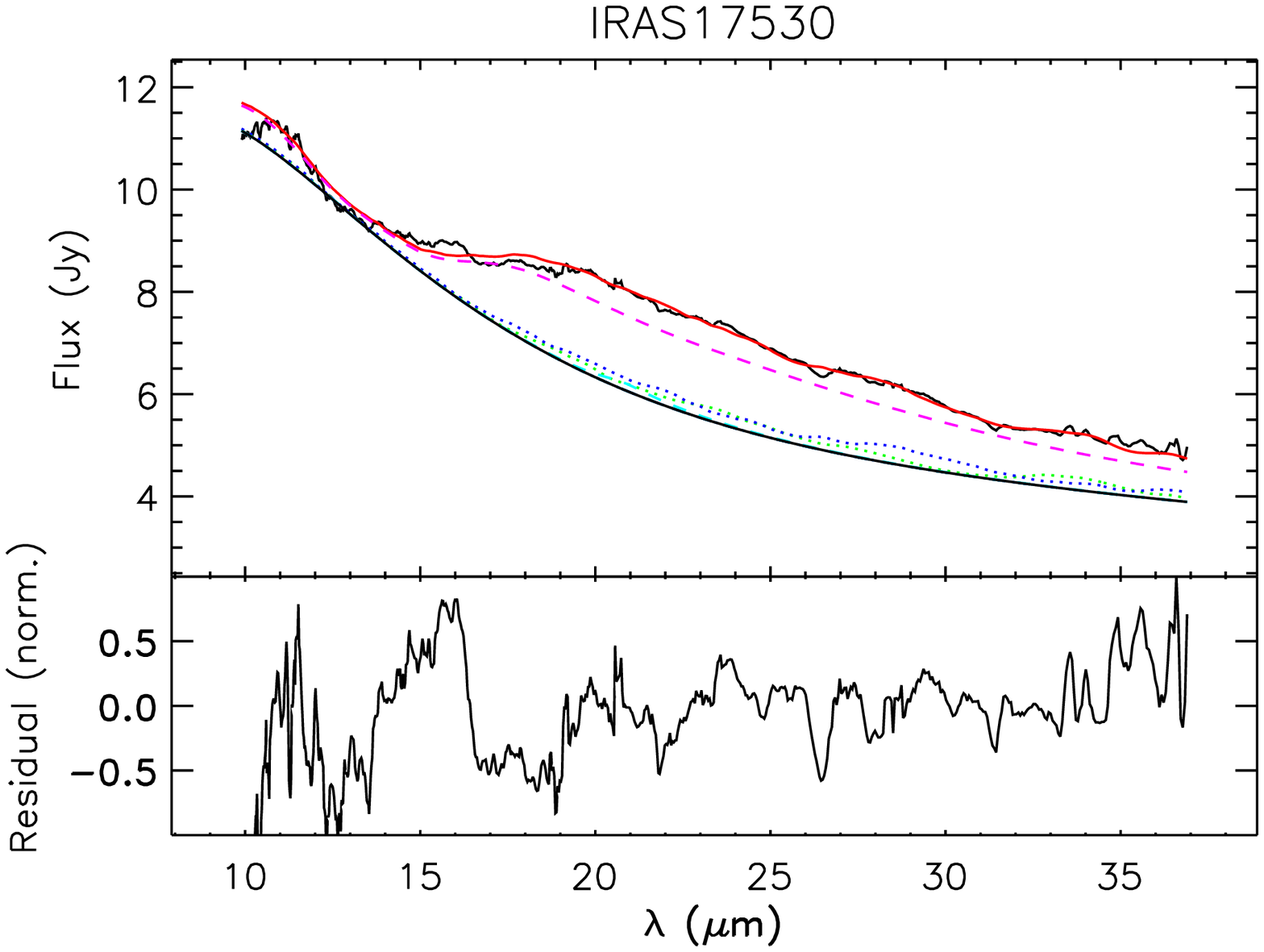}}
\vspace{0.3cm}
\hspace{0.3cm}
\resizebox{6cm}{!}{\includegraphics{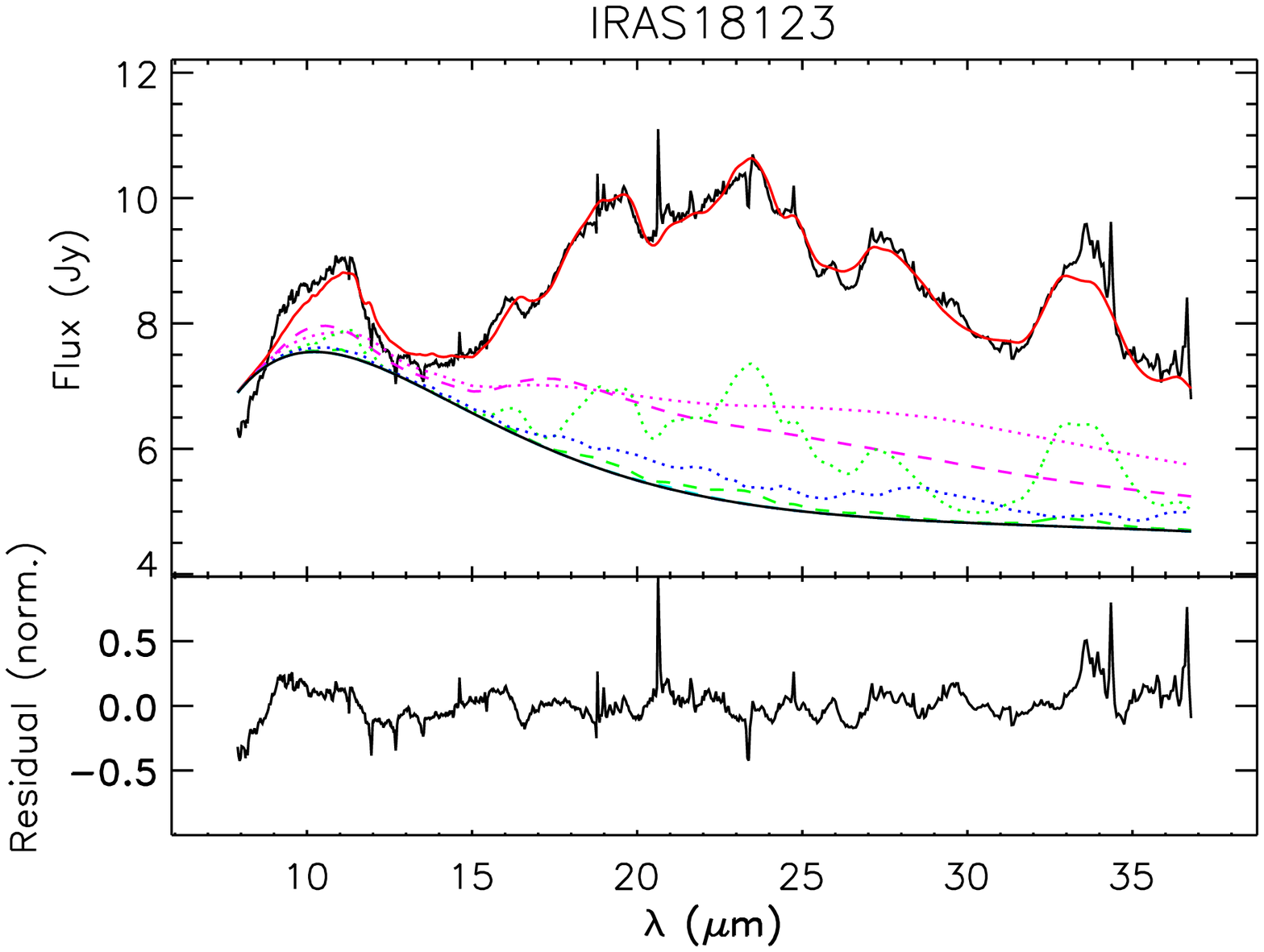}}
\caption{Same as Fig.~\ref{fits1}.}
\label{fits2}
\end{figure}

\begin{figure}

\resizebox{6cm}{!}{\includegraphics{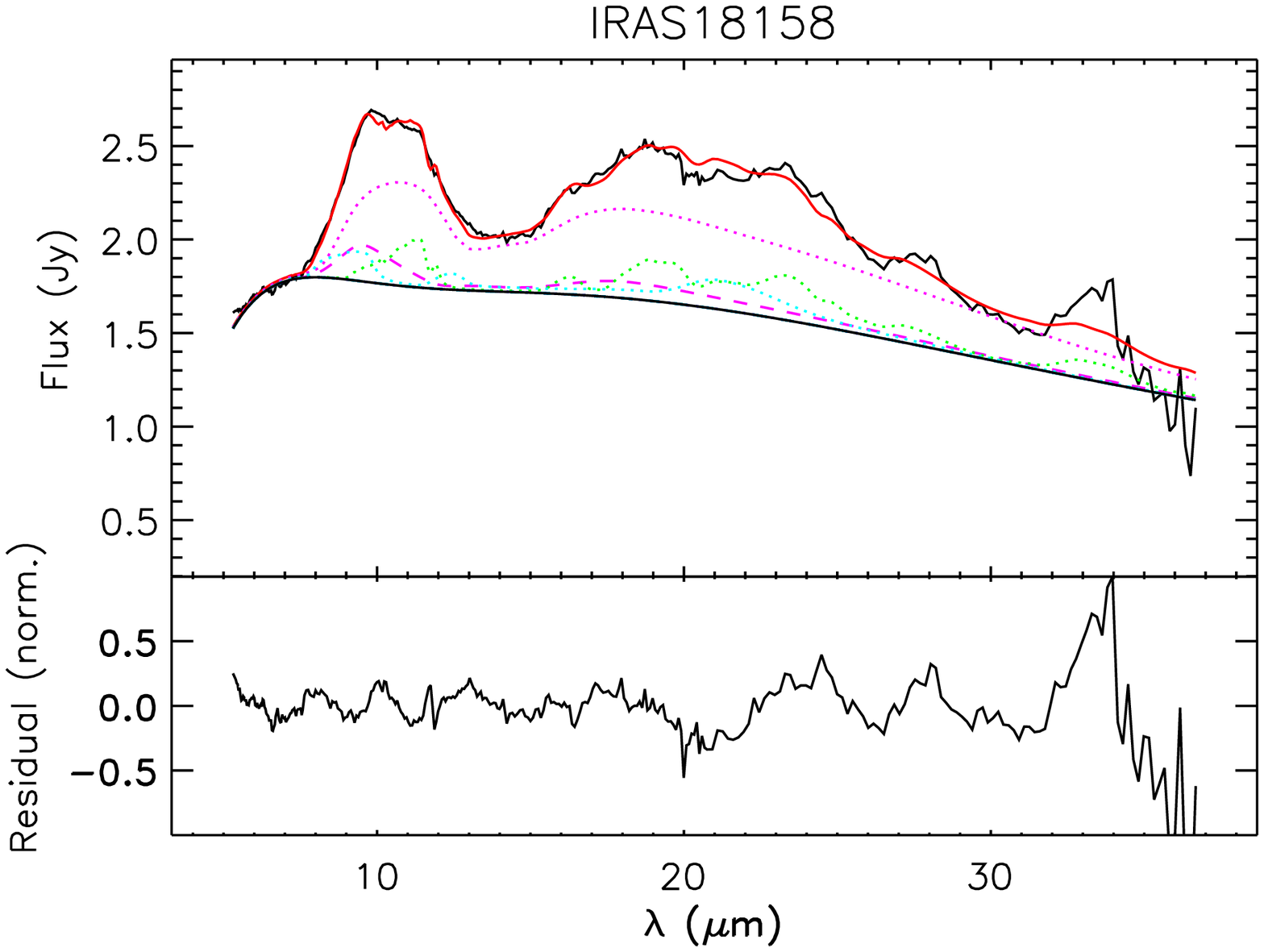}}
\vspace{0.3cm}
\hspace{0.3cm}
\resizebox{6cm}{!}{\includegraphics{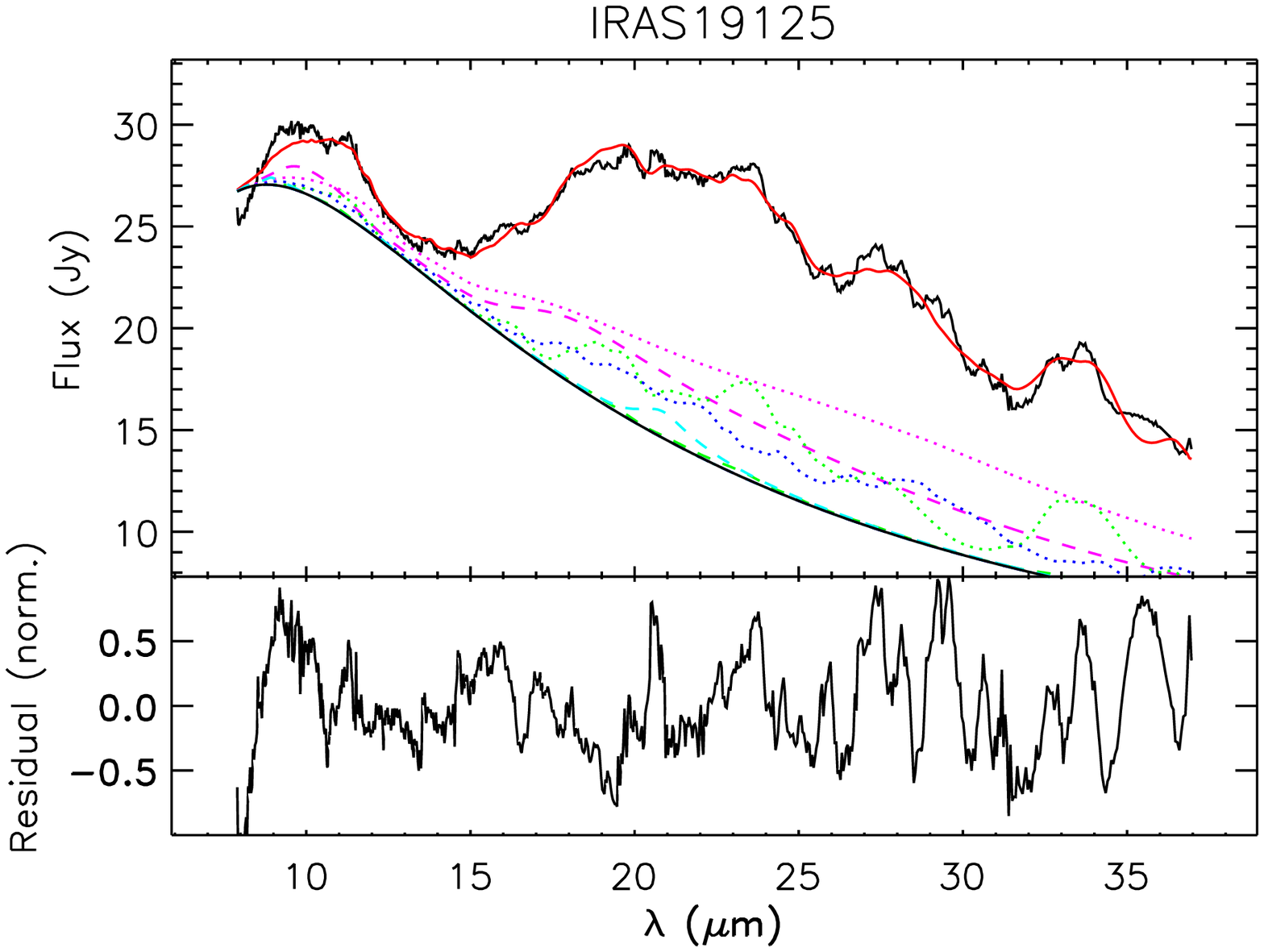}}
\vspace{0.3cm}
\hspace{0.3cm}
\resizebox{6cm}{!}{\includegraphics{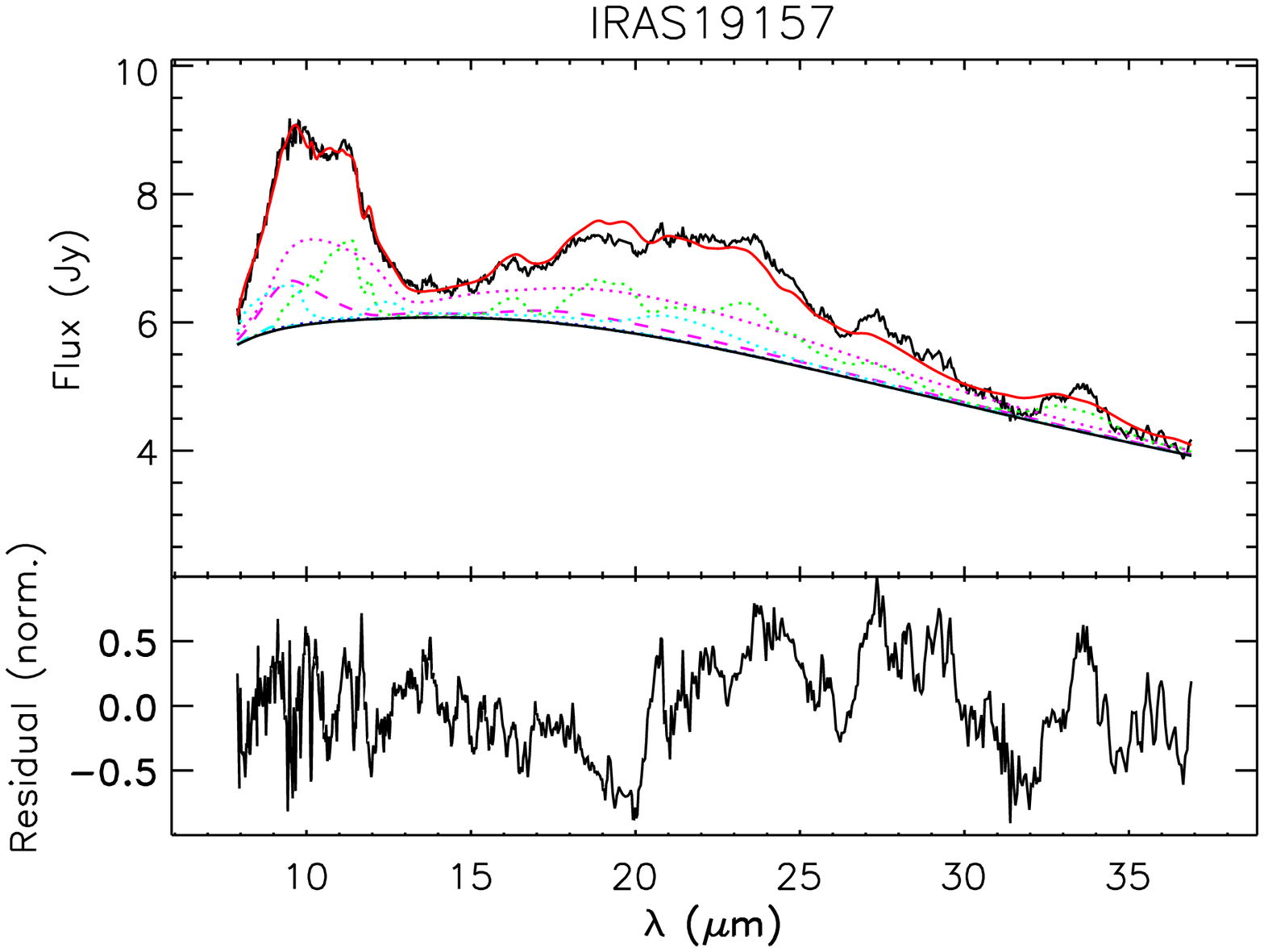}}
\resizebox{6cm}{!}{\includegraphics{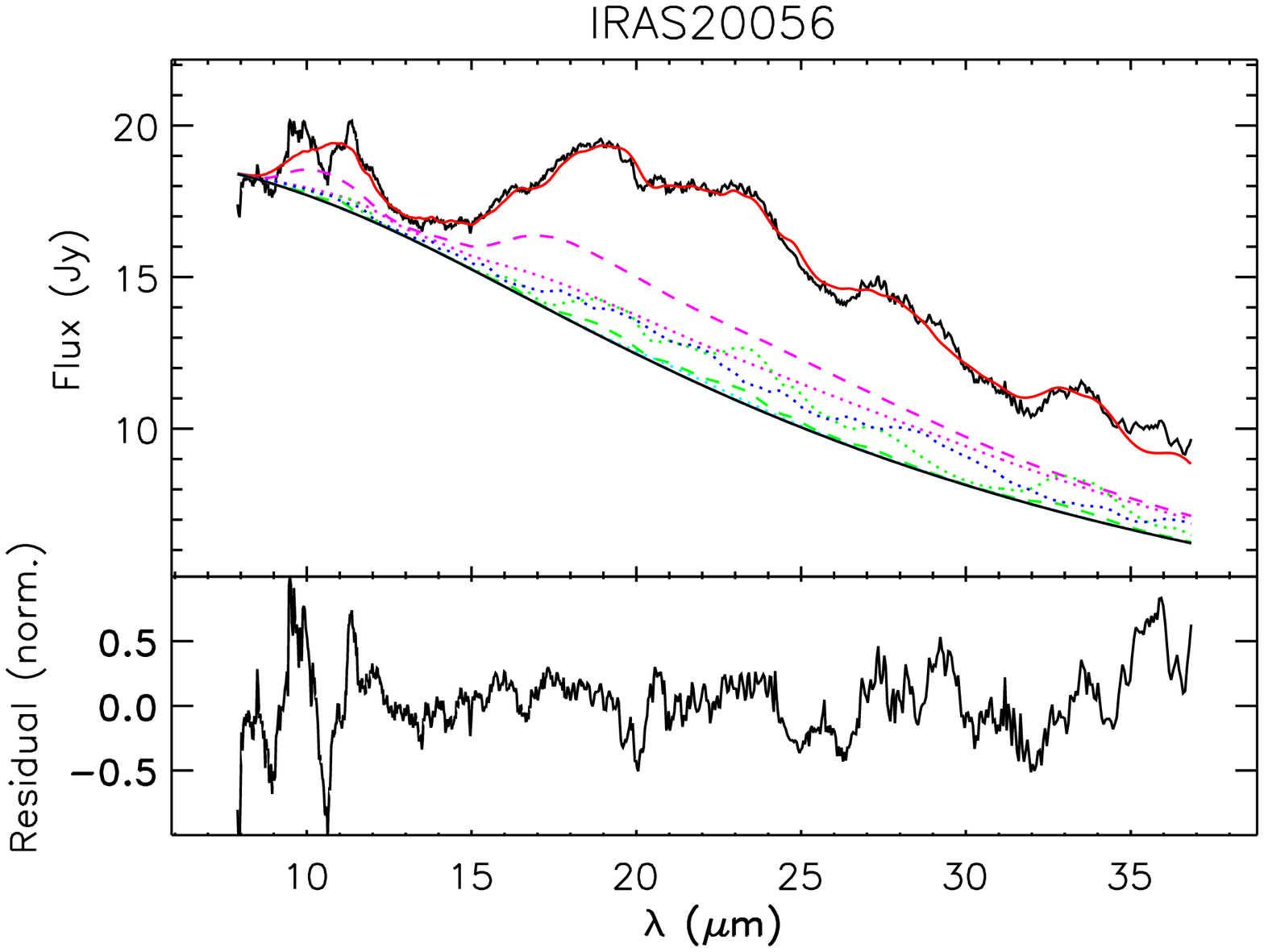}}
\vspace{0.3cm}
\hspace{0.3cm}
\resizebox{6cm}{!}{\includegraphics{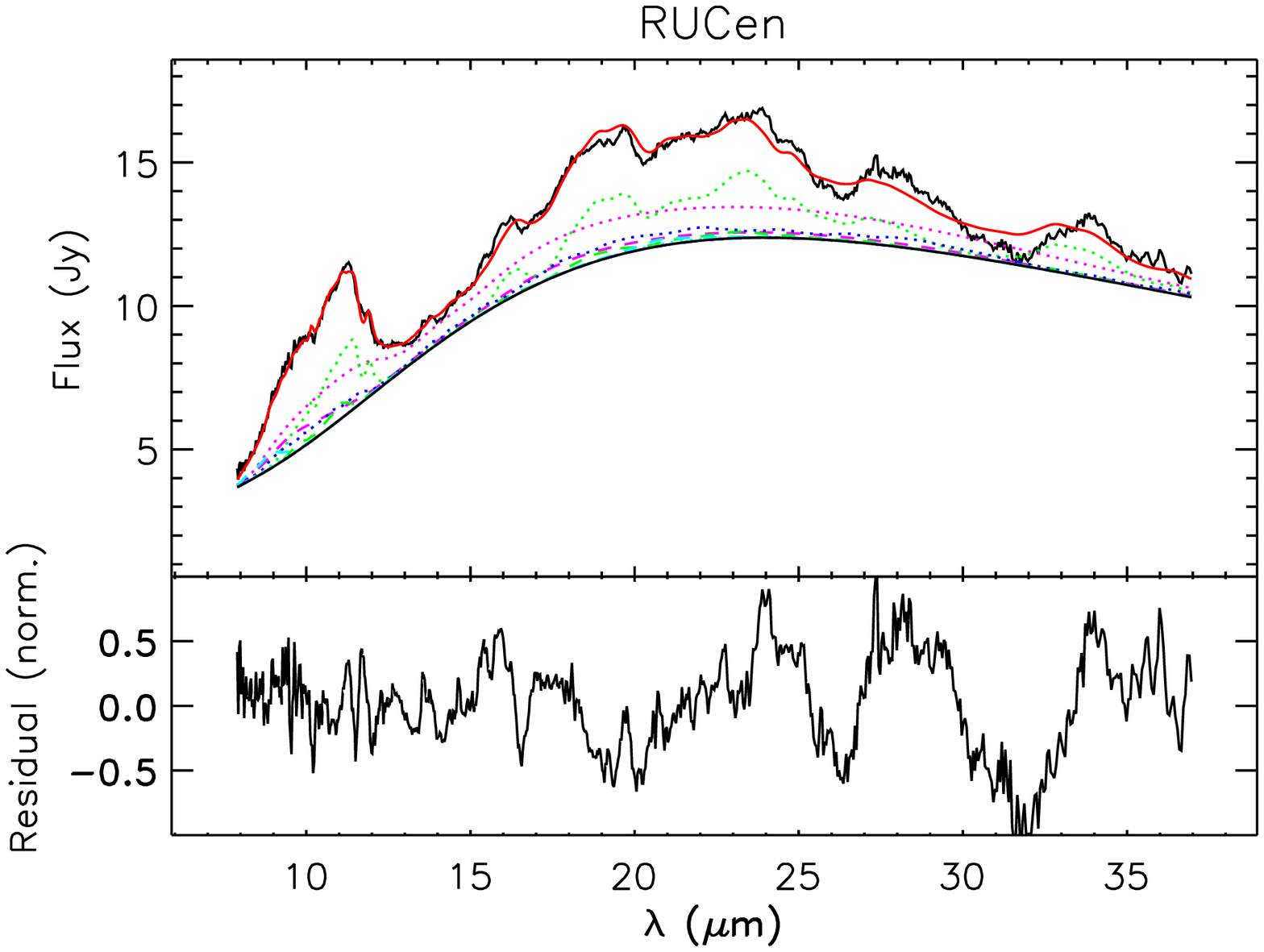}}
\vspace{0.3cm}
\hspace{0.3cm}
\resizebox{6cm}{!}{\includegraphics{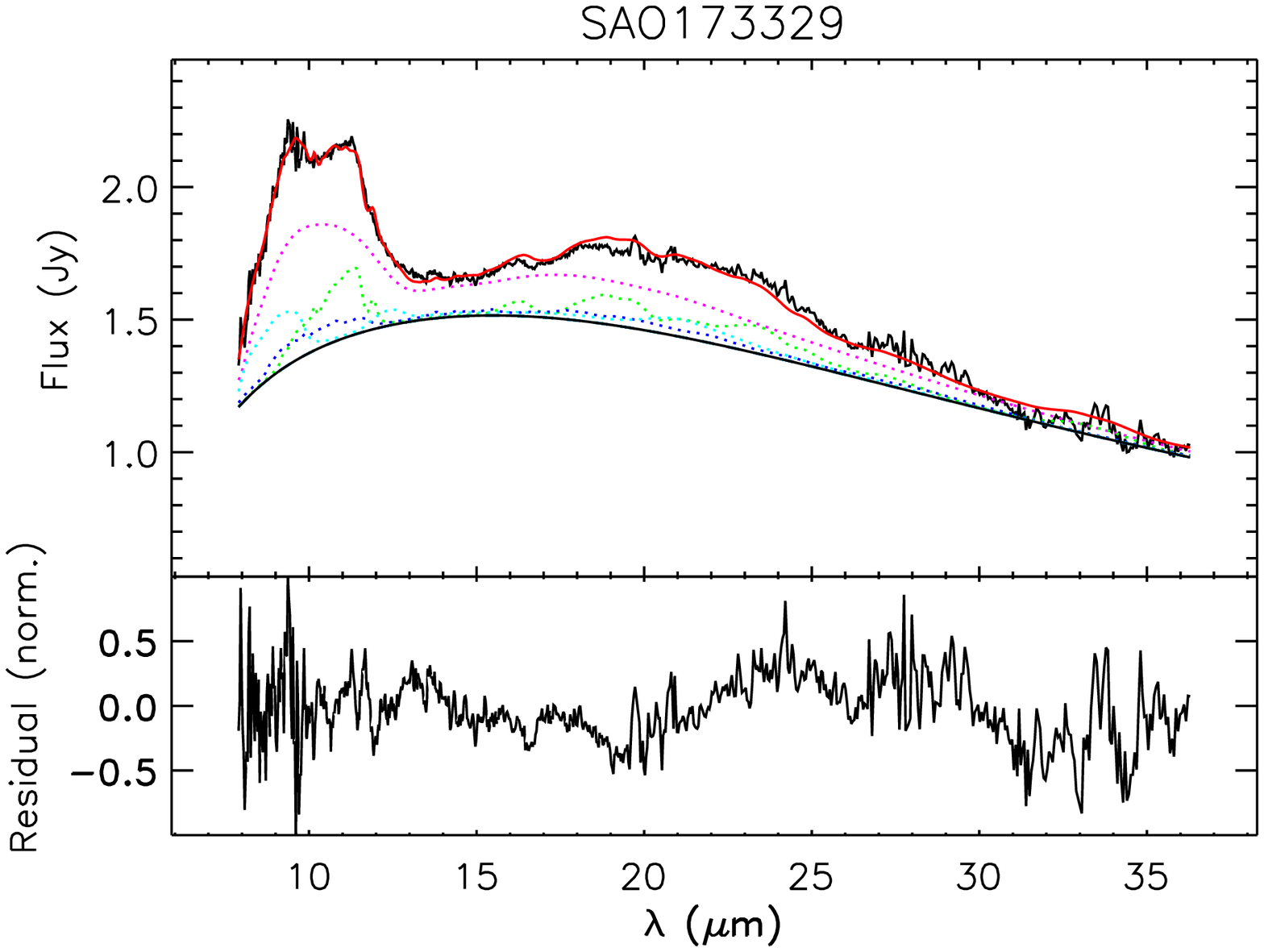}}

\resizebox{6cm}{!}{\includegraphics{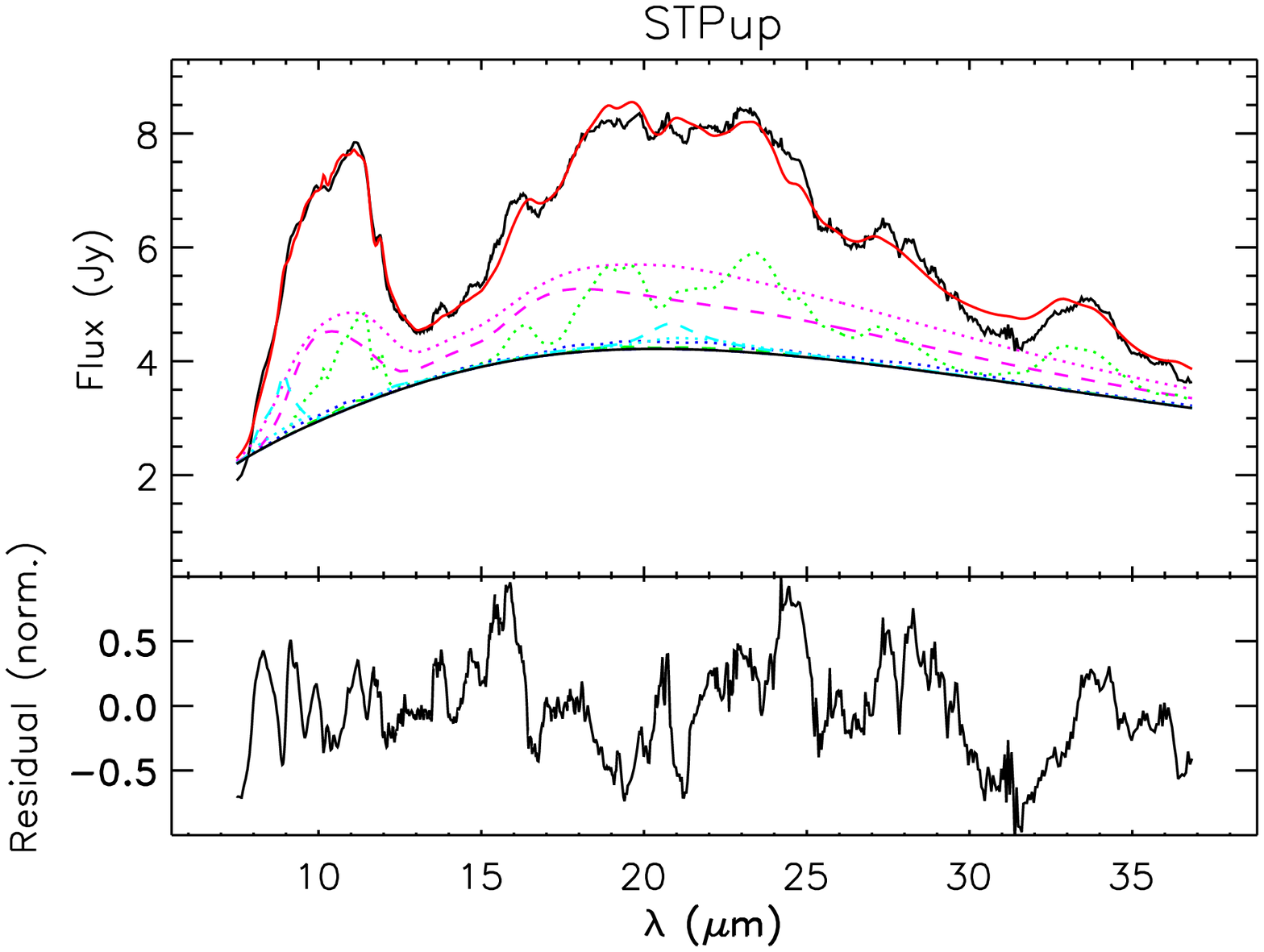}}
\vspace{0.3cm}
\hspace{0.3cm}
\resizebox{6cm}{!}{\includegraphics{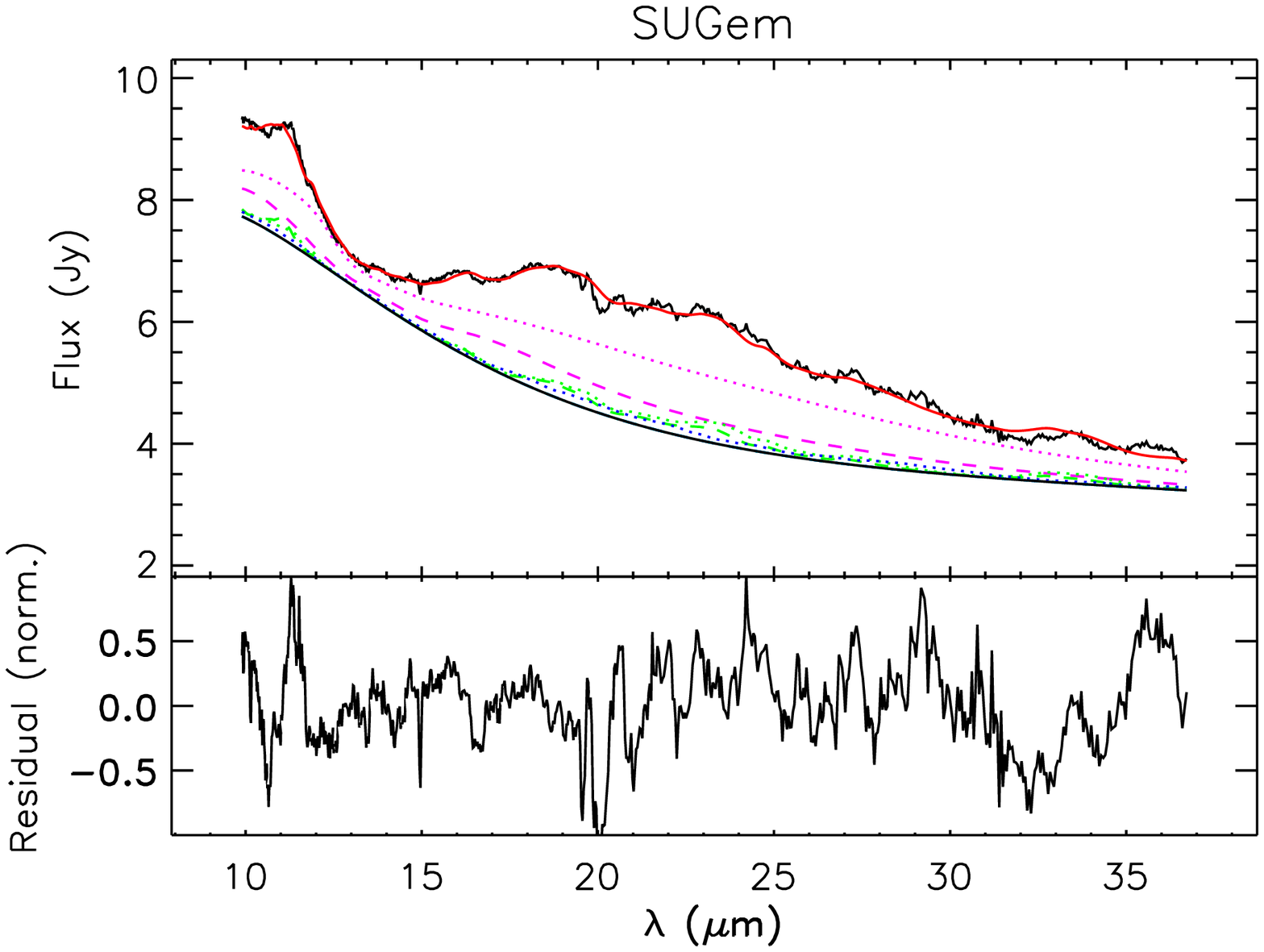}}
\vspace{0.3cm}
\hspace{0.3cm}
\resizebox{6cm}{!}{\includegraphics{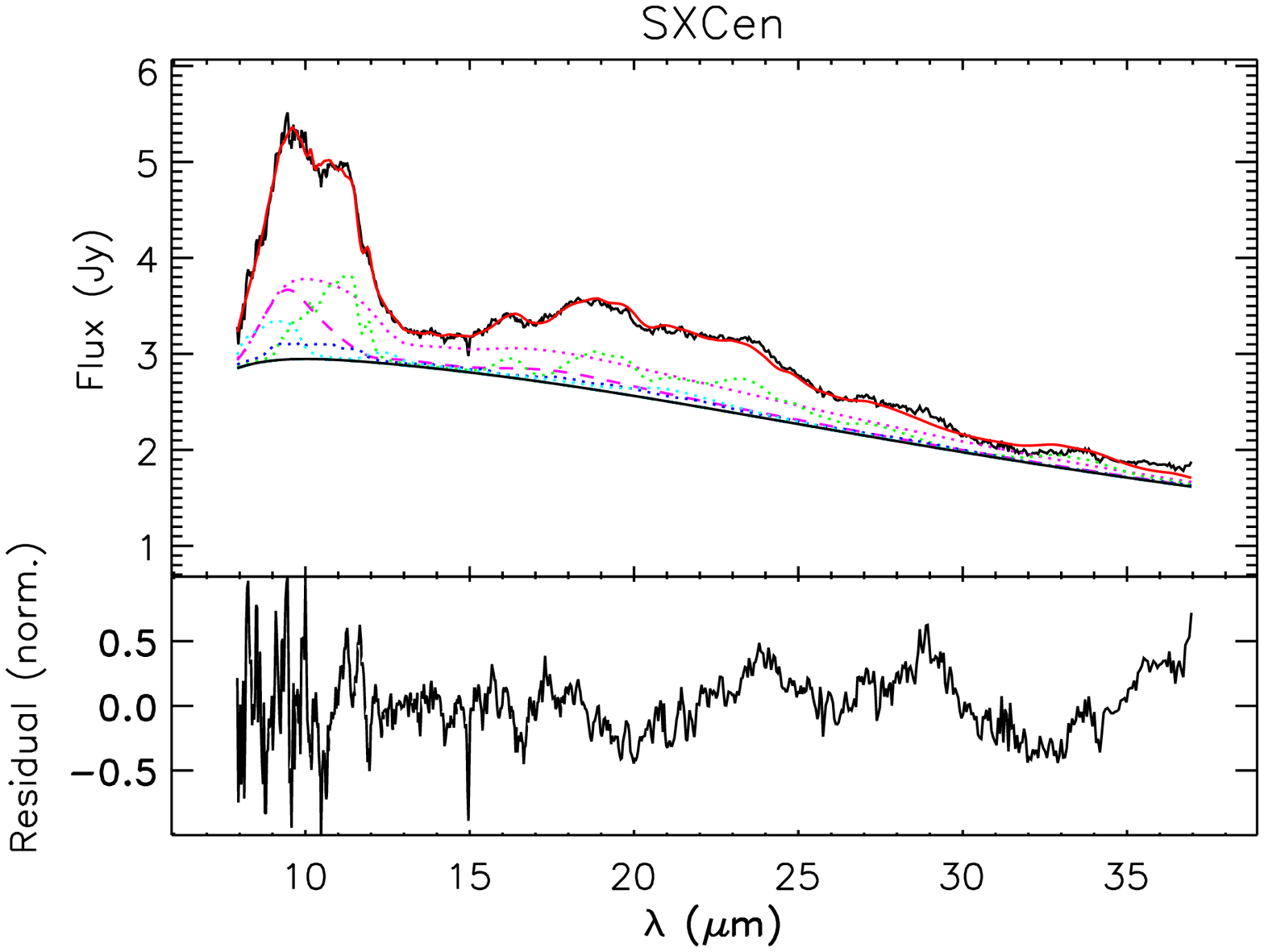}}

\resizebox{6cm}{!}{\includegraphics{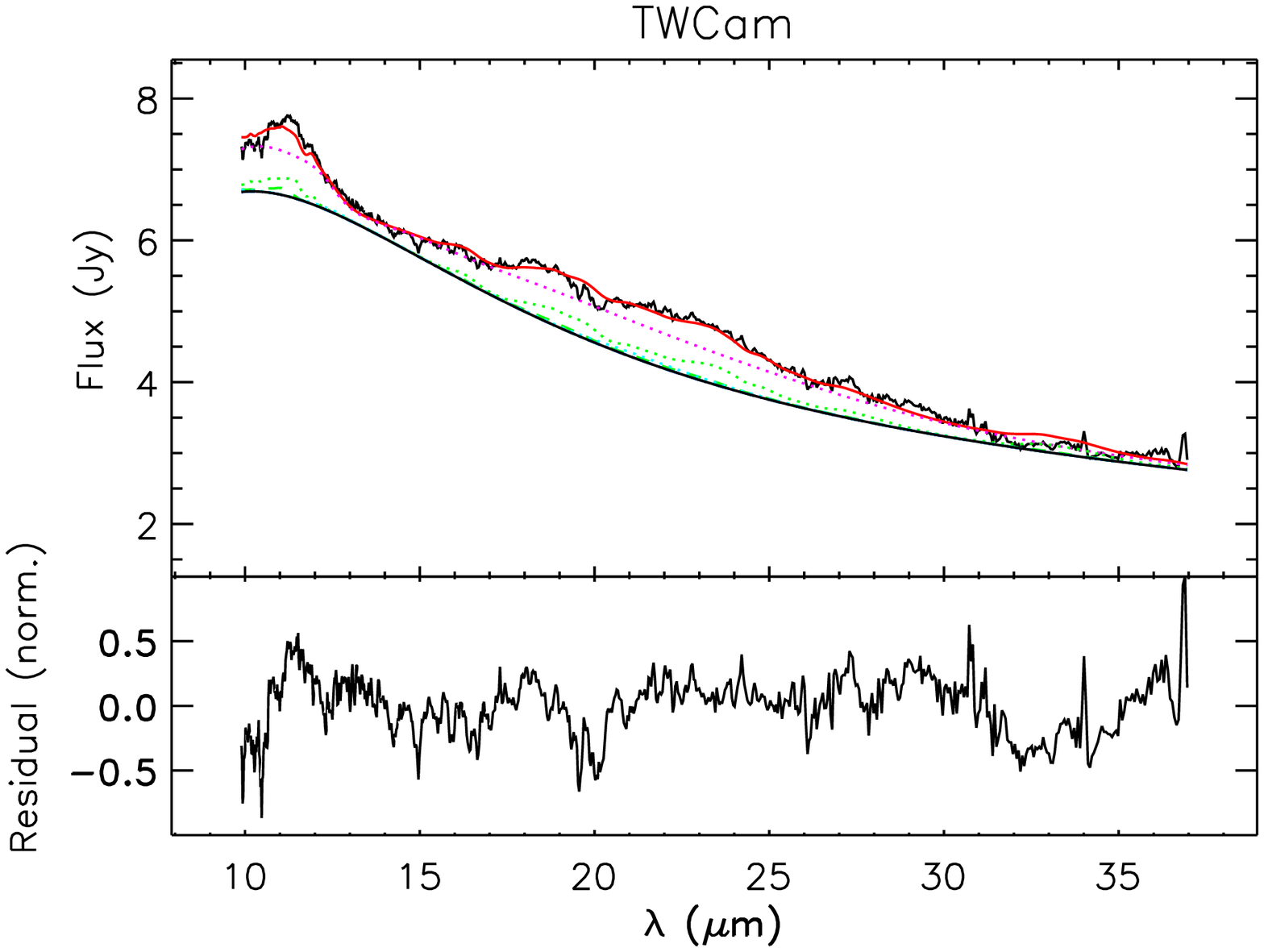}}
\vspace{0.3cm}
\hspace{0.3cm}
\resizebox{6cm}{!}{\includegraphics{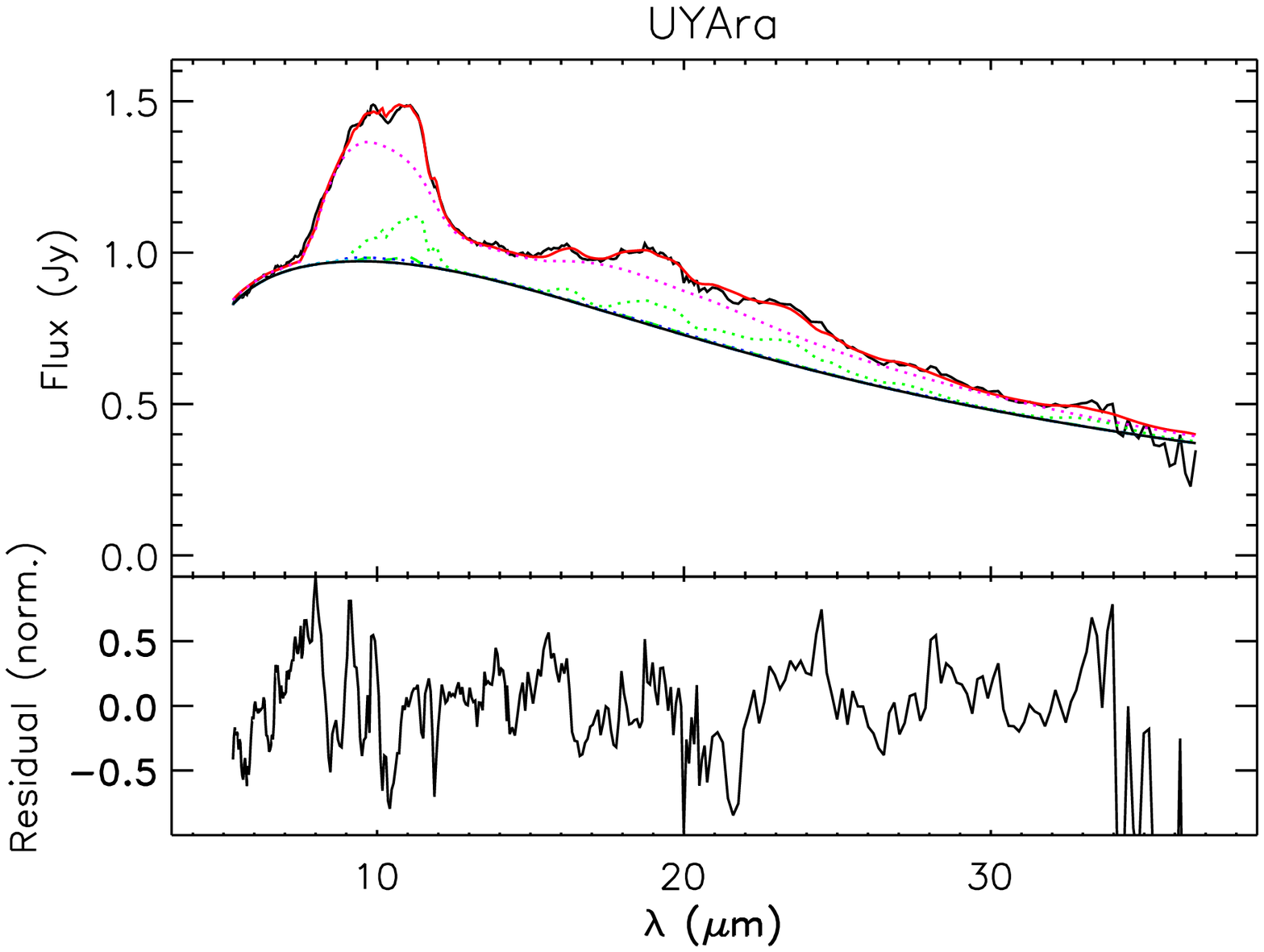}}
\vspace{0.3cm}
\hspace{0.3cm}
\resizebox{6cm}{!}{\includegraphics{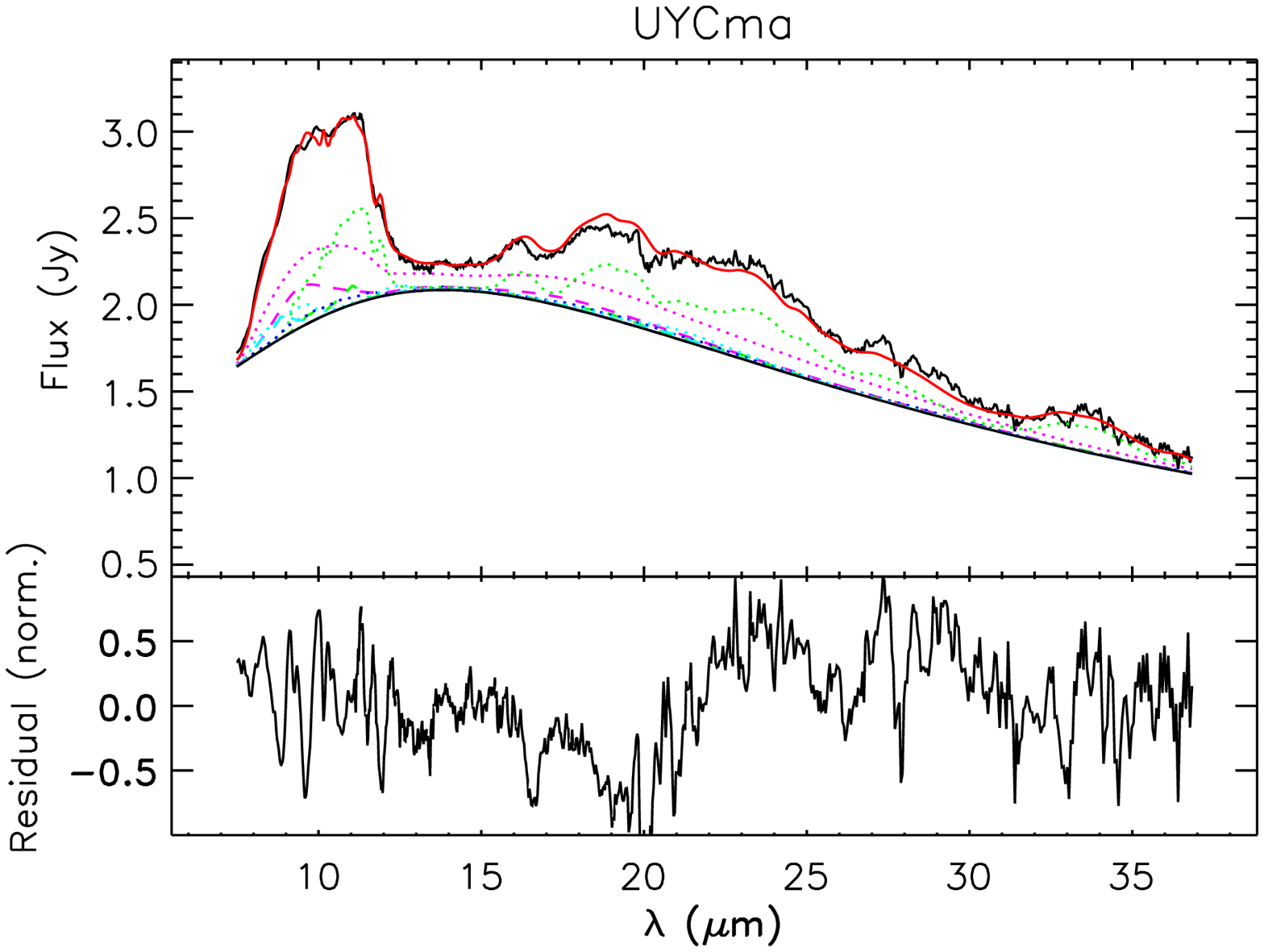}}
\caption{Same as Fig.~\ref{fits1}.}
\label{fits3}
\end{figure}

\begin{figure}
\resizebox{6cm}{!}{\includegraphics{plots/bestfit/HV12631.ps}}
\vspace{0.3cm}
\hspace{0.3cm}
\resizebox{6cm}{!}{\includegraphics{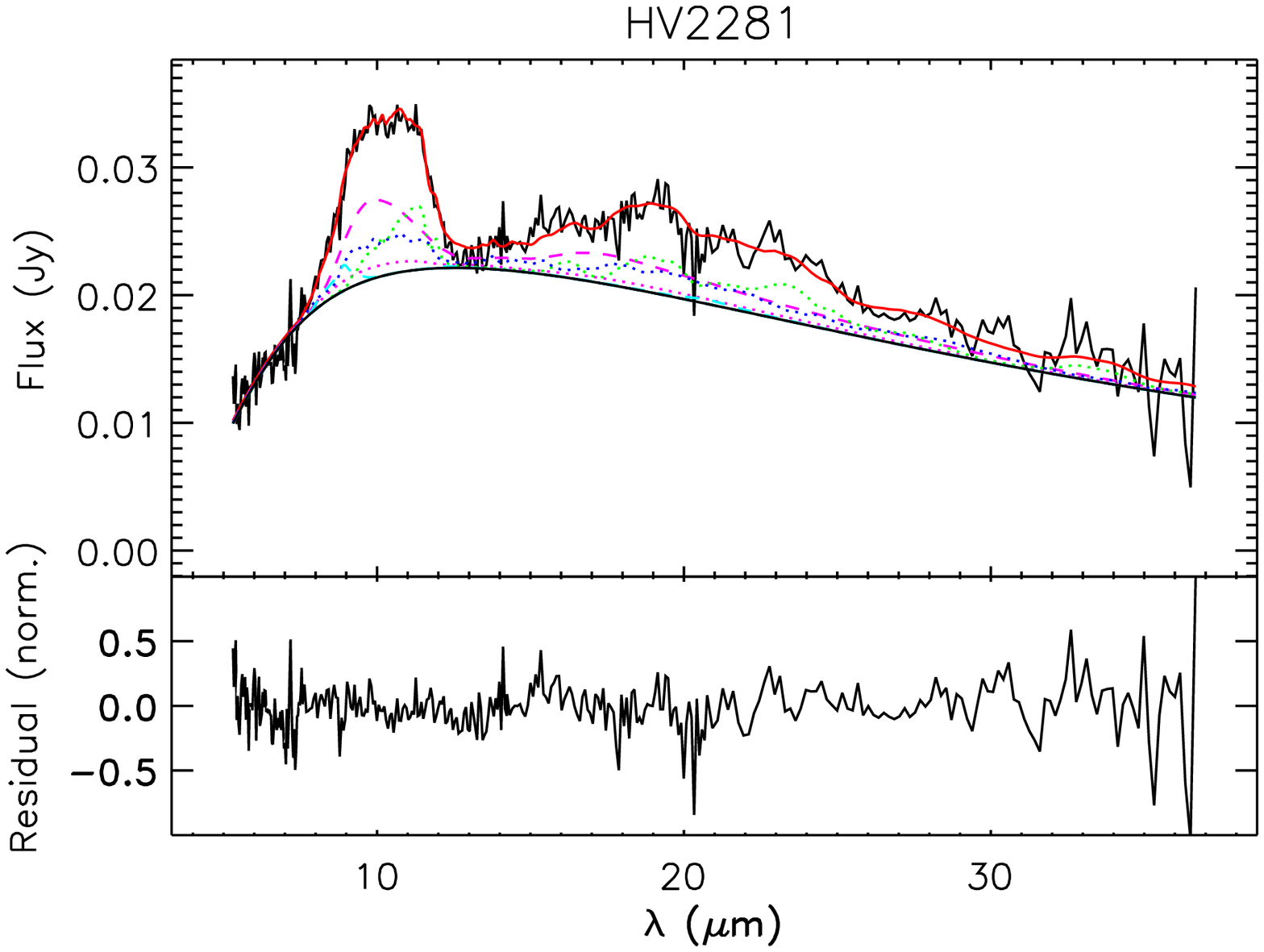}}
\vspace{0.3cm}
\hspace{0.3cm}
\resizebox{6cm}{!}{\includegraphics{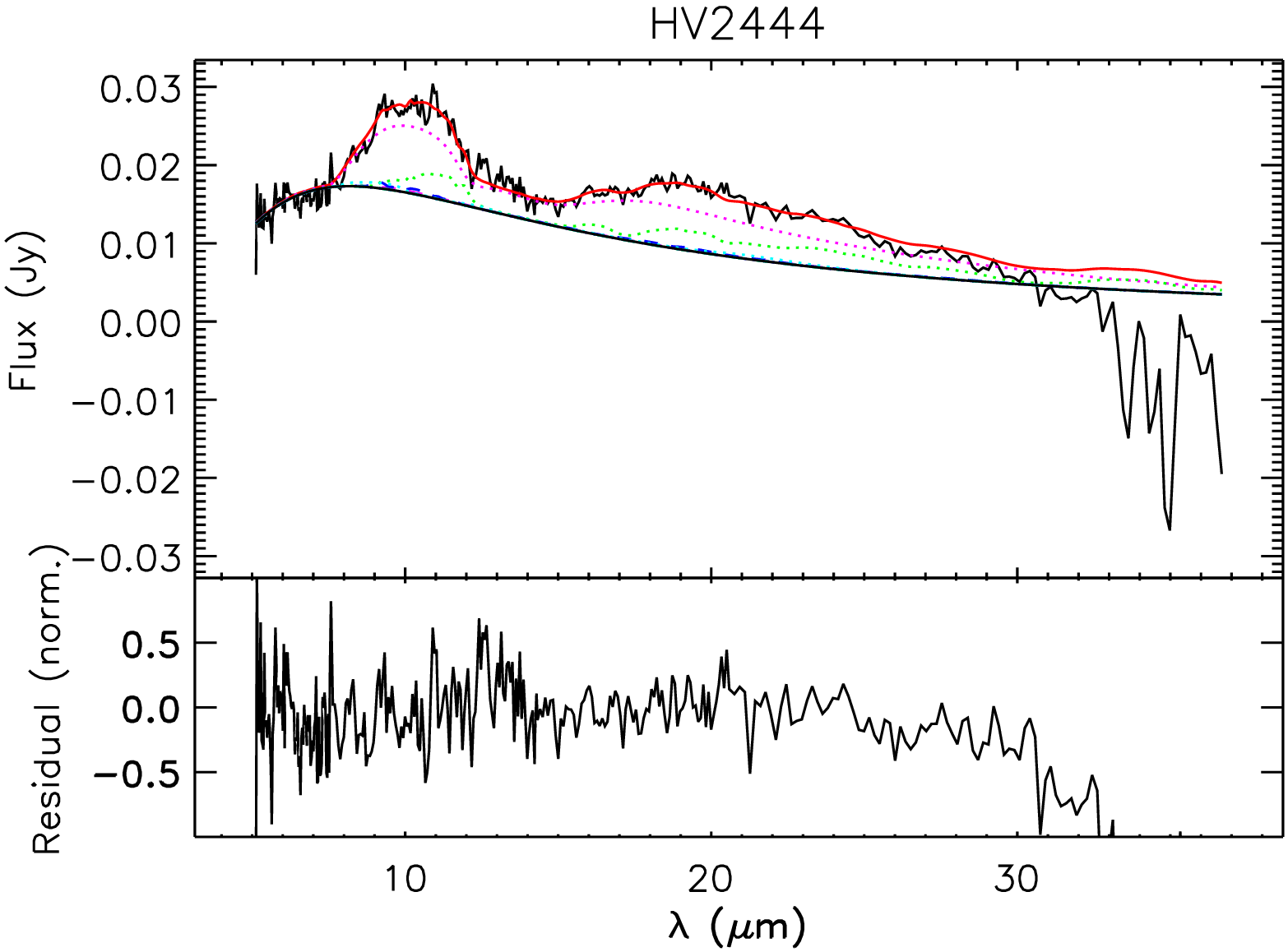}}

\resizebox{6cm}{!}{\includegraphics{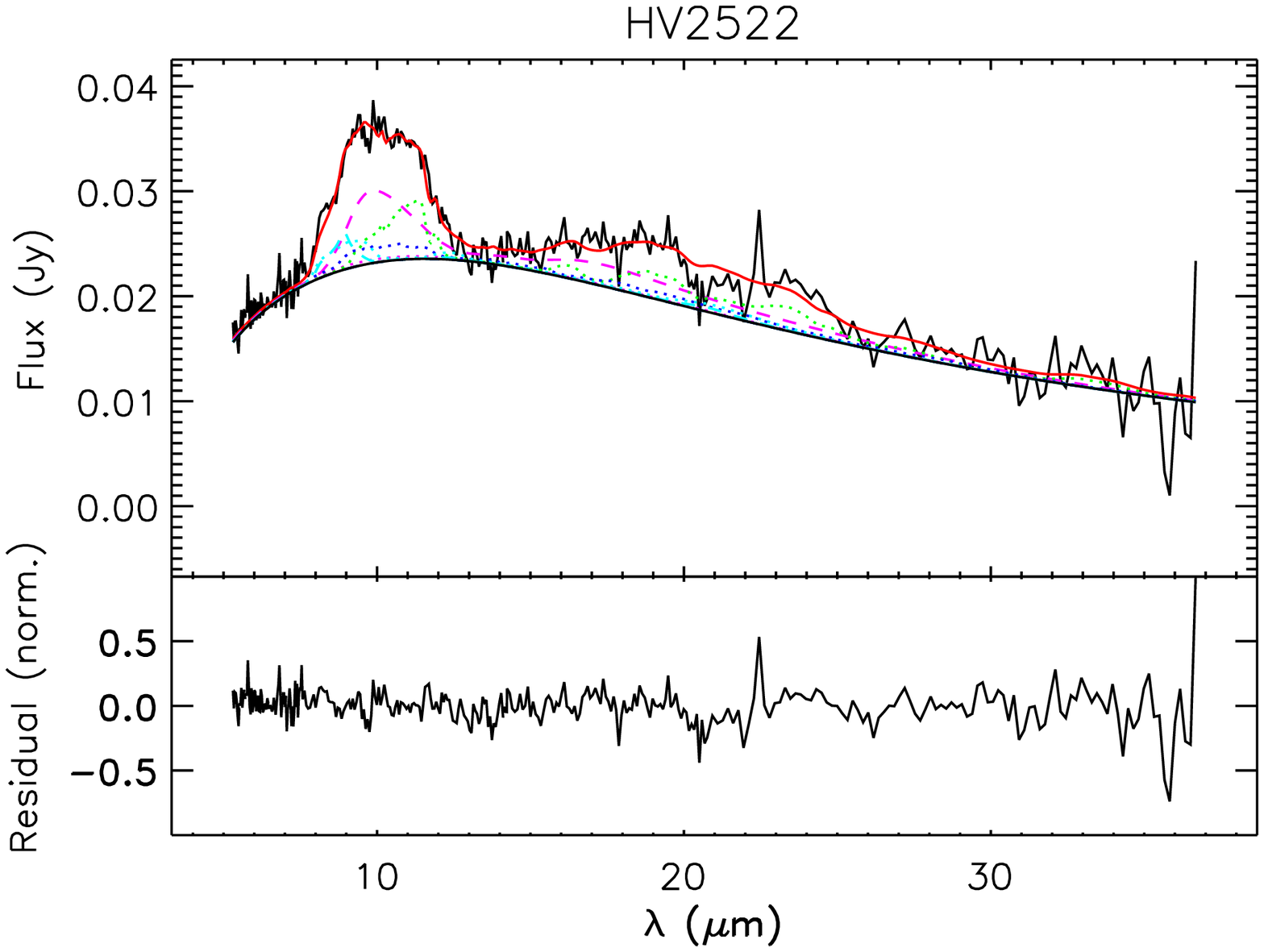}}
\vspace{0.3cm}
\hspace{0.3cm}
\resizebox{6cm}{!}{\includegraphics{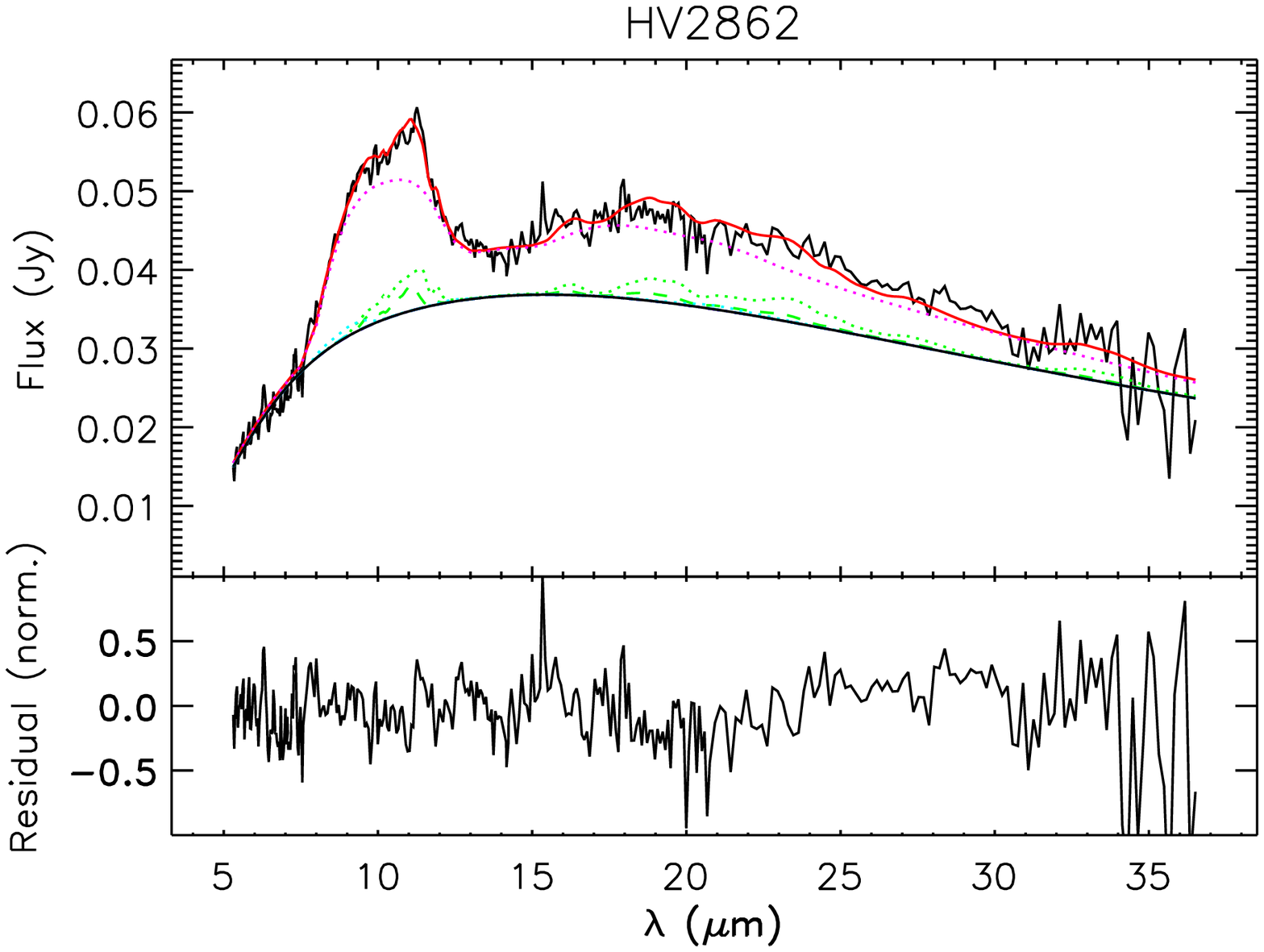}}
\vspace{0.3cm}
\hspace{0.3cm}
\resizebox{6cm}{!}{\includegraphics{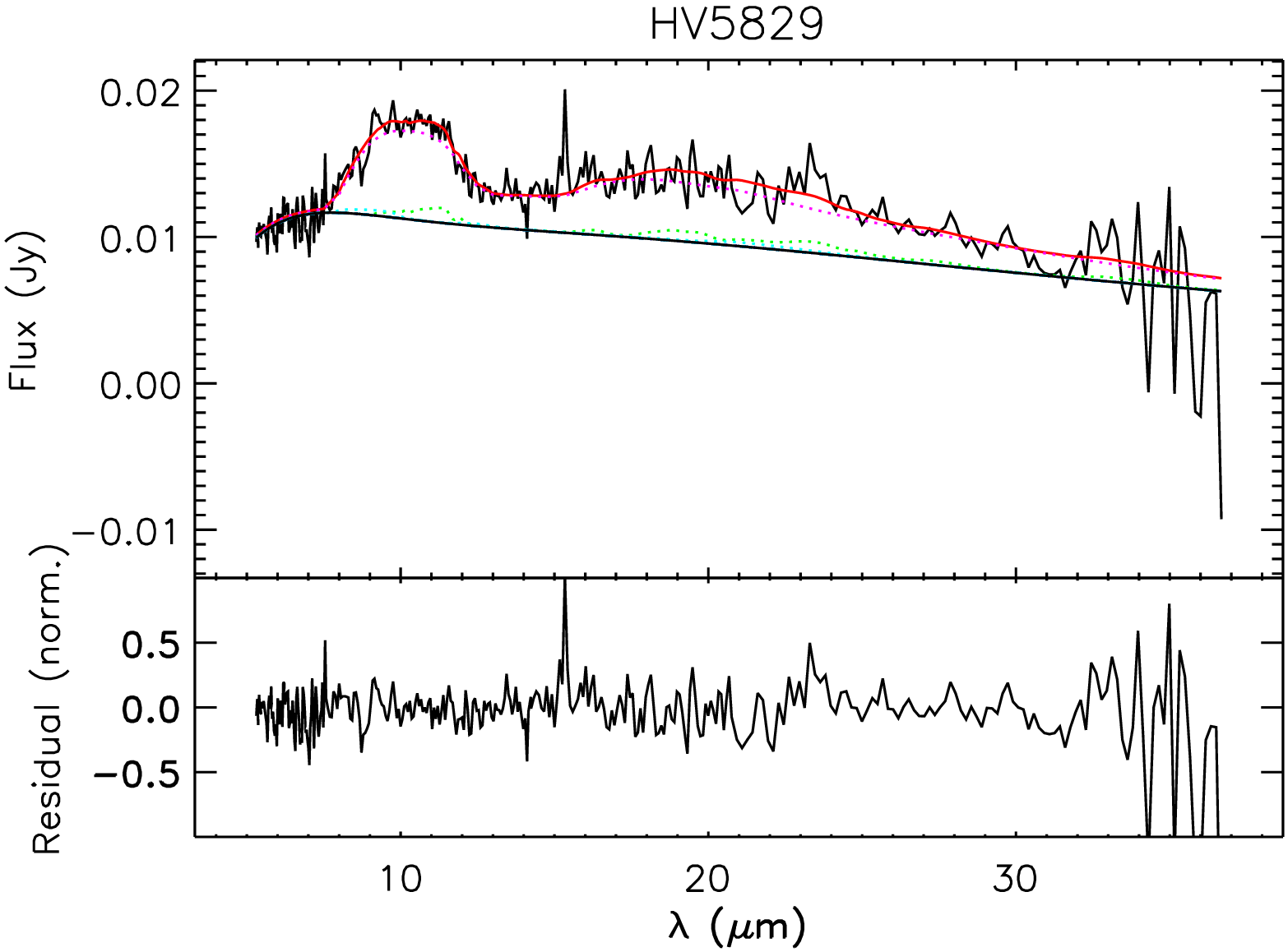}}

\resizebox{6cm}{!}{\includegraphics{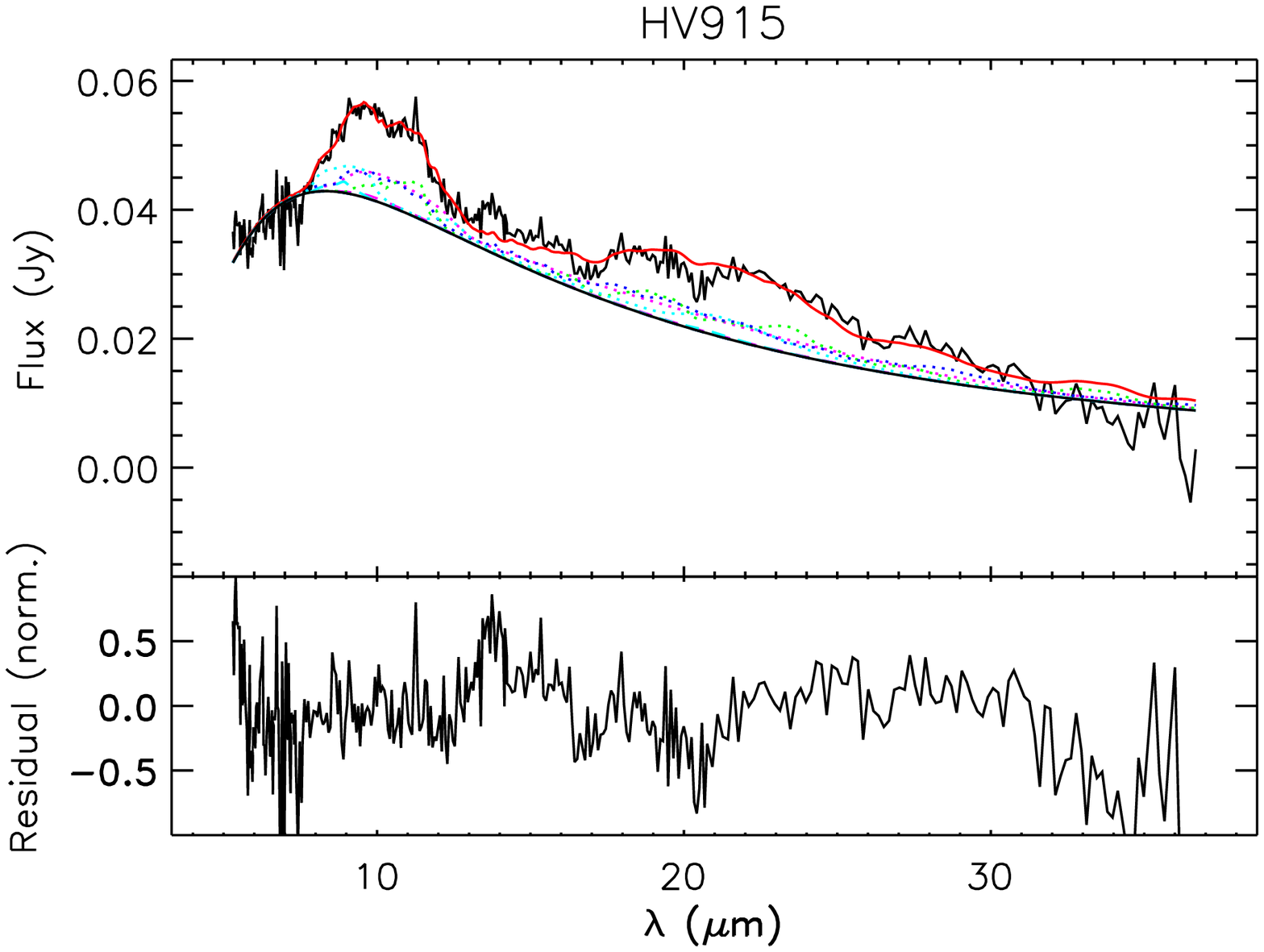}}
\vspace{0.3cm}
\hspace{0.3cm}
\resizebox{6cm}{!}{\includegraphics{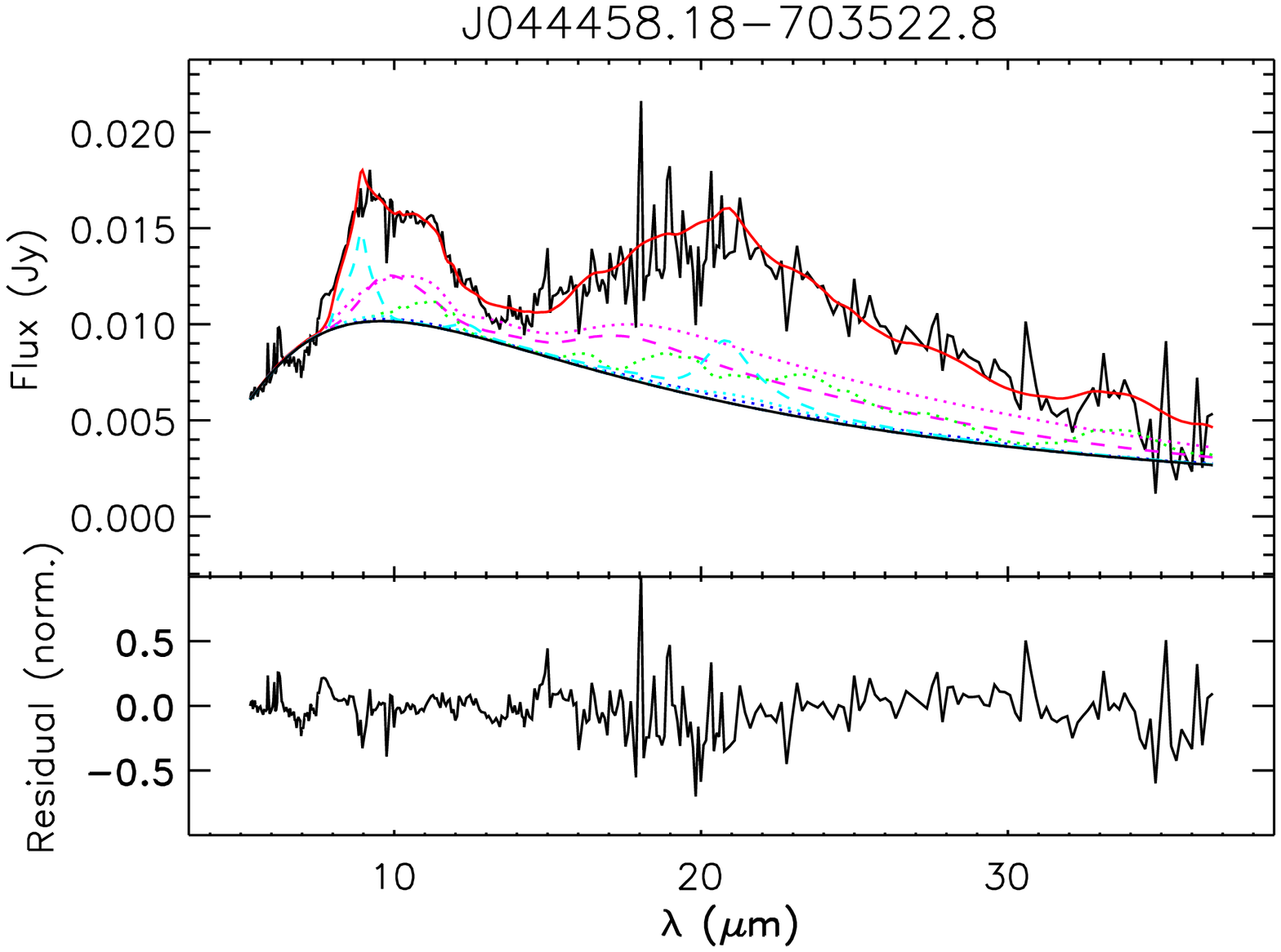}}
\vspace{0.3cm}
\hspace{0.3cm}
\resizebox{6cm}{!}{\includegraphics{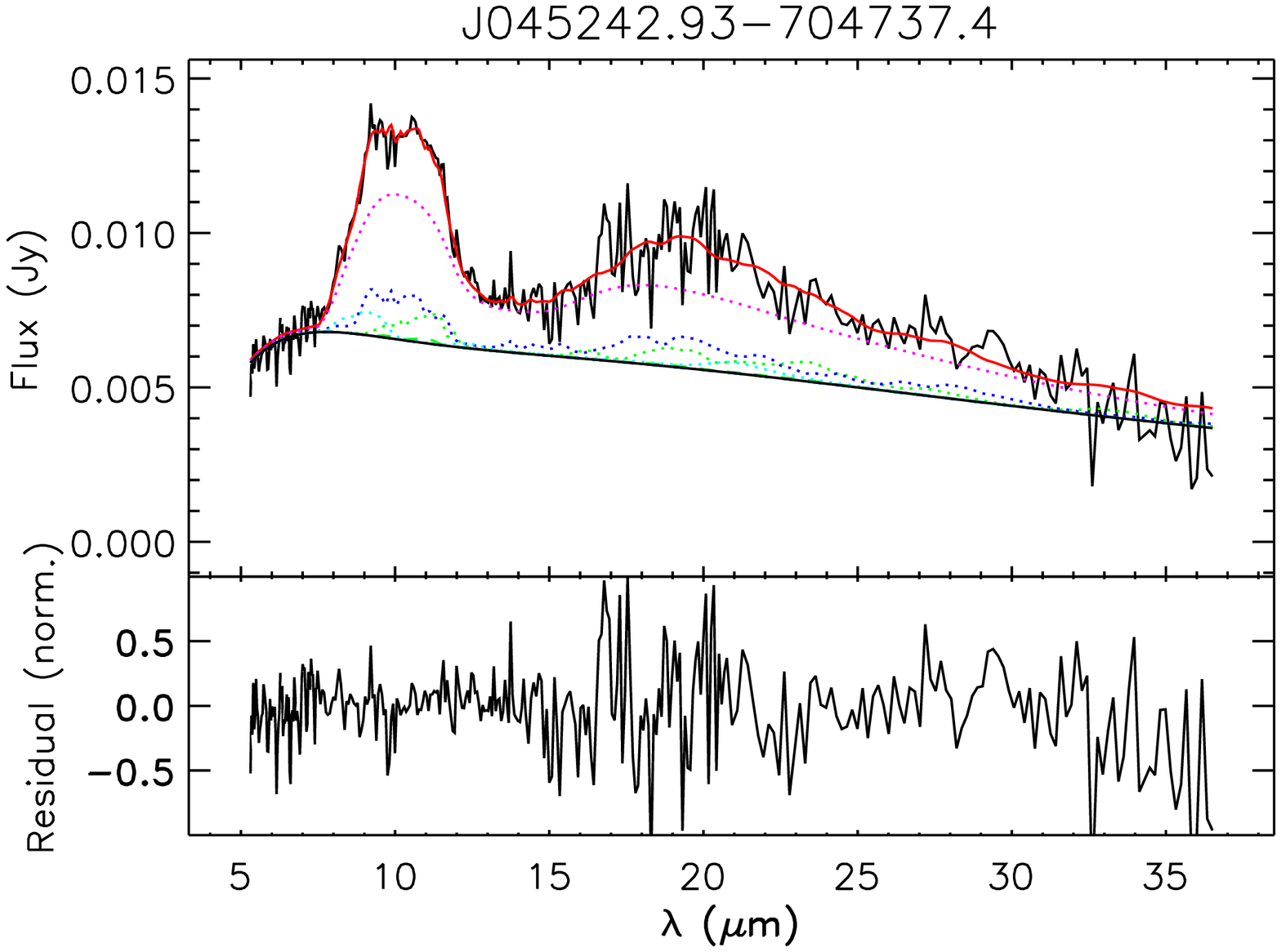}}

\resizebox{6cm}{!}{\includegraphics{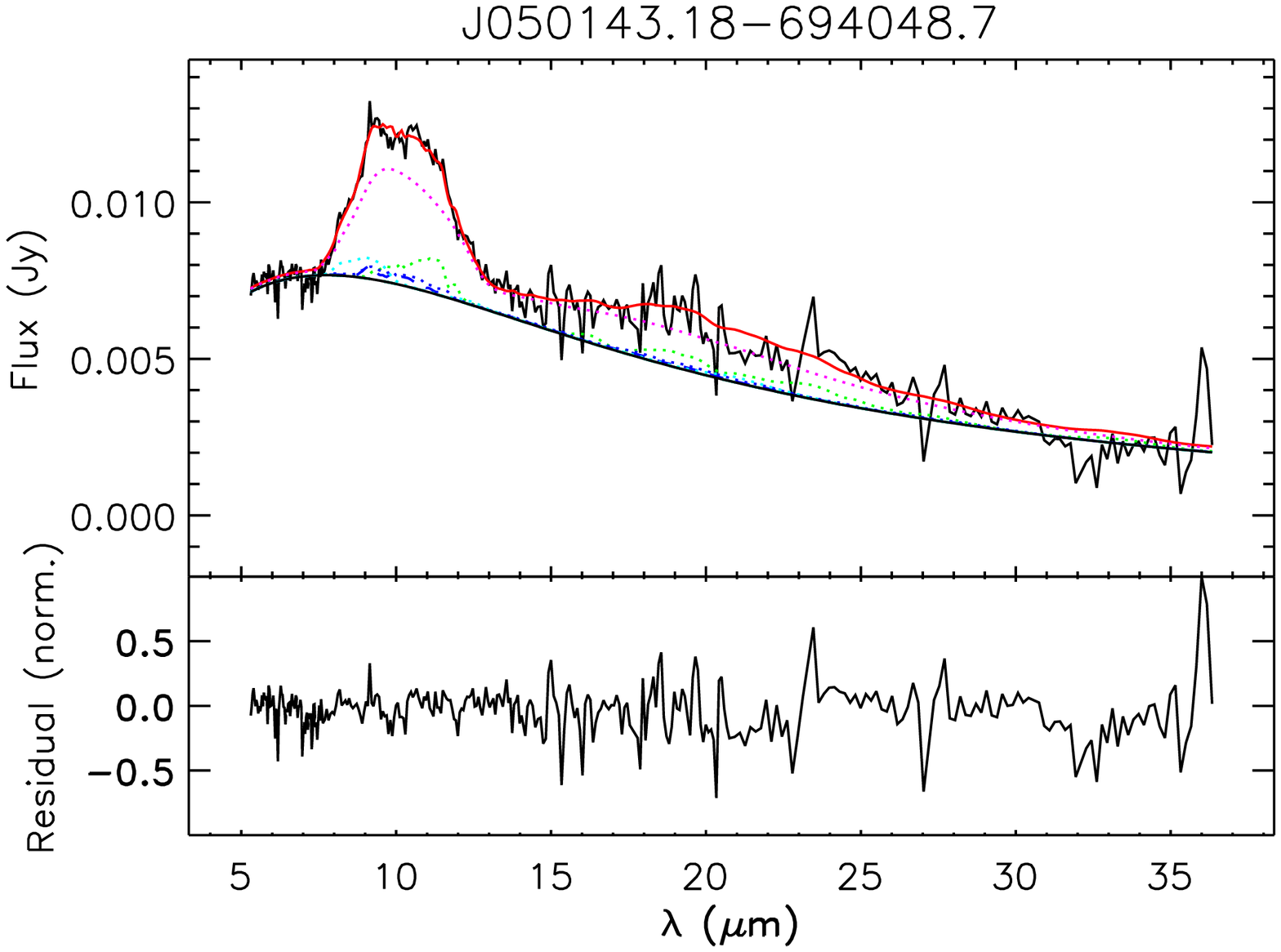}}
\vspace{0.3cm}
\hspace{0.3cm}
\resizebox{6cm}{!}{\includegraphics{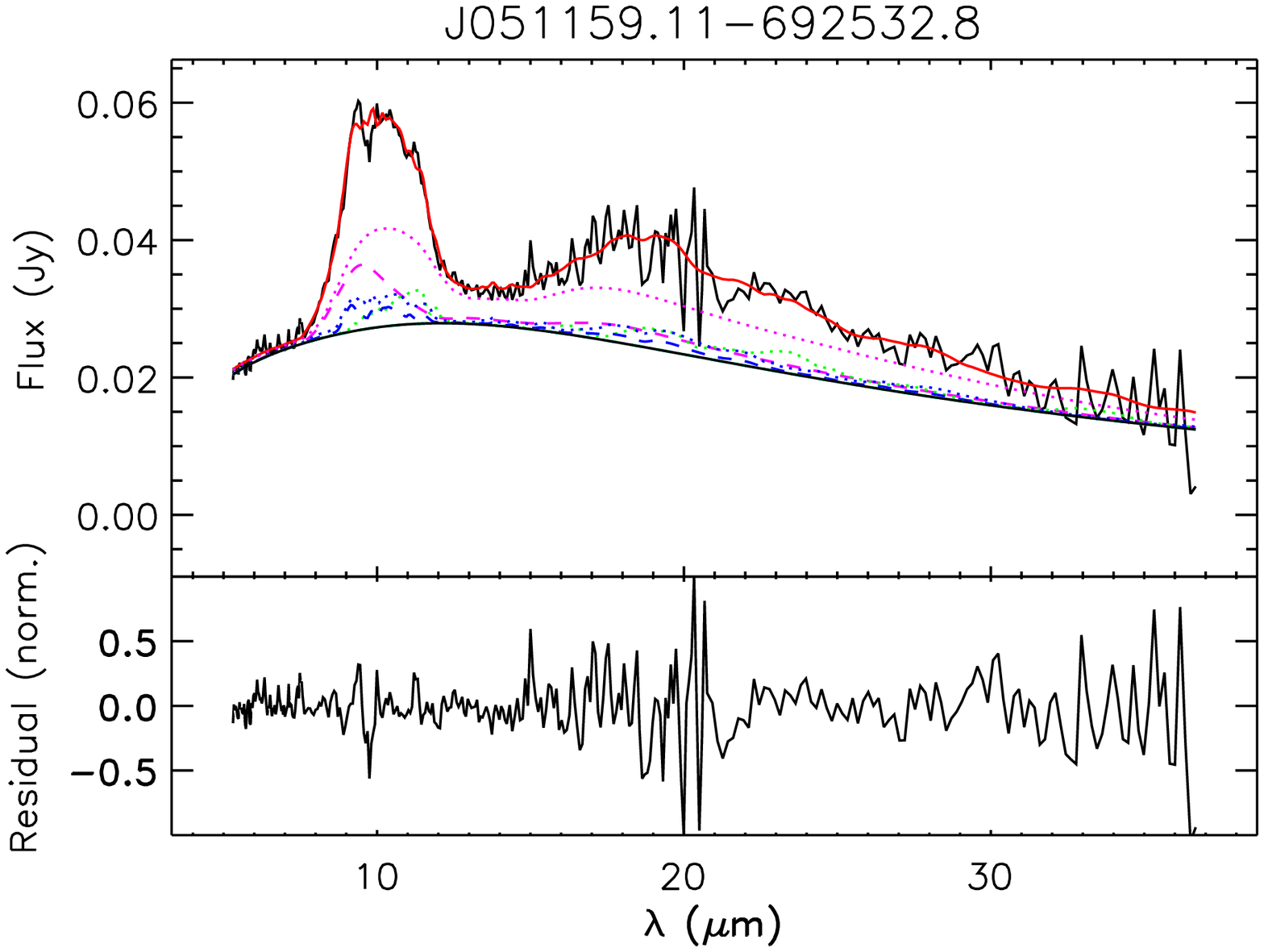}}
\vspace{0.3cm}
\hspace{0.3cm}
\resizebox{6cm}{!}{\includegraphics{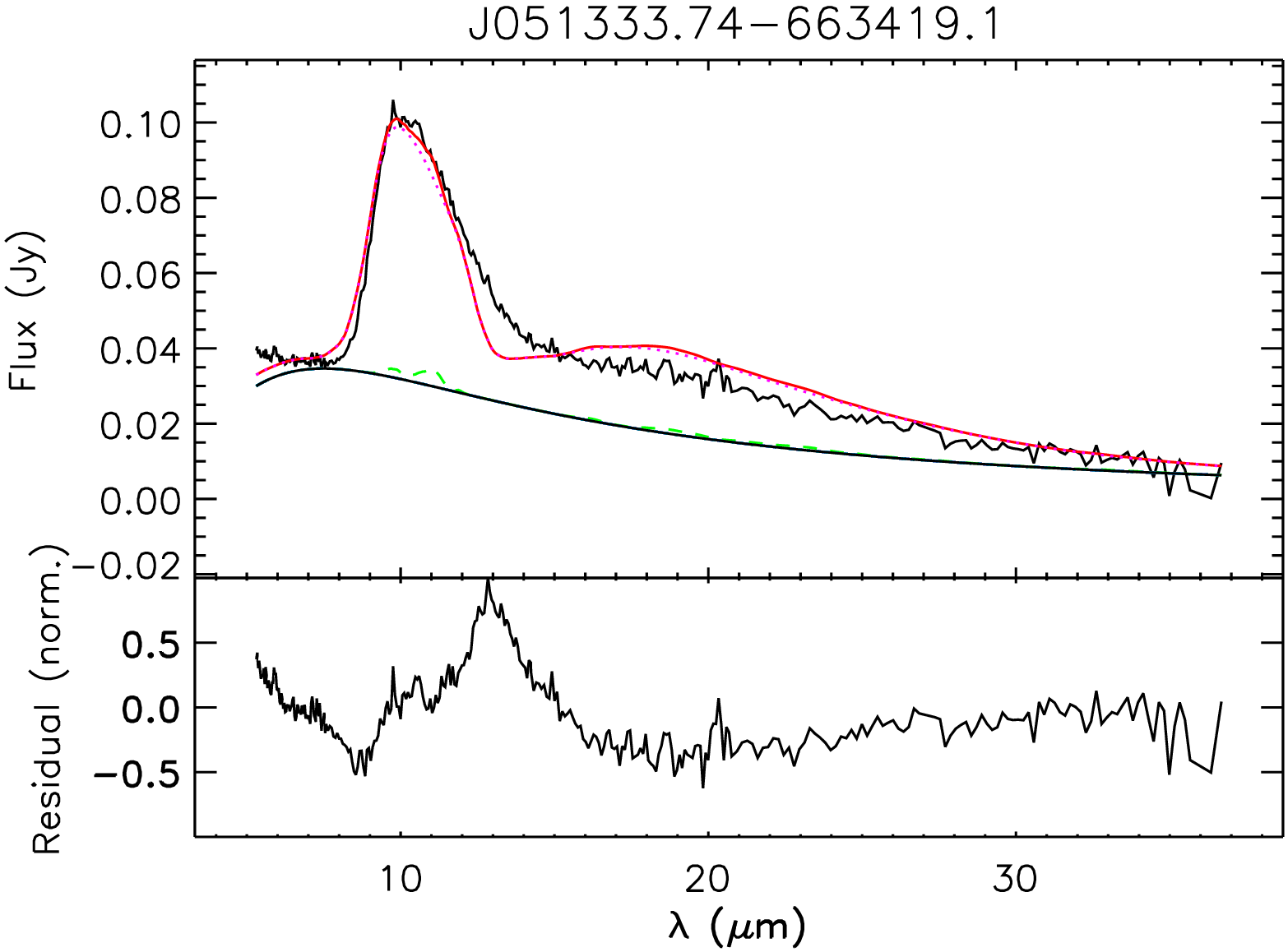}}
\caption{Best model fits for our LMC sample stars, showing the contribution of the different dust species.
Top: The observed spectrum (black curve) is plotted together with the best model fit (red curve) and the continuum (black solid line).
Forsterite is plotted in green, enstatite in blue, silica in cyan and amorphous olivine and pyroxene in magenta.
Small grains (0.1\,$\mu$m) are plotted as dashed lines and larger grains (2 and 4\,$\mu$m) as dotted lines.
Bottom: The normalised residuals after subtraction of our best model of the observed spectra.}
\label{fits4}
\end{figure}

\begin{figure}
\resizebox{6cm}{!}{\includegraphics{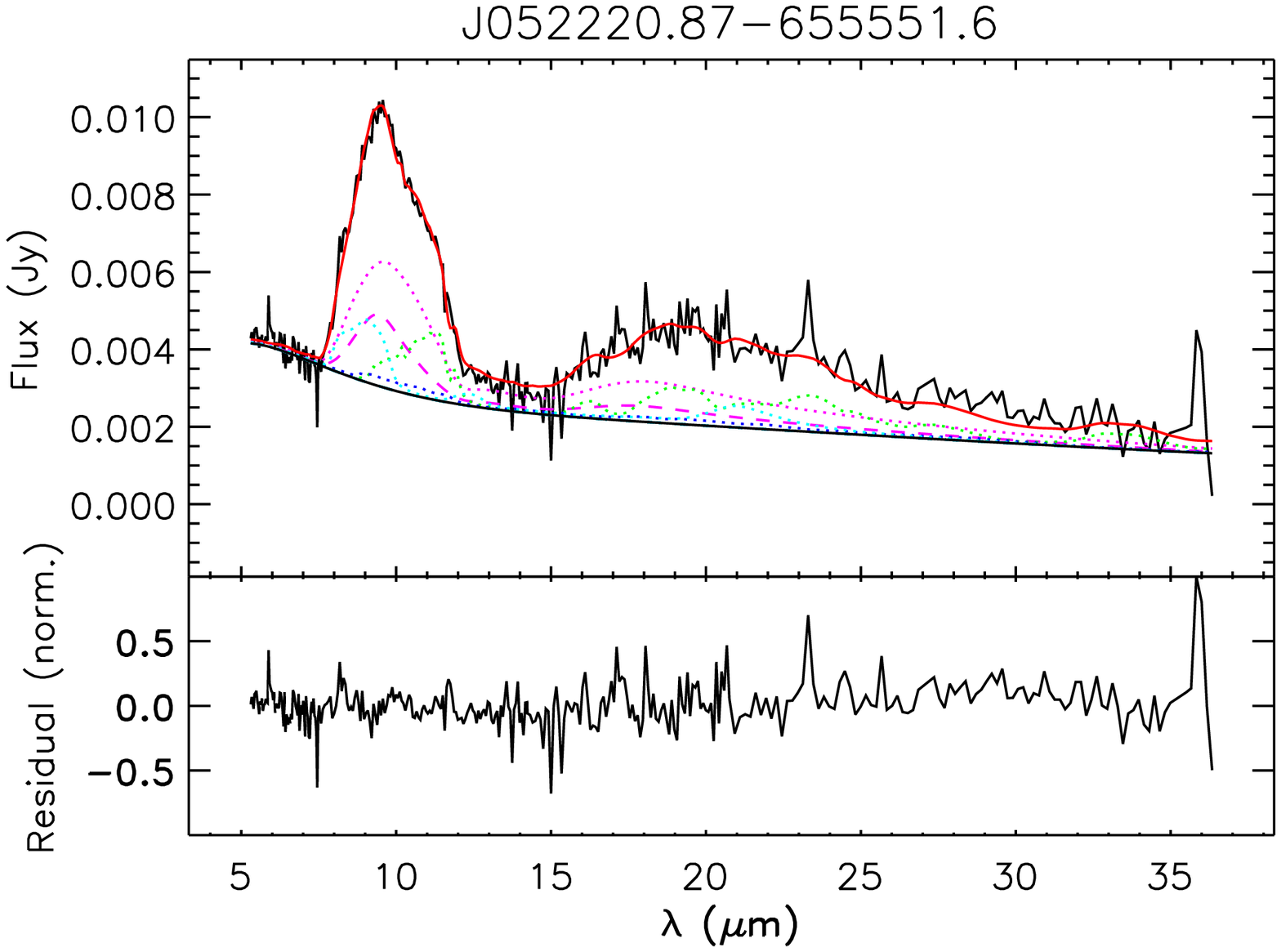}}
\vspace{0.3cm}
\hspace{0.3cm}
\resizebox{6cm}{!}{\includegraphics{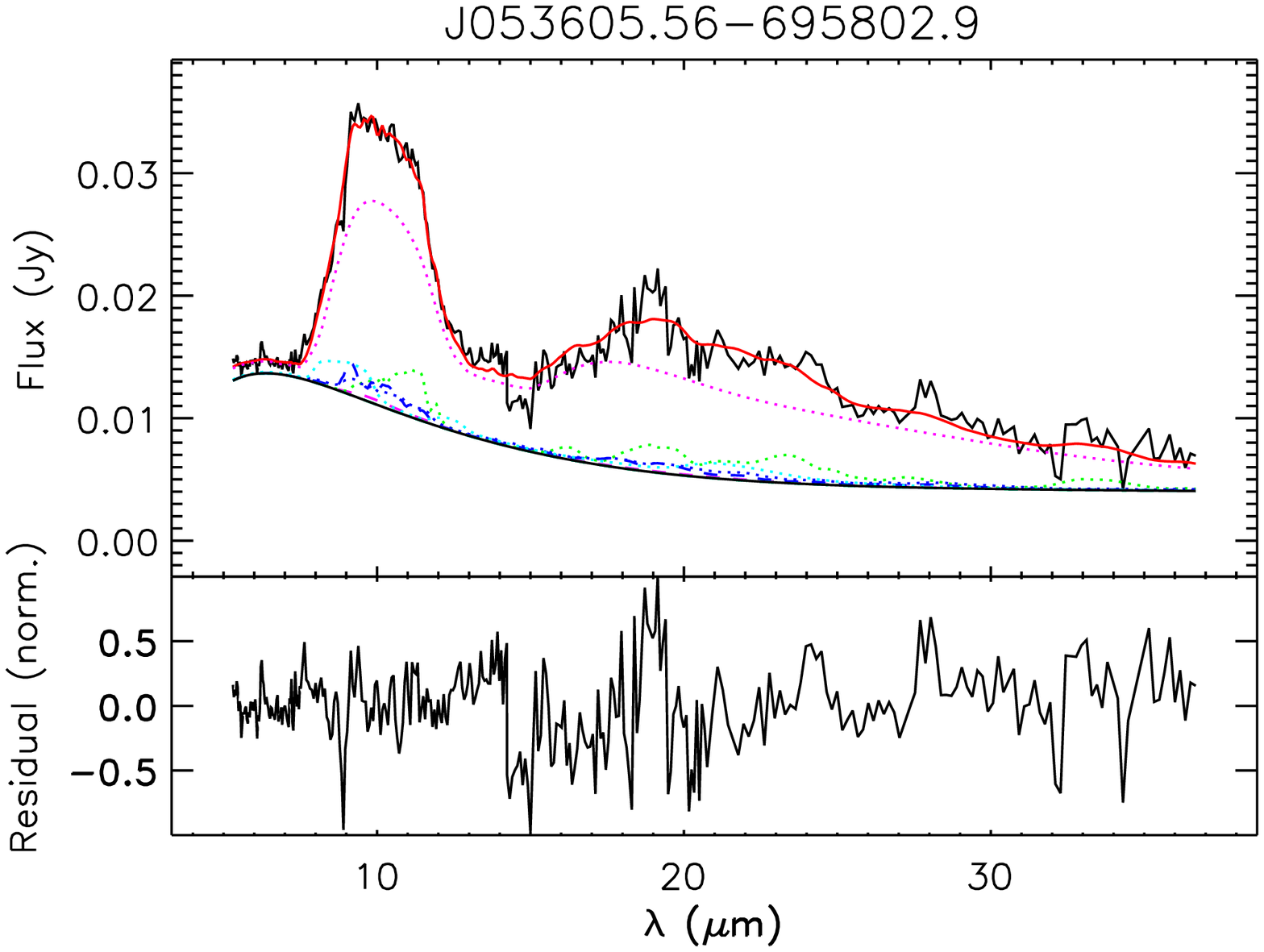}}
\vspace{0.3cm}
\hspace{0.3cm}
\resizebox{6cm}{!}{\includegraphics{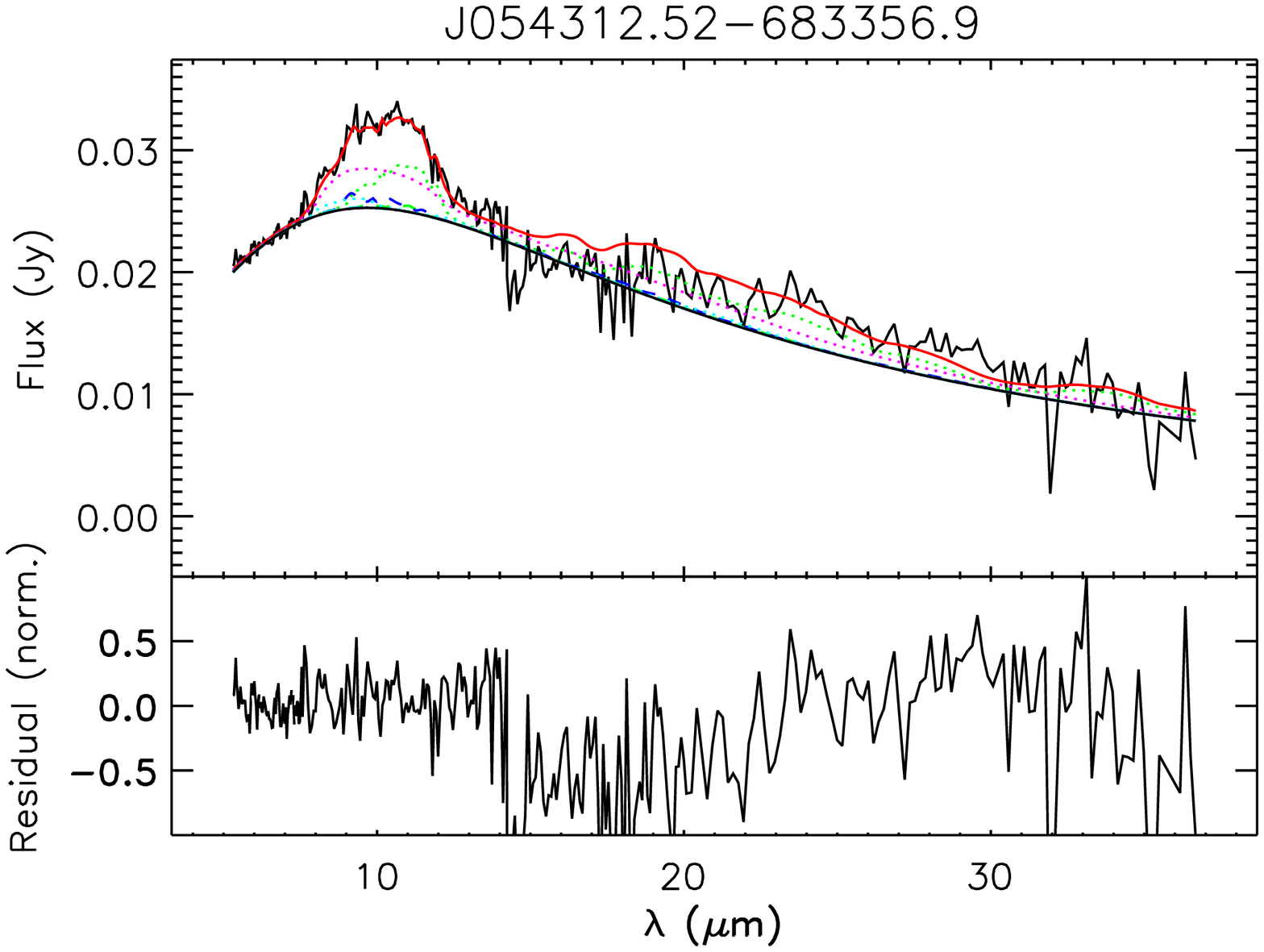}}

\resizebox{6cm}{!}{\includegraphics{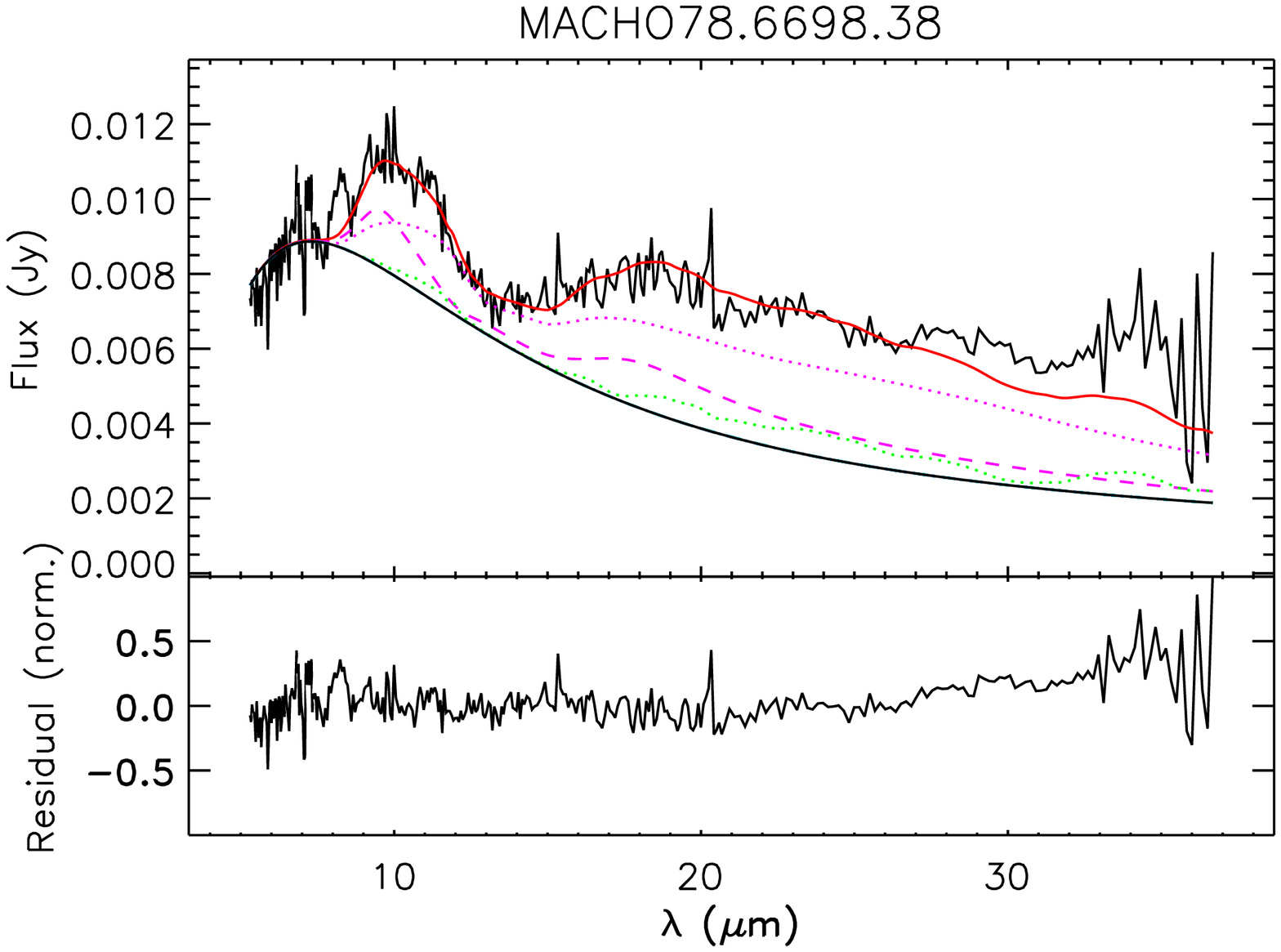}}
\vspace{0.3cm}
\hspace{0.3cm}
\resizebox{6cm}{!}{\includegraphics{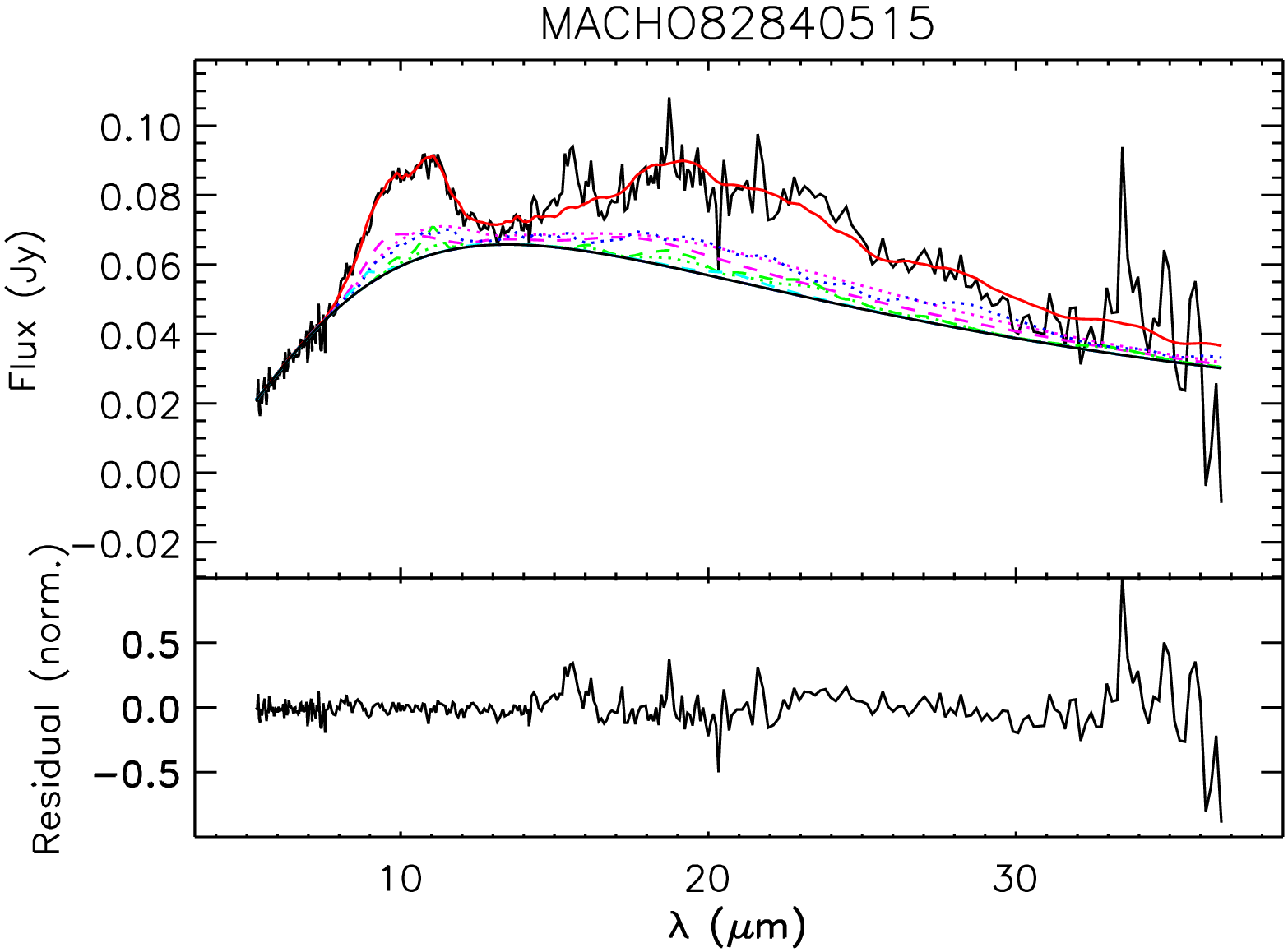}}
\vspace{0.3cm}
\hspace{0.3cm}
\resizebox{6cm}{!}{\includegraphics{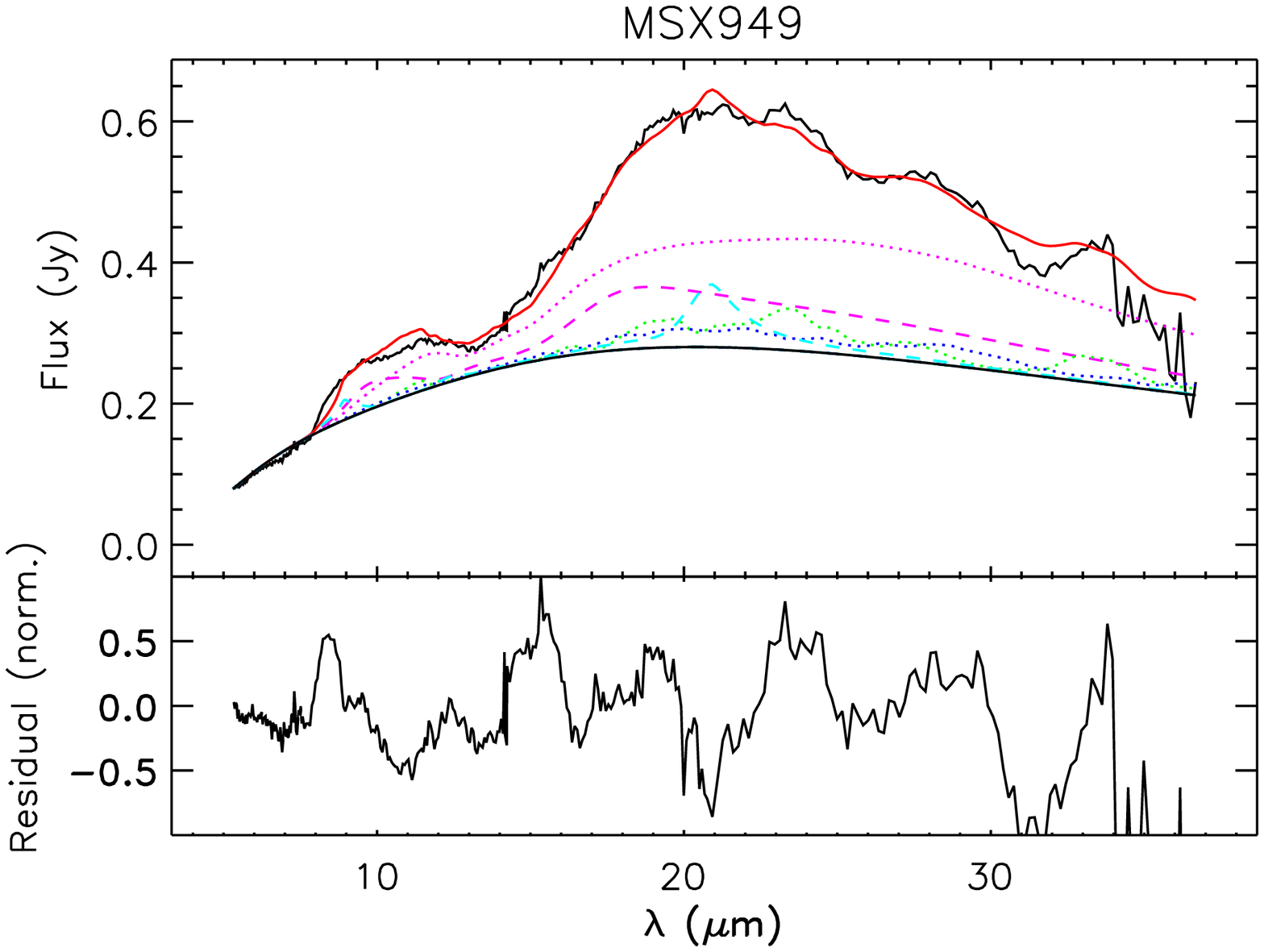}}

\resizebox{6cm}{!}{\includegraphics{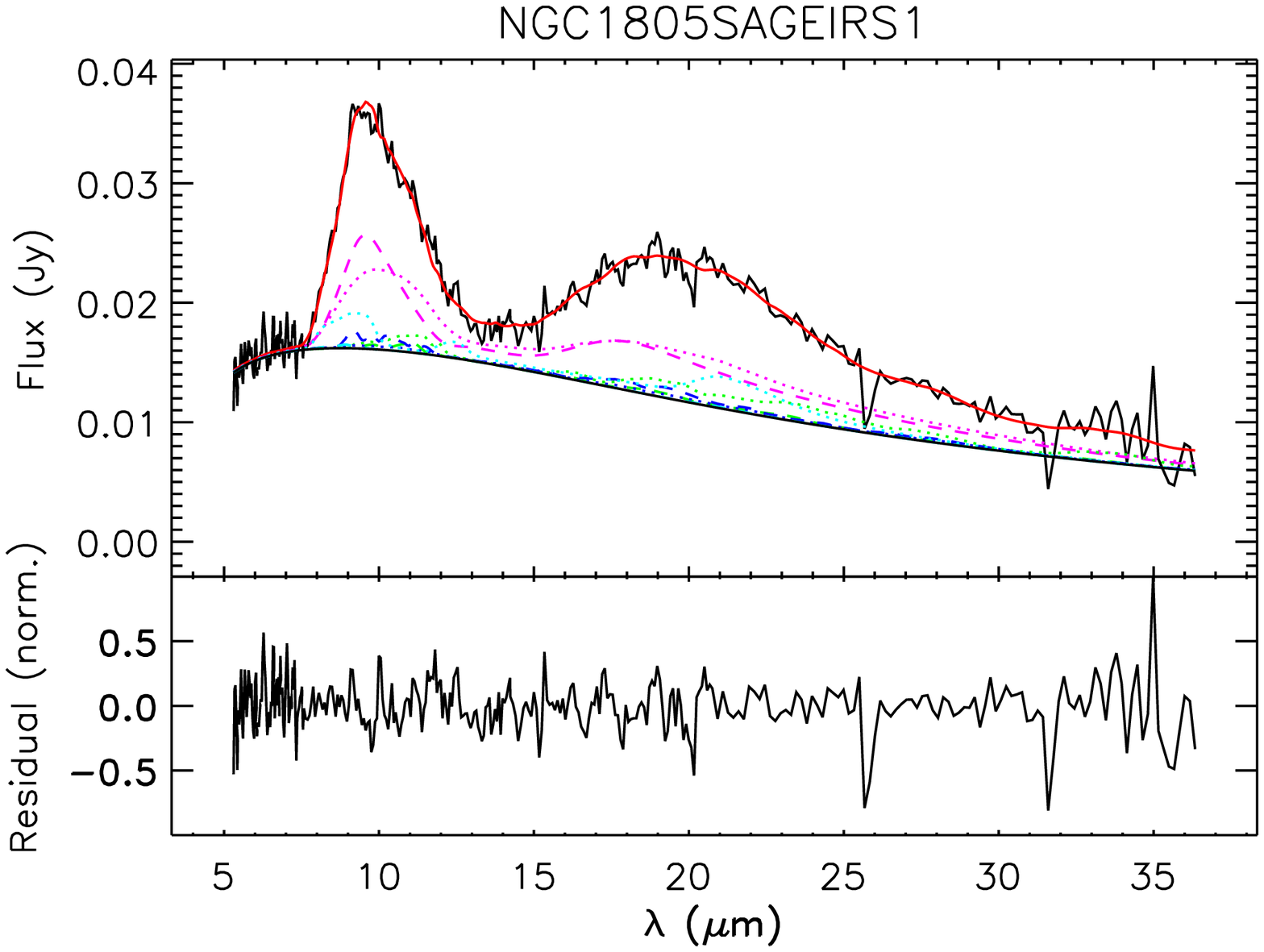}}
\vspace{0.3cm}
\hspace{0.3cm}
\resizebox{6cm}{!}{\includegraphics{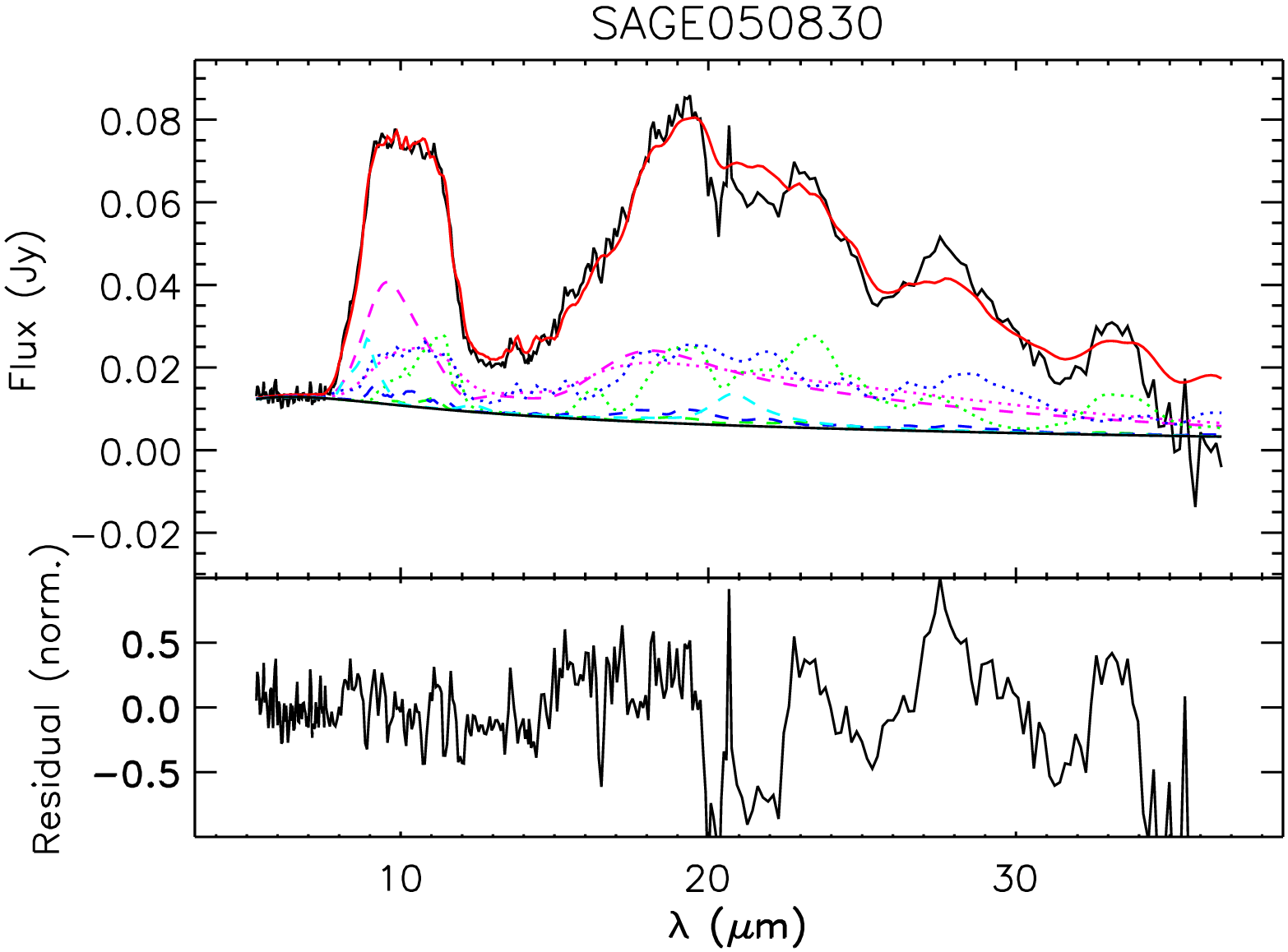}}
\vspace{0.3cm}
\hspace{0.3cm}
\resizebox{6cm}{!}{\includegraphics{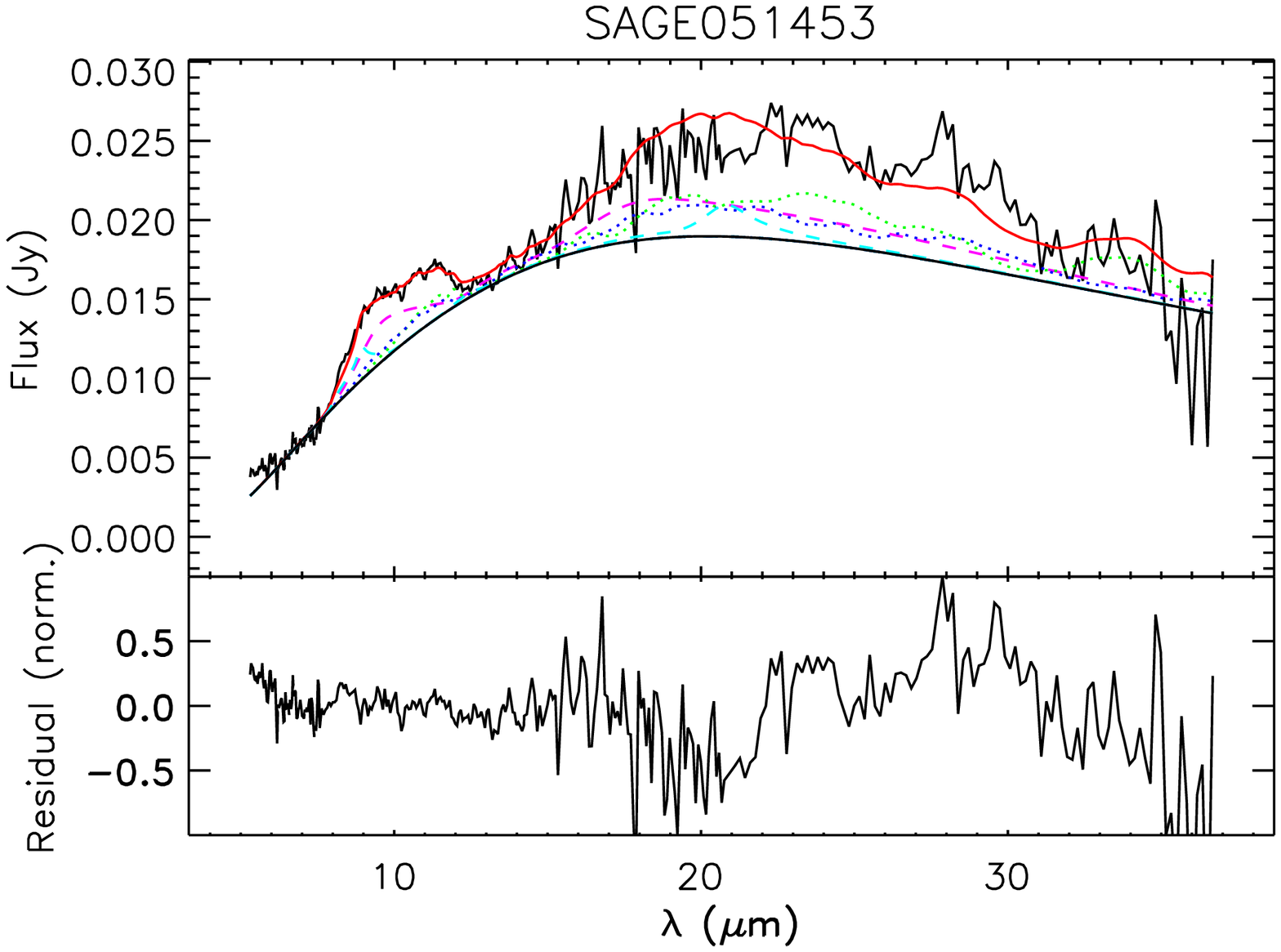}}

\resizebox{6cm}{!}{\includegraphics{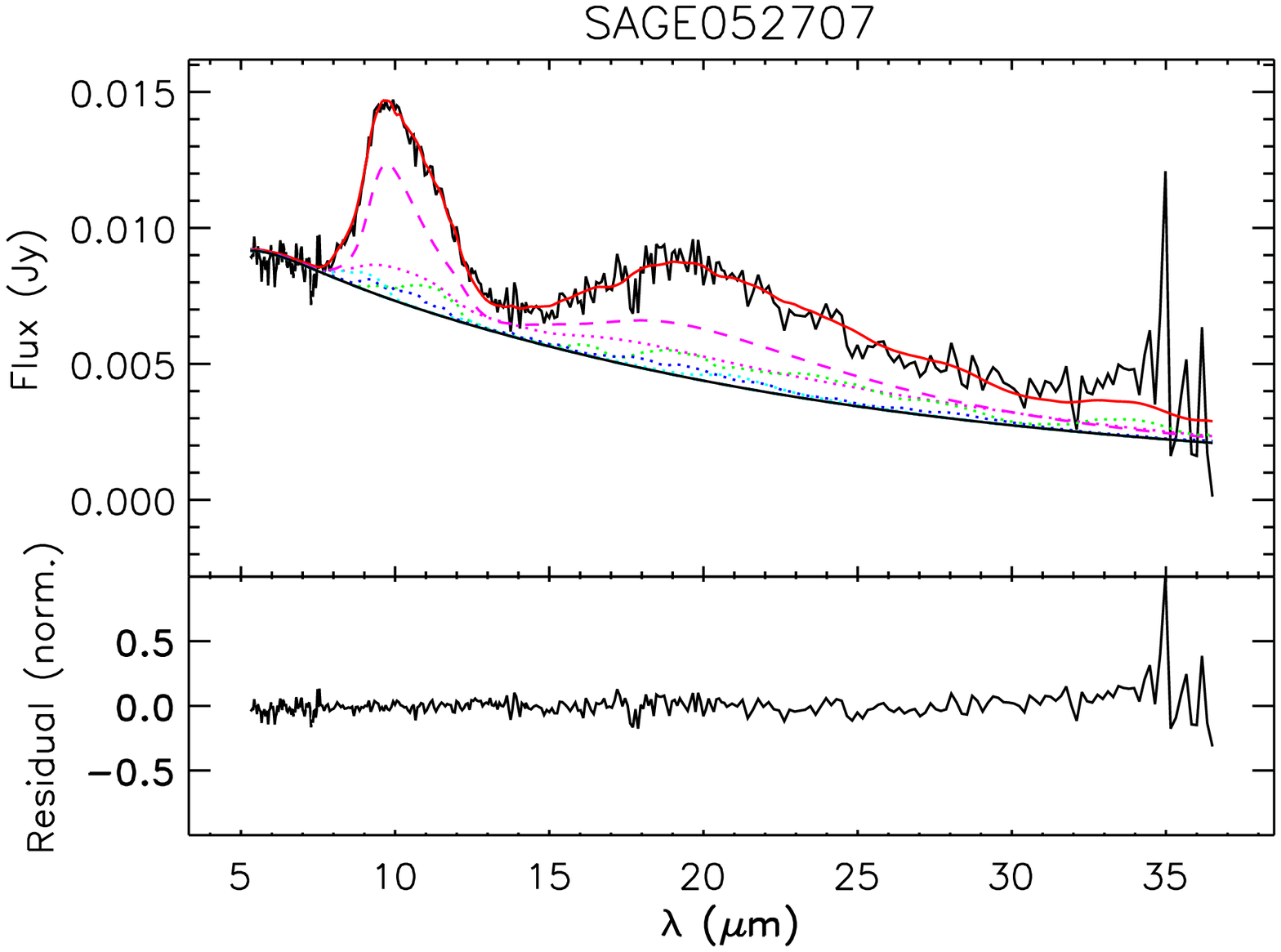}}
\vspace{0.3cm}
\hspace{0.3cm}
\resizebox{6cm}{!}{\includegraphics{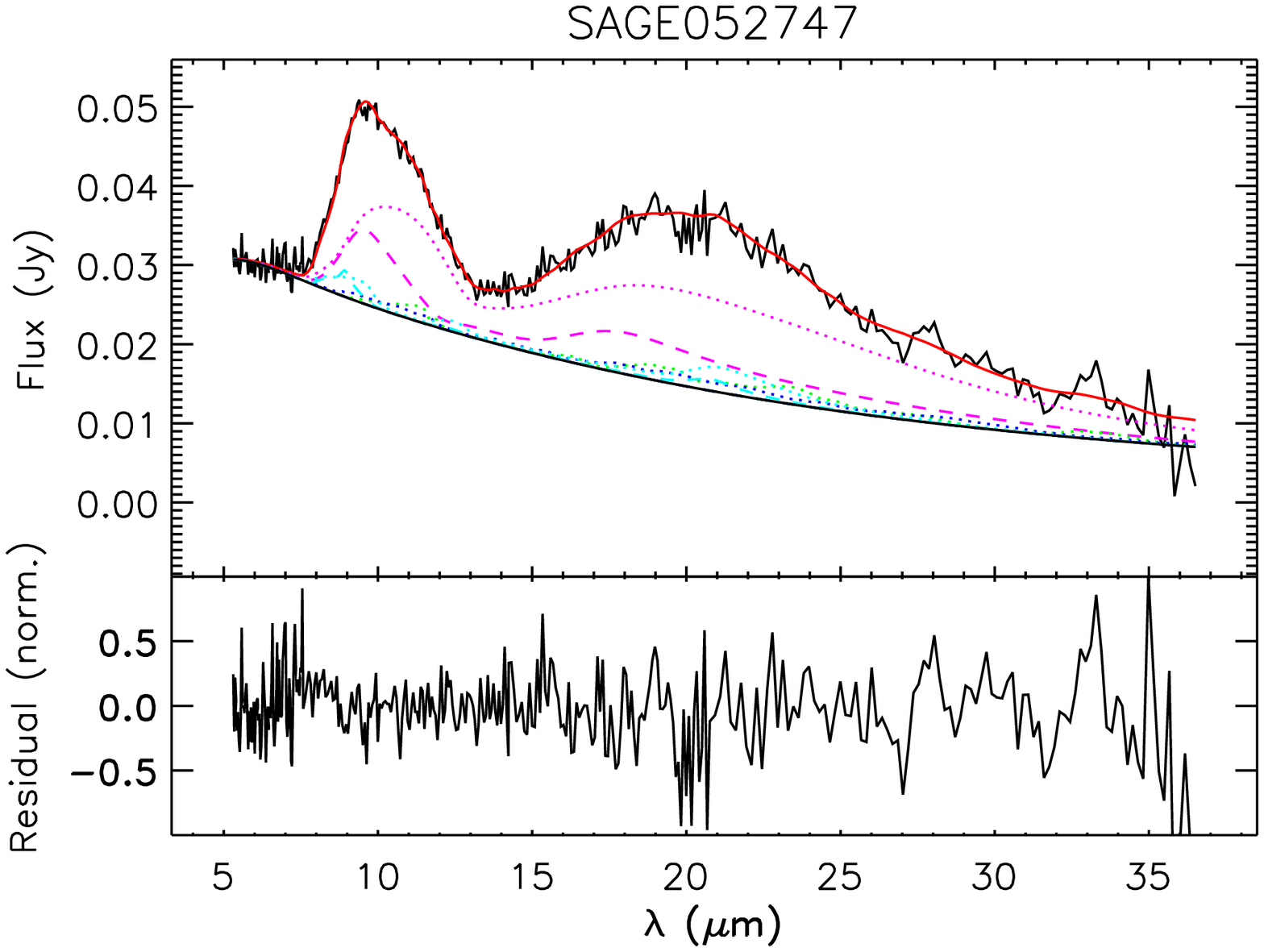}}
\vspace{0.3cm}
\hspace{0.3cm}
\resizebox{6cm}{!}{\includegraphics{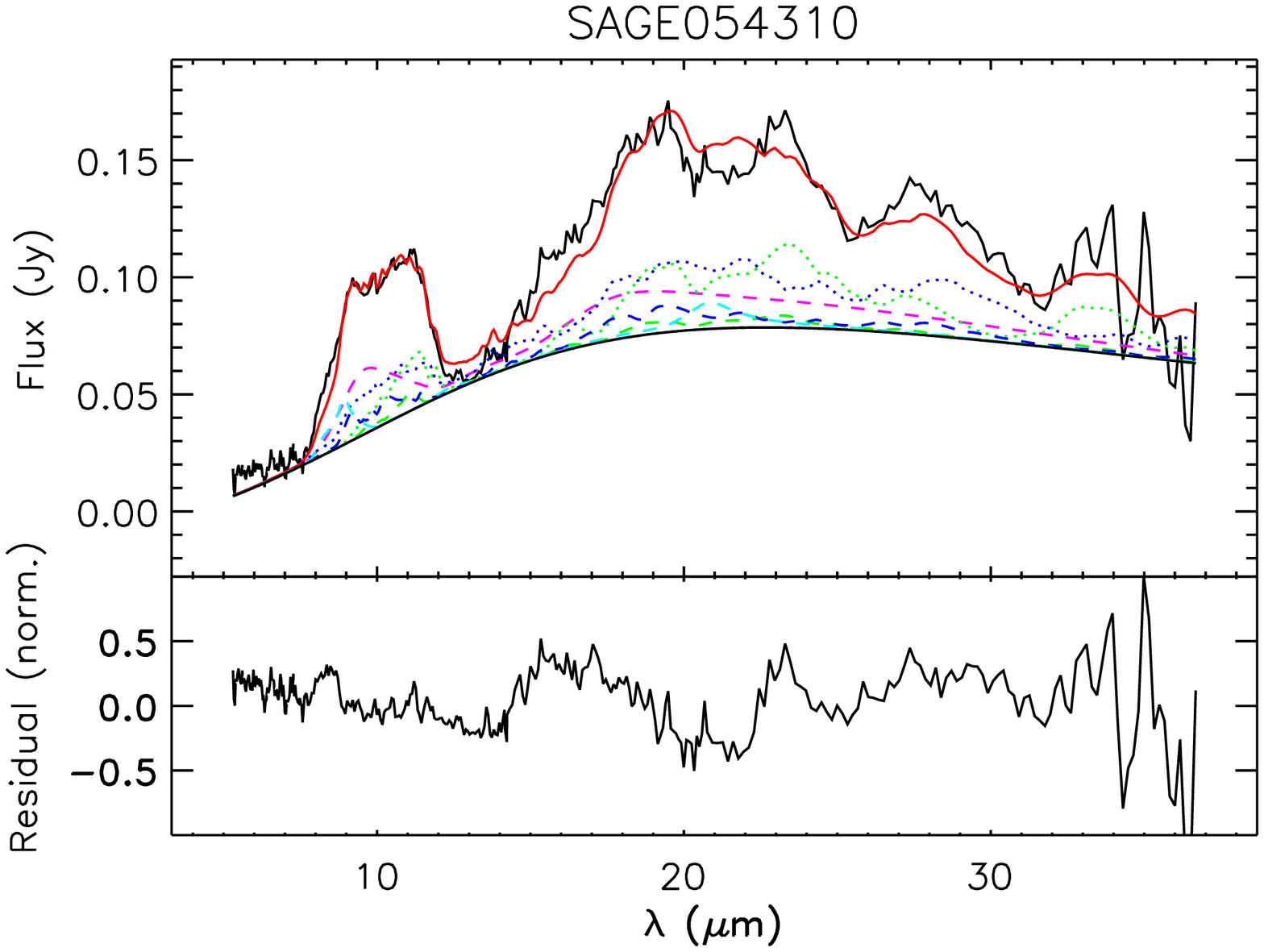}}
\caption{Same as Fig.~\ref{fits4}.}
\label{fits5}
\end{figure}

\begin{figure}[ht]
\vspace{0cm}
\hspace{0cm}
\resizebox{8cm}{!}{\includegraphics{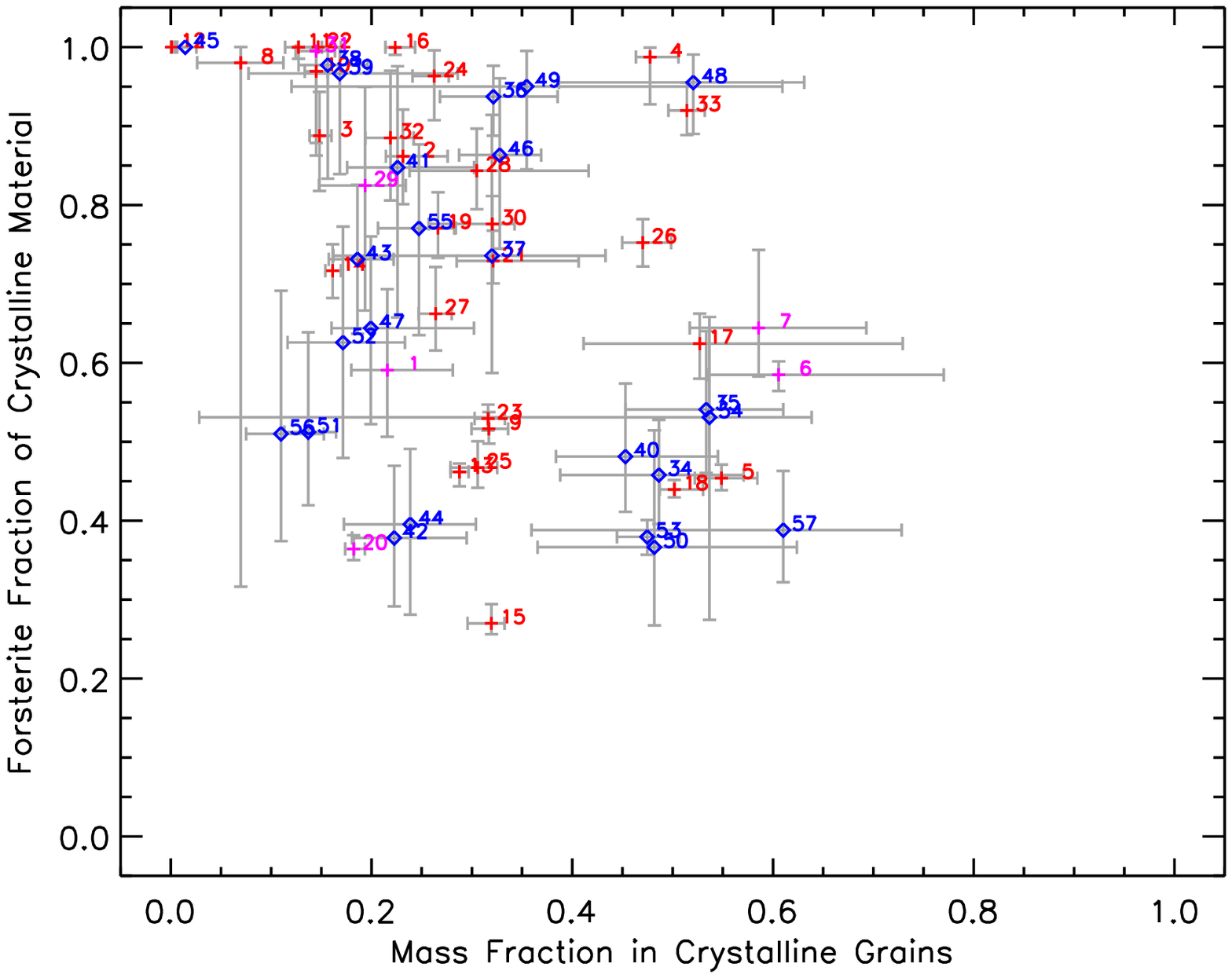}}
\resizebox{8cm}{!}{\includegraphics{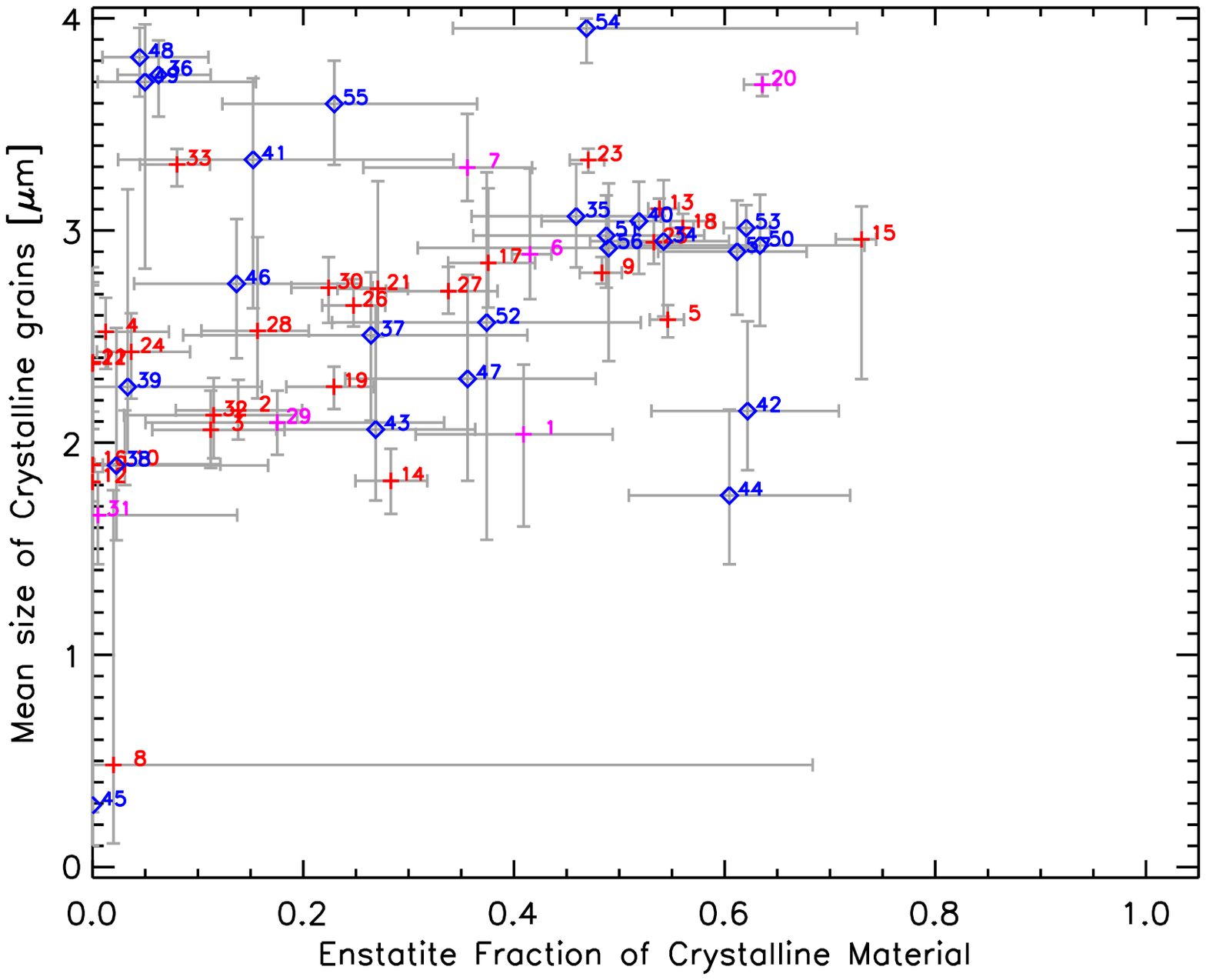}}
\caption{Left: The mass fraction in crystalline grains versus the forsterite fraction in the crystalline material.
Right: The mean size of crystalline grains versus the enstatite fraction in the crystalline material.
Galactic sources are
given in red plus signs and LMC sources in blue diamonds. The magenta symbols depict Galactic sources for which the infrared spectra
only start from 9.9\,$\mu$m. The numbers correspond to numbers given in Tables~\ref{galsterren} and \ref{lmcsterren}.}
\label{forst_cryst}
\label{enstfrac_meansizecryst}

\end{figure}

\begin{figure}[ht]
\vspace{0cm}
\hspace{0cm}
\resizebox{8cm}{!}{\includegraphics{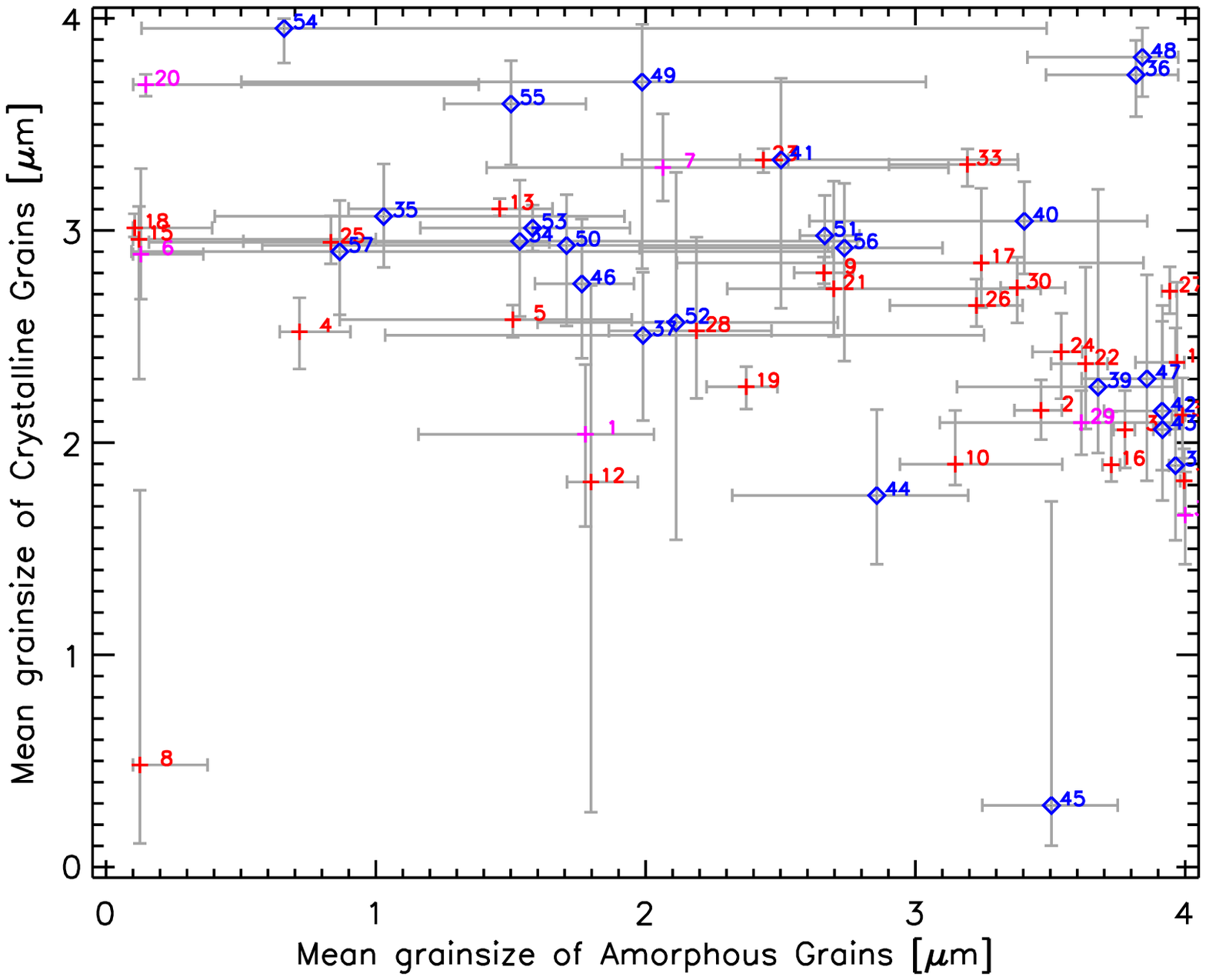}}
\resizebox{8cm}{!}{\includegraphics{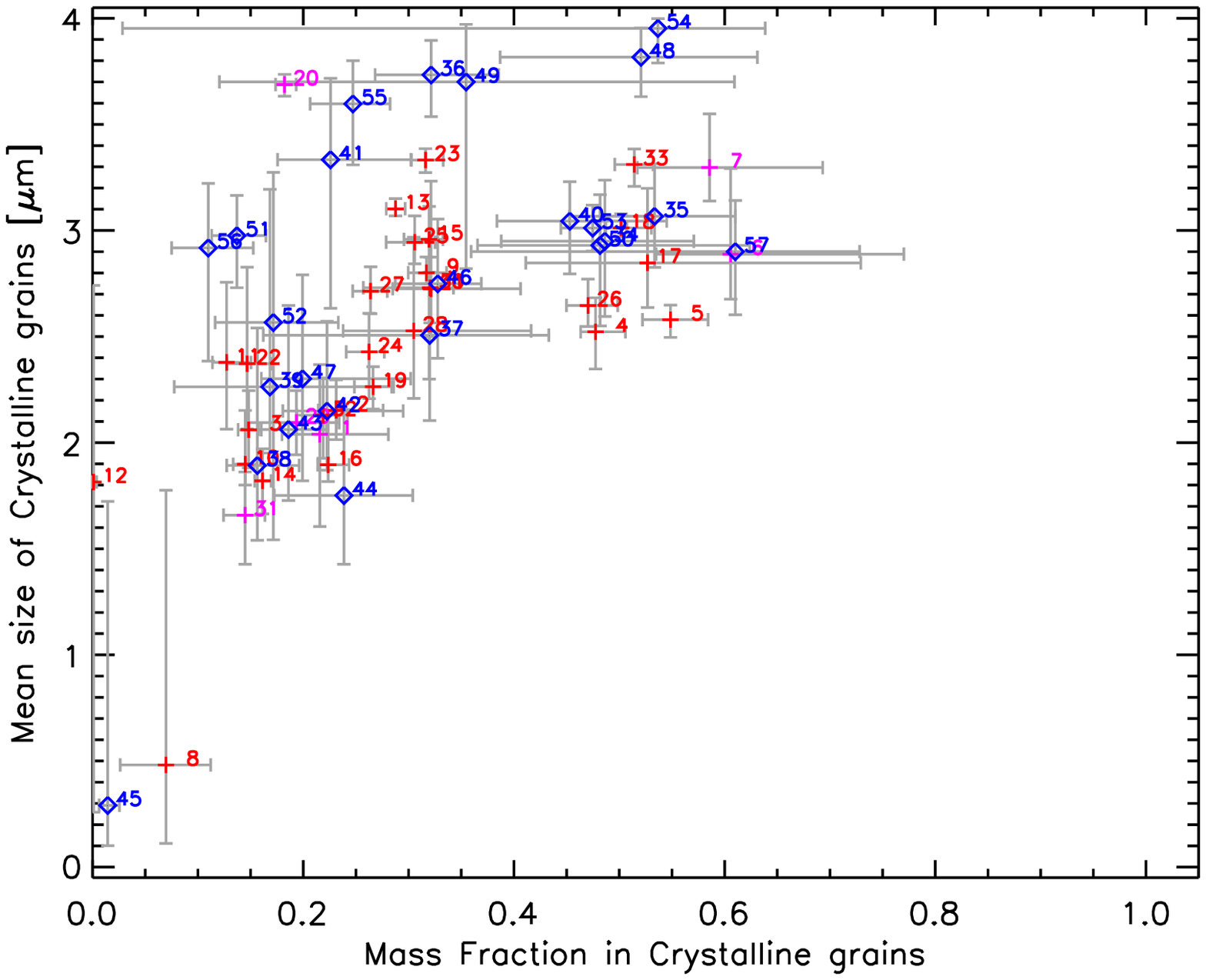}}
\caption{Left: The mean size of crystalline grains versus the mean size of amorphous grains.
Right: The mean size of the crystalline grains versus crystallinity fraction. 
Galactic sources are
given in red plus signs and LMC sources in blue diamonds. The magenta symbols depict Galactic sources for which the infrared spectra
only start from 9.9\,$\mu$m. The numbers correspond to numbers given in Tables~\ref{galsterren} and \ref{lmcsterren}.}
\label{meansizecryst_meansizeamorf}
\label{cryst_meansizecryst}

\end{figure}

\begin{figure}[ht]
\vspace{0cm}
\hspace{0cm}
\resizebox{8cm}{!}{\includegraphics{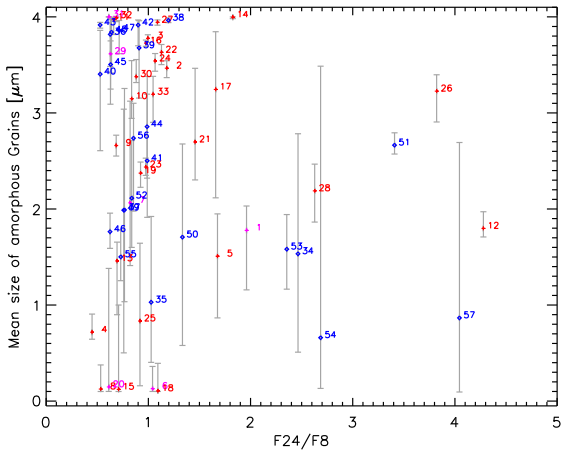}}
\caption{The mean size of the amorphous grains versus the disc flaring, determined by the $F_{24}/F_8$ flux ratio.
Galactic sources are
given in red plus signs and LMC sources in blue diamonds. The magenta symbols depict Galactic sources for which the infrared spectra
only start from 9.9\,$\mu$m. The numbers correspond to numbers given in Tables~\ref{galsterren} and \ref{lmcsterren}.}
\label{amorf_flaring}
\end{figure}


\begin{table}[h]
\caption{ Best fit parameters deduced from our full spectral fitting.
Listed the dust and continuum temperatures and their relative fractions.}
\label{fitresults1}
\centering
\begin{tabular}{llllllll}
\hline \hline
N$^\circ$ & Name &  $T_{dust1}$ & $T_{dust2}$ & Fraction & $T_{cont1}$ & $T_{cont2}$ & Fraction \\
       &              &     (K)     & (K)         & $T_{dust1}$- $T_{dust2}$    & (K)         & (K)         & $T_{cont1}$-$T_{cont2}$   \\
\hline
  1 &EPLyr &$ 100._{   0.}^{   0.}$ &$ 200._{   0.}^{   0.}$ &$ 0.90_{ 0.00}^{ 0.00}- 0.10_{ 0.00}^{ 0.00}$ &$ 200._{   0.}^{   0.}$ &$ 996._{ 111.}^{   4.}$ &$ 0.98_{ 0.00}^{ 0.01}- 0.02_{ 0.01}^{ 0.00}$\\
  2 &HD131356 &$ 200._{   0.}^{   0.}$ &$1000._{   0.}^{   0.}$ &$ 0.90_{ 0.10}^{ 0.00}- 0.10_{ 0.00}^{ 0.10}$ &$ 200._{   0.}^{   0.}$ &$ 500._{   0.}^{   0.}$ &$ 0.90_{ 0.00}^{ 0.01}- 0.10_{ 0.01}^{ 0.00}$\\
  3 &HD213985 &$ 100._{   0.}^{   0.}$ &$1000._{   0.}^{   0.}$ &$ 0.90_{ 0.00}^{ 0.00}- 0.10_{ 0.00}^{ 0.00}$ &$ 200._{   0.}^{   0.}$ &$ 800._{   0.}^{   0.}$ &$ 0.98_{ 0.00}^{ 0.00}- 0.02_{ 0.00}^{ 0.00}$\\
  4 &HD52961 &$ 200._{   0.}^{   0.}$ &$ 704._{   4.}^{ 111.}$ &$ 0.90_{ 0.00}^{ 0.00}- 0.10_{ 0.00}^{ 0.00}$ &$ 100._{   0.}^{   0.}$ &$1000._{   0.}^{   0.}$ &$ 0.99_{ 0.00}^{ 0.00}- 0.01_{ 0.00}^{ 0.00}$\\
  5 &IRAS05208 &$ 426._{ 109.}^{  75.}$ &$ 771._{ 153.}^{ 204.}$ &$ 0.70_{ 0.30}^{ 0.20}- 0.30_{ 0.20}^{ 0.30}$ &$ 200._{   0.}^{   0.}$ &$ 500._{   0.}^{   0.}$ &$ 0.93_{ 0.01}^{ 0.00}- 0.07_{ 0.00}^{ 0.01}$\\
  6 &IRAS06034 &$ 200._{   0.}^{   0.}$ &$ 530._{  30.}^{  71.}$ &$ 0.90_{ 0.00}^{ 0.00}- 0.10_{ 0.00}^{ 0.00}$ &$ 200._{   0.}^{   0.}$ &$ 500._{   0.}^{   0.}$ &$ 0.94_{ 0.00}^{ 0.00}- 0.06_{ 0.00}^{ 0.00}$\\
  7 &IRAS06072 &$ 200._{   0.}^{   0.}$ &$ 680._{ 182.}^{ 296.}$ &$ 0.90_{ 0.00}^{ 0.00}- 0.10_{ 0.00}^{ 0.00}$ &$ 178._{  79.}^{ 124.}$ &$ 564._{ 165.}^{ 263.}$ &$ 0.94_{ 0.02}^{ 0.04}- 0.06_{ 0.04}^{ 0.02}$\\
  8 &IRAS06338 &$ 179._{  80.}^{ 459.}$ &$1000._{   0.}^{   0.}$ &$ 0.60_{ 0.50}^{ 0.30}- 0.40_{ 0.30}^{ 0.50}$ &$ 277._{  79.}^{  23.}$ &$ 859._{ 157.}^{  58.}$ &$ 0.94_{ 0.01}^{ 0.01}- 0.06_{ 0.01}^{ 0.01}$\\
  9 &IRAS09060 &$ 200._{   0.}^{   0.}$ &$ 600._{   0.}^{   0.}$ &$ 0.90_{ 0.00}^{ 0.00}- 0.10_{ 0.00}^{ 0.00}$ &$ 100._{   0.}^{   0.}$ &$ 600._{   0.}^{   0.}$ &$ 0.95_{ 0.01}^{ 0.00}- 0.05_{ 0.00}^{ 0.01}$\\
 10 &IRAS09144 &$ 209._{   9.}^{  96.}$ &$ 627._{  27.}^{ 290.}$ &$ 0.90_{ 0.00}^{ 0.00}- 0.10_{ 0.00}^{ 0.00}$ &$ 200._{   0.}^{   0.}$ &$ 700._{   0.}^{   0.}$ &$ 0.93_{ 0.01}^{ 0.00}- 0.07_{ 0.00}^{ 0.01}$\\
 11 &IRAS09538 &$ 274._{  76.}^{ 319.}$ &$ 699._{ 140.}^{  10.}$ &$ 0.80_{ 0.40}^{ 0.10}- 0.20_{ 0.10}^{ 0.40}$ &$ 376._{  78.}^{  24.}$ &$ 976._{  78.}^{  24.}$ &$ 0.91_{ 0.01}^{ 0.01}- 0.09_{ 0.01}^{ 0.01}$\\
 12 &IRAS10174 &$ 100._{   0.}^{   0.}$ &$ 300._{   0.}^{   0.}$ &$ 0.90_{ 0.00}^{ 0.00}- 0.10_{ 0.00}^{ 0.00}$ &$ 100._{   0.}^{   0.}$ &$ 423._{  23.}^{  79.}$ &$ 0.98_{ 0.00}^{ 0.01}- 0.02_{ 0.01}^{ 0.00}$\\
 13 &IRAS11000 &$ 177._{  79.}^{  23.}$ &$ 426._{  88.}^{  75.}$ &$ 0.80_{ 0.20}^{ 0.10}- 0.20_{ 0.10}^{ 0.20}$ &$ 100._{   0.}^{   0.}$ &$ 603._{   3.}^{ 119.}$ &$ 0.96_{ 0.00}^{ 0.02}- 0.04_{ 0.02}^{ 0.00}$\\
 14 &IRAS13258 &$ 100._{   0.}^{   0.}$ &$ 200._{   0.}^{   0.}$ &$ 0.80_{ 0.00}^{ 0.00}- 0.20_{ 0.00}^{ 0.00}$ &$ 100._{   0.}^{   0.}$ &$ 500._{   0.}^{   0.}$ &$ 0.99_{ 0.00}^{ 0.00}- 0.01_{ 0.00}^{ 0.00}$\\
 15 &IRAS15556 &$ 100._{   0.}^{   0.}$ &$ 200._{   0.}^{   0.}$ &$ 0.20_{ 0.10}^{ 0.50}- 0.80_{ 0.50}^{ 0.10}$ &$ 100._{   0.}^{   0.}$ &$ 689._{  93.}^{  11.}$ &$ 0.99_{ 0.01}^{ 0.00}- 0.01_{ 0.00}^{ 0.01}$\\
 16 &IRAS16230 &$ 200._{   0.}^{   0.}$ &$ 500._{   0.}^{   0.}$ &$ 0.90_{ 0.00}^{ 0.00}- 0.10_{ 0.00}^{ 0.00}$ &$ 100._{   0.}^{   0.}$ &$ 500._{   0.}^{   0.}$ &$ 0.95_{ 0.01}^{ 0.00}- 0.05_{ 0.00}^{ 0.01}$\\
 17 &IRAS17038 &$ 218._{  18.}^{ 135.}$ &$ 952._{  53.}^{  48.}$ &$ 0.90_{ 0.10}^{ 0.00}- 0.10_{ 0.00}^{ 0.10}$ &$ 200._{   0.}^{   0.}$ &$ 513._{  13.}^{  91.}$ &$ 0.95_{ 0.01}^{ 0.01}- 0.05_{ 0.01}^{ 0.01}$\\
 18 &IRAS17233 &$ 320._{  20.}^{  82.}$ &$ 563._{  64.}^{ 137.}$ &$ 0.80_{ 0.10}^{ 0.10}- 0.20_{ 0.10}^{ 0.10}$ &$ 200._{   0.}^{   0.}$ &$ 600._{   0.}^{   0.}$ &$ 0.92_{ 0.00}^{ 0.00}- 0.08_{ 0.00}^{ 0.00}$\\
 19 &IRAS17243 &$ 200._{   0.}^{   0.}$ &$ 500._{   0.}^{   0.}$ &$ 0.90_{ 0.00}^{ 0.00}- 0.10_{ 0.00}^{ 0.00}$ &$ 200._{   0.}^{   0.}$ &$ 600._{   0.}^{   0.}$ &$ 0.90_{ 0.00}^{ 0.00}- 0.10_{ 0.00}^{ 0.00}$\\
 20 &IRAS17530 &$ 100._{   0.}^{   0.}$ &$ 200._{   0.}^{   0.}$ &$ 0.90_{ 0.10}^{ 0.00}- 0.10_{ 0.00}^{ 0.10}$ &$ 100._{   0.}^{   0.}$ &$ 600._{   0.}^{   0.}$ &$ 0.97_{ 0.00}^{ 0.00}- 0.03_{ 0.00}^{ 0.00}$\\
 21 &IRAS18123 &$ 136._{  36.}^{  65.}$ &$ 240._{  40.}^{  79.}$ &$ 0.80_{ 0.10}^{ 0.10}- 0.20_{ 0.10}^{ 0.10}$ &$ 100._{   0.}^{   0.}$ &$ 464._{  65.}^{  36.}$ &$ 0.98_{ 0.00}^{ 0.00}- 0.02_{ 0.00}^{ 0.00}$\\
 22 &IRAS18158 &$ 201._{  66.}^{ 101.}$ &$ 414._{  14.}^{ 201.}$ &$ 0.60_{ 0.20}^{ 0.30}- 0.40_{ 0.30}^{ 0.20}$ &$ 200._{   0.}^{   0.}$ &$ 700._{   0.}^{   0.}$ &$ 0.96_{ 0.00}^{ 0.00}- 0.04_{ 0.00}^{ 0.00}$\\
 23 &IRAS19125 &$ 100._{   0.}^{   0.}$ &$ 200._{   0.}^{   0.}$ &$ 0.90_{ 0.00}^{ 0.00}- 0.10_{ 0.00}^{ 0.00}$ &$ 500._{   0.}^{   0.}$ &$ 900._{   0.}^{   0.}$ &$ 0.91_{ 0.01}^{ 0.00}- 0.09_{ 0.00}^{ 0.01}$\\
 24 &IRAS19157 &$ 200._{   0.}^{   0.}$ &$ 799._{   0.}^{   1.}$ &$ 0.90_{ 0.00}^{ 0.00}- 0.10_{ 0.00}^{ 0.00}$ &$ 200._{   0.}^{   0.}$ &$ 601._{   1.}^{   0.}$ &$ 0.94_{ 0.00}^{ 0.00}- 0.06_{ 0.00}^{ 0.00}$\\
 25 &IRAS20056 &$ 100._{   0.}^{   0.}$ &$ 200._{   0.}^{   0.}$ &$ 0.90_{ 0.10}^{ 0.00}- 0.10_{ 0.00}^{ 0.10}$ &$ 304._{   4.}^{ 111.}$ &$ 850._{  50.}^{  66.}$ &$ 0.91_{ 0.01}^{ 0.01}- 0.09_{ 0.01}^{ 0.01}$\\
 26 &RUCen &$ 277._{ 155.}^{  23.}$ &$ 576._{ 160.}^{  24.}$ &$ 0.90_{ 0.00}^{ 0.00}- 0.10_{ 0.00}^{ 0.00}$ &$ 200._{   0.}^{   0.}$ &$ 596._{ 111.}^{   4.}$ &$ 0.99_{ 0.01}^{ 0.00}- 0.01_{ 0.00}^{ 0.01}$\\
 27 &SAO173329 &$ 101._{   1.}^{   0.}$ &$ 998._{   0.}^{   2.}$ &$ 0.90_{ 0.00}^{ 0.00}- 0.10_{ 0.00}^{ 0.00}$ &$ 200._{   0.}^{   0.}$ &$ 501._{   1.}^{   0.}$ &$ 0.90_{ 0.00}^{ 0.00}- 0.10_{ 0.00}^{ 0.00}$\\
 28 &STPup &$ 203._{   3.}^{ 119.}$ &$ 487._{  90.}^{  25.}$ &$ 0.80_{ 0.10}^{ 0.10}- 0.20_{ 0.10}^{ 0.10}$ &$ 200._{   0.}^{   0.}$ &$ 472._{  73.}^{  28.}$ &$ 0.94_{ 0.04}^{ 0.01}- 0.06_{ 0.01}^{ 0.04}$\\
 29 &SUGem &$ 213._{  18.}^{ 158.}$ &$ 506._{  83.}^{ 176.}$ &$ 0.80_{ 0.10}^{ 0.10}- 0.20_{ 0.10}^{ 0.10}$ &$ 158._{  59.}^{  42.}$ &$ 776._{ 178.}^{ 130.}$ &$ 0.97_{ 0.01}^{ 0.01}- 0.03_{ 0.01}^{ 0.01}$\\
 30 &SXCen &$ 171._{  71.}^{ 494.}$ &$ 990._{  95.}^{  10.}$ &$ 0.80_{ 0.60}^{ 0.10}- 0.20_{ 0.10}^{ 0.60}$ &$ 200._{   0.}^{   0.}$ &$ 617._{  17.}^{  86.}$ &$ 0.93_{ 0.01}^{ 0.03}- 0.07_{ 0.03}^{ 0.01}$\\
 31 &TWCam &$ 206._{  31.}^{  98.}$ &$ 400._{   0.}^{   0.}$ &$ 0.70_{ 0.10}^{ 0.10}- 0.30_{ 0.10}^{ 0.10}$ &$ 100._{   0.}^{   0.}$ &$ 500._{   0.}^{   0.}$ &$ 0.95_{ 0.00}^{ 0.00}- 0.05_{ 0.00}^{ 0.00}$\\
 32 &UYAra &$ 219._{  70.}^{ 437.}$ &$ 869._{  70.}^{  53.}$ &$ 0.70_{ 0.30}^{ 0.10}- 0.30_{ 0.20}^{ 0.20}$ &$ 300._{   0.}^{   0.}$ &$ 800._{   0.}^{   0.}$ &$ 0.91_{ 0.00}^{ 0.00}- 0.09_{ 0.00}^{ 0.00}$\\
 33 &UYCma &$ 200._{   0.}^{   0.}$ &$1000._{   0.}^{   0.}$ &$ 0.90_{ 0.00}^{ 0.00}- 0.10_{ 0.00}^{ 0.00}$ &$ 300._{   0.}^{   0.}$ &$ 900._{   0.}^{   0.}$ &$ 0.97_{ 0.00}^{ 0.00}- 0.03_{ 0.00}^{ 0.00}$\\
 34 &HV12631 &$ 189._{  93.}^{  16.}$ &$ 382._{  84.}^{  40.}$ &$ 0.80_{ 0.50}^{ 0.10}- 0.20_{ 0.10}^{ 0.50}$ &$ 232._{  32.}^{  69.}$ &$ 605._{  81.}^{ 147.}$ &$ 0.93_{ 0.03}^{ 0.02}- 0.07_{ 0.02}^{ 0.03}$\\
 35 &HV2281 &$ 301._{  56.}^{ 109.}$ &$ 984._{ 165.}^{  16.}$ &$ 0.90_{ 0.00}^{ 0.10}- 0.10_{ 0.10}^{ 0.00}$ &$ 203._{   3.}^{ 119.}$ &$ 504._{   4.}^{ 169.}$ &$ 0.85_{ 0.00}^{ 0.00}- 0.15_{ 0.00}^{ 0.00}$\\
 36 &HV2444 &$ 323._{ 227.}^{  86.}$ &$ 505._{   5.}^{ 106.}$ &$ 0.50_{ 0.30}^{ 0.40}- 0.50_{ 0.40}^{ 0.30}$ &$ 575._{ 183.}^{  25.}$ &$ 733._{  39.}^{ 126.}$ &$ 0.81_{ 0.20}^{ 0.10}- 0.19_{ 0.10}^{ 0.20}$\\
 37 &HV2522 &$ 218._{  67.}^{ 126.}$ &$ 916._{ 171.}^{  85.}$ &$ 0.60_{ 0.40}^{ 0.30}- 0.40_{ 0.30}^{ 0.40}$ &$ 304._{   4.}^{ 111.}$ &$ 714._{  44.}^{ 239.}$ &$ 0.87_{ 0.10}^{ 0.04}- 0.13_{ 0.04}^{ 0.10}$\\
 38 &HV2862 &$ 117._{  17.}^{ 135.}$ &$ 613._{  13.}^{ 104.}$ &$ 0.90_{ 0.10}^{ 0.00}- 0.10_{ 0.00}^{ 0.10}$ &$ 200._{   0.}^{   0.}$ &$ 500._{   0.}^{   0.}$ &$ 0.90_{ 0.00}^{ 0.00}- 0.10_{ 0.00}^{ 0.00}$\\
 39 &HV5829 &$ 250._{  60.}^{  86.}$ &$ 783._{ 136.}^{ 141.}$ &$ 0.90_{ 0.10}^{ 0.00}- 0.10_{ 0.00}^{ 0.10}$ &$ 203._{   3.}^{ 119.}$ &$ 656._{  57.}^{  74.}$ &$ 0.92_{ 0.03}^{ 0.03}- 0.08_{ 0.03}^{ 0.03}$\\
 40 &HV915 &$ 230._{  30.}^{ 105.}$ &$ 783._{ 150.}^{ 119.}$ &$ 0.90_{ 0.10}^{ 0.00}- 0.10_{ 0.00}^{ 0.10}$ &$ 520._{ 213.}^{  81.}$ &$ 719._{  56.}^{ 101.}$ &$ 0.75_{ 0.54}^{ 0.17}- 0.25_{ 0.17}^{ 0.54}$\\
 41 &J044458.18-703522.8 &$ 209._{   9.}^{  96.}$ &$ 452._{  52.}^{ 191.}$ &$ 0.90_{ 0.00}^{ 0.00}- 0.10_{ 0.00}^{ 0.00}$ &$ 431._{ 236.}^{  70.}$ &$ 681._{  83.}^{  33.}$ &$ 0.91_{ 0.01}^{ 0.01}- 0.09_{ 0.01}^{ 0.01}$\\
 42 &J045242.93-704737.4 &$ 243._{  49.}^{ 139.}$ &$ 712._{ 104.}^{ 207.}$ &$ 0.80_{ 0.20}^{ 0.10}- 0.20_{ 0.10}^{ 0.20}$ &$ 196._{ 111.}^{   4.}$ &$ 695._{ 106.}^{   5.}$ &$ 0.95_{ 0.02}^{ 0.01}- 0.05_{ 0.01}^{ 0.02}$\\
 43 &J050143.18-694048.7 &$ 100._{   0.}^{   0.}$ &$1000._{   0.}^{   0.}$ &$ 0.00_{ 0.00}^{ 0.00}- 1.00_{ 0.00}^{ 0.00}$ &$ 400._{   0.}^{   0.}$ &$1000._{   0.}^{   0.}$ &$ 0.90_{ 0.00}^{ 0.00}- 0.10_{ 0.00}^{ 0.00}$\\
 44 &J051159.11-692532.8 &$ 204._{  27.}^{ 186.}$ &$ 709._{ 114.}^{ 120.}$ &$ 0.90_{ 0.10}^{ 0.00}- 0.10_{ 0.00}^{ 0.10}$ &$ 313._{  31.}^{  90.}$ &$ 806._{  58.}^{ 120.}$ &$ 0.93_{ 0.02}^{ 0.01}- 0.07_{ 0.01}^{ 0.02}$\\
 45 &J051333.74-663419.1 &$ 226._{ 127.}^{ 684.}$ &$ 990._{  95.}^{  10.}$ &$ 0.10_{ 0.10}^{ 0.50}- 0.90_{ 0.50}^{ 0.10}$ &$ 625._{  44.}^{  82.}$ &$ 999._{   0.}^{   1.}$ &$ 0.90_{ 0.00}^{ 0.04}- 0.10_{ 0.04}^{ 0.00}$\\
 46 &J052220.87-655551.6 &$ 200._{   0.}^{   0.}$ &$ 986._{  89.}^{  14.}$ &$ 0.90_{ 0.00}^{ 0.00}- 0.10_{ 0.00}^{ 0.00}$ &$ 200._{   0.}^{   0.}$ &$1000._{   0.}^{   0.}$ &$ 0.97_{ 0.00}^{ 0.00}- 0.03_{ 0.00}^{ 0.00}$\\
 47 &J053605.56-695802.9 &$ 183._{  85.}^{  50.}$ &$ 812._{ 172.}^{ 117.}$ &$ 0.90_{ 0.10}^{ 0.00}- 0.10_{ 0.00}^{ 0.10}$ &$ 123._{  23.}^{ 166.}$ &$ 815._{  15.}^{ 152.}$ &$ 0.98_{ 0.07}^{ 0.01}- 0.02_{ 0.01}^{ 0.07}$\\
 48 &J054312.52-683356.9 &$ 200._{   0.}^{   0.}$ &$ 950._{ 153.}^{  50.}$ &$ 0.90_{ 0.10}^{ 0.00}- 0.10_{ 0.00}^{ 0.10}$ &$ 400._{   0.}^{   0.}$ &$ 996._{ 111.}^{   4.}$ &$ 0.94_{ 0.02}^{ 0.00}- 0.06_{ 0.00}^{ 0.02}$\\
 49 &MACHO78.6698.38 &$ 125._{  25.}^{  76.}$ &$ 533._{ 234.}^{ 329.}$ &$ 0.90_{ 0.00}^{ 0.00}- 0.10_{ 0.00}^{ 0.00}$ &$ 151._{  52.}^{  50.}$ &$ 701._{   1.}^{   0.}$ &$ 0.94_{ 0.04}^{ 0.01}- 0.06_{ 0.01}^{ 0.04}$\\
 50 &MACHO82840515 &$ 208._{   8.}^{  98.}$ &$ 501._{  74.}^{ 100.}$ &$ 0.90_{ 0.10}^{ 0.00}- 0.10_{ 0.00}^{ 0.10}$ &$ 300._{   0.}^{   0.}$ &$ 500._{   0.}^{   0.}$ &$ 0.82_{ 0.02}^{ 0.03}- 0.18_{ 0.03}^{ 0.02}$\\
 51 &MSX949 &$ 100._{   0.}^{   0.}$ &$ 200._{   0.}^{   0.}$ &$ 0.70_{ 0.10}^{ 0.00}- 0.30_{ 0.00}^{ 0.10}$ &$ 200._{   0.}^{   0.}$ &$ 500._{   0.}^{   0.}$ &$ 0.95_{ 0.00}^{ 0.00}- 0.05_{ 0.00}^{ 0.00}$\\
 52 &NGC1805SAGEIRS1 &$ 242._{ 115.}^{  88.}$ &$ 582._{ 142.}^{ 243.}$ &$ 0.80_{ 0.40}^{ 0.10}- 0.20_{ 0.10}^{ 0.40}$ &$ 289._{  93.}^{  20.}$ &$ 786._{  89.}^{  30.}$ &$ 0.89_{ 0.06}^{ 0.01}- 0.11_{ 0.01}^{ 0.06}$\\
 53 &SAGE050830 &$ 218._{  18.}^{  84.}$ &$ 418._{  18.}^{  84.}$ &$ 0.80_{ 0.00}^{ 0.10}- 0.20_{ 0.10}^{ 0.00}$ &$ 248._{  48.}^{ 320.}$ &$ 877._{  78.}^{  69.}$ &$ 0.89_{ 0.24}^{ 0.05}- 0.12_{ 0.05}^{ 0.24}$\\
 54 &SAGE051453 &$ 184._{  87.}^{  16.}$ &$ 376._{ 107.}^{  24.}$ &$ 0.90_{ 0.10}^{ 0.00}- 0.10_{ 0.00}^{ 0.10}$ &$ 200._{   0.}^{   0.}$ &$ 412._{  12.}^{  92.}$ &$ 0.90_{ 0.13}^{ 0.02}- 0.10_{ 0.02}^{ 0.13}$\\
 55 &SAGE052707 &$ 203._{  11.}^{ 279.}$ &$ 681._{ 112.}^{ 120.}$ &$ 0.90_{ 0.10}^{ 0.00}- 0.10_{ 0.00}^{ 0.10}$ &$ 298._{ 139.}^{   2.}$ &$ 998._{ 139.}^{   2.}$ &$ 0.90_{ 0.00}^{ 0.07}- 0.10_{ 0.07}^{ 0.00}$\\
 56 &SAGE052747 &$ 248._{  48.}^{  53.}$ &$ 524._{ 116.}^{ 111.}$ &$ 0.90_{ 0.10}^{ 0.10}- 0.10_{ 0.10}^{ 0.10}$ &$ 331._{  31.}^{  70.}$ &$ 990._{  95.}^{  10.}$ &$ 0.88_{ 0.03}^{ 0.02}- 0.12_{ 0.02}^{ 0.03}$\\
 57 &SAGE054310 &$ 197._{ 119.}^{   3.}$ &$ 465._{  81.}^{  35.}$ &$ 0.90_{ 0.00}^{ 0.00}- 0.10_{ 0.00}^{ 0.00}$ &$ 200._{   0.}^{   0.}$ &$ 430._{  30.}^{  71.}$ &$ 0.95_{ 0.00}^{ 0.00}- 0.05_{ 0.00}^{ 0.00}$\\

\hline
\end{tabular}
\end{table}

\begin{table*}
\caption{Best fit parameters deduced from our full spectral fitting. The abundances of small, medium and large grains of the various
dust species are given as fractions of the total mass, excluding the dust responsible for the continuum emission.}
\label{fitresults2}
\centering
\begin{tabular}{lcccc}
\hline \hline   
 N$^\circ$ & MgOlivine & MgPyroxene & MgFeOlivine & MgFePyroxene \\
                & Small - Medium -  Large & Small - Medium -  Large & Small -Medium  -  Large  &  Small - Medium -   Large \\
\hline
  1    &$ 0.00_{ 0.00}^{ 0.00}    -	 0.00_{ 0.00}^{ 0.00}    -   0.00_{ 0.00}^{ 0.00}$    &$ 0.05_{ 0.05}^{ 0.00}    -	 0.00_{ 0.00}^{ 0.00}    -   0.00_{ 0.00}^{ 0.00}$    &$44.26_{ 4.57}^{ 6.75}    -	 0.00_{ 0.00}^{ 0.00}    -  34.12_{13.02}^{ 5.73}$    &$ 0.00_{ 0.00}^{ 0.00}    -	 0.00_{ 0.00}^{ 0.00}    -   0.00_{ 0.00}^{ 0.00}$\\
  2    &$ 1.24_{ 1.08}^{ 3.32}    -	 0.00_{ 0.00}^{ 0.00}    -  14.39_{ 5.46}^{ 3.41}$    &$ 9.29_{ 1.89}^{ 0.94}    -	 0.00_{ 0.00}^{ 0.00}    -   3.31_{ 2.94}^{ 7.60}$    &$ 0.00_{ 0.00}^{ 0.00}    -	 0.00_{ 0.00}^{ 0.00}    -   0.62_{ 0.62}^{ 6.21}$    &$ 0.00_{ 0.00}^{ 0.00}    -	 0.00_{ 0.00}^{ 0.00}    -  44.60_{17.07}^{ 5.33}$\\
  3    &$ 0.00_{ 0.00}^{ 0.00}    -	 0.00_{ 0.00}^{ 0.00}    -  16.25_{ 2.57}^{ 2.46}$    &$ 4.75_{ 0.83}^{ 0.92}    -	 0.00_{ 0.00}^{ 0.00}    -  11.16_{ 5.11}^{ 4.11}$    &$ 0.00_{ 0.00}^{ 0.00}    -	 0.00_{ 0.00}^{ 0.00}    -   0.01_{ 0.02}^{ 0.00}$    &$ 0.00_{ 0.00}^{ 0.00}    -	 0.00_{ 0.00}^{ 0.00}    -  49.23_{ 4.85}^{ 6.12}$\\
  4    &$ 0.00_{ 0.00}^{ 0.00}    -	 0.00_{ 0.00}^{ 0.00}    -   0.00_{ 0.00}^{ 0.00}$    &$36.80_{ 3.63}^{ 1.40}    -	 0.00_{ 0.00}^{ 0.00}    -   0.00_{ 0.00}^{ 0.00}$    &$ 0.00_{ 0.00}^{ 0.00}    -	 0.00_{ 0.00}^{ 0.00}    -   0.00_{ 0.00}^{ 0.00}$    &$ 0.00_{ 0.00}^{ 0.00}    -	 0.00_{ 0.00}^{ 0.00}    -   0.00_{ 0.00}^{ 0.00}$\\
  5    &$ 0.00_{ 0.00}^{ 0.00}    -	 0.00_{ 0.00}^{ 0.00}    -   0.00_{ 0.00}^{ 0.00}$    &$11.06_{ 2.61}^{ 1.79}    -	 0.00_{ 0.00}^{ 0.00}    -   1.18_{ 1.15}^{ 3.16}$    &$17.15_{ 5.15}^{ 7.62}    -	 0.00_{ 0.00}^{ 0.00}    -  15.48_{ 7.14}^{ 5.25}$    &$ 0.00_{ 0.00}^{ 0.00}    -	 0.00_{ 0.00}^{ 0.00}    -   0.00_{ 0.00}^{ 0.00}$\\
  6    &$34.95_{26.48}^{11.28}    -	 0.00_{ 0.00}^{ 0.00}    -   0.00_{ 0.00}^{ 0.00}$    &$ 4.06_{ 3.88}^{ 9.11}    -	 0.00_{ 0.00}^{ 0.00}    -   0.00_{ 0.00}^{ 0.00}$    &$ 0.00_{ 0.00}^{ 0.00}    -	 0.00_{ 0.00}^{ 0.00}    -   0.00_{ 0.00}^{ 0.00}$    &$ 0.00_{ 0.00}^{ 0.00}    -	 0.00_{ 0.00}^{ 0.00}    -   0.00_{ 0.00}^{ 0.00}$\\
  7    &$ 2.71_{ 2.58}^{ 5.19}    -	 0.00_{ 0.00}^{ 0.00}    -  10.13_{ 9.95}^{ 6.89}$    &$ 0.42_{ 0.41}^{ 1.62}    -	 0.00_{ 0.00}^{ 0.00}    -   0.18_{ 0.18}^{ 3.08}$    &$19.04_{19.29}^{12.34}    -	 0.00_{ 0.00}^{ 0.00}    -   8.80_{ 8.80}^{14.46}$    &$ 0.00_{ 0.00}^{ 0.00}    -	 0.00_{ 0.00}^{ 0.00}    -   0.00_{ 0.00}^{ 0.00}$\\
  8    &$ 0.00_{ 0.00}^{ 0.00}    -	 0.00_{ 0.00}^{ 0.00}    -   0.00_{ 0.00}^{ 0.00}$    &$61.88_{29.71}^{10.53}    -	 0.00_{ 0.00}^{ 0.00}    -   0.35_{ 0.35}^{11.55}$    &$10.11_{10.18}^{36.34}    -	 0.00_{ 0.00}^{ 0.00}    -   0.00_{ 0.00}^{ 0.00}$    &$ 0.00_{ 0.00}^{ 0.00}    -	 0.00_{ 0.00}^{ 0.00}    -   0.00_{ 0.00}^{ 0.00}$\\
  9    &$ 0.00_{ 0.00}^{ 0.00}    -	 0.00_{ 0.00}^{ 0.00}    -  17.23_{ 5.91}^{ 2.92}$    &$21.53_{ 0.94}^{ 1.05}    -	 0.00_{ 0.00}^{ 0.00}    -   6.72_{ 3.20}^{ 3.37}$    &$ 1.89_{ 1.44}^{ 1.78}    -	 0.00_{ 0.00}^{ 0.00}    -  18.43_{ 3.13}^{ 6.89}$    &$ 0.00_{ 0.00}^{ 0.00}    -	 0.00_{ 0.00}^{ 0.00}    -   0.00_{ 0.00}^{ 0.00}$\\
 10    &$ 0.00_{ 0.00}^{ 0.00}    -	 0.00_{ 0.00}^{ 0.00}    -  12.05_{ 9.01}^{ 4.80}$    &$10.29_{ 4.30}^{ 1.57}    -	 0.00_{ 0.00}^{ 0.00}    -  16.91_{ 5.12}^{ 8.84}$    &$ 0.11_{ 0.11}^{ 4.84}    -	12.48_{ 9.12}^{ 5.73}    -  21.69_{ 5.87}^{19.58}$    &$ 0.00_{ 0.00}^{ 0.00}    -	 0.00_{ 0.00}^{ 0.00}    -   2.85_{ 2.61}^{ 5.75}$\\
 11    &$ 0.00_{ 0.00}^{ 0.00}    -	 0.00_{ 0.00}^{ 0.00}    -   5.66_{ 5.66}^{19.27}$    &$ 0.65_{ 0.60}^{ 3.11}    -	 0.00_{ 0.00}^{ 0.00}    -  22.03_{ 4.83}^{ 4.15}$    &$ 0.00_{ 0.00}^{ 0.34}    -	 0.00_{ 0.00}^{ 0.00}    -  48.58_{30.81}^{10.07}$    &$ 0.00_{ 0.00}^{ 0.00}    -	 0.00_{ 0.00}^{ 0.00}    -   3.06_{ 3.07}^{15.32}$\\
 12    &$24.17_{ 5.77}^{ 3.02}    -	 0.79_{ 0.79}^{ 3.98}    -   0.28_{ 0.28}^{ 2.49}$    &$ 0.01_{ 0.01}^{ 0.00}    -	13.60_{ 6.42}^{ 4.15}    -   8.45_{ 5.02}^{ 6.89}$    &$24.29_{ 3.53}^{ 4.79}    -	 1.14_{ 1.15}^{ 4.54}    -  27.00_{ 3.73}^{ 3.34}$    &$ 0.00_{ 0.00}^{ 0.00}    -	 0.00_{ 0.00}^{ 0.00}    -   0.00_{ 0.00}^{ 0.00}$\\
 13    &$ 6.79_{ 4.62}^{ 8.22}    -	 0.84_{ 0.84}^{ 7.81}    -   1.71_{ 1.69}^{ 8.93}$    &$20.96_{ 4.30}^{ 3.29}    -	 0.00_{ 0.00}^{ 0.00}    -   0.03_{ 0.03}^{ 1.29}$    &$17.07_{ 4.78}^{ 6.63}    -	 0.16_{ 0.16}^{14.54}    -  22.17_{12.91}^{ 3.65}$    &$ 0.03_{ 0.03}^{ 2.32}    -	 0.00_{ 0.00}^{ 0.00}    -   0.00_{ 0.00}^{ 0.00}$\\
 14    &$ 0.00_{ 0.00}^{ 0.00}    -	 0.00_{ 0.00}^{ 0.00}    -   0.00_{ 0.00}^{ 0.00}$    &$ 0.00_{ 0.00}^{ 0.00}    -	 0.00_{ 0.00}^{ 0.00}    -   0.00_{ 0.00}^{ 0.00}$    &$ 0.00_{ 0.00}^{ 0.00}    -	 0.00_{ 0.00}^{ 0.00}    -  83.47_{ 0.88}^{ 0.75}$    &$ 0.00_{ 0.00}^{ 0.00}    -	 0.00_{ 0.00}^{ 0.00}    -   0.00_{ 0.00}^{ 0.00}$\\
 15    &$ 0.00_{ 0.00}^{ 0.00}    -	 0.00_{ 0.00}^{ 0.00}    -   0.00_{ 0.00}^{ 0.00}$    &$ 1.64_{ 1.64}^{14.18}    -	 0.00_{ 0.00}^{ 0.00}    -   0.00_{ 0.00}^{ 0.00}$    &$66.37_{ 9.89}^{ 1.96}    -	 0.00_{ 0.00}^{ 0.00}    -   0.13_{ 0.13}^{ 0.00}$    &$ 0.07_{ 0.07}^{ 3.02}    -	 0.00_{ 0.00}^{ 0.00}    -   0.00_{ 0.00}^{ 0.00}$\\
 16    &$ 0.00_{ 0.00}^{ 0.00}    -	 0.00_{ 0.00}^{ 0.00}    -  23.00_{ 6.45}^{ 3.41}$    &$ 5.21_{ 0.71}^{ 0.61}    -	 0.00_{ 0.00}^{ 0.00}    -   0.00_{ 0.00}^{ 0.00}$    &$ 0.00_{ 0.00}^{ 0.00}    -	 0.00_{ 0.00}^{ 0.00}    -  41.98_{ 2.96}^{ 4.35}$    &$ 0.00_{ 0.00}^{ 0.00}    -	 0.00_{ 0.00}^{ 0.00}    -   0.00_{ 0.00}^{ 0.00}$\\
 17    &$ 0.00_{ 0.00}^{ 0.00}    -	 0.00_{ 0.00}^{ 0.00}    -  19.60_{10.76}^{ 8.83}$    &$ 3.29_{ 3.02}^{ 5.27}    -	 0.00_{ 0.00}^{ 0.00}    -  10.87_{10.38}^{10.28}$    &$ 1.21_{ 1.17}^{ 3.42}    -	 0.00_{ 0.00}^{ 0.00}    -   7.86_{ 6.97}^{28.56}$    &$ 0.00_{ 0.00}^{ 0.00}    -	 0.00_{ 0.00}^{ 0.00}    -   1.32_{ 1.31}^{ 5.80}$\\
 18    &$ 0.92_{ 0.88}^{ 2.77}    -	 0.00_{ 0.00}^{ 0.00}    -   0.05_{ 0.05}^{ 3.11}$    &$ 8.28_{ 2.02}^{ 1.61}    -	 0.00_{ 0.00}^{ 0.00}    -   0.00_{ 0.00}^{ 0.00}$    &$38.44_{ 2.93}^{ 3.41}    -	 0.00_{ 0.00}^{ 0.00}    -   0.00_{ 0.00}^{ 0.00}$    &$ 0.00_{ 0.00}^{ 0.00}    -	 0.00_{ 0.00}^{ 0.00}    -   0.00_{ 0.00}^{ 0.00}$\\
 19    &$ 0.00_{ 0.00}^{ 0.00}    -	 0.00_{ 0.00}^{ 0.00}    -  14.95_{ 3.36}^{ 3.03}$    &$ 6.52_{ 1.17}^{ 1.16}    -	 0.00_{ 0.00}^{ 0.00}    -   0.00_{ 0.00}^{ 0.00}$    &$24.04_{ 2.70}^{ 3.25}    -	 0.02_{ 0.02}^{ 0.00}    -  17.42_{ 3.22}^{ 3.40}$    &$ 0.00_{ 0.00}^{ 0.00}    -	 0.00_{ 0.00}^{ 0.00}    -   0.00_{ 0.00}^{ 0.00}$\\
 20    &$20.41_{ 6.32}^{ 2.77}    -	 0.00_{ 0.00}^{ 0.00}    -   0.00_{ 0.00}^{ 0.00}$    &$21.91_{ 6.34}^{ 3.86}    -	 0.69_{ 0.69}^{11.80}    -   0.00_{ 0.00}^{ 0.00}$    &$36.06_{ 5.44}^{ 4.86}    -	 0.25_{ 0.25}^{20.56}    -   0.51_{ 0.51}^{20.09}$    &$ 0.00_{ 0.00}^{ 0.00}    -	 0.00_{ 0.00}^{ 0.00}    -   0.00_{ 0.00}^{ 0.00}$\\
 21    &$ 0.00_{ 0.00}^{ 0.00}    -	 0.18_{ 0.18}^{ 2.12}    -  14.92_{14.64}^{32.27}$    &$ 0.00_{ 0.00}^{ 0.00}    -	 0.00_{ 0.00}^{ 0.00}    -   1.61_{ 1.59}^{ 8.25}$    &$22.95_{14.24}^{ 7.77}    -	 0.00_{ 0.00}^{ 0.00}    -  26.96_{24.87}^{13.46}$    &$ 0.00_{ 0.00}^{ 0.00}    -	 0.00_{ 0.00}^{ 0.00}    -   0.00_{ 0.00}^{ 0.00}$\\
 22    &$ 0.43_{ 0.44}^{ 3.40}    -	 0.00_{ 0.00}^{ 0.00}    -   6.47_{ 5.14}^{ 6.06}$    &$ 6.55_{ 2.12}^{ 1.81}    -	 0.35_{ 0.35}^{ 5.38}    -   5.98_{ 4.17}^{ 4.68}$    &$ 0.90_{ 0.90}^{ 4.91}    -	 0.00_{ 0.00}^{ 0.00}    -  52.59_{ 5.85}^{ 5.37}$    &$ 0.00_{ 0.00}^{ 0.00}    -	 0.00_{ 0.00}^{ 0.00}    -   0.34_{ 0.34}^{ 5.03}$\\
 23    &$ 3.57_{ 2.61}^{ 2.24}    -	 0.00_{ 0.00}^{ 0.00}    -   0.35_{ 0.35}^{ 2.94}$    &$14.90_{ 2.41}^{ 2.59}    -	 0.00_{ 0.00}^{ 0.00}    -   0.00_{ 0.00}^{ 0.00}$    &$ 6.19_{ 3.63}^{ 4.37}    -	 0.00_{ 0.00}^{ 0.00}    -  40.44_{ 1.67}^{ 1.90}$    &$ 0.00_{ 0.00}^{ 0.00}    -	 0.00_{ 0.00}^{ 0.00}    -   0.00_{ 0.00}^{ 0.00}$\\
 24    &$ 0.56_{ 0.55}^{ 4.14}    -	 0.00_{ 0.00}^{ 0.00}    -  47.31_{ 3.41}^{ 2.76}$    &$ 7.82_{ 1.52}^{ 1.09}    -	 0.00_{ 0.00}^{ 0.00}    -   3.83_{ 2.98}^{ 3.22}$    &$ 0.00_{ 0.00}^{ 0.00}    -	 0.00_{ 0.00}^{ 0.00}    -   0.33_{ 0.33}^{12.32}$    &$ 0.00_{ 0.00}^{ 0.00}    -	 0.00_{ 0.00}^{ 0.00}    -   0.00_{ 0.00}^{ 0.00}$\\
 25    &$ 5.12_{ 4.82}^{ 6.22}    -	 0.00_{ 0.00}^{ 0.00}    -   0.00_{ 0.00}^{ 0.00}$    &$ 1.57_{ 1.50}^{ 3.46}    -	 0.00_{ 0.00}^{ 0.00}    -   0.00_{ 0.00}^{ 0.00}$    &$49.34_{ 6.66}^{ 6.75}    -	 0.00_{ 0.00}^{ 0.00}    -  10.66_{10.57}^{14.52}$    &$ 0.01_{ 0.01}^{ 0.73}    -	 0.00_{ 0.00}^{ 0.00}    -   1.44_{ 1.43}^{ 9.16}$\\
 26    &$ 0.00_{ 0.00}^{ 0.00}    -	 0.00_{ 0.00}^{ 0.00}    -   2.86_{ 2.68}^{ 5.64}$    &$ 4.32_{ 2.02}^{ 1.30}    -	 0.00_{ 0.00}^{ 0.00}    -   0.00_{ 0.00}^{ 0.00}$    &$ 4.14_{ 3.01}^{ 5.16}    -	 0.29_{ 0.29}^{26.96}    -  38.25_{ 7.80}^{ 4.60}$    &$ 0.00_{ 0.00}^{ 0.00}    -	 0.00_{ 0.00}^{ 0.00}    -   0.00_{ 0.00}^{ 0.00}$\\
 27    &$ 0.00_{ 0.00}^{ 0.00}    -	 0.00_{ 0.00}^{ 0.00}    -  20.26_{ 7.02}^{ 4.30}$    &$ 0.01_{ 0.01}^{ 0.00}    -	 0.00_{ 0.00}^{ 0.00}    -   5.61_{ 2.40}^{ 2.65}$    &$ 0.00_{ 0.00}^{ 0.00}    -	 0.00_{ 0.00}^{ 0.00}    -   6.28_{ 4.89}^{11.78}$    &$ 0.00_{ 0.00}^{ 0.00}    -	 0.00_{ 0.00}^{ 0.00}    -  29.29_{ 6.57}^{ 4.04}$\\
 28    &$ 0.00_{ 0.00}^{ 0.00}    -	 0.00_{ 0.00}^{ 0.00}    -  10.32_{ 6.89}^{ 8.18}$    &$ 0.67_{ 0.67}^{ 3.39}    -	 0.00_{ 0.00}^{ 0.00}    -   7.62_{ 4.37}^{ 5.90}$    &$25.20_{ 4.23}^{ 3.63}    -	 4.95_{ 4.70}^{ 8.99}    -   7.79_{ 7.00}^{13.98}$    &$ 0.00_{ 0.00}^{ 0.00}    -	 0.00_{ 0.00}^{ 0.00}    -   4.76_{ 4.54}^{ 8.42}$\\
 29    &$ 0.00_{ 0.00}^{ 0.00}    -	 0.00_{ 0.00}^{ 0.00}    -  18.98_{13.72}^{14.04}$    &$ 0.04_{ 0.04}^{ 0.67}    -	 0.00_{ 0.00}^{ 0.00}    -   0.00_{ 0.00}^{ 0.00}$    &$ 8.32_{ 8.29}^{11.53}    -	 0.06_{ 0.06}^{ 1.73}    -  52.42_{20.47}^{16.56}$    &$ 0.00_{ 0.00}^{ 0.00}    -	 0.00_{ 0.00}^{ 0.00}    -   0.00_{ 0.00}^{ 0.00}$\\
 30    &$ 0.00_{ 0.00}^{ 0.00}    -	 0.00_{ 0.00}^{ 0.00}    -   7.67_{ 5.27}^{ 6.21}$    &$10.14_{ 2.08}^{ 0.70}    -	 0.00_{ 0.00}^{ 0.00}    -   0.00_{ 0.00}^{ 0.00}$    &$ 0.00_{ 0.00}^{ 0.00}    -	 0.00_{ 0.00}^{ 0.00}    -  36.42_{ 7.28}^{17.15}$    &$ 0.00_{ 0.00}^{ 0.00}    -	 0.00_{ 0.00}^{ 0.00}    -   4.61_{ 3.32}^{ 2.50}$\\
 31    &$ 0.00_{ 0.00}^{ 0.00}    -	 0.00_{ 0.00}^{ 0.00}    -  79.45_{11.38}^{ 4.73}$    &$ 0.00_{ 0.00}^{ 0.00}    -	 0.00_{ 0.00}^{ 0.00}    -   0.00_{ 0.00}^{ 0.00}$    &$ 0.00_{ 0.00}^{ 0.00}    -	 0.00_{ 0.00}^{ 0.00}    -   2.56_{ 2.46}^{ 9.19}$    &$ 0.00_{ 0.00}^{ 0.00}    -	 0.00_{ 0.00}^{ 0.00}    -   0.00_{ 0.00}^{ 0.00}$\\
 32    &$ 0.00_{ 0.00}^{ 0.00}    -	 0.00_{ 0.00}^{ 0.00}    -   8.47_{ 4.70}^{ 4.59}$    &$ 0.14_{ 0.14}^{ 1.07}    -	 0.00_{ 0.00}^{ 0.00}    -  20.70_{ 6.47}^{ 5.91}$    &$ 0.00_{ 0.00}^{ 0.00}    -	 0.00_{ 0.00}^{ 0.00}    -   0.24_{ 0.24}^{ 4.08}$    &$ 0.00_{ 0.00}^{ 0.00}    -	 0.00_{ 0.00}^{ 0.00}    -  47.64_{ 7.17}^{ 8.11}$\\
 33    &$ 2.28_{ 2.04}^{ 4.14}    -	 0.07_{ 0.07}^{ 3.41}    -   5.40_{ 3.33}^{ 3.30}$    &$ 6.36_{ 1.06}^{ 1.13}    -	 0.00_{ 0.00}^{ 0.00}    -  26.60_{ 2.71}^{ 2.80}$    &$ 0.00_{ 0.00}^{ 0.00}    -	 0.00_{ 0.00}^{ 0.00}    -   0.00_{ 0.00}^{ 0.00}$    &$ 0.00_{ 0.00}^{ 0.00}    -	 0.00_{ 0.00}^{ 0.00}    -   0.00_{ 0.00}^{ 0.00}$\\
 34    &$ 0.00_{ 0.00}^{ 0.00}    -	 0.00_{ 0.00}^{ 0.00}    -   0.01_{ 0.01}^{ 0.00}$    &$ 7.20_{ 4.20}^{ 4.04}    -	 0.00_{ 0.00}^{ 0.00}    -  10.02_{ 8.77}^{15.05}$    &$18.35_{11.11}^{11.41}    -	 2.03_{ 1.99}^{12.79}    -   9.83_{ 8.75}^{13.55}$    &$ 0.00_{ 0.00}^{ 0.00}    -	 0.00_{ 0.00}^{ 0.00}    -   0.12_{ 0.12}^{ 9.23}$\\
 35    &$ 0.01_{ 0.01}^{ 0.00}    -	 0.00_{ 0.00}^{ 0.00}    -   0.00_{ 0.00}^{ 0.00}$    &$ 1.18_{ 1.16}^{ 5.47}    -	 0.43_{ 0.43}^{ 6.74}    -   6.50_{ 6.00}^{11.11}$    &$20.83_{10.49}^{10.49}    -	 1.86_{ 1.86}^{13.23}    -   1.04_{ 1.04}^{12.42}$    &$ 9.47_{ 8.10}^{10.78}    -	 1.04_{ 1.05}^{14.14}    -   2.58_{ 2.53}^{ 9.53}$\\
 36    &$ 2.28_{ 2.06}^{ 6.41}    -	 0.28_{ 0.28}^{ 5.70}    -   0.00_{ 0.00}^{ 0.00}$    &$ 0.00_{ 0.00}^{ 0.00}    -	 0.37_{ 0.37}^{ 7.03}    -  52.93_{ 9.53}^{ 8.96}$    &$ 0.54_{ 0.54}^{ 8.36}    -	 0.00_{ 0.00}^{ 0.00}    -   7.49_{ 6.11}^{11.67}$    &$ 0.00_{ 0.00}^{ 0.00}    -	 0.00_{ 0.00}^{ 0.00}    -   0.00_{ 0.00}^{ 0.00}$\\
 37    &$ 4.23_{ 4.06}^{14.86}    -	 0.00_{ 0.00}^{ 0.00}    -   0.10_{ 0.10}^{ 4.27}$    &$ 6.05_{ 5.00}^{ 5.65}    -	 0.10_{ 0.10}^{ 4.00}    -   1.38_{ 1.38}^{12.17}$    &$17.53_{13.72}^{14.68}    -	 1.21_{ 1.21}^{13.48}    -   4.61_{ 4.57}^{13.99}$    &$ 0.09_{ 0.09}^{ 3.77}    -	 0.20_{ 0.20}^{ 6.82}    -  25.25_{18.98}^{36.91}$\\
 38    &$ 0.00_{ 0.00}^{ 0.00}    -	 0.00_{ 0.00}^{ 0.00}    -   0.49_{ 0.48}^{ 7.59}$    &$ 0.00_{ 0.00}^{ 0.00}    -	 0.00_{ 0.00}^{ 0.00}    -   2.66_{ 2.50}^{ 6.42}$    &$ 0.00_{ 0.00}^{ 0.00}    -	 0.00_{ 0.00}^{ 0.00}    -  16.68_{ 7.45}^{ 8.27}$    &$ 0.00_{ 0.00}^{ 0.00}    -	 0.00_{ 0.00}^{ 0.00}    -  63.03_{ 8.76}^{ 7.73}$\\
 39    &$ 0.00_{ 0.00}^{ 0.00}    -	 0.00_{ 0.00}^{ 0.00}    -   9.69_{ 8.62}^{16.73}$    &$ 3.76_{ 3.58}^{ 7.16}    -	 0.03_{ 0.03}^{ 0.00}    -  17.81_{12.93}^{17.76}$    &$ 1.89_{ 1.84}^{ 8.67}    -	 0.40_{ 0.40}^{17.45}    -  18.43_{12.96}^{16.10}$    &$ 0.00_{ 0.00}^{ 0.00}    -	 0.00_{ 0.00}^{ 0.00}    -  27.97_{24.79}^{30.26}$\\
 40    &$ 6.12_{ 5.25}^{10.41}    -	 0.61_{ 0.62}^{18.84}    -  28.64_{15.94}^{11.62}$    &$ 0.21_{ 0.21}^{ 2.09}    -	 0.00_{ 0.00}^{ 0.00}    -   0.00_{ 0.00}^{ 0.00}$    &$ 0.00_{ 0.00}^{ 0.00}    -	 0.00_{ 0.00}^{ 0.00}    -   0.00_{ 0.00}^{ 0.00}$    &$ 0.00_{ 0.00}^{ 0.00}    -	 0.00_{ 0.00}^{ 0.00}    -   0.00_{ 0.00}^{ 0.00}$\\
 41    &$ 0.04_{ 0.04}^{ 2.22}    -	 0.00_{ 0.00}^{ 0.00}    -   0.66_{ 0.66}^{ 7.67}$    &$ 7.21_{ 5.86}^{ 5.22}    -	 0.10_{ 0.10}^{ 0.00}    -  35.09_{11.60}^{ 9.85}$    &$12.16_{ 8.05}^{ 9.49}    -	 0.00_{ 0.00}^{ 0.00}    -   0.12_{ 0.12}^{10.54}$    &$ 0.00_{ 0.00}^{ 0.00}    -	 0.08_{ 0.08}^{ 5.82}    -   5.19_{ 5.04}^{13.38}$\\
 42    &$ 0.00_{ 0.00}^{ 0.00}    -	 0.00_{ 0.00}^{ 0.00}    -  11.09_{ 8.23}^{ 9.64}$    &$ 0.32_{ 0.32}^{ 9.42}    -	 0.15_{ 0.16}^{ 0.00}    -  32.89_{ 8.83}^{ 7.39}$    &$ 0.00_{ 0.00}^{ 0.32}    -	 0.02_{ 0.02}^{ 0.00}    -  17.20_{11.06}^{11.74}$    &$ 0.02_{ 0.02}^{ 0.00}    -	 0.00_{ 0.00}^{ 0.00}    -  12.24_{ 8.19}^{ 9.53}$\\
 43    &$ 0.00_{ 0.00}^{ 0.00}    -	 0.02_{ 0.02}^{ 2.18}    -  46.94_{ 6.01}^{ 6.01}$    &$ 0.00_{ 0.00}^{ 0.00}    -	 0.00_{ 0.00}^{ 0.00}    -   2.12_{ 2.05}^{ 6.24}$    &$ 0.00_{ 0.00}^{ 0.00}    -	 0.00_{ 0.00}^{ 0.00}    -   2.51_{ 2.40}^{ 8.33}$    &$ 0.00_{ 0.00}^{ 0.00}    -	 0.10_{ 0.10}^{ 4.43}    -  26.29_{ 7.56}^{ 5.95}$\\
 44    &$ 0.02_{ 0.02}^{ 0.00}    -	 0.01_{ 0.01}^{ 0.00}    -   2.63_{ 2.53}^{10.05}$    &$13.38_{ 3.55}^{ 4.23}    -	 1.23_{ 1.24}^{10.70}    -  11.41_{ 8.03}^{ 8.54}$    &$ 2.28_{ 2.18}^{ 6.87}    -	10.68_{ 9.80}^{13.91}    -  34.38_{16.03}^{17.63}$    &$ 0.01_{ 0.01}^{ 0.00}    -	 0.05_{ 0.05}^{ 0.00}    -   0.00_{ 0.00}^{ 0.00}$\\
 45    &$ 0.00_{ 0.00}^{ 0.00}    -	24.43_{12.09}^{12.61}    -  30.16_{18.98}^{15.65}$    &$ 0.00_{ 0.00}^{ 0.00}    -	 0.00_{ 0.00}^{ 0.00}    -   0.00_{ 0.00}^{ 0.00}$    &$ 0.00_{ 0.00}^{ 0.00}    -	 0.00_{ 0.00}^{ 0.00}    -  43.97_{12.83}^{11.56}$    &$ 0.00_{ 0.00}^{ 0.00}    -	 0.00_{ 0.00}^{ 0.00}    -   0.00_{ 0.00}^{ 0.00}$\\
 46    &$ 0.03_{ 0.03}^{ 1.63}    -	 0.00_{ 0.00}^{ 0.00}    -   0.00_{ 0.00}^{ 0.00}$    &$17.13_{ 4.41}^{ 4.70}    -	33.24_{11.13}^{ 8.62}    -   5.30_{ 4.91}^{ 9.62}$    &$ 0.00_{ 0.00}^{ 0.00}    -	 0.00_{ 0.00}^{ 0.00}    -   0.00_{ 0.00}^{ 0.00}$    &$ 0.00_{ 0.00}^{ 0.00}    -	 0.00_{ 0.00}^{ 0.00}    -   0.00_{ 0.00}^{ 0.00}$\\
 47    &$ 2.08_{ 1.65}^{ 3.16}    -	 0.00_{ 0.00}^{ 0.00}    -   1.46_{ 1.46}^{14.73}$    &$ 0.55_{ 0.55}^{ 2.31}    -	 0.00_{ 0.00}^{ 0.00}    -  26.40_{ 4.21}^{ 4.50}$    &$ 0.00_{ 0.00}^{ 0.00}    -	 0.00_{ 0.00}^{ 0.00}    -  41.92_{15.60}^{ 7.92}$    &$ 0.00_{ 0.00}^{ 0.00}    -	 0.00_{ 0.00}^{ 0.00}    -   0.38_{ 0.38}^{ 8.40}$\\
 48    &$ 0.93_{ 0.93}^{ 5.18}    -	 0.22_{ 0.22}^{ 4.06}    -   0.81_{ 0.81}^{ 8.74}$    &$ 0.13_{ 0.13}^{ 2.73}    -	 0.38_{ 0.38}^{ 3.60}    -   2.42_{ 2.42}^{11.63}$    &$ 0.00_{ 0.00}^{ 0.00}    -	 0.00_{ 0.00}^{ 0.00}    -   5.76_{ 5.41}^{14.43}$    &$ 0.00_{ 0.00}^{ 0.00}    -	 0.00_{ 0.00}^{ 0.00}    -  32.11_{14.78}^{15.20}$\\
 49    &$ 0.96_{ 0.96}^{14.89}    -	 0.00_{ 0.00}^{ 0.00}    -   0.00_{ 0.00}^{ 0.00}$    &$24.35_{ 5.46}^{11.53}    -	 0.51_{ 0.51}^{16.67}    -   1.10_{ 1.10}^{28.26}$    &$ 0.00_{ 0.00}^{ 0.00}    -	 0.00_{ 0.00}^{ 0.00}    -  36.95_{31.06}^{29.60}$    &$ 0.00_{ 0.00}^{ 0.00}    -	 0.00_{ 0.00}^{ 0.00}    -   0.00_{ 0.00}^{ 0.00}$\\
 50    &$ 1.70_{ 1.68}^{10.63}    -	 0.00_{ 0.00}^{ 0.00}    -  15.44_{13.59}^{13.45}$    &$17.59_{ 6.25}^{ 5.42}    -	 0.90_{ 0.90}^{10.62}    -   5.66_{ 5.36}^{15.67}$    &$ 4.99_{ 4.69}^{ 9.62}    -	 0.00_{ 0.00}^{ 0.00}    -   0.77_{ 0.77}^{ 9.36}$    &$ 0.28_{ 0.28}^{23.77}    -	 0.00_{ 0.00}^{ 0.00}    -   0.71_{ 0.71}^{15.36}$\\
 51    &$ 0.00_{ 0.00}^{ 0.00}    -	 0.00_{ 0.00}^{ 0.00}    -   0.01_{ 0.01}^{ 0.00}$    &$21.94_{ 2.72}^{ 2.21}    -	 0.00_{ 0.00}^{ 0.00}    -   0.54_{ 0.54}^{12.00}$    &$ 0.00_{ 0.00}^{ 0.00}    -	 0.00_{ 0.00}^{ 0.00}    -  56.16_{ 4.07}^{ 2.92}$    &$ 0.00_{ 0.00}^{ 0.00}    -	 0.00_{ 0.00}^{ 0.00}    -   0.00_{ 0.00}^{ 0.00}$\\
 52    &$ 3.68_{ 3.47}^{ 8.63}    -	 4.39_{ 4.28}^{10.09}    -   0.88_{ 0.89}^{ 7.18}$    &$17.63_{ 8.55}^{ 8.46}    -	 9.19_{ 8.55}^{15.60}    -   5.39_{ 5.05}^{13.11}$    &$ 2.49_{ 2.41}^{ 8.27}    -	 0.53_{ 0.54}^{ 7.85}    -   1.08_{ 1.07}^{ 9.51}$    &$ 5.27_{ 5.13}^{14.71}    -	 3.65_{ 3.56}^{11.34}    -  15.34_{14.18}^{26.80}$\\
 53    &$ 0.00_{ 0.00}^{ 0.00}    -	 0.00_{ 0.00}^{ 0.00}    -   0.00_{ 0.00}^{ 0.00}$    &$22.95_{ 2.05}^{ 2.69}    -	 0.00_{ 0.00}^{ 0.00}    -   8.08_{ 5.72}^{ 6.47}$    &$ 0.05_{ 0.05}^{ 0.00}    -	10.41_{ 6.91}^{ 7.33}    -   6.98_{ 5.56}^{ 6.77}$    &$ 0.06_{ 0.06}^{ 2.17}    -	 0.00_{ 0.00}^{ 0.00}    -   0.01_{ 0.01}^{ 0.00}$\\
 54    &$ 0.00_{ 0.00}^{ 0.00}    -	 0.00_{ 0.00}^{ 0.00}    -   0.00_{ 0.00}^{ 0.00}$    &$24.94_{17.43}^{ 6.47}    -	 0.00_{ 0.00}^{ 0.00}    -   4.44_{ 4.44}^{31.90}$    &$ 0.00_{ 0.00}^{ 0.00}    -	 0.00_{ 0.00}^{ 0.00}    -   7.72_{ 7.77}^{43.37}$    &$ 0.00_{ 0.00}^{ 0.00}    -	 0.00_{ 0.00}^{ 0.00}    -   0.33_{ 0.33}^{ 9.63}$\\
 55    &$40.18_{ 5.63}^{ 5.98}    -	 0.00_{ 0.00}^{ 0.00}    -   0.03_{ 0.03}^{ 0.00}$    &$ 4.85_{ 2.89}^{ 3.38}    -	 1.96_{ 1.90}^{ 4.75}    -   4.32_{ 3.59}^{ 6.57}$    &$ 1.26_{ 1.26}^{ 9.65}    -	 0.00_{ 0.00}^{ 0.00}    -  17.98_{ 6.74}^{ 6.01}$    &$ 0.00_{ 0.00}^{ 0.00}    -	 0.03_{ 0.03}^{ 0.00}    -   0.12_{ 0.12}^{ 3.99}$\\
 56    &$ 2.03_{ 2.00}^{12.71}    -	 0.00_{ 0.00}^{ 0.00}    -  33.82_{13.76}^{10.93}$    &$13.31_{ 6.93}^{ 5.13}    -	 0.00_{ 0.00}^{ 0.00}    -   8.85_{ 7.77}^{ 9.82}$    &$ 9.80_{ 6.56}^{14.84}    -	 0.00_{ 0.00}^{ 0.00}    -   9.71_{ 7.77}^{10.73}$    &$ 0.25_{ 0.25}^{ 3.42}    -	 0.00_{ 0.00}^{ 0.00}    -   1.02_{ 1.02}^{14.62}$\\
 57    &$ 0.00_{ 0.00}^{ 0.00}    -	 0.00_{ 0.00}^{ 0.00}    -   0.00_{ 0.00}^{ 0.00}$    &$18.84_{ 3.40}^{ 3.17}    -	 0.00_{ 0.00}^{ 0.00}    -   2.68_{ 2.70}^{13.40}$    &$ 3.20_{ 3.05}^{ 6.98}    -	 0.00_{ 0.00}^{ 0.00}    -   9.92_{ 9.92}^{25.50}$    &$ 0.03_{ 0.03}^{ 0.00}    -	 0.00_{ 0.00}^{ 0.00}    -   0.00_{ 0.00}^{ 0.00}$\\

\hline
\end{tabular}
\end{table*}

\begin{table*}
\caption{Best fit parameters deduced from our full spectral fitting. 
The last column gives the continuum flux contribution, listed as a percentage of the total integrated flux over the 
full wavelength range.}
\label{fitresults3}
\centering
\begin{tabular}{lcccc}
\hline \hline   
 N$^\circ$ & Silica & Forsterite & Enstatite & Continuum\\
                & 
           Small  - Medium -   Large & Small  - Medium -   Large & Small  - Medium -   Large & \\
\hline
  1    &$ 0.00_{ 0.00}^{ 0.00}    -	 0.00_{ 0.00}^{ 0.00}    -   0.00_{ 0.00}^{ 0.00}$    &$ 8.66_{ 2.37}^{ 3.64}    -	 3.85_{ 2.44}^{ 2.56}    -   0.05_{ 0.05}^{ 4.58}$    &$ 0.01_{ 0.01}^{ 1.19}    -	 0.00_{ 0.00}^{ 0.00}    -   9.00_{ 3.18}^{ 3.30}$    &$52.84_{ 0.85}^{ 4.99}$\\
  2    &$ 0.00_{ 0.00}^{ 0.09}    -	 0.00_{ 0.00}^{ 0.00}    -   3.44_{ 0.84}^{ 2.41}$    &$ 1.44_{ 0.56}^{ 0.39}    -	18.35_{ 1.10}^{ 2.77}    -   0.00_{ 0.00}^{ 0.00}$    &$ 0.02_{ 0.02}^{ 0.26}    -	 0.00_{ 0.00}^{ 0.00}    -   3.31_{ 1.49}^{ 2.17}$    &$77.81_{ 0.31}^{ 1.03}$\\
  3    &$ 0.00_{ 0.00}^{ 0.11}    -	 0.22_{ 0.22}^{ 0.75}    -   3.55_{ 1.08}^{ 0.57}$    &$ 0.50_{ 0.30}^{ 0.37}    -	12.61_{ 0.89}^{ 0.86}    -   0.00_{ 0.00}^{ 0.00}$    &$ 0.37_{ 0.29}^{ 0.40}    -	 0.00_{ 0.00}^{ 0.00}    -   1.33_{ 1.01}^{ 1.33}$    &$74.46_{ 0.21}^{ 0.20}$\\
  4    &$ 0.01_{ 0.01}^{ 0.36}    -	14.03_{ 1.86}^{ 0.96}    -   1.41_{ 1.27}^{ 3.11}$    &$13.45_{ 1.49}^{ 1.38}    -	 8.92_{ 3.66}^{ 4.67}    -  24.74_{ 4.66}^{ 4.29}$    &$ 0.00_{ 0.00}^{ 0.00}    -	 0.00_{ 0.00}^{ 0.00}    -   0.64_{ 0.60}^{ 3.30}$    &$67.56_{ 0.49}^{ 0.37}$\\
  5    &$ 0.27_{ 0.20}^{ 0.25}    -	 0.00_{ 0.00}^{ 0.00}    -   0.00_{ 0.00}^{ 0.00}$    &$ 4.99_{ 0.50}^{ 0.47}    -	19.90_{ 1.58}^{ 1.57}    -   0.00_{ 0.00}^{ 0.00}$    &$ 4.77_{ 0.68}^{ 0.87}    -	 0.00_{ 0.00}^{ 0.00}    -  25.19_{ 2.13}^{ 1.94}$    &$69.00_{ 0.64}^{ 0.68}$\\
  6    &$ 0.25_{ 0.24}^{ 1.24}    -	 0.00_{ 0.00}^{ 0.00}    -   0.17_{ 0.17}^{ 1.33}$    &$ 6.07_{ 1.97}^{ 1.10}    -	12.27_{ 2.25}^{ 2.88}    -  17.12_{ 4.71}^{ 9.58}$    &$ 2.90_{ 2.29}^{ 2.15}    -	 2.48_{ 2.17}^{ 3.60}    -  19.74_{ 4.43}^{ 7.76}$    &$79.35_{ 0.47}^{ 0.64}$\\
  7    &$ 0.00_{ 0.00}^{ 0.00}    -	 0.00_{ 0.00}^{ 0.00}    -   0.15_{ 0.15}^{ 0.83}$    &$ 4.63_{ 0.51}^{ 0.53}    -	 9.81_{ 5.69}^{ 3.73}    -  23.91_{11.62}^{18.41}$    &$ 0.50_{ 0.49}^{ 1.66}    -	 0.00_{ 0.00}^{ 0.00}    -  19.72_{ 4.34}^{ 2.16}$    &$82.26_{ 0.95}^{ 1.50}$\\
  8    &$20.14_{ 3.37}^{ 2.56}    -	 0.55_{ 0.55}^{ 7.15}    -   0.00_{ 0.00}^{ 0.00}$    &$ 5.16_{ 3.40}^{ 3.60}    -	 1.53_{ 1.49}^{ 7.41}    -   0.00_{ 0.00}^{ 0.00}$    &$ 0.00_{ 0.00}^{ 0.00}    -	 0.00_{ 0.00}^{ 0.00}    -   0.27_{ 0.27}^{13.44}$    &$81.87_{ 1.68}^{ 1.15}$\\
  9    &$ 0.00_{ 0.00}^{ 0.00}    -	 0.00_{ 0.00}^{ 0.00}    -   2.50_{ 0.49}^{ 0.49}$    &$ 3.21_{ 0.30}^{ 0.37}    -	12.70_{ 0.88}^{ 0.78}    -   0.46_{ 0.45}^{ 2.57}$    &$ 0.00_{ 0.00}^{ 0.00}    -	 0.00_{ 0.00}^{ 0.00}    -  15.33_{ 1.16}^{ 1.13}$    &$70.18_{ 0.34}^{ 0.29}$\\
 10    &$ 1.85_{ 0.71}^{ 0.38}    -	 0.00_{ 0.00}^{ 0.00}    -   7.29_{ 0.83}^{ 2.42}$    &$ 1.11_{ 0.70}^{ 0.44}    -	12.91_{ 0.87}^{ 1.05}    -   0.00_{ 0.00}^{ 0.00}$    &$ 0.06_{ 0.06}^{ 0.46}    -	 0.00_{ 0.00}^{ 0.00}    -   0.41_{ 0.40}^{ 1.72}$    &$73.51_{ 2.02}^{ 0.84}$\\
 11    &$ 0.00_{ 0.00}^{ 0.14}    -	 0.00_{ 0.00}^{ 0.00}    -   7.29_{ 1.26}^{ 1.30}$    &$ 0.01_{ 0.01}^{ 0.27}    -	10.21_{ 1.89}^{ 2.80}    -   2.50_{ 2.09}^{ 2.80}$    &$ 0.00_{ 0.00}^{ 0.00}    -	 0.00_{ 0.00}^{ 0.00}    -   0.00_{ 0.00}^{ 0.22}$    &$79.83_{ 1.06}^{ 2.44}$\\
 12    &$ 0.00_{ 0.00}^{ 0.00}    -	 0.00_{ 0.00}^{ 0.00}    -   0.20_{ 0.19}^{ 0.77}$    &$ 0.01_{ 0.01}^{ 0.09}    -	 0.07_{ 0.06}^{ 0.26}    -   0.01_{ 0.01}^{ 0.29}$    &$ 0.00_{ 0.00}^{ 0.00}    -	 0.00_{ 0.00}^{ 0.00}    -   0.00_{ 0.00}^{ 0.00}$    &$28.11_{ 3.30}^{ 1.07}$\\
 13    &$ 1.06_{ 0.40}^{ 0.37}    -	 0.00_{ 0.00}^{ 0.00}    -   0.44_{ 0.42}^{ 0.93}$    &$ 0.03_{ 0.04}^{ 0.43}    -	12.83_{ 0.75}^{ 0.50}    -   0.42_{ 0.38}^{ 0.80}$    &$ 0.00_{ 0.00}^{ 0.00}    -	 0.00_{ 0.00}^{ 0.00}    -  15.47_{ 0.56}^{ 0.58}$    &$74.30_{ 0.75}^{ 1.04}$\\
 14    &$ 0.04_{ 0.04}^{ 0.42}    -	 0.07_{ 0.07}^{ 0.45}    -   0.29_{ 0.26}^{ 0.46}$    &$ 0.00_{ 0.00}^{ 0.00}    -	11.44_{ 0.33}^{ 0.28}    -   0.09_{ 0.09}^{ 0.68}$    &$ 3.12_{ 0.35}^{ 0.38}    -	 0.00_{ 0.00}^{ 0.00}    -   1.47_{ 0.86}^{ 1.01}$    &$52.14_{ 0.43}^{ 0.48}$\\
 15    &$ 0.02_{ 0.02}^{ 0.67}    -	 0.00_{ 0.00}^{ 0.00}    -   0.00_{ 0.00}^{ 0.00}$    &$ 0.50_{ 0.48}^{ 2.26}    -	 5.61_{ 1.60}^{ 0.86}    -   2.43_{ 1.30}^{ 1.29}$    &$ 4.55_{ 2.23}^{ 2.80}    -	 1.47_{ 1.39}^{ 3.62}    -  17.21_{ 5.70}^{ 1.96}$    &$78.00_{0.27}^{0.33}$\\
 16    &$ 0.23_{ 0.17}^{ 0.27}    -	 0.00_{ 0.00}^{ 0.00}    -   7.21_{ 0.74}^{ 0.66}$    &$ 2.03_{ 0.57}^{ 0.67}    -	19.48_{ 1.80}^{ 1.34}    -   0.85_{ 0.83}^{ 4.55}$    &$ 0.00_{ 0.00}^{ 0.00}    -	 0.00_{ 0.00}^{ 0.00}    -   0.00_{ 0.00}^{ 0.21}$    &$75.33_{ 0.36}^{ 0.27}$\\
 17    &$ 1.64_{ 0.48}^{ 0.40}    -	 0.00_{ 0.00}^{ 0.00}    -   1.51_{ 1.25}^{ 1.96}$    &$ 1.71_{ 0.63}^{ 0.69}    -	23.99_{ 1.97}^{ 2.80}    -   7.55_{ 7.60}^{14.59}$    &$ 0.58_{ 0.47}^{ 0.69}    -	 0.00_{ 0.00}^{ 0.00}    -  18.86_{ 4.09}^{ 5.28}$    &$82.47_{ 1.05}^{ 1.08}$\\
 18    &$ 1.61_{ 0.51}^{ 0.40}    -	 0.00_{ 0.00}^{ 0.00}    -   0.00_{ 0.00}^{ 0.00}$    &$ 2.87_{ 1.05}^{ 0.71}    -	19.12_{ 1.03}^{ 1.78}    -   0.41_{ 0.41}^{ 1.92}$    &$ 0.00_{ 0.00}^{ 0.00}    -	 0.03_{ 0.03}^{ 1.12}    -  28.26_{ 1.28}^{ 1.48}$    &$74.24_{0.24}^{0.21}$\\
 19    &$ 0.00_{ 0.00}^{ 0.00}    -	 0.00_{ 0.00}^{ 0.00}    -  10.42_{ 0.77}^{ 0.75}$    &$ 2.72_{ 0.53}^{ 0.60}    -	17.76_{ 1.15}^{ 1.07}    -   0.00_{ 0.00}^{ 0.00}$    &$ 0.00_{ 0.00}^{ 0.00}    -	 0.00_{ 0.00}^{ 0.00}    -   6.16_{ 1.46}^{ 1.42}$    &$83.65_{ 0.21}^{ 0.19}$\\
 20    &$ 1.96_{ 0.35}^{ 0.39}    -	 0.00_{ 0.00}^{ 0.00}    -   0.00_{ 0.00}^{ 0.00}$    &$ 0.00_{ 0.00}^{ 0.00}    -	 2.84_{ 0.43}^{ 0.48}    -   3.80_{ 0.69}^{ 0.80}$    &$ 0.00_{ 0.00}^{ 0.00}    -	 0.00_{ 0.00}^{ 0.00}    -  11.58_{ 0.62}^{ 0.60}$    &$86.71_{ 1.02}^{ 0.92}$\\
 21    &$ 0.14_{ 0.13}^{ 0.33}    -	 0.00_{ 0.00}^{ 0.00}    -   1.11_{ 1.11}^{ 4.29}$    &$ 1.37_{ 0.57}^{ 0.65}    -	16.62_{ 3.51}^{ 2.16}    -   5.51_{ 4.54}^{11.41}$    &$ 0.00_{ 0.00}^{ 0.00}    -	 0.33_{ 0.33}^{ 3.89}    -   8.29_{ 2.43}^{ 1.65}$    &$68.29_{ 0.31}^{ 0.28}$\\
 22    &$ 0.00_{ 0.00}^{ 0.00}    -	 0.00_{ 0.00}^{ 0.00}    -  11.71_{ 1.20}^{ 1.66}$    &$ 0.00_{ 0.00}^{ 0.00}    -	11.63_{ 2.13}^{ 1.78}    -   3.04_{ 2.54}^{ 4.20}$    &$ 0.00_{ 0.00}^{ 0.00}    -	 0.00_{ 0.00}^{ 0.00}    -   0.00_{ 0.00}^{ 0.00}$    &$79.96_{ 0.29}^{ 0.26}$\\
 23    &$ 2.75_{ 0.36}^{ 0.30}    -	 0.00_{ 0.00}^{ 0.00}    -   0.18_{ 0.17}^{ 0.76}$    &$ 0.74_{ 0.54}^{ 0.83}    -	 9.09_{ 1.23}^{ 1.07}    -   6.89_{ 1.10}^{ 1.01}$    &$ 0.00_{ 0.00}^{ 0.00}    -	 0.00_{ 0.00}^{ 0.00}    -  14.88_{ 1.04}^{ 1.03}$    &$69.41_{ 0.14}^{ 0.15}$\\
 24    &$ 0.28_{ 0.20}^{ 0.32}    -	 0.00_{ 0.00}^{ 0.00}    -  13.61_{ 0.88}^{ 0.71}$    &$ 0.13_{ 0.12}^{ 0.40}    -	20.26_{ 2.25}^{ 1.84}    -   4.88_{ 2.74}^{ 2.82}$    &$ 0.00_{ 0.00}^{ 0.00}    -	 0.00_{ 0.00}^{ 0.00}    -   0.99_{ 0.87}^{ 1.60}$    &$82.50_{ 0.36}^{ 0.32}$\\
 25    &$ 0.00_{ 0.00}^{ 0.00}    -	 0.12_{ 0.12}^{ 0.70}    -   1.16_{ 0.81}^{ 0.96}$    &$ 3.42_{ 1.58}^{ 1.72}    -	 9.49_{ 1.92}^{ 1.70}    -   1.35_{ 1.15}^{ 1.90}$    &$ 0.00_{ 0.00}^{ 0.11}    -	 0.00_{ 0.00}^{ 0.00}    -  16.32_{ 1.97}^{ 1.69}$    &$79.60_{ 1.47}^{ 0.95}$\\
 26    &$ 1.82_{ 0.34}^{ 0.36}    -	 0.00_{ 0.00}^{ 0.00}    -   1.28_{ 0.93}^{ 1.69}$    &$ 2.53_{ 0.51}^{ 0.49}    -	26.80_{ 2.29}^{ 2.18}    -   6.03_{ 2.75}^{ 3.50}$    &$ 0.00_{ 0.00}^{ 0.00}    -	 0.00_{ 0.00}^{ 0.00}    -  11.68_{ 1.61}^{ 1.91}$    &$79.95_{ 0.47}^{ 0.49}$\\
 27    &$ 0.07_{ 0.07}^{ 0.29}    -	 1.95_{ 1.07}^{ 1.12}    -  10.13_{ 2.08}^{ 1.85}$    &$ 0.12_{ 0.11}^{ 0.40}    -	16.58_{ 1.21}^{ 0.86}    -   0.72_{ 0.70}^{ 2.45}$    &$ 0.04_{ 0.04}^{ 0.34}    -	 0.00_{ 0.00}^{ 0.00}    -   8.93_{ 1.82}^{ 1.78}$    &$82.74_{ 0.52}^{ 0.25}$\\
 28    &$ 3.44_{ 0.37}^{ 0.34}    -	 0.00_{ 0.00}^{ 0.00}    -   4.75_{ 2.22}^{ 3.90}$    &$ 0.37_{ 0.31}^{ 0.45}    -	20.22_{ 1.61}^{ 1.83}    -   4.86_{ 4.68}^{ 7.95}$    &$ 0.00_{ 0.00}^{ 0.00}    -	 0.00_{ 0.00}^{ 0.00}    -   5.04_{ 2.32}^{ 3.01}$    &$60.03_{ 1.52}^{ 2.05}$\\
 29    &$ 0.00_{ 0.00}^{ 0.00}    -	 0.00_{ 0.00}^{ 0.00}    -   0.84_{ 0.81}^{ 2.53}$    &$ 2.22_{ 1.38}^{ 1.53}    -	14.13_{ 8.09}^{ 6.01}    -   0.03_{ 0.03}^{ 0.68}$    &$ 0.00_{ 0.00}^{ 0.00}    -	 0.00_{ 0.00}^{ 0.00}    -   2.96_{ 2.12}^{ 2.00}$    &$83.09_{ 3.99}^{ 2.90}$\\
 30    &$ 0.00_{ 0.00}^{ 0.00}    -	 1.20_{ 0.87}^{ 0.90}    -   7.92_{ 1.51}^{ 1.32}$    &$ 0.06_{ 0.06}^{ 0.28}    -	19.99_{ 1.67}^{ 1.60}    -   4.85_{ 3.49}^{ 2.84}$    &$ 0.00_{ 0.00}^{ 0.00}    -	 0.00_{ 0.00}^{ 0.00}    -   7.14_{ 1.29}^{ 1.35}$    &$76.51_{ 1.26}^{ 0.55}$\\
 31    &$ 0.00_{ 0.00}^{ 0.00}    -	 0.00_{ 0.00}^{ 0.00}    -   3.51_{ 1.54}^{ 2.29}$    &$ 2.54_{ 1.33}^{ 1.34}    -	11.85_{ 3.04}^{ 2.80}    -   0.01_{ 0.01}^{ 0.00}$    &$ 0.00_{ 0.00}^{ 0.00}    -	 0.00_{ 0.00}^{ 0.00}    -   0.08_{ 0.08}^{ 2.05}$    &$90.67_{ 0.35}^{ 0.20}$\\
 32    &$ 0.06_{ 0.06}^{ 0.26}    -	 0.00_{ 0.00}^{ 0.00}    -   0.87_{ 0.71}^{ 1.07}$    &$ 1.06_{ 0.60}^{ 0.87}    -	18.16_{ 1.80}^{ 1.66}    -   0.00_{ 0.00}^{ 0.00}$    &$ 0.03_{ 0.03}^{ 0.75}    -	 0.00_{ 0.00}^{ 0.00}    -   2.64_{ 2.00}^{ 2.06}$    &$83.69_{ 0.20}^{ 0.22}$\\
 33    &$ 1.33_{ 0.27}^{ 0.32}    -	 0.00_{ 0.00}^{ 0.00}    -   6.53_{ 0.73}^{ 0.73}$    &$ 2.04_{ 0.40}^{ 0.51}    -	13.43_{ 1.87}^{ 2.34}    -  31.80_{ 2.67}^{ 2.62}$    &$ 0.13_{ 0.13}^{ 0.52}    -	 0.00_{ 0.00}^{ 0.00}    -   4.01_{ 2.03}^{ 1.76}$    &$80.38_{ 0.26}^{ 0.35}$\\
 34    &$ 3.80_{ 0.93}^{ 1.00}    -	 0.00_{ 0.00}^{ 0.00}    -   0.00_{ 0.00}^{ 0.00}$    &$ 3.10_{ 1.08}^{ 1.33}    -	14.11_{ 4.57}^{ 4.57}    -   5.03_{ 4.33}^{ 7.94}$    &$ 0.14_{ 0.14}^{ 1.50}    -	 4.75_{ 3.52}^{ 5.80}    -  21.50_{ 8.52}^{ 7.07}$    &$47.88_{ 4.76}^{ 3.51}$\\
 35    &$ 1.62_{ 1.06}^{ 1.11}    -	 0.08_{ 0.08}^{ 1.72}    -   0.00_{ 0.00}^{ 0.00}$    &$ 0.39_{ 0.37}^{ 1.45}    -	23.24_{ 5.75}^{ 5.20}    -   5.04_{ 4.50}^{11.84}$    &$ 0.06_{ 0.06}^{ 1.35}    -	 0.27_{ 0.27}^{ 5.56}    -  24.35_{ 7.19}^{ 6.07}$    &$83.15_{ 1.70}^{ 0.92}$\\
 36    &$ 0.00_{ 0.00}^{ 0.00}    -	 0.06_{ 0.06}^{ 2.35}    -   3.91_{ 1.85}^{ 2.00}$    &$ 0.00_{ 0.00}^{ 0.00}    -	 0.30_{ 0.30}^{ 5.14}    -  29.73_{ 5.13}^{ 5.71}$    &$ 2.11_{ 1.32}^{ 2.03}    -	 0.00_{ 0.00}^{ 0.00}    -   0.00_{ 0.00}^{ 0.00}$    &$80.36_{ 6.69}^{ 8.37}$\\
 37    &$ 2.17_{ 1.41}^{ 1.26}    -	 2.01_{ 1.88}^{ 3.81}    -   3.07_{ 3.01}^{ 5.46}$    &$ 0.51_{ 0.47}^{ 0.99}    -	21.49_{10.46}^{ 7.85}    -   0.69_{ 0.69}^{ 7.31}$    &$ 0.03_{ 0.03}^{ 0.90}    -	 0.11_{ 0.11}^{ 3.15}    -   9.17_{ 6.43}^{ 7.17}$    &$82.01_{ 4.34}^{ 2.87}$\\
 38    &$ 0.22_{ 0.21}^{ 0.56}    -	 1.07_{ 0.88}^{ 1.09}    -   0.23_{ 0.23}^{ 1.65}$    &$ 3.07_{ 1.09}^{ 1.12}    -	 9.80_{ 3.99}^{ 3.15}    -   2.31_{ 2.20}^{ 7.26}$    &$ 0.00_{ 0.00}^{ 0.00}    -	 0.00_{ 0.00}^{ 0.00}    -   0.45_{ 0.45}^{ 3.24}$    &$79.88_{ 0.84}^{ 0.66}$\\
 39    &$ 0.30_{ 0.30}^{ 1.28}    -	 0.13_{ 0.13}^{ 2.10}    -   2.76_{ 2.23}^{ 3.60}$    &$ 0.03_{ 0.03}^{ 1.16}    -	11.81_{ 5.56}^{ 7.76}    -   4.13_{ 4.03}^{14.24}$    &$ 0.57_{ 0.56}^{ 3.17}    -	 0.00_{ 0.00}^{ 0.00}    -   0.29_{ 0.29}^{ 6.29}$    &$81.79_{ 2.67}^{ 2.88}$\\
 40    &$ 1.45_{ 0.90}^{ 1.33}    -	 0.00_{ 0.00}^{ 0.00}    -  17.67_{ 4.02}^{ 4.03}$    &$ 0.08_{ 0.08}^{ 0.69}    -	20.62_{ 3.98}^{ 4.81}    -   0.97_{ 0.97}^{ 7.41}$    &$ 0.30_{ 0.30}^{ 1.90}    -	 0.00_{ 0.00}^{ 0.00}    -  23.33_{ 6.42}^{ 5.47}$    &$86.15_{ 0.97}^{ 1.09}$\\
 41    &$ 9.57_{ 5.60}^{ 2.45}    -	 0.66_{ 0.65}^{ 3.31}    -   6.53_{ 4.85}^{ 9.81}$    &$ 0.05_{ 0.05}^{ 1.26}    -	 6.46_{ 3.95}^{ 4.19}    -  12.14_{ 7.13}^{ 6.90}$    &$ 0.04_{ 0.04}^{ 0.74}    -	 0.17_{ 0.17}^{ 2.62}    -   3.73_{ 3.22}^{ 6.58}$    &$68.59_{ 1.53}^{ 2.84}$\\
 42    &$ 0.01_{ 0.01}^{ 0.57}    -	 2.26_{ 1.46}^{ 1.36}    -   1.53_{ 1.44}^{ 3.34}$    &$ 0.79_{ 0.59}^{ 1.11}    -	 6.97_{ 2.80}^{ 2.91}    -   0.83_{ 0.82}^{ 7.21}$    &$ 0.09_{ 0.09}^{ 2.10}    -	11.51_{ 2.95}^{ 3.72}    -   2.07_{ 2.02}^{ 5.63}$    &$70.33_{ 3.01}^{ 1.81}$\\
 43    &$ 0.01_{ 0.01}^{ 0.34}    -	 3.26_{ 1.02}^{ 0.92}    -   0.14_{ 0.14}^{ 1.85}$    &$ 0.44_{ 0.40}^{ 0.98}    -	10.04_{ 4.33}^{ 2.84}    -   3.14_{ 2.99}^{ 6.48}$    &$ 1.80_{ 1.16}^{ 1.38}    -	 3.18_{ 2.47}^{ 2.61}    -   0.00_{ 0.00}^{ 0.00}$    &$81.10_{ 0.48}^{ 0.50}$\\
 44    &$ 0.01_{ 0.01}^{ 0.41}    -	 0.03_{ 0.03}^{ 0.77}    -   0.00_{ 0.00}^{ 0.00}$    &$ 0.17_{ 0.17}^{ 0.99}    -	 9.47_{ 4.20}^{ 4.08}    -   0.02_{ 0.02}^{ 0.60}$    &$ 4.42_{ 1.80}^{ 2.05}    -	 7.92_{ 5.23}^{ 5.59}    -   1.87_{ 1.85}^{ 7.37}$    &$69.58_{ 4.47}^{ 2.67}$\\
 45    &$ 0.00_{ 0.00}^{ 0.00}    -	 0.00_{ 0.00}^{ 0.00}    -   0.00_{ 0.00}^{ 0.00}$    &$ 1.23_{ 0.76}^{ 0.73}    -	 0.18_{ 0.18}^{ 2.12}    -   0.04_{ 0.04}^{ 2.67}$    &$ 0.00_{ 0.00}^{ 0.00}    -	 0.00_{ 0.00}^{ 0.00}    -   0.00_{ 0.00}^{ 0.00}$    &$55.22_{ 3.28}^{ 2.53}$\\
 46    &$ 0.21_{ 0.19}^{ 0.62}    -	 7.93_{ 2.96}^{ 1.97}    -   3.39_{ 2.66}^{ 4.03}$    &$ 0.03_{ 0.03}^{ 0.86}    -	19.91_{ 3.83}^{ 3.42}    -   8.11_{ 5.42}^{ 6.68}$    &$ 0.00_{ 0.00}^{ 0.00}    -	 0.01_{ 0.01}^{ 0.00}    -   4.71_{ 3.45}^{ 4.56}$    &$60.58_{ 0.63}^{ 0.61}$\\
 47    &$ 0.00_{ 0.00}^{ 0.00}    -	 0.10_{ 0.10}^{ 1.00}    -   7.17_{ 1.33}^{ 1.63}$    &$ 0.18_{ 0.17}^{ 0.53}    -	 8.60_{ 2.45}^{ 1.89}    -   4.30_{ 3.83}^{ 9.81}$    &$ 2.89_{ 0.98}^{ 0.88}    -	 1.24_{ 1.14}^{ 2.95}    -   2.73_{ 2.40}^{ 3.56}$    &$52.99_{ 1.40}^{ 5.73}$\\
 48    &$ 0.33_{ 0.31}^{ 0.80}    -	 0.02_{ 0.02}^{ 0.66}    -   4.82_{ 3.06}^{ 4.54}$    &$ 0.84_{ 0.75}^{ 1.95}    -	 0.42_{ 0.42}^{ 4.47}    -  48.19_{11.43}^{ 9.63}$    &$ 1.47_{ 1.31}^{ 2.15}    -	 0.34_{ 0.35}^{ 6.62}    -   0.79_{ 0.79}^{ 6.39}$    &$89.87_{ 3.05}^{ 1.16}$\\
 49    &$ 0.65_{ 0.55}^{ 1.39}    -	 0.01_{ 0.01}^{ 0.00}    -   0.00_{ 0.00}^{ 0.00}$    &$ 0.19_{ 0.19}^{ 4.10}    -	 3.72_{ 3.69}^{30.02}    -  29.42_{18.01}^{25.88}$    &$ 1.96_{ 1.84}^{ 4.45}    -	 0.00_{ 0.00}^{ 0.00}    -   0.18_{ 0.18}^{ 4.20}$    &$77.89_{ 6.02}^{ 6.70}$\\
 50    &$ 2.25_{ 1.19}^{ 1.26}    -	 0.95_{ 0.91}^{ 3.87}    -   0.57_{ 0.57}^{ 4.67}$    &$ 7.20_{ 2.19}^{ 2.22}    -	10.92_{ 7.36}^{ 9.89}    -   0.01_{ 0.01}^{ 0.00}$    &$ 0.20_{ 0.20}^{ 2.32}    -	 0.59_{ 0.59}^{ 6.44}    -  29.27_{ 8.99}^{ 8.70}$    &$81.33_{ 1.88}^{ 2.61}$\\
 51    &$ 7.61_{ 0.64}^{ 0.53}    -	 0.00_{ 0.00}^{ 0.00}    -   0.03_{ 0.03}^{ 0.71}$    &$ 0.00_{ 0.00}^{ 0.00}    -	 6.76_{ 0.79}^{ 0.83}    -   0.05_{ 0.05}^{ 4.10}$    &$ 0.02_{ 0.02}^{ 0.78}    -	 0.00_{ 0.00}^{ 0.00}    -   6.87_{ 2.46}^{ 2.54}$    &$63.78_{ 0.42}^{ 0.40}$\\
 52    &$ 0.36_{ 0.34}^{ 0.87}    -	 2.13_{ 1.77}^{ 4.21}    -  10.83_{ 5.82}^{ 4.57}$    &$ 0.94_{ 0.90}^{ 1.85}    -	 3.09_{ 2.47}^{ 3.28}    -   6.74_{ 4.96}^{ 5.41}$    &$ 3.05_{ 1.83}^{ 1.91}    -	 0.07_{ 0.07}^{ 2.22}    -   3.27_{ 2.91}^{ 4.67}$    &$65.99_{ 4.44}^{ 2.10}$\\
 53    &$ 3.98_{ 0.48}^{ 0.46}    -	 0.00_{ 0.00}^{ 0.00}    -   0.00_{ 0.00}^{ 0.00}$    &$ 0.96_{ 0.39}^{ 0.48}    -	17.04_{ 1.35}^{ 1.57}    -   0.00_{ 0.00}^{ 0.00}$    &$ 2.31_{ 0.91}^{ 1.14}    -	 0.00_{ 0.00}^{ 0.00}    -  27.17_{ 2.83}^{ 2.80}$    &$21.98_{ 1.93}^{ 1.72}$\\
 54    &$ 7.91_{ 3.02}^{ 1.04}    -	 0.51_{ 0.51}^{ 2.27}    -   0.49_{ 0.49}^{ 4.11}$    &$ 0.16_{ 0.16}^{ 2.02}    -	 0.95_{ 0.94}^{ 5.09}    -  29.41_{22.43}^{ 8.83}$    &$ 0.02_{ 0.02}^{ 2.01}    -	 0.00_{ 0.00}^{ 0.00}    -  23.13_{13.30}^{ 9.06}$    &$80.83_{12.75}^{ 2.74}$\\
 55    &$ 0.01_{ 0.01}^{ 0.22}    -	 1.84_{ 1.45}^{ 1.57}    -   2.72_{ 2.23}^{ 3.09}$    &$ 0.01_{ 0.01}^{ 0.29}    -	 0.27_{ 0.27}^{ 1.81}    -  18.74_{ 4.10}^{ 3.95}$    &$ 0.21_{ 0.21}^{ 1.04}    -	 4.18_{ 2.42}^{ 2.92}    -   1.30_{ 1.29}^{ 4.36}$    &$72.36_{ 0.58}^{ 0.56}$\\
 56    &$ 2.07_{ 0.99}^{ 1.15}    -	 2.33_{ 2.13}^{ 3.08}    -   5.83_{ 4.30}^{ 4.36}$    &$ 0.08_{ 0.08}^{ 0.79}    -	 5.14_{ 1.60}^{ 2.05}    -   0.08_{ 0.08}^{ 7.26}$    &$ 0.12_{ 0.12}^{ 1.20}    -	 0.01_{ 0.01}^{ 0.00}    -   5.53_{ 2.77}^{ 3.25}$    &$66.15_{ 1.17}^{ 2.56}$\\
 57    &$ 4.32_{ 1.60}^{ 1.03}    -	 0.00_{ 0.00}^{ 0.00}    -   0.00_{ 0.00}^{ 0.00}$    &$ 2.76_{ 1.56}^{ 1.44}    -	14.26_{ 6.04}^{ 6.32}    -   7.32_{ 6.06}^{ 8.52}$    &$ 6.36_{ 2.14}^{ 2.08}    -	 0.00_{ 0.00}^{ 0.00}    -  30.32_{10.15}^{ 8.87}$    &$53.30_{10.45}^{ 4.72}$\\
\hline
\end{tabular}
\end{table*}

\end{appendix}

\end{document}